\documentclass[preprint]{revtex4-1}
\usepackage{amsmath}
\usepackage{amssymb}
\usepackage{amsthm}
\usepackage{booktabs}
\usepackage{color}      
\usepackage{float}
\usepackage{geometry} 
\usepackage{graphicx}   
\usepackage{multirow}
\usepackage{longtable}
\usepackage{setspace}
\usepackage{subfigure}  
\usepackage{tabu}
\usepackage{verbatim}

\newtheorem{question}{Question}
\newtheorem{answer}{Answer}
\newtheorem{stage}{Stage}
\newtheorem{theorem}{Theorem}
\newtheorem*{remark}{Remark}

\begin{document}
\bibliographystyle{alpha}  

\begin{abstract}
This article is intended to an introductory lecture in material physics, in which the modern computational group theory and the electronic structure calculation are in collaboration. The effort of mathematicians in the field of group theory have ripened as a new trend, called ``computer algebra'', outcomes of which now can be available as handy computational packages, and would also be useful to physicists with practical purposes. This article, in the former part, explains how to use the computer algebra for the applications in the solid-state simulation, by means of one of the computer algebra package, the GAP system. The computer algebra enables us to obtain various group theoretical properties with ease, such as the representations, the character tables, the subgroups, etc. Furthermore, it would grant us a new perspective of material design, which could be executed in mathematically rigorous and systematic way.
Some technical details and some computations which require the knowledge of a little higher mathematics (but computable easily by the computer algebra) are also given. The selected topics will provide the reader with some insights toward the dominating role of the symmetry in crystal, or, the ``mathematical first principles'' in it.  
In the latter part of the article, we analyze the relation between the structural symmetry and the electronic structure in C$_{60}$ (as an example to the system without periodicity).
The principal object of the study is to illustrate the hierarchical change of the quantum-physical properties of the molecule, in accordance with the reduction of the symmetry (as it descends down in the ladder of subgroups). 
As an application, this article also presents the computation of the vibrational modes of the C$_{60}$ by means of the computer algebra. 
In order to serve the common interest of the researchers, the details of the computations (the required initial data and the small programs developed for the purpose) are explained as minutely as possible.

\end{abstract}

\title{Computer Algebra and Material Design}
\author{Akihito Kikuchi}
\affiliation{Canon Inc, Tokyo, Japan}
\email{akihito\_kikuchi@gakushikai.jp (The corresponding author)} 
\maketitle


\section{Introduction}
In the history of physics, the cooperation of physicists and mathematicians has yielded a great harvest in the first half of the twentieth century, as the installation of group theory in quantum mechanics\cite{Weyl}. One of the main applications is realized in the field of solid-state physics\cite{INUI, TINK, DRESS}. In subsequent years, however, such productive relationship between physics and mathematics became enfeebled, as physicists and mathematicians were pursuing their own interests separately. In material science, the typical tool of study has turned into ``first principle electronic structure computation'', in which rapid computers are intensively used so that the quantitative simulation could be achieved. In contrast, the group theoretical view in the quantum physics is rather a qualitative one, which could explain the likeness in similar material structures but could not illuminate the origin of subtle but distinct differences. The standpoints of group theoretical analysis, and of first principles simulation, are located at cross-purposes. This is one of the reasons which brought about the breaking-off between the group theory and the first principles electronic structure computation. There is another hardship which hinders the collaboration of mathematics and physics. The computation of the ``representation'' in group theory\cite{BURNSIDE, ABSTGRP, Lam, Ser, Fulton}, useful in physical applications, requires special arts, which is recondite to non-experts. Thus, for the purpose of general use, the character tables are listed up in literature and textbooks (recently, moreover in databases). Traditionally we are obliged to consult with the non-electric data, which is prone to troublesome errors. Such a circumstance has little affinity to the modern style of computational physics, where necessary data should easily be accessible on the computer or computed anew by the researcher. 
Nevertheless, it must be stressed here that such a cumbersome situation has already become surmountable. The development of the effective computer architectures and the program packages for computational group theory in recent decades have benefited mathematicians so well that they could actually obtain definite symbolic or numeric solutions to the problems of their own, not only in proving the existence of the solution\cite{COMPTGRPTHEO, CGTNOTES}. These computational tools are not restricted to mathematicians, but available to every scientific researcher. Powerful outcomes by mathematicians, also, will confer a modernized viewpoint to the material science, so that the conventional computational tools could be assisted and improved, although, in the present state, most of the material scientists seem to be content in having old-fashioned knowledge of mathematics. 
This article takes several examples in the group theoretical analysis in the electronic structure calculation and presents the details of the computation so that the readers would know how to use modern mathematical packages of computational group theory. The necessary group theoretical data are computed by the desktop computer, without references to other resources, and are applicable to the analysis of the quantum physics in materials which is principally governed by the symmetry of the system. The article furthermore expounds the possible style of the systematic material design, by which the electronic structure might be controlled artificially from the viewpoint of the structural symmetry. From these topics, the potentiality of the cooperation of computer algebra and first-principles electronic structure (which are legitimate heirs of group theory and quantum mechanics) may be observed.

To be fair, there are a lot of works of physicists who make use of computer algebra, although the works of pedagogical purpose are not so many. My article is only one of them. And here I list up some of such works for the benefit of readers.

The vibrational spectrum of C$_{60}$ is analyzed in the thesis of Mooij\cite{MOOIJ}. The representation of the group is computed by the computer algebra system Maple; the vibrational modes are computed by the reduction of the size of the dynamical matrix (of the empirical potential model); the parameters of these models are fit against the experimental data for the optically active modes. My article also treats the vibrational modes of this molecule, but it does not arrive at the quantitative computation of vibrational frequencies. My article, by means of GAP, classifies geometrically possible deformations with respect to irreducible representations and generates the basis set of deformations in the symbolic formulas, although the obtained results could be applied to more quantitative studies.

The book by El-Batanouny and Wooten treats the wide range of topics related to the symmetry in the condensed matter\cite{BATAN}. The interest of the research is almost the same as my article. One of the principal difference to my article is this: the authors of this book developed the programs for the group theoretical study by means of Mathematica, while I adopt the GAP system. The authors of this book developed Mathematica programs which compute the characters of groups by means of Dixon's method and applied them toward various topics. (I should say this: the GAP package computes the characters of groups by the same method, so we can directly step into applications in physics, as end-users of mathematics.) I recommend the readers to read the book along with my article, if possible, in order to solidify their understanding.

The book by Stauffer et al treats the general topics on the computer simulation and the computer algebra\cite{STAUF}. The adopted computer algebra system is "reduce", for the purpose of symbolic computations, and the application to the physics is rather introductory.

Hergert and coworkers made the group-theoretical analysis of electronic and photonic band structure\cite{HERG1,HERG2}. The symmetry properties of Schr\"{o}dinger equation and Maxwell equation are investigated. The computer algebra system Mathematica enables them to simplify group theoretical study. 
 
As for the introduction to the computational physics, the book by Thijssen will be useful\cite{THIJ}.

In the study of quantum chemistry, Sakiyama and Waki utilized the computer algebra system GAP\cite{CONFORM}.  The possible conformations (for a hexakis-methylamine nickel(ii) complex cation) are generated by the programs built on GAP, and the stability of possible formations is computed and compared through the quantitative electronic structure computation.

In the thesis of Rykhlinskaya, the author developed a computer algebra package "Bethe" in the framework of Maple, and applied it to the symmetric molecular geometry and symmetries, the vibrational analysis of the molecule, and the analysis of the atomic behavior and the atoms in the crystal field\cite{RYKL}.The applications of Bethe are demonstrated in the article of Fritzshche\cite{FRITZ}.  

Barnett gave us the review article in the application of computer algebra, not only in the life science but in the material science\cite{BARN}. The interest of the author is centered on the symbolic computation (especially in chemistry), not of the computer group theory. However, we can find useful references there, if we extend our researches beyond group theory.  

``Bilbao Crystallographic Server'' is an online database of crystallography\cite{AROYO1,AROYO2,AROYO3}. From this server, we can find data of crystallographic point groups and space groups and useful programs.

Also, there are studies of the group theoretical property of C$_{60}$ by  mathematicians\cite{Chung, KOSTANT}. The authors of these articles (Chung, Kostant, and Sternberg) discuss the embedding of the icosahedral group (of C$_{60}$ symmetry) into a larger group, which is a realization of the statement of Galois. And they analyze the energy spectrum of the molecule by the representation theory, with the hope that their results will find applications in the physical properties of the molecule.

Eick and Souvignier give us a survey of the algorithms for space groups and crystallographic groups available in the computer algebra system GAP and in the software packages Carat and Cryst\cite{CRYST}. 

It seems to me, that the application of the GAP system to the group theoretical analysis in the material science is still undeveloped, although a great deal of the products in the pure mathematics can be attainable easily from GAP. Some of these related works explain to us how to use the computational packages developed on some computer algebra system, and provide us with those packages, but they are not kind enough to explain their implementation. So in my article, I intend to present "the raw programs" and "the raw computed results" so that the readers may have experience on the actual computations by GAP. Again, I must say this: the examples taken in the following chapters are not novel, and a lot of group theoretical analysis is already done, without the aid of computer algebra. I only show an easier way of these analyses by means of a desktop PC. If readers thought that some of the methods are unfamiliar or novel, they are not of my invention, in the strict sense: they would be well-known things in the field of pure mathematics, and they are only interpretations in the viewpoint of physics. 

﻿\section{Computation of group theoretical properties using ``GAP''}
The computational discrete algebra package GAP is one of the powerful tools for the computation in the group theory\cite{GAPSYSTEM}. It can be applied to the determination of the group theoretical properties which are necessary for solid-state electronic structure computation. For this purpose, the point group of the crystal must be determined. The group operations are to be listed up; the multiplication table is to be prepared; through which the character table is computed. 

\begin{figure}[ht]
\centering
\includegraphics[width=0.5\linewidth]{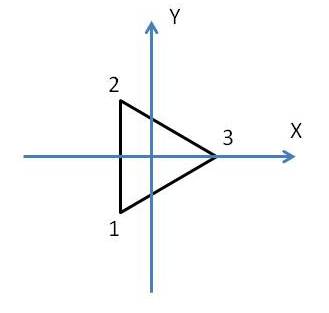}
\caption{A triangle, as the object of the symmetric operation. The symmetric operation can be through the permutations of three symbols, 1,2,3, or the rotations and reflections in the x-y plane.}
\label{fig:TRIANGLE}
\end{figure}

At first, the concept of the character of the group is explained here. Let us consider an equilateral triangle as in Fig. \ref{fig:TRIANGLE}. There are operations in the plane, such as, the rotation by 60 degrees at the center of mass, and the reflection at an axis which connects the center of mass and one of the vertexes. These operations exchange vertexes among themselves but fix the triangle unmoved. In the mathematical terminology, these operations consist a group $C_{3v}$.
These operations are the permutation of the vertexes, numbered as 1,2, and 3 in the clockwise direction.

They are expressed as
\begin{eqnarray}
E&=&\left(
\begin{array}{ccc}
1 & 2 & 3  \nonumber\\ 
1 & 2 & 3
\end{array}\right) ,\nonumber
A=\left(
\begin{array}{ccc}
1 & 2 & 3 \\ 
2 & 1 & 3
\end{array}\right) ,\nonumber\\
B&=&\left(
\begin{array}{ccc}
1 & 2 & 3 \\
1 & 3 & 2
\end{array}\right) ,\nonumber
C=\left(
\begin{array}{ccc}
1 & 2 & 3 \\
3 & 2 & 1
\end{array}\right) ,\nonumber\\
D&=&\left(
\begin{array}{ccc}
1 & 2 & 3 \\
3 & 1 & 2
\end{array}\right) ,\nonumber
F=\left(
\begin{array}{ccc}
1 & 2 & 3 \\
2 & 3 & 1
\end{array}\right) .
\end{eqnarray}
In the above the top and the bottom low denote the initial and the final arrangements of the vertexes respectively.

These operations can also be expressed by six rotation matrices in the x-y plane:
\begin{eqnarray}
E&=&\left(
\begin{array}{cc}
1 & 0 \\ 
0 & 1
\end{array}\right) ,\nonumber
A=\left(
\begin{array}{cc}
1 & 0 \\ 
0 & -1
\end{array}\right) ,\nonumber\\
B&=&\left(
\begin{array}{cc}
-\frac{1}{2} & \frac{\sqrt{3}}{2}  \\
\frac{\sqrt{3}}{2} & \frac{1}{2}
\end{array}\right) ,\nonumber
C=\left(
\begin{array}{cc}
-\frac{1}{2} & -\frac{\sqrt{3}}{2} \\ 
-\frac{\sqrt{3}}{2} & \frac{1}{2}
\end{array}\right) ,\nonumber\\
D&=&\left(
\begin{array}{cc}
-\frac{1}{2} & \frac{\sqrt{3}}{2} \\ 
-\frac{\sqrt{3}}{2} & -\frac{1}{2}
\end{array}\right) ,\nonumber
F=\left(
\begin{array}{cc}
-\frac{1}{2} & -\frac{\sqrt{3}}{2} \\ 
\frac{\sqrt{3}}{2} & -\frac{1}{2}
\end{array}\right).
\end{eqnarray}

The multiplication table of these elements is written in table \ref{mltblc3v}.

\begin{table}[h!]
\centering
\begin{tabular}{c|cccccc}
   & E & A & B & C & D & F \\ \hline
 E & E & A & B & C & D & F \\ 
 A & A & E & D & F & B & C \\ 
 B & B & F & E & D & C & A \\ 
 C & C & D & F & E & A & B \\ 
 D & D & C & A & B & F & E \\ 
 F & F & B & C & A & E & D \\ 
\end{tabular} 
\caption{The multiplication table. The multiplications $xy$ between the elements in the left column $x$ and in the top row $y$ are shown.}
\label{mltblc3v}
\end{table}

Let us examine these maps:
\begin{eqnarray}
\Gamma_1&:& E,A,B,C,D,F\rightarrow 1 \nonumber\\
\Gamma_{1'}&:& E,D,F \rightarrow 1;A,B,C\rightarrow -1 {\rm \:(Determinant\; of \; matrices
)}\nonumber\\
\Gamma_2&:& E,A,B,C,D,F \rightarrow E,A,B,C,D,F {\rm\:(Identity\; map\; as\; matrices)}\nonumber\\
\end{eqnarray}

One can see that the multiplication table is kept unaltered by these maps with the replacement of the six symbols to corresponding targets. The geometrical operations in the triangle, forming a group, are represented by these maps, even if not always faithfully; the viewpoint can be switched from geometrical one to numerical one. This is an example of the representation of the group; all elements in the group are represented by proxies of scalars or matrices; the multiplications among them are subject to the same rule as the original group elements. The trace of the matrix representation, such as $\Gamma_2$, is called ``character'', having certain properties favorable to the application of quantum physics, as will be seen later. The character table of $C_{3v}$ is given in table~\ref{CHTBLC3V}.

\begin{table}[h!]
\centering
\begin{tabular}{c|cccccc}
  & E & A & B & C & D & F \\ 
\hline 
$\Gamma_1$ & 1 & 1 & 1 & 1 & 1 & 1 \\ 
$\Gamma_{1'}$ &1  &-1  & -1 &-1  & 1 & 1 \\ 
$\Gamma_2$ & 2 & 0 & 0 & 0 & -1 & -1 \\ 
 
\end{tabular}
\caption{The character table of C$_{3v}$.}
\label{CHTBLC3V} 
\end{table}

As can be seen in the multiplication table, the group elements are divided into subsets such as
\begin{equation}
  {\rm Cl}(a) = \{ g \in G|{\rm \;there\; exists\;} x \in G {\rm \;such\;as\;} g=xax^{-1} \}.
\end{equation}
These subsets are called conjugacy classes. The elements in the same conjugacy class have a common value of the character; it is customary for character tables to be labeled by conjugacy classes, not by each group elements. The conjugacy classes for $C_{3v}$ are given by three subsets: $\rm \{E\},\{A,B,C\}\; and\; \{D,F\}$.

The gap computation for this example proceeds as follows. The group is defined by the minimal set of generators including a rotation and a reflection:
\begin{verbatim}
gap> G:=Group((1,2,3),(1,2));
Group([ (1,2,3), (1,2) ])
\end{verbatim}
The program returns the results in the shorthand notation preferred by mathematicians. 
\begin{verbatim}
gap> Elements(G);
[ (), (2,3), (1,2), (1,2,3), (1,3,2), (1,3) ]
gap> M:=MultiplicationTable(G);
[ [ 1, 2, 3, 4, 5, 6 ], [ 2, 1, 4, 3, 6, 5 ], [ 3, 5, 1, 6, 2, 4 ], 
  [ 4, 6, 2, 5, 1, 3 ], [ 5, 3, 6, 1, 4, 2 ], [ 6, 4, 5, 2, 3, 1 ] ]
\end{verbatim}

The multiplication table is given by a matrix, which is represented by a doubly nested list; each of the inner lists should be read as one of the lows in the matrix. The entry $M_{ij}$ shows the result of the multiplication $g_i\cdot g_j$ between elements $g_i$ and $g_j$.  

Subgroups and generators are computed by:
\begin{verbatim}
gap> AllSubgroups(G);
[ Group(()), Group([ (2,3) ]), Group([ (1,2) ]), Group([ (1,3) ]), 
  Group([ (1,2,3) ]), Group([ (1,2,3), (2,3) ]) ]
gap> GeneratorsOfGroup(G);
[ (1,2,3), (1,2) ]
gap> G.1;
(1,2,3)
gap> G.2;
(1,2)
gap> S:=Subgroup(G,[(1,2)]);
Subgroup(G,[(1,2)]);
gap> Elements(S);
[ (), (1,2) ]
gap> A:=Group((1,2,3));
Group([ (1,2,3) ])
gap> IsSubgroup(G,A);
true
gap> IsSubgroup(A,G);
false
gap> A:=Group((1,2,3,4));
Group([ (1,2,3,4) ])
gap> IsSubgroup(G,A);
false
\end{verbatim}    

The group can also be defined from the multiplication table as
\begin{verbatim}
gap> G2:=GroupByMultiplicationTable(M);
<group of size 6 with 6 generators>
gap> Elements(G2);
[ m1, m2, m3, m4, m5, m6 ]
\end{verbatim}
The elements in the group constructed from the multiplication tables are represented by abstract symbols m1,m2,...,m6. 

In order to construct the group from the rotation in the Euclidean space, the generating set of the rotation matrices should be given. 
\begin{verbatim}
gap> M1:=[[-1/2,ER(3)/2],[-ER(3)/2,-1/2]];
[ [ -1/2, -1/2*E(12)^7+1/2*E(12)^11 ], 
  [ 1/2*E(12)^7-1/2*E(12)^11, -1/2 ] ]
gap> M2:=[[1,0],[0,-1]];
[ [ 1, 0 ], [ 0, -1 ] ]
gap> G3:=Group(M1,M2);
Group(
[ 
  [ [ -1/2, -1/2*E(12)^7+1/2*E(12)^11 ], 
      [ 1/2*E(12)^7-1/2*E(12)^11, -1/2 ] ], [ [ 1, 0 ], [ 0, -1 ] ] 
 ])
\end{verbatim}
Here ER(n) is $\sqrt{n}$ and E(n) is the primitive n-th root of unity $\exp(2\pi i/n)$.

Let us inquire of the GAP package whether the three different definitions should generate equivalent groups.
\begin{verbatim}
gap> G=G;
true
gap> G2=G;
false
gap> G3=G;
false
\end{verbatim}
The GAP program decides that the three groups are not identical in the strict sense, because they are composed from different resources, i.e. permutations, abstract symbols, and matrices. However, we can construct isomorphisms among them. The following GAP command returns the isomorphism on the generators of the groups.
\begin{verbatim}
gap> IsomorphismGroups(G,G);
[ (1,2,3), (1,2) ] -> [ (1,2,3), (1,2) ]
gap> IsomorphismGroups(G2,G);
[ m1, m2, m3, m4, m5, m6 ] -> [ (), (2,3), (1,2), (1,2,3), (1,3,2), 
  (1,3) ]
gap> IsomorphismGroups(G,G2);
[ (1,2,3), (1,2) ] -> [ m4, m3 ]
gap> G4:=Group((1,2,3,4))
gap> IsomorphismGroups(G,G4);
fail
\end{verbatim}
The existence of the isomorphism between G and G3 can also be verified in this way.

The conjugacy classes are computed by this command:
\begin{verbatim}
gap> ConjugacyClasses(G);
[ ()^G, (2,3)^G, (1,2,3)^G ]
\end{verbatim}

To access entries in the conjugacy classes, a special method is needed. The following cannot work well:
\begin{verbatim}
gap> Elements(ConjugacyClasses(G));
[ ()^G, (2,3)^G, (1,2,3)^G ]
\end{verbatim}

Instead, using ``List'' command and anonymous functions in $\lambda$ calculus, we can access each entry in the list:
\begin{verbatim}
gap> List(Elements(G),x->x); 
[ (), (2,3), (1,2), (1,2,3), (1,3,2), (1,3) ]
gap> List(Elements(G),x->x^-1); 
[ (), (2,3), (1,2), (1,3,2), (1,2,3), (1,3) ]
gap> List(ConjugacyClasses(G),Elements);
[ [ () ], [ (2,3), (1,2), (1,3) ], [ (1,2,3), (1,3,2) ] ]
gap> List(ConjugacyClasses(G),Representative);
[ (), (2,3), (1,2,3) ]
\end{verbatim}

The character table is computed by this command:
\begin{verbatim}
gap> irrg:=Irr(G);
[ Character( CharacterTable( Sym( [ 1 .. 3 ] ) ), [ 1, 1, 1 ] ), 
  Character( CharacterTable( Sym( [ 1 .. 3 ] ) ), [ 1, -1, 1 ] ), 
  Character( CharacterTable( Sym( [ 1 .. 3 ] ) ), [ 2, 0, -1 ] ) ]
\end{verbatim}

The character table can be accessible as a numeral list, or as a map on the group elements:
\begin{verbatim}
gap> List([1,2,3],y->List([1,2,3],x->irrg[y][x]));
[ [ 1, 1, 1 ], [ 1, -1, 1 ], [ 2, 0, -1 ] ]
gap> List([1,2,3],y->List(Elements(G),x->x^irrg[y]));
[ [ 1, 1, 1, 1, 1, 1 ], [ 1, -1, -1, 1, 1, -1 ], 
  [ 2, 0, 0, -1, -1, 0 ] ]
\end{verbatim}

The irreducible representation is computed by
\begin{verbatim}
gap> rep:=IrreducibleRepresentations(G);
[ Pcgs([ (2,3), (1,2,3) ]) -> [ [ [ 1 ] ], [ [ 1 ] ] ], 
  Pcgs([ (2,3), (1,2,3) ]) -> [ [ [ -1 ] ], [ [ 1 ] ] ], 
  Pcgs([ (2,3), (1,2,3) ]) -> 
    [ [ [ 0, 1 ], [ 1, 0 ] ], [ [ E(3), 0 ], [ 0, E(3)^2 ] ] ] ]
gap> List(Elements(G),x->x^rep[3]);
[ [ [ 1, 0 ], [ 0, 1 ] ], [ [ 0, 1 ], [ 1, 0 ] ], 
  [ [ 0, E(3)^2 ], [ E(3), 0 ] ], [ [ E(3), 0 ], [ 0, E(3)^2 ] ], 
  [ [ E(3)^2, 0 ], [ 0, E(3) ] ], [ [ 0, E(3) ], [ E(3)^2, 0 ] ] ]
\end{verbatim}
This command returns the irreducible representations, giving the relations between the generators of the group and the matrix representation. The characters are computed by taking traces of the matrix representation. 

The character table is also computed by 
\begin{verbatim}
gap> tbl:=CharacterTable(G);
CharacterTable( Sym( [ 1 .. 3 ] ) )
gap> Display(tbl);
CT1

     2  1  1  .
     3  1  .  1

       1a 2a 3a
    2P 1a 1a 3a
    3P 1a 2a 1a

X.1     1  1  1
X.2     1 -1  1
X.3     2  . -1
\end{verbatim}
To make use of the character table computed in this way, various subsidiary commands are prepared in GAP system. (N.B. The orderings of representations or characters by these commands do not always coincide with each other.)
\begin{verbatim}
gap> ConjugacyClasses(tbl);
[ ()^G, (2,3)^G, (1,2,3)^G ]
gap> Irr(tbl);
[ Character( CharacterTable( Sym( [ 1 .. 3 ] ) ), [ 1, -1, 1 ] ), 
  Character( CharacterTable( Sym( [ 1 .. 3 ] ) ), [ 2, 0, -1 ] ), 
  Character( CharacterTable( Sym( [ 1 .. 3 ] ) ), [ 1, 1, 1 ] ) ]
\end{verbatim}

\begin{remark}
There are various algorithms for the computation of characters. The simplest one, the Burnside algorithm, starts from the counting in the elements in the conjugacy classes, sets up a matrix, and the characters are computed through the eigenvalue problem. The Dixon (or Dixon-Schneider) algorithm goes in a more advanced way, employing the computation in the prime fields and the other techniques of the symbolic computation, but it is effective in the large groups\cite{DIXON, SCH,HULP93}. The latter algorithm is adopted in GAP. 
\end{remark}
\section{Some preliminaries}
\subsection{Projection operator}
Once the characters of the group are obtained, ``projection operators''  are constructed and assigned to each of the irreducible representation. The definition is as follows:
\begin{equation}
P^{(p)}=\frac{l_p}{|G|}\sum_{T \in G} \chi^{(p)*}_T \cdot O_T,
\end{equation}
where $l_p$ is the dimension of the irreducible representation,$|G|$ is the order of the group, and $\chi^{(p)}_T$ is the character of the group element $T$, which is allowed to be complex-valued. $O_T$ is the operator allotted to each $T$ in the space in which the group action is defined. This operator acts on functions in the Euclidean space as follows:
\begin{equation}
O_T f(r)= f(T^{-1}(r)).
\end{equation}   
By this construction, we can distinguish whether a function belongs to the corresponding irreducible representation or not. If the projector is applied to the basis functions of the corresponding irreducible representations, these functions are invariant as a set: on the other hand, the projector is applied to the basis functions of different irreducible representations, the result of the projection goes to zero. (The epithet ``invariant'' should be understood in the following way: in many cases, plural basis functions belong to one irreducible representation. Through the group operations, these basis functions are interchanged with each other or re-expressed as the linear combination of them; moreover, they are never transformed into functions belonging to other representations. In a word, they are invariant as a subspace; if the subspace corresponding to a representation is minimal, no more being divisible, it is called irreducible. The whole of representations and characters of the arbitrary finite group could be computed exactly, without omission, by the principle of the group theory.)

If we know the explicit forms of the irreducible matrix representation $D_{ij}^{(p)}(R)$, the projection operator of the following form is useful:
\begin{equation}
P^{(p)}_{kl}=\frac{l_p}{|G|}\sum_{T \in G} D^{(p)*}_{kl}(O_T) \cdot O_T.
\end{equation}
Owing to the definition of the partner functions 
\begin{equation}
O_T|p;l\rangle=\sum_j |p;j\rangle D^{(p)}_{jl}(T)  
\end{equation}
and the orthogonality theorem of the representation
\begin{equation}
\sum_T D_{ik}^{(p)*}(T)D_{lm}^{(q)}(T)=\frac{|G|}{l_p}\delta_{pq}\delta_{il}\delta_{km}
\end{equation}
this projector transforms one basis vector into another basis in one irreducible representation: 
\begin{equation}
P^{(p)}_{kl}|p;l\rangle=|p;k\rangle.
\end{equation}
Moreover it can project out k-th partner function of the irreducible representation from an arbitrary function
$F=\sum_{q}\sum_j f_j^{(q)}|q;j\rangle$
as
\begin{equation}
P^{(p)}_{kk}F= f_k^{(p)}|p;k\rangle.
\end{equation}

At this junction, the connection between the group theory and the energy spectrum of the Schrodinger equations arises. Let the potential term be invariant under some symmetry operations. With this potential term, certain eigenfunctions should exist. When the symmetry operations are applied to the Schrodinger equation, the Laplacian, the potential, and the energy spectrum are invariant: the changeable is only the wavefunction, which is expressed by the basis functions in the irreducible representation. If the irreducible representation is one-dimensional, the wavefunction will be invariant up to a certain phase factor by the symmetry operation: on the other-hand, if the representation is multi-dimensional, the symmetry operation may generate different wave-functions which are other solutions to the same equation with the same eigenvalue. Such circumstances lead to the degeneracy of the energy spectrum. The identification of each wavefunction to its proper irreducible representation enables us to clarify the relation between the energy spectrum and the symmetry in the system, which is helpful in the analysis of the electronic properties of materials.   

\subsection{Space group}
An n-dimensional space group S is a discrete subgroup in a group of Euclidean motions in $\mathbb{R}^n$, such that the subgroup T in S, composed of pure translations(without rotations or reflections) is a free Abelian subgroup of rank n, having finite index in S. Thus an exact sequence of groups exists as
\begin{equation}
0\rightarrow T \rightarrow S \rightarrow P \rightarrow 1,    
\end{equation}
in which $S$ acts on $T$ by means of the conjugation action $t\rightarrow s\cdot t\cdot s^{-1}$, which represents the action of the point group $P$, defined as a factor group $S/T$, on the lattice translations.

In a more intuitive expression, the space group $S$ is the set of symmetry operations in a periodic system; it is composed of rotations and reflections which fix components in the unit cell, combined with parallel movements along crystal axes; therefore $S$ is of infinite order; the parallel movements, by themselves, compose the discrete subgroup $T$ of infinite order, in which the direction and the stride of the movements are determined by the primitive lattice vectors. Meanwhile, the point group $P$ is the set of the reminders of operations in $S$, taken modulo of the crystalline periodicity, i.e. with the parallel movements in $T$ being nullified; it is a finite group and may include fractional translations, which are represented by linear combinations of fractions of primitive lattice vectors.

\subsection{Crystallographic group}
In the case of crystallographic group, the symmetry operation is the affine mapping, composed of the linear parts $R_T$, including the reflection, rotation, and inversion, and the translation parts $\tau_T$\cite{CRYST}. The definition is
\begin{equation}
O_T=\{R_T|\tau_T\}: r \rightarrow R_T\cdot r+\tau_T,
\end{equation}
\begin{equation}
O_T\cdot O_S=\{R_T\cdot R_S|\tau_T + R_T\cdot \tau_s\},
\end{equation}
\begin{equation}
O_T^{-1}=\{R_T^{-1}|-R_T^{-1}\cdot \tau_T\}.
\end{equation}
The operation on the plane wave is 
\begin{equation}
O_T \exp(ikr)=\exp(ik(R_T^{-1}r-R_T^{-1}\tau))=\exp(i(R_T\cdot k)(r-\tau)).
\end{equation}

The affine mapping gives rise to the phase factor $\exp(-i(R_T\cdot k)\tau))$, which is not unity at a general k-point. On the other hand, the matrix representation of the affine mapping is subject to the following relation:
\begin{equation}
D^{(p,k)}( \{R_k|f_{R_k}+\tau_n    \})
=e^{-ik\cdot \tau_n}D^{(p,k)}(\{R_k|f_R\})
\end{equation}
The trace of the matrix $D^{(p,k)}$ is the character: therefore the phase-factors in the character and the wave-number part in the Bloch-type wavefunction ($e^{ikr}$ in $e^{ikr}u_k(r)$ ) are canceled with each other as complex conjugates; in the general k-point, the phase shift in the character can actually be negligible in the projection operator.(Indeed there are textbooks or lectures omitting these terms.) However, if the Bloch-type wavefunction is expressed by plane-wave expansion $\sum C_G e^{i(k+G)r}$ , one should be cautious. The phase shift in the periodic part is $e^{iR_k(G)\tau_n}$, which remains without cancellation in the projection operation.

\subsection{Theoretical set-up for Wyckoff positions}
The mathematical definition of the Wyckoff positions is stated as follows.

Let $G$ be a space group, $T$ be its translation lattice, $K$ be its point group, $t_g$ be fractional translations. Each element in $G$ has a form of $\{k|t_k+t\}$ for $k\in K$ and $t \in T $, and acts on a vector $v$ in $V=\mathbb{R}^n$ as
\begin{equation}
\{k|t_k+y\}(v)=k\cdot v+t_k+t.
\end{equation}  
The stabilizer of $v\in V$ under this action is denoted as Stab$_G(v)$=$\{g\in G|g(v)=v\}$. 
Let $v\in V$. An equivalence relations $\sim$ is set up if Stab$_G(v)$ is conjugate to Stab$_G(w)$,i.e. Stab$_G(w)$=Stab$_G(g(v))$=$g$ Stab$_G(w)$$g^{-1}$ for some $g\in G$. This equivalence classes of $\sim$ is called the Wyckoff positions of $G$.

This definition, given in the mathematicians' terminology, is a little recondite for physicists. For a more intuitive understanding, it can be restated as:  

\begin{quote}
The Wyckoff positions are a set of coordinate points, composed from two  subsets:

[Type 1] The coordinate points which may be fixed by a certain symmetry operation of the point group. They are regarded as generators of the Wyckoff positions.

[Type 2] The coordinate points generated from those in Type 1 subset by all of the symmetry operations.

\end{quote}

The command in GAP ``WyckoffPositions(S)'' returns the type 1 set for a space group S; the equivalent coordinate points to a generator ``W'' in type 1 set by the symmetry operations are computed from an another command ``WyckoffOrbit(W)''.  

\subsection{Units in the computation}
Throughout the computations of quantum physics in this article, we mainly use atomic units, abbreviated as a.u. if necessary; for the length, the Bohr radius(a$_0 \approxeq 5.292\times 10^{-11}$m); for the energy, electron volt unit (eV) or the Hartree unit(E$_h\approxeq 27.211$eV). However, some computations adopt arbitrary unit in the energy.


\section{Application 1: Identification of wavefunctions to irreducible representations}

The application in this section is the classical example of group theory in quantum physics.
\subsection{The simplest case: at $\Gamma$ point}
In this section, the group-theoretical analysis of the wavefunction is exemplified. The wavefunctions at the $\Gamma$ point in the diamond crystal are classified to corresponding irreducible representations. The treatment for the general k-point ($k\ne 0$) shall be discussed later.(The knowledge of the distinction between symmorphic or non-symmorphic crystal is necessary; the existence of these two types of crystal makes the discussion not a little complicated.)

At first, the character table should be computed. The symmetric operations in the diamond structure, whose unit cell is the minimal one, including two carbon atoms, are given in the appendix, as well as the multiplication table of these operations.

When one uses the GAP packages, there are several options for the preparation of the point group; of which three types can be used.

First: the symmetry operations are given as a set of the three-dimensional matrix. 

\begin{verbatim}
gap> MT[  1]:=[[  1,  0,  0],[  0,  1,  0],[  0,  0,  1]];;
gap> MT[  2]:=[[  1,  0,  0],[  0, -1,  0],[  0,  0, -1]];;
..........................
gap> MT[ 48]:=[[  0,  0,  1],[  0,  1,  0],[ -1,  0,  0]];;
gap> G:=Group(MT);
\end{verbatim}

Second: the multiplication table in the group is supplied.

\begin{verbatim}
gap> M:=[[1,2,3,...,48],[...],...,[...]];;
gap> G:=GroupByMultiplicationTable(M);
gap> Elements(G);
[ m1, m2, m3, m4,............,m45, m46, m47, m48 ]
\end{verbatim}

In this case, the group elements are denoted by the symbolical way, m1,m2,...,m48.
 
Actually, all of the group elements are not necessarily provided to define a group: it is enough to give the smallest generating set, from which other elements are constructed. The group should be remade by means of the smallest generating set. (In the present implementation of GAP, especially when the multiplication table is supplied, the following tendency in the computation is observed: the computations for groups, constructed from the minimal generating set, are much quicker than those for groups where the all elements are stored as generators.) For the case of the diamond, the computation goes as:

\begin{verbatim}
GS:=SmallGeneratingSet(G);
[ m36, m48 ]
G:=Group(GS);
<group with 2 generators>
\end{verbatim}

The command ``SmallGeneratingSet()'' yields a reasonably small generating set. In the cases of finite solvable groups (a typical example of this is the crystal point group) and of finitely generated nilpotent groups, ``MinimalGeneratingSet()'' command is also available to get minimal generators, but the computation is time-consuming. In addition, each element ``elm'' in the group ``G`` can be expressed by generators, by means of Factorization(G,elm):

\begin{verbatim}
gap> List(Elements(G),x->Factorization(G,x));
[ <identity ...>, x2^-1*x1*x2*x1, x2^2, (x2*x1)^2, x2^2*x1^2, 
  x1*x2^2*x1, x1^2, x2^-1*x1^-1*x2^-1*x1, x1^-1*x2^-1*x1*x2, 
  x2^-1*x1^2*x2, x2^-1*x1*x2*x1^-1, x1^-2, x2^-1*x1, x1^-1*x2^-1, 
  x2*x1, x2*x1*x2^2, x2^2*x1*x2, x2*x1^-1, x1*x2, x2^-1*x1^-1, 
  x1^-1*x2^-1*x1^2, x2^-1*x1^3, x2*x1*x2^2*x1^2, x2*x1^3, x1^3, 
  x2^-1*x1^-1*x2^-1*x1^2, x2^2*x1^3, x1*x2^2*x1^2, x2^2*x1^-1, 
  x2*x1*x2, x1^-1, x2^-1*x1*x2, x1*x2^2, x2^2*x1, x2^-1*x1^-1*x2^-1,
  x1, x1^2*x2, x1*x2*x1^-1, x2*x1^-2, x1*x2^2*x1*x2, x2*x1*x2^2*x1, 
  x2*x1^2, x1^-1*x2^-1*x1, x2^-1*x1^2, x1*x2*x1, x2^-1, 
  x2*(x2*x1)^2, x2 ]
\end{verbatim}
 
Third: the crystallographic groups are defined by four-dimensional augmented matrices, in which both of the point group operations and the translations are inscribed. The notations for the augmented can be given in the following form

\begin{equation}
\left(
\begin{array}{cc}
 A & v  \\ 
 0 & 1 
\end{array}
\right) 
\end{equation} 
acting on column vectors $(x,1)$ from the right 
or, alternatively,
\begin{equation}
\left(\begin{array}{cc}
 A^T & 0  \\ 
 v & 1
\end{array} \right) 
\end{equation} 
acting on low vectors $(x,1)$ from the left, which represent the affine mapping $ x\rightarrow A \cdot x + v$.(Crystallographers prefer the later notation.) 

Once the crystallographic group is defined, the point group can easily be deduced. 

As for the diamond case, in the GAP computation, the crystallographic group is defined as follows. (The minimal generating set is used for simplicity.)

\begin{verbatim}
gap> M1:=[[0,0,1,0],[1,0,0,0],[0,-1,0,0],[1/4,1/4,1/4,1]];;
gap> M2:=[[0,0,-1,0],[0,-1,0,0],[1,0,0,0],[0,0,0,1]];;
gap> S:=AffineCrystGroup([M1,M2]);
<matrix group with 2 generators>
gap> P:=PointGroup(S);
Group([ [ [ 0, 0, 1 ], [ 1, 0, 0 ], [ 0, -1, 0 ] ], 
  [ [ 0, 0, -1 ], [ 0, -1, 0 ], [ 1, 0, 0 ] ] ])
\end{verbatim} 

By these preparations the conjugacy classes and the character table of the diamond crystal are computed as tables \ref{tab:conjgacy_dia_fcc} and \ref{tab:character_tbl_fcc_dia_G}.(The group elements for the symmetry operation are given in a table in the appendix. The entries in the conjugacy classes are given by the numbering of the table of the group elements.)

\begin{table}[h!]
  \centering
  \begin{tabular}{l|l}
    \hline
    Class   & Group Elements\\\hline
    C.1     & 1  (The identity)     \\
    C.2     & 2     , 3     , 4       \\
    C.3     & 5     , 6     , 7     , 8     , 9     , 10    , 11    , 12 \\
    C.4     & 13    , 16    , 17    , 18    , 21    , 23      \\
    C.5     & 14    , 15    , 19    , 20    , 22    , 24      \\
    C.6     & 25      \\
    C.7     & 26    , 27    , 28      \\
    C.8     & 29    , 30    , 31    , 32    , 33    , 34    , 35    , 36 \\
    C.9     & 37    , 40    , 41    , 42    , 45    , 47      \\
    C.10    & 38    , 39    , 43    , 44    , 46    , 48      \\
    \hline
    \end{tabular}%
  \caption{Conjugacy classes of the point group in the diamond structure. The elements are indexed in the same way as the table of the symmetry operations. )  }
   \label{tab:conjgacy_dia_fcc}%
\end{table}%

\begin{table}[h!]
  \centering
    \begin{tabular}{r|rrrrrrrrrr} %
    {} & 
    {C.1} & 
    {C.2} & {C.3} & {C.4} & {C.5} & {C.6} & {C.7} & {C.8} & {C.9} & {C.10} \\
\hline
    x.1   & 1     & 1     & 1     & 1     & 1     & 1     & 1     & 1     & 1     & 1 \\
    x.2   & 1     & 1     & 1     & -1    & -1    & -1    & -1    & -1    & 1     & 1 \\
    x.3   & 1     & 1     & 1     & -1    & -1    & 1     & 1     & 1     & -1    & -1 \\
    x.4   & 1     & 1     & 1     & 1     & 1     & -1    & -1    & -1    & -1    & -1 \\
    x.5   & 2     & 2     & -1    & 0     & 0     & -2    & -2    & 1     & 0     & 0 \\
    x.6   & 2     & 2     & -1    & 0     & 0     & 2     & 2     & -1    & 0     & 0 \\
    x.7   & 3     & -1    & 0     & -1    & 1     & -3    & 1     & 0     & 1     & -1 \\
    x.8   & 3     & -1    & 0     & -1    & 1     & 3     & -1    & 0     & -1    & 1 \\
    x.9   & 3     & -1    & 0     & 1     & -1    & -3    & 1     & 0     & -1    & 1 \\
    x.10  & 3     & -1    & 0     & 1     & -1    & 3     & -1    & 0     & 1     & -1 \\
    \end{tabular}%
      \caption{The character table of the basic diamond structure at $\Gamma$ point.}
  \label{tab:character_tbl_fcc_dia_G}%
\end{table}%

We can apportion the wavefunction at the $\Gamma$ point to each irreducible representations by applying the projection operator. (It is enough to check whether the numerical result of the projection remains almost as it is, or it nearly vanishes to zero.)  The identification in the example of the diamond is given in table \ref{tab:wf_and_irrep_dia}. In the numerical expansion coefficients of the wavefunctions, we can imagine the existence of a hidden mechanism, as is suggested by the succession of zeros, or the interchanges of specific values with the alteration of signs. The hierarchy and the symmetry in them can be unraveled and classified by the group theoretical analysis. 

\begin{table}[h!]
  \centering
  \begin{tabular}{c|rrr|r|r|r|r|r|r|r|r|r}   \hline
          &   $k_1$    & $k_2$      & $k_3$      & $\|G\|^2$      &  1 & 2     & 3     & 4     & 5     & 6     & 7     & 8 \\\hline
  1   & 0     & 0     & 0     & 0     & -0.9765 & 0     & 0     & 0     & 0     & 0     & 0     & 0 \\
    2     & -1    & -1    & -1    & 1.6243 & 0.0742 & 0.3519 & -0.0007 & 0.2934 & -0.3562 & -0.1637 & -0.3913 & -0.3212 \\
    3     & 1     & 1     & 1     & 1.6243 & 0.0742 & 0.3519 & -0.0007 & 0.2934 & 0.3562 & 0.1637 & 0.3913 & 0.3212 \\
    4     & 0     & -1    & 0     & 1.6243 & -0.0742 & -0.1488 & 0.1173 & 0.4172 & -0.0785 & 0.5376 & 0.1078 & -0.3212 \\
    5     & 0     & 1     & 0     & 1.6243 & -0.0742 & -0.1488 & 0.1173 & 0.4172 & 0.0785 & -0.5376 & -0.1078 & 0.3212 \\
    6     & 0     & 0     & 1     & 1.6243 & -0.0742 & 0.3159 & 0.301 & -0.1397 & 0.084 & 0.295 & -0.4612 & 0.3212 \\
    7     & 0     & 0     & -1    & 1.6243 & -0.0742 & 0.3159 & 0.301 & -0.1397 & -0.084 & -0.295 & 0.4612 & -0.3212 \\
    8     & 1     & 0     & 0     & 1.6243 & -0.0742 & 0.1848 & -0.419 & 0.0159 & -0.5186 & 0.0788 & 0.1777 & 0.3212 \\
    9     & -1    & 0     & 0     & 1.6243 & -0.0742 & 0.1848 & -0.419 & 0.0159 & 0.5186 & -0.0788 & -0.1777 & -0.3212 \\
    10    & -1    & -1    & 0     & 1.8756 & 0     & 0.0285 & -0.2391 & 0.3432 & 0     & 0     & 0     & -0.1671 \\
    11    & 1     & 1     & 0     & 1.8756 & 0     & 0.0285 & -0.2391 & 0.3432 & 0     & 0     & 0     & 0.1671 \\
    12    & 0     & -1    & -1    & 1.8756 & 0     & 0.1325 & 0.3315 & 0.2199 & 0     & 0     & 0     & -0.1671 \\
    13    & 0     & 1     & 1     & 1.8756 & 0     & 0.1325 & 0.3315 & 0.2199 & 0     & 0     & 0     & 0.1671 \\
    14    & 1     & 0     & 1     & 1.8756 & 0     & 0.3968 & -0.0935 & -0.0981 & 0     & 0     & 0     & 0.1671 \\
    15    & -1    & 0     & -1    & 1.8756 & 0     & 0.3968 & -0.0935 & -0.0981 & 0     & 0     & 0     & -0.1671 \\\hline
      Ene.    &       &       &       &       & -2.1091 & 19.5799 & 19.5799 & 19.5799 & 25.1394 & 25.1394 & 25.1394 & 33.1443 \\\hline
      Rep.    &       &     &     & & x.1     & x.10    & x.10    & x.10    & x.7     & x.7     & x.7     & x.2 \\
\hline
    \end{tabular}%
      \caption{The identification between the irreducible representation and the wavefunctions in diamond crystal at $\Gamma$ point.  Wavefunctions are expressed by coefficients ($C_i, k_1, k_2, k_3$) in the plane wave expansion $C_i\exp(iGr);G=k_1 G_1 + k_2 G_2 + k_3 G_3$, where $G_1, G_2,G_3$ are reciprocal vectors. The left columns (from the second to the fifth ) indicate ($k_1,k_2,k_3, |G|^2$) and the remaining parts give the coefficients $C_i$. The energies and the irreducible representations are given in last two columns indicated by ``Ene.'' and ``Rep.'', respectively.
The carbon atoms are placed at $\pm(1/8\cdot a_1+1/8\cdot a_2+1/8\cdot a_3)$ so that the coefficients of the wavefunctions in the plane wave expansion are real-valued.
}
  \label{tab:wf_and_irrep_dia}%
\end{table}%

﻿
\subsection{Character table computation in super-cell}
It is a standard way to use the minimal unit cell. In the diamond structure, the minimal primitive cell is chosen to be the same as the face-centered one, in which two carbon atoms are located with the point-group G(I $\times$ O, C$_2$$ \times$ S$_4$). Meanwhile, one can choose the cubic unit cell; the volume of which is four times as large as the minimal cell, because the cubic cell is constructed from four of the simplest unit cell stuck together obliquely. The unit cell enlarged in a similar way, (if possible, in which extra atoms are embedded) is called super-cell. The point groups of the supercells are inevitably altered by the presence of extra fractional translations. The group theoretical computations for such super-cells will be difficult for bare human powers since the point group becomes larger and more complicated. Nevertheless, if one can use the computer algebra system, the character tables are easily computed. We should note that it is more advisable to compute characters of groups by ourselves if we can. Although the possible super-cells are infinite, the character tables available in textbooks and databases are limited, possibly only for the minimal unit cell.



For the super-cell, actually, the point group of the minimal crystal can be extended by the semidirect product with a finite Abelian group, by which extra fractional translations in the enlarged cell are represented. (The construction of the semidirect product will be stated elsewhere in the following section of this article.) The easiest and straightest way for this is to remake the multiplication table. This can be done in the following way: first, apply possible fractional translations to group elements in symmetry operations of the minimal cell, so that all symmetry operations in the supercell are prepared in the following form: 
\begin{equation}
\{E|\tau_i\}\{R_j|\sigma_j\}.
\end{equation}

By means of these new set of operations, we must construct the multiplication table in the super-cell:
\begin{equation}
\{R_j|\sigma_j+\tau_i\}\{R_k|\sigma_k+\tau_l\}=\{R_p|\sigma_p+\tau_q\}.
\end{equation}
In the evaluation, we can make use of the multiplication table in the minimal cell:
\begin{equation}
\{R_j|\sigma_j\}\{R_k|\sigma_k\}=\{R_j\cdot R_k|\sigma_j+R_j\sigma_k\}\equiv\{R_p|\sigma_p\}
\label{PS1}
\end{equation}
There is a relation:
\begin{equation}
\{R_j|\sigma_j+\tau_i\}\{R_k|\sigma_k+\tau_l\}=
\{R_j\cdot R_k|\sigma_j+\tau_i+R_j\sigma_k+R_j\tau_l\}=\{R_p|\sigma_p+\tau_q\}.
\label{PS2}
\end{equation}
From the equations (\ref{PS1}) and (\ref{PS2}), it leads that
\begin{equation}
\tau_q\equiv\sigma_j+\tau_i+R_j\sigma_k+R_j\tau_l-\sigma_p {\rm \:(modulo\;translational\;vectors)}
\end{equation}
Using this $\tau_q$, the extended multiplication table can be constructed with ease.

The exemplary calculation for the diamond structure shall be executed in this section. The basic unit cell in the diamond structure is the face-centered one, including two atoms,  The lattice vectors are defined as

\begin{equation}
a_1=(0,1/2,1/2),a_2=(1/2,0,1/2),a_3=(1/2,1/2,0)
\end{equation}

This basic unit cell could be extended four-fold, in which atomic coordinates are given by translations such as $x\rightarrow x+\tau_i\; (i=0,1,2,3)$. The vectors $\tau_i$ are defined as:
\begin{equation}
\tau_0=(0,0,0),\tau_1=(0,1/2,1/2),\tau_2=(1/2,0,1/2),\tau_3=(1/2,1/2,0)
\end{equation}
The extended unit cell is the cubic one, including eight atoms, to which the lattice vectors are defined as:
\begin{equation}
a'_1=(1,0,0),a'_2=(0,1,0),a'_3=(0,0,1)
\end{equation}

Let F be the group composed of the four fractional translation vectors $\tau_i(i=0,1,2,3)$. And let T be the group of primitive lattice translations, composed of $a'_1,a'_2,a'_3$ and the zero vector. The factor group N(:=F/T) is a finite group, whose generators $\hat{\tau}$ are given by the projection of $\tau_i$ onto N. This factor group is the Klein four group, or $Z_2 \oplus Z_2$. The generators satisfy the following relations.  

\begin{eqnarray}
2\cdot\hat{\tau}_i = \hat{\tau}_0 \\\nonumber
\hat{\tau}_1+\hat{\tau}_2+\hat{\tau}_3 
= \hat{\tau}_0
\end{eqnarray}

The point-group $G'$in the cubic unit cell is the semidirect product $N \rtimes G$, between the group of the fractional translations N and the point-group $G$ in the basic unit cell. The new group includes 192 $(=48\times 4)$ elements. The textbooks or databases, however,  only show the point group of the basic, minimal unit cells, but not those of the extended super-cells. In the latter cases, the point group is made of the semidirect product; one can compute its character tables by straightforwardly constructing the multiplication table, or, in a more complicated and elegant way, i.e. by making use of the induced representation of the group, when the group N is Abelian. The computation of the induced representation will be explained later.

The group elements in the extended group are renumbered as follows:
\begin{equation}
\{E|\tau_i\}\{R_j|s_j\}\rightarrow{\rm\;(i\times48+j)th\; element\; of\; the\; extended\; group} 
\end{equation} 

The conjugacy classes and the character tables for the extended group are given in table 
\ref{tab:cng_class_cub_dia} and \ref{tab:character_tbl_cubic_dia_G}.

  \begin{longtable}{c|l}
\hline
    Class &  Group Elements 
\\\hline
{$\hat{\rm C}$.1} & [0,1] 
\\\hline
{$\hat{\rm C}$.2} &  [0,2],[0,3],[0,4],[1,2],[2,3],[3,4] 
\\\hline
\multirow{4}{*}{$\hat{\rm C}$.3} &  [0,5],[0,6],[0,7],[0,8],[0,9],[0,10],[0,11],[0,12],
                               \\&[1,5],[1,6],[1,7],[1,8],[1,9],[1,10],[1,11],[1,12],
                               \\&[2,5],[2,6],[2,7],[2,8],[2,9],[2,10],[2,11],[2,12],
                               \\&[3,5],[3,6],[3,7],[3,8],[3,9],[3,10],[3,11],[3,12] 
\\\hline
\multirow{2}{*}{$\hat{\rm C}$.4} &  [0,13],[0,17],[0,21],[1,16],[1,17],[1,23],
                                \\&[2,16],[2,18],[2,21],[3,13],[3,18],[3,23] 
\\\hline
\multirow{4}{*}{$\hat{\rm C}$.5} &  [0,14],[0,15],[0,19],[0,20],[0,22],[0,24],\\&[1,14],[1,15],[1,19],[1,20],[1,22],[1,24],\\&[2,14],[2,15],[2,19],[2,20],[2,22],[2,24],\\&[3,14],[3,15],[3,19],[3,20],[3,22],[3,24] 
\\\hline
\multirow{2}{*}{$\hat{\rm C}$.6} &  [0,16],[0,18],[0,23],[1,13],[1,18],[1,21],\\&[2,13],[2,17],[2,23],[3,16],[3,17],[3,21] 
\\\hline
{$\hat{\rm C}$.7} &  [0,25],[1,25],[2,25],[3,25] 
\\\hline
\multirow{2}{*}{$\hat{\rm C}$.8} &  [0,26],[0,27],[0,28],[1,26],[1,27],[1,28],\\&[2,26],[2,27],[2,28],[3,26],[3,27],[3,28] 
\\\hline
\multirow{4}{*}{$\hat{\rm C}$.9} &  [0,29],[0,30],[0,31],[0,32],[0,33],[0,34],[0,35],[0,36],\\&[1,29],[1,30],[1,31],[1,32],[1,33],[1,34],[1,35],[1,36],\\&[2,29],[2,30],[2,31],[2,32],[2,33],[2,34],[2,35],[2,36],\\&[3,29],[3,30],[3,31],[3,32],[3,33],[3,34],[3,35],[3,36] 
\\\hline
\multirow{2}{*}{$\hat{\rm C}$.10} &  [0,37],[0,40],[0,41],[0,42],[0,45],[0,47],\\&[1,41],[1,42],[2,45],[2,47],[3,37],[3,40] 
\\\hline
\multirow{4}{*}{$\hat{\rm C}$.11} &  [0,38],[0,39],[0,43],[0,44],[0,46],[0,48],\\&[1,38],[1,39],[1,43],[1,44],[1,46],[1,48],\\&[2,38],[2,39],[2,43],[2,44],[2,46],[2,48],\\&[3,38],[3,39],[3,43],[3,44],[3,46],[3,48] 
\\\hline
{$\hat{\rm C}$.12} &  [1,1],[2,1],[3,1] 
\\\hline
{$\hat{\rm C}$.13} &  [1,3],[1,4],[2,2],[2,4],[3,2],[3,3] 
\\\hline
\multirow{2}{*}{$\hat{\rm C}$.14} &  [1,37],[1,40],[1,45],[1,47],[2,37],[2,40],[2,41],[2,42],\\&[3,41],[3,42],[3,45],[3,47] 
\\\hline
  \caption{Conjugacy classes for the cubic diamond structure. The element $\{1|\tau_j\}\{R_i|\sigma_i\}$ is denoted as [j,i].}
    \label{tab:cng_class_cub_dia}%
\end{longtable}

\begin{table}[h!]
  \centering
    \begin{tabular}{r|rrrrrrrrrrrrrr}
    {} &
     {$\hat{\rm C}$.1}   & {$\hat{\rm C}$.2} & {$\hat{\rm C}$.3} & {$\hat{\rm C}$.4} & {$\hat{\rm C}$.5} & {$\hat{\rm C}$.6} & {$\hat{\rm C}$.7} & {$\hat{\rm C}$.8} & {$\hat{\rm C}$.9} & {$\hat{\rm C}$.10} & {$\hat{\rm C}$.11} & {$\hat{\rm C}$.12} & {$\hat{\rm C}$.13} & {$\hat{\rm C}$.14} \\\hline
    y.1   & 1     & 1     & 1     & 1     & 1     & 1     & 1     & 1     & 1     & 1     & 1     & 1     & 1     & 1 \\
    y.2   & 1     & 1     & 1     & -1    & -1    & -1    & -1    & -1    & -1    & 1     & 1     & 1     & 1     & 1 \\
    y.3   & 1     & 1     & 1     & -1    & -1    & -1    & 1     & 1     & 1     & -1    & -1    & 1     & 1     & -1 \\
    y.4   & 1     & 1     & 1     & 1     & 1     & 1     & -1    & -1    & -1    & -1    & -1    & 1     & 1     & -1 \\
    y.5   & 2     & 2     & -1    & 0     & 0     & 0     & -2    & -2    & 1     & 0     & 0     & 2     & 2     & 0 \\
    y.6   & 2     & 2     & -1    & 0     & 0     & 0     & 2     & 2     & -1    & 0     & 0     & 2     & 2     & 0 \\
    y.7   & 3     & -1    & 0     & -1    & 1     & -1    & -3    & 1     & 0     & 1     & -1    & 3     & -1    & 1 \\
    y.8   & 3     & -1    & 0     & -1    & 1     & -1    & 3     & -1    & 0     & -1    & 1     & 3     & -1    & -1 \\
    y.9   & 3     & -1    & 0     & 1     & -1    & 1     & -3    & 1     & 0     & -1    & 1     & 3     & -1    & -1 \\
    y.10  & 3     & -1    & 0     & 1     & -1    & 1     & 3     & -1    & 0     & 1     & -1    & 3     & -1    & 1 \\
    y.11  & 6     & -2    & 0     & -2    & 0     & 2     & 0     & 0     & 0     & 0     & 0     & -2    & 2     & 0 \\
    y.12  & 6     & -2    & 0     & 2     & 0     & -2    & 0     & 0     & 0     & 0     & 0     & -2    & 2     & 0 \\
    y.13  & 6     & 2     & 0     & 0     & 0     & 0     & 0     & 0     & 0     & -2    & 0     & -2    & -2    & 2 \\
    y.14  & 6     & 2     & 0     & 0     & 0     & 0     & 0     & 0     & 0     & 2     & 0     & -2    & -2    & -2 \\
    \end{tabular}%
      \caption{The character table of the cubic diamond unit cell at $\Gamma$ point.}
    \label{tab:character_tbl_cubic_dia_G}%
\end{table}%

The eight lowest eigenvalues in energy spectra in the minimal and cubic diamond cell are shown in table \ref{tab:ENERGYOFFCCANDCUBIC}. There are two sets of threefold degeneracy in the former case. On the other hand, in the latter, there is one sixfold degeneracy, which is attributable to the overlap of the spectra at the three $X$ points in the Brillouin zone of the minimal unit cell. However, the character tables of the $\Gamma$ point in the minimal diamond cell could not assign them in suitable representation; the character tables rebuilt for the cubic cell can do this.
 
\begin{table}[h!]
    \begin{tabular}{lrrrrrrrr}\hline\hline
    Minimal   & -2.11 & 19.58 & 19.58 & 19.58 & 25.14 & 25.14 & 25.14 & 33.14 \\
    Representation& x.1     & x.10    & x.10    & x.10    & x.7     & x.7     & x.7     & x.2 \\
    \hline
    Cubic  & -2.11 & 6.74  & 6.74  & 6.74  & 6.74  & 6.74  & 6.74  & 13.03 \\
    Representation & y.1     & y.14    & y.14    & y.14    & y.14     & y.14     & y.14     & y.12 \\\hline\hline
        \end{tabular}%
  \caption{The energy spectra at the $\Gamma$ point in the minimal and cubic diamond unit cell (unit in eV). The assigned irreducible representations are also shown. }
  \label{tab:ENERGYOFFCCANDCUBIC}%
\end{table}%

﻿\section{Application 2: A systematic way of the material designing} 
\subsection{How to manage quantum-dynamics in crystal?}
The group theoretical view will be able to bring about a systematic way of material design. In this section, although in the level of table-top exercise, a practicable form of material design is presented. In a nutshell, through the systematic reduction of the crystal symmetry, the band-structure could tactically be modulated; the degeneracy of the energy spectrum can be artificially split into different levels and the new energy-gap can be opened. Toward the systematic reduction of the crystal symmetry, the crystallographic concept, called Wyckoff positions, is utilized\cite{WYCKOFF, CRYST}. This mathematical idea is not an abstract one, but actually computable by means of the computer algebra. 

One should note this: the degeneracy of the energy spectrum is governed by the symmetry of the crystal. 
From the mathematical viewpoint, the point group of the crystal system is the manifestation of the symmetry; the subgroups of the point group describe the possible style with the reduced symmetry.(It is well known that a material has general tendency to reduce its own symmetry in order to release itself from the energetically unstable situation owing to the existence of the degeneracy of occupied energy levels.) By tracing the tree structure of the inclusion relations of the subgroup of the finite groups, one can enumerate every possibility of reduced symmetry. For each of reduced symmetries, one can compute group representations. The comparison of the character tables in the subgroups and the parent groups (by means of analyzing compatibility relations) one can rummage the possibility of the breaking-down of the energy spectrum degeneracy. The opening of the energy-gaps in the reduced symmetry will be speculated, sorted and classified with their group theoretical origins. This plan, therefore, will lead to a kind of systematic way of material design. It must be admitted that such a tactic still lacks in the sufficiently quantitative argument, since the only possible existence of energy-gap, together with their relationship with crystal symmetry reduction, is discussed. In order to evaluate the stability of the reduced structure or the actual possibility of its existence, one must employ massive first-principles computation. However, the group theoretical inspection has its own merit; it can be achieved by a light and quick computation, frugal in hardware usage; maybe simple computations in the level of elementary perturbation theory will do. Hence it can be applied to screening purposes for picking up candidates before serious massive first-principles computation. 

As an example, consider the crystallographic group of a cube, including following operations.


\begin{itemize}
\item Three rotations along x, y, or z axes by 90 degrees.
\item Exchange of x, y, and z axes ( Rotations along the diagonal in the cube by 120 degrees).
\item The inversion and reflections are omitted for simplicity.
\end{itemize}

\begin{figure}
\centering
\includegraphics[width=1.0\linewidth]{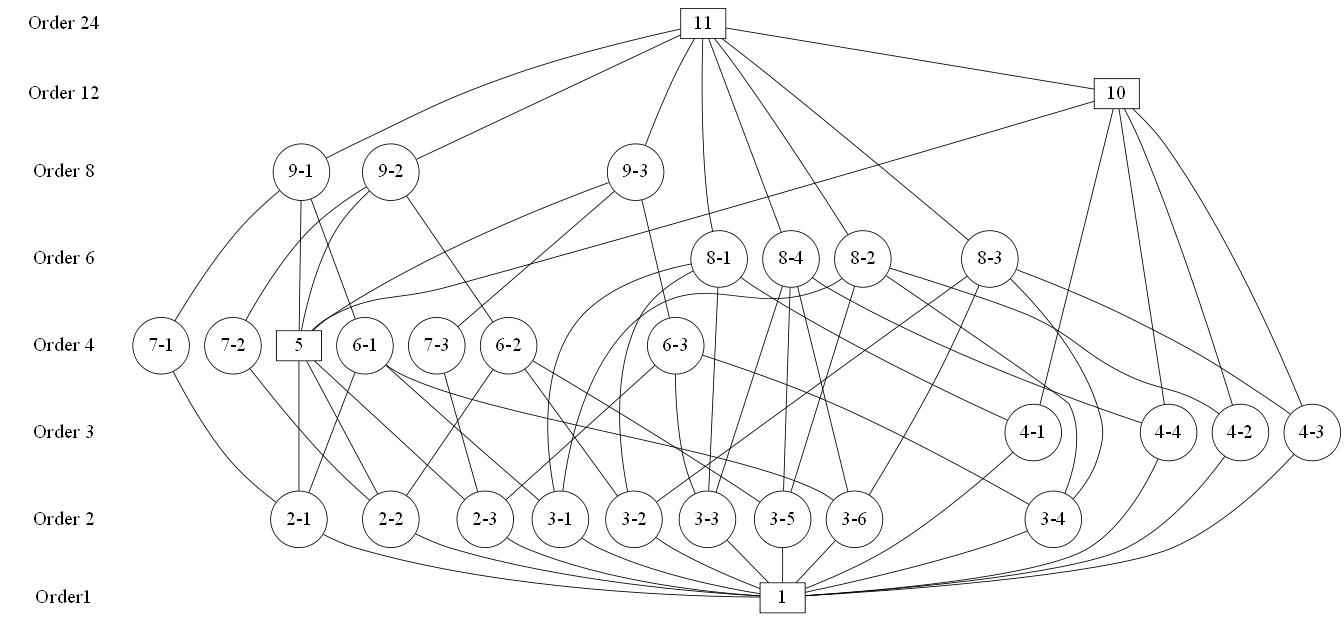}
\caption{The paradigm of the crystal symmetry, or the subgroup lattice.}
\label{fig:paradigm}
\end{figure}

These rotations generate 24 operations in the cubic lattice to construct the point group. Using GAP, The relationship of the inclusion in subgroups (the lattice subgroup) can be inquired and visualized(Fig.\ref{fig:paradigm}). The point group of the cubic lattice is located at the topmost node of the tree. We can trace the subgroups of lower symmetries in descending along the branches and finally arrive at the trivial group (the identity group), which is located at the root of the tree. This means that we can explore and predict the alternation of the band structure (viz, the splitting of the degenerated energy levels, or the opening of new band gaps) systematically, by consulting with the ``paradigm'' of the crystal symmetry.  

Now there arise following questions.

\begin{question}
How to make up actual crystal structures in correspondence with the subgroups of lower symmetries?
\end{question}

\begin{answer}
\rm The reduction of the symmetry may be occasioned, for example, by the presence of alien atoms, or the occurrence of point defects in the perfect crystal. The change in the electronic structure will be investigated by simple perturbation theory with sufficient accuracy.
\end{answer}

\begin{question}
In what positions alien atoms or point defects can be located? It is easy to cause the great change the symmetry, or, to abolish it completely. However, for the case with the small change in the symmetry, for example, that between neighboring connected entries in the subgroup lattice, can we set up favorable atomic dispositions?
\end{question}

\begin{answer}
\rm Choose an arbitrary point. Apply all of the operations of the subgroup in consideration to this coordinate. The number of the generated points is equal to the order of the subgroup. The set of these points is invariant with respect to the operation of the subgroup. By putting alien atoms or defects in this set of points, we can reduce the symmetry of the crystal, and, by continuing this way, descend through the subgroup lattice.
\end{answer}

\begin{question}
It is impossible for us to put alien atoms at arbitrary points in the crystal. With the view of the material design, a certain restriction should be imposed on the number of alien atoms.  We would like to know the set of atomic sites, with a limitation in number, and invariant by the subgroup operation.
\end{question}

\begin{answer}
\rm Make use of the concept of the Wyckoff position.
\end{answer}

The mathematical definition is given in the appendix. In a familiar word of material sciences, the Wyckoff positions are the set of equivalent points in the unit cell, which are transformed among themselves by symmetry operations. In general, the total number of such points is the same as that of the symmetry operations (the order of the point group). However, if the generating point of the Wyckoff position remains fixed by certain operations of the point group, the total number of Wyckoff positions will be fewer, which will be a factor of the order of the point group. In a mathematical word, it is equal to the order of some factor group. In GAP program, the Wyckoff position is easily computed from a ``Cryst'' package.

In this example of the cubic lattice, the Wyckoff positions are classified as follows:

\begin{itemize}
\item A discrete point (0,0,0).
\item A discrete point (1/2,1/2,1/2).
\item The centers of the edges of the cube:(1/2,0,0),(0,1/2,0),(0,0,1/2).
\item The centers of the faces:(1/2,1/2,0) (1/2,0,1/2),(0,1/2,1/2)
\item A segment (x,0,0) (0$<$x$<$1),invariant by 90-degrees rotation along the x axis.
\item A segment (0,y,0) (0$<$y$<$1), invariant by 90-degrees rotation along the y axis.
\item A segment (0,0,z) (0$<$z$<$1), invariant by 90-degrees rotation along the z axis.
\item A segment (x,1/2,1/2) (0$<$x$<$1),invariant by 90-degrees rotation along the x axis.
\item A segment (1/2,y,1/2) (0$<$y$<$1), invariant by 90-degrees rotation along the y axis.
\item A segment (1/2,1/2,z) (0$<$z$<$1), invariant 90-degrees rotation along the z axis.
\item A segment (t,t,t), invariant 120-degrees rotations which exchange three axes.(In a similar way, all of the diagonal lines of the cube belong to this type.)
\item Several segments (not discreet points), connected to the vertexes, the centers of the edges.
\item Arbitrary points in the unit cell.
\end{itemize}

\begin{figure}
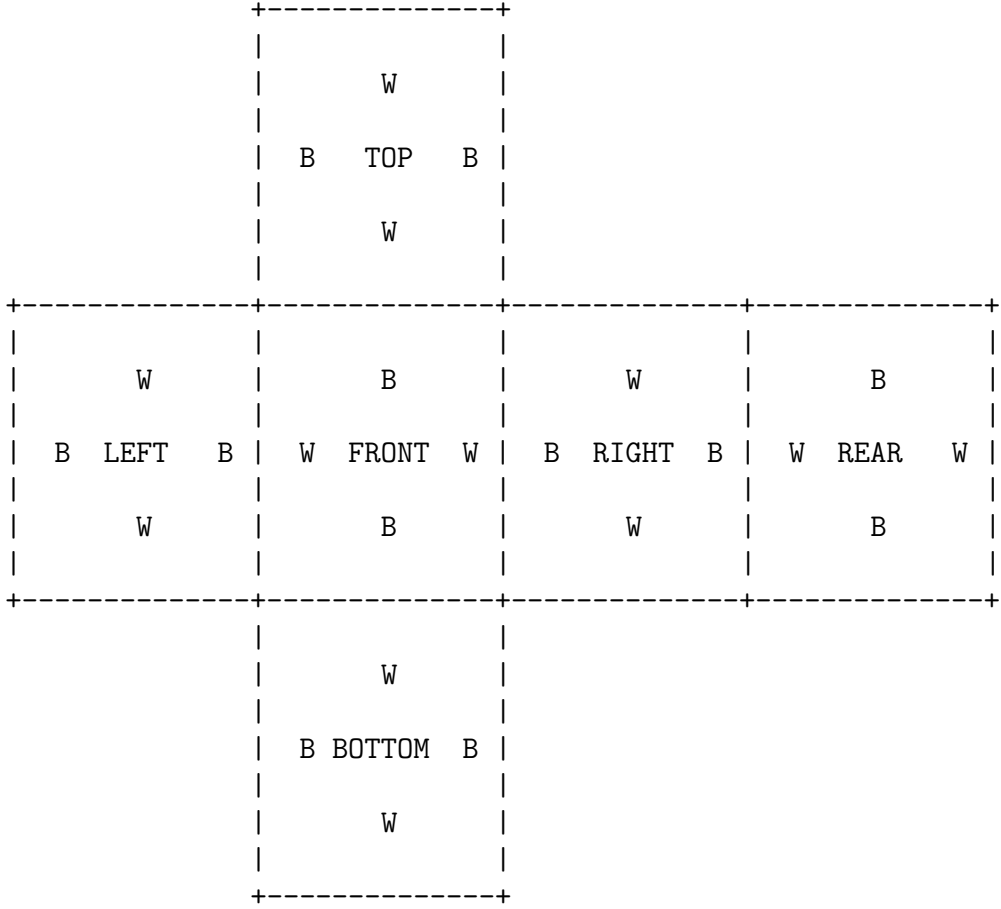

\centering
\begin{verbatim}
                     +--------------+
                     |              |
                     |       W      |
                     |              |
                     |  B   TOP   B |
                     |              |
                     |       W      |
                     |              |
      +--------------+--------------+--------------+--------------+
      |              |              |              |              |
      |       W      |       B      |       W      |       B      |
      |              |              |              |              |
      |  B  LEFT   B |  W  FRONT  W |  B  RIGHT  B |  W  REAR   W |
      |              |              |              |              |
      |       W      |       B      |       W      |       B      |
      |              |              |              |              |
      +--------------+--------------+--------------+--------------+
                     |              |
                     |       W      |
                     |              |
                     |  B BOTTOM  B |
                     |              |
                     |       W      |
                     |              |
                     +--------------+

\end{verbatim}
\caption{An illustration of split Wyckoff Orbit in the six faces of the cube.}
\label{fig:WyckoffOrbit}
\end{figure}

In table \ref{tab:wycoffpositioncube}, the Wyckoff positions of the group of full symmetry of the cubic cell and those of one of the subgroups are compared. These two groups have common generators of the Wyckoff positions. However, in some of them, the number of equivalent positions differs. By making use of such positions and stationing atoms in them, we can switch the crystal symmetry. For example, take the Wyckoff positions given in the 7-th low. They are in the orbit of the generating point $(t,0,1/2)$. The group of No.11 and No.10 are isomorphic to the symmetric group S$_4$ and the alternative group A$_4$ respectively. In the full symmetry, given by the group of No.11, the orbit is composed of 12 points. On the other hand, in the reduced symmetry of the group No.10, these 12 points are split into two orbits, each of which includes 6 points. If atoms (say, of type W) are positioned in all of 12 points in the cubic lattice, the crystal symmetry remains unaltered as that of the group of No.11. However, if, in the half part of them, at the Wyckoff positions of the Group No.10, the atoms are replaced by those of type B, the crystal symmetry is reduced to the group of No.10. Those Wyckoff positions are illustrated in Fig. \ref{fig:WyckoffOrbit}, where two split orbits are shown in B(black) and W(white) atoms.

\begin{table}[h!]
  \begin{tabular}{c|c|c|c}
     \hline
     Group NO.11  & \multirow{2}[0]{*}{Size}  & Group NO.10  & \multirow{2}[0]{*}{Size}  \\
     (\#G=24)     &   & (\#G=12)        &   \\\hline   
    [0,0,0]       &  1 & [0,0,0]       & 1 \\\hline
    [1/2,1/2,1/2] & 1 & [1/2,1/2,1/2] & 1 \\\hline
    [0,1/2,1/2]   & 3 & [0,1/2,1/2]   & 3 \\\hline
    [1/2,0,0]     & 3 & [1/2,0,0]     & 3 \\\hline
    [t,0,0]       & 6 & [t,0,0]       & 6 \\\hline
    [t,1/2,1/2]   & 6 & [t,1/2,1/2]   & 6 \\\hline
    \multirow{3}[0]{*}{[t,0,1/2]} & \multirow{3}[0]{*}{12} & [0,t,1/2] & 6 \\
                                  &                        & [1/2,t,0] & 6 \\
                                  &                        &  (Split in two) & \\\hline
    [t,t,0]       & 12& not existent & 0 \\\hline
    [t,t,1/2]     & 12& not existent  & 0 \\\hline
    [t,t,t]       & 8 & [t,t,t]       & 4 \\\hline
    [t1,t2,t3]    & 24& [t1,t2,t3]    & 12 \\\hline
    \end{tabular}%
  \centering
  \caption{The Wyckoff positions in the cubic lattice. The left three columns gives the generators and the right two show the number of equivalent points.}
    \label{tab:wycoffpositioncube}%
\end{table}%

\subsection{Control of band gaps by means of the Wyckoff positions}
In this section, the example of the material design discussed in the previous section is presented; in which ab-initio electronic structure calculations are executed, where the symmetry of the crystal is artificially altered and the band gaps are controlled. 

The exemplary calculation goes in the following way: in order to alter the crystal symmetry, certain coordinate points are chosen; and at these localities, we put impurities, which interact to electrons with short-ranged potentials; these impurities are assumed to be neutral; therefore they do not alter the net charge.  The change in the symmetry of the crystal will be caused by the presence of impurities, the effects of which are expressed by model potentials with adjustable parameters. We use the following form of the scattering potential. It is given in the real space as
\begin{equation}
V(r)=C\exp\left(-\frac{1}{2}\left(\frac{r}{r_{loc}}\right)^2 \right).
\end{equation}
The reciprocal space representation is
\begin{equation}
V(K)=\frac{1}{\Omega}Z\exp\left(-\frac{1}{2}\left(r_{loc}K\right)^2\right),
\label{IMPURITYPOT}
\end{equation}
where $\Omega$ is the volume, required in the normalization of the plane wave $\frac{1}{\sqrt{\Omega}}\exp\left(i K r\right)$. 
and $Z=C\sqrt{(2\pi)^3}(r_{loc}^3)$. There are two parameters: Z is the strength of the potential and $r_{loc} $ is the effective range.

A simple example is taken from diamond band structure. The coordinate of the carbon atoms are 
\begin{eqnarray}
(0\cdot {V}_1+0\cdot {V}_3+0\cdot {V}_3)\\\nonumber
(1/4\cdot {V}_1+1/4\cdot {V}_3+1/4\cdot {V}_3).
\end{eqnarray}

The lattice vectors are
\begin{eqnarray}
{V}_1&=&\left( \frac{a}{2},\frac{a}{2},0\right),\\\nonumber
{V}_2&=&\left( \frac{a}{2},0,\frac{a}{2}\right),\\\nonumber
{V}_3&=&\left( 0,\frac{a}{2},\frac{a}{2}\right)    .
\end{eqnarray}

In this structure, the lowest eight eigenvalues at $\Gamma$ point are shown in table \ref{tab:lowest_eight_eig_standard}.
\begin{table}[h!]
\centering
\begin{tabular}{ccccccccc}\hline\hline
$\Gamma$ &-2.11 & 19.58 & 19.58    & 19.58 &    25.14 & 25.14    & 25.14     &33.14\\
Representation& x.1     & x.10    & x.10    & x.10    & x.7     & x.7     & x.7     & x.2 \\\hline\hline
\end{tabular} 
\caption{The lowest eight eigenvalues at $\Gamma$ point in the diamond structure (unit in eV).}
\label{tab:lowest_eight_eig_standard}
\end{table}
There are two set of threefold degeneracy above and below the band gap, the energy levels of which are at 19.58 eV and 25.14 eV. (The irreducible representations to them are x.10 and x.7 respectively, as given in the previous section.)

The band gap (in exact terms, the degeneracy of the eigenvalues) is modified by the insertion of a single impurity in the crystal. Its position is chosen at  $(1/8\cdot {V}_1+1/8\cdot {V}_3+1/8\cdot {V}_3)$. The point group of the genuine diamond structure ( of the order of 48) is reduced to its subgroup (of the order of 12.) The symmetry operations of the subgroup are shown in table \ref{tab:symopr_dia_imp}.

\begin{table}[h!]
\centering
  \caption{The symmetry operations with the presence of an impurity. The rows (from the second to the bottom ) show the rotation matrix R (R11, R12,..., R33) and the translation T (T1, T2, T3). }
    \begin{tabular}{r|rrrrrrrrr|rrr}
\hline
          & R11   &  R12   &  R13   & R21   & R22   & R23    & R31   & R32   & R33   & T1    & T2    & T3 \\
\hline
    1     & 1     & 0     & 0     & 0     & 1     & 0     & 0     & 0     & 1     & 0     & 0     & 0 \\
    2     & 0     & 1     & 0     & 0     & 0     & 1     & 1     & 0     & 0     & 0     & 0     & 0 \\
    3     & 0     & 0     & 1     & 1     & 0     & 0     & 0     & 1     & 0     & 0     & 0     & 0 \\
    4     & 0     & -1    & 0     & -1    & 0     & 0     & 0     & 0     & -1    & 1/4 & 1/4 & 1/4 \\
    5     & -1    & 0     & 0     & 0     & 0     & -1    & 0     & -1    & 0     & 1/4 & 1/4 & 1/4 \\
    6     & 0     & 0     & -1    & 0     & -1    & 0     & -1    & 0     & 0     & 1/4 & 1/4 & 1/4 \\
    7     & -1    & 0     & 0     & 0     & -1    & 0     & 0     & 0     & -1    & 1/4 & 1/4 & 1/4 \\
    8     & 0     & -1    & 0     & 0     & 0     & -1    & -1    & 0     & 0     & 1/4 & 1/4 & 1/4 \\
    9     & 0     & 0     & -1    & -1    & 0     & 0     & 0     & -1    & 0     & 1/4 & 1/4 & 1/4 \\
    10    & 0     & 1     & 0     & 1     & 0     & 0     & 0     & 0     & 1     & 0     & 0     & 0 \\
    11    & 1     & 0     & 0     & 0     & 0     & 1     & 0     & 1     & 0     & 0     & 0     & 0 \\
    12    & 0     & 0     & 1     & 0     & 1     & 0     & 1     & 0     & 0     & 0     & 0     & 0 \\
\hline
    \end{tabular}%
  \label{tab:symopr_dia_imp}%
\end{table}%

The character table of this subgroup is given in table \ref{tab:chr_tbl_dia_imp}. The rows S.1,..., S.6 show the character values of the subgroup, and the rows x.7 and x.10 shows those of the group of the full diamond symmetry (without impurity). 
\begin{table}[h!]
  \centering
  \caption{The character table of the diamond structure with an impunity. The characters for each operation are shown here.}
    \begin{tabular}{r|rrrrrrrrrrrr} %
         & 1 & 2      &  3     &    4   &   5    &   6    &    7   &    8   &  9     &   10    &  11     &  12\\\hline
    S.1    & 1     & 1     & 1     & 1     & 1     & 1     & 1     & 1     & 1     & 1     & 1     & 1 \\
    S.2    & 1     & 1     & 1     & -1    & -1    & -1    & -1    & -1    & -1    & 1     & 1     & 1 \\
    S.3    & 1     & 1     & 1     & -1    & -1    & -1    & 1     & 1     & 1     & -1    & -1    & -1 \\
    S.4    & 1     & 1     & 1     & 1     & 1     & 1     & -1    & -1    & -1    & -1    & -1    & -1 \\
    S.5    & 2     & -1    & -1    & 0     & 0     & 0     & -2    & 1     & 1     & 0     & 0     & 0 \\
    S.6    & 2     & -1    & -1    & 0     & 0     & 0     & 2     & -1    & -1    & 0     & 0     & 0 \\\hline
    x.7    & 3     & 0     & 0     & -1    & -1    & -1    & -3    & 0     & 0     & 1     & 1     & 1 \\
    x.10   & 3     & 0     & 0     & 1     & 1     & 1     & 3     & 0     & 0     & 1     & 1     & 1 \\ 
    \end{tabular}%
  \label{tab:chr_tbl_dia_imp}%
\end{table}%

In the rows of the table, there are two relations: x.7=S.2+S.5 and x.10=S.1+S.6; which are called matching relations.(In some mathematical context, they are referred as branching rules.) These relations indicate the threefold degeneracy in the energy levels in the full diamond symmetry split into two irreducible representations S.2 and S.5, or S.1 and S.6 with the presence of an impurity. The character values of the identity element, shown in the second columns from the left, are the degeneracy of the energy levels; so we can predict the threefold-degeneracy split into two different levels, viz. twofold and single levels. 

To inspect this, the actual computation is executed; where the impurity potential is set to be a weak Gaussian form. For example, set the parameters in Eq.\ref{IMPURITYPOT} as $Z=-0.1 \rm{a.u.},r_{loc}=0.1\rm{a.u}$. The lowest eight eigenvalues are given in table \ref{tab:eig_dia_imp}. The prediction by group theory is realized there.
\begin{table}[h!]
\begin{tabular}{ccccccccc}\hline\hline
$\Gamma$ & -2.18 &    18.86    & 19.67    & 19.67    & 25.04    & 25.18&    25.18    &33.13\\\hline\hline
\end{tabular}
\caption{The lowest eight eigenvalues at $\Gamma$ point in the diamond structure with an impurity (unit in eV).}
\label{tab:eig_dia_imp} 
\end{table}

In the practice of the artificial material design, for example, the following demand will arise:

\begin{quote}
Demand: How to substantiate the situation, in which the band gap is altered, but the degeneracy of the energy levels is unaffected?     
\end{quote}

\begin{figure}[h!]
\centering
\includegraphics[scale=1.0]{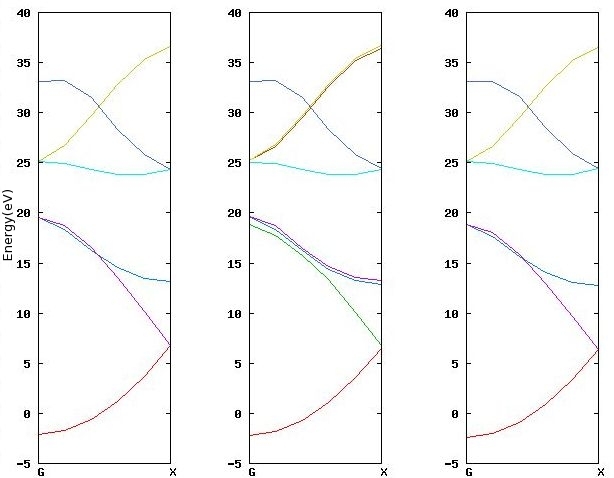}
\caption{The band dispersions in diamond structures altered by the presence of impurities, along $\Gamma$-L direction. The left figure is that without impurity: the center, with a single impurity: the right, with four impurities. }
\label{fig:GXdispersions}
\end{figure}

Let us proceed with the example of the diamond. In the diamond structure, there are plural possible Wyckoff positions.(The computation can be done by GAP.) For example, chose the position of $(1/8 V_1+1/8 V_2+1/8 V_3 )$. By the symmetry operations, it is transformed into three other points such as:
$(5/8 V_1+1/8 V_2+1/8 V_3 ),(1/8 V_1+5/8 V_2+1/8 V_3 ),(1/8 V_1+1/8 V_2+5/8 V_3 )$.
If impurities are stationed at these four points, there is no change in the point group; thus the degeneracy of the energy spectrum remains unchanged. The computed energy spectrum is given in table \ref{tab:eig_dia_imp_4}, which guarantees the validity of the group theory. The band dispersions along $\Gamma(0,0,0)-X(2\pi/a,0,0)$ are shown in Fig. \ref{fig:GXdispersions}, in which that of the basic diamond structure, and that with single impurity, and that with four impurities are compared.  

\begin{table}[h!]
\begin{tabular}{ccccccccc}\hline\hline
$\Gamma$ & -2.41 &    18.83    & 18.83    & 18.83 &    25.11     & 25.11&    25.11    &33.07\\\hline\hline
\end{tabular}
\caption{The lowest eight eigenvalues at $\Gamma$ point in the diamond structure with four impurities (unit in eV).}
\label{tab:eig_dia_imp_4}
\end{table}

There is a certain noteworthy point in these two cases (of the single impurity, and of the four impurities). In the former, the single impurity is located at$(1/8 V_1+1/8 V_2+1/8 V_3 )$; this point is one of the Wyckoff positions in the point group of the crystal with the deteriorated symmetry; it remains fixed by any symmetry operation and not transformed into other positions; however, this point also belongs to the Wyckoff position of the perfect diamond crystal and there transformed to four points. We can interpret this circumstance as follows: the former example corresponds to a situation, where the only one of the four Wyckoff position in the latter case is chosen so that the symmetry of the crystal could be reduced.  In a similar way, (but in a more general and liberal way, by inserting extra atoms or excavating vacancy site) we can reduce the crystal symmetry for the purpose of controlling the band structure; by doing this we can cleave into the degeneracy of the energy spectra and open extra energy gaps.

\subsection{A possible plan of material design in the super-cell} 
Now we have seen that the band structure is controllable by the presence of impurities or defects, although the presented example is that of instructive purpose, and not realistic, as an only model case for the sake of illustrating a possible form of theoretical material design. The diamond structure is compressed in such a compact cell that the supernumerary atoms could not be actually stationed at arbitrary places, such as at Wyckoff positions. Meanwhile, in recent decades, crystalline materials which have a large unit cell such as In$_3$O$_2$ have been put into practical use. At a glance, In$_3$O$_2$ seems to have a huge complex structure. However, actually, it is a super-cell composed from the stacking of body-centered lattice of In-O, with oxygen vacancies. The band dispersion is almost free-electron-like; its behavior under the presence of doped Sn, viz. impurities in the cell, can be understood by the group theoretical view\cite{ITOBAND}. Furthermore, the modern material composing technologies, such as epitaxial growth, or micro atom-scale probes, may enable us to realize such microscopic disposition of atoms in the supercell structure, which will lead to an actual phase of material design. In the super-cell, there is a more broad freedom in the atomic arrangement, in which certain atoms could be replaced or removed so that the periodical arrangement of impurities or defects should be permitted. 

Now a principle of material design based on group theoretical view can be proposed.

\begin{itemize}

\item[1] Chose a certain crystal as a starting point: we may start from simple atomic- and electronic- structures, such that we can execute the electronic structure computation easily, as the primitive module of the design.

\item[2] Consider an enlarged cell, composed of the stacking of the minimal unit cell. (The crystal axes may be cut newly and a supercell should be created.) 

\item[3] We can determine the band structure and the band gap of the enlarged cell, simply by folding the band structure of the minimal cell.(The folding of the band structure is to re-distribute k-points of the old Brillouin zone to the new one in accordance with the new periodicity and to redraw the band dispersion by means of new k-points.)

\item[4] Suppose a situation where the symmetry is reduced by the formation of the supercell with atomic displacements. Check the possibility of the opening of the extra band gaps (or the splitting of the degeneracy) at k-points with high symmetries, such as at $\Gamma$ points and the boundary of the Brillouin zone. The group theoretical view, such as the character theory, the compatible relation, the subgroup lattices, or the Wyckoff positions, should be fully made use of. The change in the band structure may be quantitatively evaluated by simple perturbation theory. These sorts of trials-and-errors, by tracing the sequence of crystal symmetry, will help us to search crystal structures with more desirable band structure.  
\end{itemize}

\section{Technical details}
In the preceding sections of the article, the basics of the usage of computer algebra, concerning the computation of group theoretical properties, are explained and applied to the analysis of the wave-functions. Moreover, the possible scheme for controlling band gaps, by means of tracing the sequence of subgroups, which corresponds to the reduction of crystal symmetry, is suggested through simple electronic structure computations. Meanwhile, in this part, the topics are rather promiscuous and technical, such as the computation of characters of the group of the supercell by means of ``semi-direct product'' which is an effective way to extend groups or the special treatment for non-symmorphic crystal, and so on. Moreover, ``mathematical first principle'' of crystal formation is demonstrated based on the cohomology theory, which shall explain to us how specific types of crystals, finite in number, can exist in accordance with the actions of certain point groups. The topics might arrive at somewhat higher mathematics, that may be not well-known to physicists. However, with the aid of computer algebra, one can conquer such hardships so that he might actually execute necessary computations. As for the fundamental knowledge of the group theory and its application, the reader should consult with standard references, such as Refs.\cite{Weyl,INUI,TINK,DRESS,BURNSIDE,ABSTGRP} for the group theory, such as Refs.\cite{Lam,Ser,Fulton} for the representation theory, such as Refs.\cite{COMPTGRPTHEO,CGTNOTES} for the algorithm in the computer algebra. We make use of the computer algebra package GAP\cite{GAPSYSTEM}, as well as in the previous sections.

\subsection{Determination of crystal symmetry}

It is essential for us to determine the crystal symmetry in order to execute the group theoretical analysis. The symmetry operations which fix the crystal axes and atomic coordinates must be listed to construct the point group. The determination algorithm of the point group is composed of the following two stages.

\begin{stage}
Determination of the point group of the lattice. 
\end{stage}

\begin{quote}
Let $A=(A_1,A_2,A_3)$ be the lattice vectors. There are two large categories of the crystal system; the one of which is hexagonal, and the other is cubic. In the former, rotations along the crystal axis by 60 degrees are allowed; in the latter, the rotations of this kind are composed of those by 90 degrees. The point group of arbitrary crystal system is given as a subgroup which is reduced from the full hexagonal or cubic symmetries. The full set of the symmetry operations (i.e. rotation matrices) for the hexagonal or cubic symmetry should be prepared, and from which admissible operations are extracted.

For each operation $R$ of the full symmetry group, compute the rotation of the lattice vectors:
\begin{equation}
R\cdot A_i (i=1,2,3).
\end{equation}
If the vector $R\cdot A_i$ protrudes out of the unit cell, it must be pulled back according to the crystal periodicity. Hereafter the pulling back of a vector $x$ into the unit cell is denoted as $P(x)$. For example, $R\cdot A_i = l\cdot A_1+m\cdot A_2+n\cdot A_3$ for integers $l, m, n$, then exactly $P(R\cdot A_i)=0$.

Then store the operation $R$ such as    
\[
  \|P(R\cdot A_i)\| \leq \epsilon 
\]
for all $A_i$(=1,2,3), with a threshold value $\epsilon$ ( This is an attention needed only in the floating-point computation.) From the stored set of admissible operations $\{R\}_C$, the point group of the crystal system is determined;  in this stage, however, the symmetry preservation in the atomic positions is not yet checked. 
\end{quote}

\begin{stage}
Determination of the point group of the crystal. 
\end{stage}
\begin{quotation}
The symmetry operations which fix atomic coordinates should be extracted now. The atom transformation table, the fractional translations associated to each rotation, and the multiplication table should be provided at the same time. Let $X$ and $Y$ be entries in the set of atomic positions $\{X\} $, and let T(X) and T(Y) be the corresponding type of atoms.

For each $R \in \{R\}_L$, and for each $X \in \{X\}$, compute $R\cdot X$; and chose every $Y\in \{X\}$ such as T(X) = T(Y). Now the vector $\tau_c=P(R\cdot X-Y)$ is a candidate for the fractional translation associated to R. If $\tau_c$ is the admissible fractional translation to $R$, the following conditions should be satisfied.

\end{quotation}

\begin{quotation}
{\bf Condition}:
for all $X_1$ in the atomic coordinates, there exist a $Y_1$ such that
\[
 \|\tau_c - P(R\cdot X_1-Y_1)\| \leq \epsilon,
\]            
and
\[
T(X_1) = T(Y_1).
\]
If this condition is satisfied, store $R$ and $\tau_c$ as a symmetry operation $\{R|\tau_c\}$ in the point group of the crystal. 
\end{quotation}
 
This algorithm works well if the unit cell is minimal, in the sense that the fractional translation is uniquely determined; i.e. with respect to one symmetry operation of $\{R|t\}$, other operations, such as $\{E|s\}\{R|t\}=\{R|t+s\}$, do not coexist in the point group operations. If the unit cell is not minimal, in other words, if it is a supercell (the accumulation of the minimal unit cell), the existence of plural $\tau$ for one $R$ cannot be negligible. The readers will be able to find and utilize subroutines or functions purposed for the identification of the crystal symmetry (possibly with more efficacy than the algorithm in this section)in the band calculation packages of one's own.

\subsection{Symmetry operations in the reciprocal space}
As for the plane wave, the symmetry operation in the real-space is given as
\begin{equation}
\{R_T|\tau_T\} \exp(ikr)=\exp\left(ik\left(R_T^{-1}r-R_T^{-1}\tau_T\right)\right)=\exp\left(i (R_T\cdot k)\cdot(r-\tau_T)\right).
\end{equation}
Consequently, it is more convenient to give operations with respect to reciprocal lattice vectors in the wave number space. If the wave-number-vector $k$ is expressed as $k_1\cdot G_1 + k_2 \cdot G_2 + k_3 \cdot G_3$, the rotation matrix for coefficients in the rotated vector $R_T\cdot k$ is given as 
\begin{equation}
\left(
\begin{array}{ccc}
R'_{11} &R'_{12}  & R'_{13}  \\ 
R'_{21} &R'_{22}  & R'_{23} \\ 
R'_{31} &R'_{31}  & R'_{33}
\end{array} 
\right)
\left(
\begin{array}{c}
k_1 \\ 
k_2 \\ 
k_3
\end{array} 
\right).
\end{equation}
Thus the rotation matrix for the coefficients is computed from that in the Euclidean space as:
\begin{equation}
R':=\big(\left(a_1,a_2,a_3\right)^{-1}R^T\left(a_1,a_2,a_3\right)\big)^T
\end{equation}
where $a_1,a_2,a_3$ are primitive lattice vectors.

For the case of the face-centered unit cell, the gap computation is done as:
\begin{verbatim}
gap> A:=[ [ 0, 1/2, 1/2 ], [ 1/2, 0, 1/2 ], [ 1/2, 1/2, 0 ] ];;
gap> RM:=[ [ 1, 0, 0 ], [ 0, -1, 0 ], [ 0, 0, -1 ] ];;
gap> TransposedMat(A^-1*TransposedMat(RM)*A);
[ [ -1, 0, 0 ], [ -1, 0, 1 ], [ -1, 1, 0 ] ]
\end{verbatim}

The rotation matrices in the reciprocal space, as well as those in the real space, are given in the appendix for cubic and hexagonal symmetries.

\subsection{The computation of the compatibility relation}

The irreducible representation of the diamond structure shows branching if the k-point moves from $\Gamma$ point to non-zero k-point. The situation of the branching (the compatibility relation) can be checked by GAP. For this purpose, the computation goes as follows:

\begin{itemize}

\item Compute the character table for the small group. Then, for each representation, evaluate the characters of elements in the small group, and store them in the lists $S_1, S_2,..., S_N$. (Let N be the number of the representations of the Small group.)

\item Compute character values of elements in the small group, using the irreducible representation at the $\Gamma$ point, and store them in the lists $G_1, G_2,..., G_M$.(Let M be the number of the representations of the point group.)

\item If the j-th irreducible representation (G.j) branches into the composition of those in the small (S.i) as follows
\begin{equation}
{\rm G_j}=\sum_{i=1}^{N} {\rm C_i^{j}\times S_i}
\end{equation}
the coefficients are computed by the inner product of the lists as
\begin{equation}
{\rm C_i^j=G_j \cdot S_i / \left|\#K\right|}
\end{equation}
where ${\rm \left|\#K\right|}$ is the order of the small group.

\end{itemize}

For this purpose, this function can be used:  
\begin{verbatim}
GetCompati:=function(A,B)
 local repA,repB,AV,BV,CV;
 repA:=Irr(A);
 repB:=Irr(B);
 if ( false=IsSubgroup(A,B) ) then
 Print("A < B ; Inverse A and B."); return;
 fi;
 AV:=List(repA,y->List(Elements(B),x->x^y));
 BV:=List(repB,y->List(Elements(B),x->x^y));
 CV:=AV*TransposedMat(BV)/Size(B);
 return CV;
end;
\end{verbatim}

This function computes the characters of the group elements in the group ``B'', in two ways, those by the irreducible representation of group ``A'' and group ``B''. The results are stored in two lists ``AV'' and ``BV''. Each row of these lists contains the characters of each irreducible representation. By the computation of the inner product between these lists, the ramification of the representations is computed.

Symmetry operations defined in wave-number space are used here. These matrices are three-dimensional ones given in Table \ref{tab:symopr_fcc_dia_reciprocal}.(They are stored in the list ``MT[~]'' in the following computation.) The coordinate of the K point along $\Gamma-X$ point is given, as the coefficients of reciprocal lattice vectors.

\begin{verbatim}
gap> kx:=[0,1,1];
[ 0, 1, 1 ]
\end{verbatim}

The symmetry operations which fix the k-point are extracted, and from them, the small group is generated.

\begin{verbatim}
gap> FL:=Filtered(MT,i->i*kx=kx);
[ [ [ 1, 0, 0 ], [ 0, 1, 0 ], [ 0, 0, 1 ] ], 
  [ [ -1, 0, 0 ], [ -1, 0, 1 ], [ -1, 1, 0 ] ], 
  [ [ 0, -1, 1 ], [ 0, 0, 1 ], [ -1, 0, 1 ] ], 
  [ [ 0, 1, -1 ], [ -1, 1, 0 ], [ 0, 1, 0 ] ], 
  [ [ 0, 1, -1 ], [ 0, 1, 0 ], [ -1, 1, 0 ] ], 
  [ [ 0, -1, 1 ], [ -1, 0, 1 ], [ 0, 0, 1 ] ], 
  [ [ 1, 0, 0 ], [ 0, 0, 1 ], [ 0, 1, 0 ] ], 
  [ [ -1, 0, 0 ], [ -1, 1, 0 ], [ -1, 0, 1 ] ] ]
\end{verbatim}

We can directly construct the small group from the above list of the output. For the purpose of the comparison of the representations, however, the point group and the small group must be expressed by common abstract symbols such as ``m1,m2,...'', which are employed to make up the multiplication table.(There is an alternative: the point group could be constructed from the 48 matrices of symmetry operations in wave-number space.)   

\begin{verbatim}
gap> FL:=List(FL,i->Position(MT,i));
[ 1, 2, 19, 20, 27, 28, 41, 42 ]
gap> FL:=List(FL,i->Elements(G)[i]);
[ m1, m2, m19, m20, m27, m28, m41, m42 ]
gap> GS:=Group(FL);
<group with 8 generators>
\end{verbatim}

The character table of the full point group GN is given as: 
\begin{verbatim}
gap> ConjugacyClasses(GN);
[ m1^G, m2^G, m5^G, m13^G, m14^G, m25^G, m26^G, m29^G, 
m37^G, m38^G  ]
gap> Display(Irr(GN));
[ [   1,   1,   1,   1,   1,   1,   1,   1,   1,   1 ],
  [   1,   1,   1,  -1,  -1,  -1,  -1,  -1,   1,   1 ],
  [   1,   1,   1,  -1,  -1,   1,   1,   1,  -1,  -1 ],
  [   1,   1,   1,   1,   1,  -1,  -1,  -1,  -1,  -1 ],
  [   2,   2,  -1,   0,   0,  -2,  -2,   1,   0,   0 ],
  [   2,   2,  -1,   0,   0,   2,   2,  -1,   0,   0 ],
  [   3,  -1,   0,  -1,   1,  -3,   1,   0,   1,  -1 ],
  [   3,  -1,   0,  -1,   1,   3,  -1,   0,  -1,   1 ],
  [   3,  -1,   0,   1,  -1,  -3,   1,   0,  -1,   1 ],
  [   3,  -1,   0,   1,  -1,   3,  -1,   0,   1,  -1 ] ]
\end{verbatim}

The character table of the small group is as follows:
\begin{verbatim}
gap> List(ConjugacyClasses(GS),Elements);
[ [ m1 ], [ m2 ], [ m19, m20 ], [ m27, m28 ], [ m41, m42 ] ]
gap> Display(Irr(GS));
[ [   1,   1,   1,   1,   1 ],
  [   1,   1,  -1,  -1,   1 ],
  [   1,   1,  -1,   1,  -1 ],
  [   1,   1,   1,  -1,  -1 ],
  [   2,  -2,   0,   0,   0 ] ]
\end{verbatim}

The branching relations are computed now: 
\begin{verbatim}
gap> Display(GetCompati(GN,GS));
[ [  1,  0,  0,  0,  0 ],
  [  0,  1,  0,  0,  0 ],
  [  0,  0,  1,  0,  0 ],
  [  0,  0,  0,  1,  0 ],
  [  0,  1,  0,  1,  0 ],
  [  1,  0,  1,  0,  0 ],
  [  1,  0,  0,  0,  1 ],
  [  0,  0,  0,  1,  1 ],
  [  0,  0,  1,  0,  1 ],
  [  0,  1,  0,  0,  1 ] ]
\end{verbatim}
This output list should be read as the indication of the following branching relation:
\begin{verbatim}
POINT GROUP    SMALL GROUP 
 G.1        ->  S.1
 G.2        ->  S.2
 G.3        ->  S.3
 G.4        ->  S.4
 G.5        ->  S.2 + S.4
 G.6        ->  S.1 + S.3
 G.7        ->  S.1 + S.5
 G.8        ->  S.4 + S.5
 G.9        ->  S.3 + S.5
 G.10       ->  S.2 + S.5
\end{verbatim}

\subsection{The character table at the boundary of the Brillouin zone}
\subsubsection{The necessity of the special treatment for the non-symmorphic crystal}

In general non-zero k-points, the wave-functions are apportioned to representations different to that of the $\Gamma$ point. The new representations are deduced from subgroups of the point group; the subgroup admissible to each k-point must be composed of operations which fix that k-point. The invariance on the k-point by symmetry operations should be kept up to the periodicity of the wave-space. Thus at the boundary of the Brillouin zone, different but equivalent k-points are mingled together in the representation. In general, it is sufficient to choose the subgroup which connects equivalent k-points and compute the representation in the same way as in $\Gamma$ point. However, in special cases, this naive approach inevitably fails, owing to the existence of two different kinds of crystal structures, i.e. the distinction between the symmorphic and non-symmorphic ones. Let us review this troublesome situation.
  
The matrix representation could be written as
\begin{equation}
D^k(\{\alpha|\tau\} )=\exp(-ik\cdot \tau)\Gamma(\alpha).
\end{equation}
Then the product is 
\begin{equation}
D^k (\{\alpha_1|\tau_1 \} ) D^k (\{\alpha_2|\tau_2 \} )
=\exp(-ik\cdot(\tau_1+\tau_2 ) )\Gamma(\alpha_1)\Gamma( \alpha_2)        
=\exp(-ik\cdot(\tau_1+\tau_2 ) )\Gamma(\alpha_1 \alpha_2).        
\end{equation}
The product should be equal to this:
\begin{equation}
D^k (\{\alpha_1\cdot \alpha_2 | \alpha_1 \cdot \tau_2+\tau_1 \} )
=\exp(-ik\cdot(\alpha_1\cdot \tau_2+ \tau_1 ) )\Gamma(\alpha_1 \alpha_2).        
\end{equation}
These equations lead to the relation:
\begin{equation}
    \exp(i(\alpha_1^{-1} \cdot k -k)\cdot \tau_2 )=1.        
\end{equation}
Certain conditions are required so that the above relation should be valid:
\begin{itemize}
\item[1] The k point is inside the Brillouin zone; $\alpha_1^{-1}$ does not move $k$ (i.e. $\alpha_1\cdot k =k$). At the $\Gamma$ point, this condition is always satisfied.
\item[2] The k point is located at the boundary of the Brillouin zone: the vector $\alpha_1\cdot k -k$ may coincide with a certain reciprocal vector due to the periodicity on the wave space. The subgroups of the operations of these kinds are called small groups. In addition, the $\tau_2$ should be zero or a lattice translation vector in the real space.
\end{itemize}
The crystal structure, in which the second condition always holds, is said to be symmorphic. In certain types of crystals, however, there exist affine mappings with non-zero $\tau_2$, which are fractions of some lattice translational vectors. This type of the crystal is called non-symmorphic. Indeed, the diamond structure is non-symmorphic one, as can be seen in the symmetry operations for this crystal. For the non-symmorphic case, therefore, the relation to be satisfied with respect to the product of the matrix representation does not hold at the boundary of the Brillouin zone. Thus it is necessary to employ special methods to treat this case. One of these treatments was proposed by Herring\cite{HERRING}. The intuitive interpretation of this method could be given in the following way. One should bear in mind these two points: First, at the $\Gamma$ point, there is no problem in the representation even in the non-symmorphic case. Secondly, the k-point at the boundary of the Brillouin zone can be pulled back at the $\Gamma$ point, if the unit cell is extended. For example, in a one-dimensional crystal with the lattice constant of $a$, the k-point $k=\pi/a$ is at the boundary of the Brillouin zone. However, when the doubled unit cell (with the lattice constant $2a$) is assumed, the k-point $k=\pi/a$  agrees with the new primitive reciprocal lattice vector; hence it is equivalent to the $\Gamma$ point. The irreducible representation at $\pi/a$ in the small original cell can be deduced from that at the $\Gamma$ point in the enlarged cell. Thus the difficulty in non-symmorphic representation can be eluded. We shall examine the actual computation of this method in the next section.

\subsubsection{Herring Method: An example}

\begin{figure}[h!]
\centering
\includegraphics[width=0.8\linewidth]{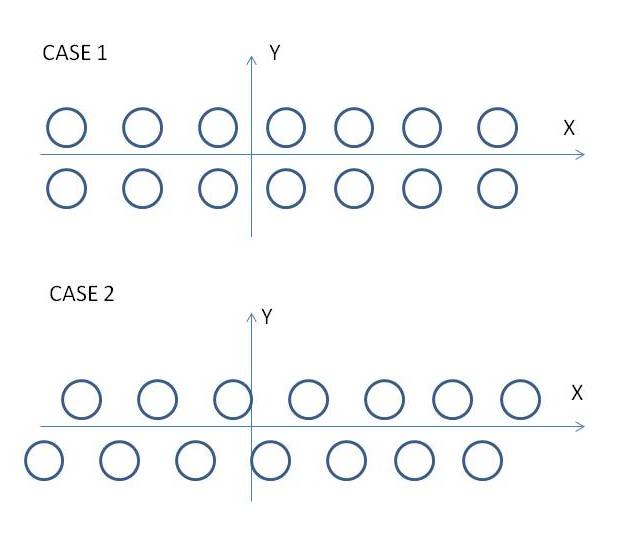}
\caption{Symmorphic and non-symmorphic one-dimensional crystals.}
\label{fig:NONSYMMORPHIC}
\end{figure}


Assume a double array of atoms, as in the Fig. \ref{fig:NONSYMMORPHIC}. This figure shows two types of one-dimensional periodic lattice, consisted by double lines of atoms. In case I, two rows forms one dimensional periodic system of squares. In case II, the two lows are shifted by half of the lattice constants. The latter is of non-symmorphic type. The symmetry operations which fix the k-point at $k=\pi/a$ for the latter case is given as
\begin{eqnarray}
e&=&\{E|0\}\\\nonumber
c&=&\{C_2|0\} {\rm\: A\;rotation\;by\; 180\; degrees\;along\;the\; origin}\\\nonumber
m&=&\{\sigma_x|a/2\}{\rm\: A\;reflection\; over\; x\; axis}\\\nonumber
g&=&\{\sigma_y|a/2\}{\rm\: A\;reflection\; over\; y\; axis}
\end{eqnarray}
From the aforementioned reasons, the irreducible representation of this group is not the proper one to represent the small group at $k=\pi/a$.

According to the Herring method, instead of this, a factor group G/I should be made use of, where G is a group of infinite order, composed of
\begin{eqnarray}
{e'}&=&\{E|2na\},\\\nonumber
{c'}&=&\{C_2 |2na\},\\\nonumber
{m'}&=&\{\sigma_x |a/2+2na\},\\\nonumber
{g'}&=&\{\sigma_y |a/2+2na\},\\\nonumber
{\bar{e}'}&=&\{E|a+2na\},\\\nonumber
{\bar{c}'}&=&\{C_2 |a+2na\},\\\nonumber
{\bar{m}'}&=&\{\sigma_x |3a/2+2na\},\\\nonumber
{\bar{g}'}&=&\{\sigma_y |3a/2+2na\},        
\end{eqnarray}
and
I is the normal subgroup composed of 
\begin{eqnarray}
{e'}=\{E|2na\}.
\end{eqnarray}

This factor group is the point group in the unit cell extended twofold in the x direction. By extending the unit cell, the k-points at the boundary of the original Brillouin zone are pulled back at the $\Gamma$ point in the new Brillouin zone. As the newly allotted $\Gamma$-point is the zero vector, the mathematical difficulty in the representation can be avoided. This situation should rather be interpreted as follows: the wavefunction at $k=\pi/a$ has the periodicity of $2a$, that is two times as large as the original lattice width. On the other hand, the wavefunction at $k=0$ in the original Brillouin zone also matches into the periodicity of $2a$.(Because originally it has periodicity of $a$) In the representation at the new $\Gamma$ point, those in the old $\Gamma$ point and the old boundary point are mingled with each other. These two different contributions, however, can be distinguishable. Let us check the character of the elements of following types $\{E|\tau\}$, viz. of the simple fractional translations in the extended cell, in which the linear operation is the identity. With respect to the representation of the boundary of the Brillouin zone of the original smaller cell, the representation matrix should be represented as $D^{k=\pi/a}(\{E|a\})=\exp(-i\frac{\pi}{a}a){\it I}^d=-{\it  I}^d$; character of which is equal to $-d$ ( d is the dimensionality of the representation). From Table \ref{tab:chr_1dimmodel} we can see the proper irreducible representation at $k=\pi/a$ is given by $\Gamma_5$ alone.

\begin{table}[h!]
\centering
\begin{tabular}{c|cccccccc}
&${e'}$&${\bar{e}'}$&${c'}$&${\bar{c}'}$&${m'}$&${\bar{m}'}$&${g'}$&${\bar{g}'}$\\\hline
${e'}$&${e'}$&${\bar{e}'}$&${c'}$&${\bar{c}'}$&${m'}$&${\bar{m}'}$&${g'}$&${\bar{g}'}$\\
${\bar{e}'}$&${\bar{e}'}$&${e'}$&${\bar{c}'}$&${c'}$&${\bar{m}'}$&${m'}$&${\bar{g}'}$&${g'}$\\
${c'}$&${c'}$&${\bar{c}'}$&${e'}$&${\bar{e}'}$&${\bar{g}'}$&${g'}$&${\bar{m}'}$&${m'}$\\
${\bar{c}'}$&${\bar{c}'}$&${c'}$&${\bar{e}'}$&${e'}$&${g'}$&${\bar{g}'}$&${m'}$&${\bar{m}'}$\\
${m'}$&${m'}$&${\bar{m}'}$&$g$&${\bar{g}'}$&${\bar{e}'}$&${e'}$&${\bar{c}'}$&${c'}$\\
${\bar{m}'}$&${\bar{m}'}$&${m'}$&${\bar{g}'}$&${g'}$&${e'}$&${\bar{e}'}$&${c'}$&${\bar{c}'}$\\
${g'}$&${g'}$&${\bar{g}'}$&${m'}$&${\bar{m}'}$&${c'}$&${\bar{c}'}$&${e'}$&${\bar{e}'}$\\
${\bar{g}'}$&${\bar{g}'}$&${g'}$&${\bar{m}'}$&${m'}$&${\bar{c}'}$&${c'}$&${\bar{e}'}$&${e'}$
\end{tabular}
\caption{The Multiplication table of G/I. The first column and the top row show the representatives in G of the elements in G/I. The elements in the table are the products between the element $a$ in the first column and the element $b$ in the top row, $a\cdot b$.}
\end{table}

\begin{table}[h!]
\centering
\begin{tabular}{c|ccccc}
&${e'}$&${\bar{e}'}$&[${c'}$,${\bar{c}'}$]&[${m'}$,${\bar{m}'}$]&$[${g'}$,{\bar{g}'}]$\\\hline
$\Gamma_1$ & 1 & 1 & 1 & 1 & 1\\
$\Gamma_2$ & 1 & 1 & 1 & -1 & -1\\
$\Gamma_3$ & 1 & 1 & -1 & 1 & -1\\
$\Gamma_4$ & 1 & 1 & -1 & -1 & -1\\
$\Gamma_5$ & 2 & -2 & 0 & 0 & 0
\end{tabular}
\caption{The character table. The top row shows the elements in the conjugacy classes. The rows below shows the characters in the five irreducible representations.}
\label{tab:chr_1dimmodel}
\end{table}

The prediction by the character table can be verified. The electronic structure calculation is executed to the mismatched double allay of Na atoms: The two atoms are placed at $ \pm(1/4 a_x+1/8 a_y) $ in the unit cell with 
$a_x=(a,0,0),a_y=(0,4a,0),a_z=(0,0,5a)$ and $a=8.1$au. This computation affirms that every pair of eigenvalues at the boundary of the Brillouin zone shows the double-fold degeneracy.

\begin{table}[h!]
\begin{tabular}{|c|c|c|c|c|c|c|c|c|}\hline\hline
k=0    &-2.775 &    -2.124 &    -0.506 &    -0.479 &    0.139 &    0.753 &    0.780 &    1.054 \\
k= $\pi$/a    &-0.717 &    -0.717 &    -0.072 &    -0.072 &    1.501 &    1.501 &    1.528 &    1.528\\\hline\hline
\end{tabular}
\caption{Energy spectrum at k=0 and $\pi$/a in the mismatched double allay of Na atoms (unit in eV). }
\end{table}

The above example is a simple one-dimensional model. For a non-symmorphic three-dimensional crystal, the irreducible representation for the k-point at the boundary of the Brillouin zone can be computed in a similar way. For this purpose we should set up a supercell structure; its periodicity should be equal to that of the wavefunction at the boundary k-point in consideration; the symmetry operations to be taken into account of must be confined to ones which fix the boundary k-point up to the periodicity of the old Brillouin zone of the small cell. (To consider the full symmetry of the new super-cell may be superfluous); the irreducible representation at the $\Gamma$ point in the supercell is the mixture of the wavefunctions at the $\Gamma$ point and the k-point in consideration, but we can distinguish them by the comparison of the character of the pure fractional translation. The example of the three-dimensional case is the irreducible representation of diamond structure in the face-centered cell at X point $k_X=[0,2\pi/a,0](=1/2G_2+1/2G_3)$. The necessary super-cell is the cubic unit cell. We can check this: the primitive lattice vectors in the cubic cell is expressed by those in the face-centered cell as $\hat{a}_1=a_1+a_2-a_3$, $\hat{a}_2=a_1-a_2+a_3$ and $\hat{a}_3=a_1+a_2-a_3$; so there are relations as $\hat{a}_1\cdot k_X=0$,$\hat{a}_2\cdot k_X=0$ and $\hat{a}_1\cdot k_X=2\pi$. These relations indicate that the wavefunction at $k_X$ in the face-centered cell fits itself into the cubic supercell, in the sense that the translations by the primitive lattice vector in the cubic cell causes no phase shift in the wavefunction. The symmetry operations for the small group at $k_X$ are constructed from those which fix $k_X$ ($k_X \rightarrow k_X$) and those which invert $k_X$ ($k_X \rightarrow -k_X$), due to the periodicity in the wave-number space. In contrast, the full symmetry operations in the cubic cell couples six wavefunctions at $[\pm 2\pi/a,0,0]$,$[0,\pm 2\pi/a,0]$,and $[0,0,\pm 2\pi/a]$. As the wave-vector $k_X$ cannot be fixed, the corresponding irreducible representation for the full symmetry is not adequate for the small group at $k_X$. (Elsewhere in this article, the point group of the full symmetry operations in the cubic unit cell is computed; from which the small group at $k_X$ can be derived as a subgroup.)
     
﻿\subsection{The semidirect product and the induced representation}
In the previous section of this article, concerning the computation of the character tables for the supercell structures, we have to remake the multiplication table of the extended group by the presence of the fractional translation. However, if we make use of the mathematical idea of the semidirect product, we can treat such a problem more effectively. In this section, this concept is explained and the possible computation GAP is presented.

The semidirect product $A \ltimes G$ is obtained by extending a group $G$ by an another group $A$ in the following way. The elements in the new group are given as the Cartesian product of each group:
\begin{equation}
    \{(a,g)  ; a\in A, g\in G\}.        
\end{equation}
The multiplication is defined as
\begin{equation}
    (a_1,g_1 )\cdot(a_2,g_2)=(a_1 g_1 (a_2),g_1 g_2),        
\end{equation}
in which the notation $g_1 (a_2)$ means the automorphism in $A$ acting on $a_2$, induced by $g_1$ in G. The typical example of this is the affine mapping in crystallographic group. The definition is given as follows:
\begin{eqnarray}
O_T=\{R_T|\tau_T \}   :  r \rightarrow R_T\cdot r+\tau_T {\rm \:(Affine\; mapping)},
\end{eqnarray}
\begin{eqnarray}
O_T\cdot O_S=\{R_T\cdot R_S|\tau_T+R_T\cdot \tau_S\}.        
\end{eqnarray}
If the group $A$ is Abelian, the representation of the semidirect product is given by
 \begin{equation}
\chi_V (a,g)=\frac{1}{|G_x |} \sum_{h\in G;hgh^{-1} \in G_x} x(h(a) ) \chi_U (hgh^{-1}).        
\end{equation}
Here G$_x$ is the set of elements in G, such that the character $x$ in A is fixed by the automorphism induced by elements in $G_x$. All of the irreducible representations of the semidirect product are exhausted by this formula through possible combinations of the character $x$ in A and the irreducible representation $\chi_U$ of $G_x$.(For the theoretical background, consult references of pure mathematics\cite{Ser,RP_SEMDIR}.)

One of the simplest example is $C_{3v}={\rm Group}((1,2,3),(1,2))\equiv S_3$. One can check the validity of this formula, taking $C_3={\rm Group}((1,2,3))$ as A and $I={\rm Group}((1,2))$ as G. Let us follow this computation.

The character tables of those three groups are given as
\begin{table}[h!]
\centering
\begin{tabular}{c|ccc}
         & [()] & [(1,2),(2,3),(3,1)]&[(1,2,3),(1,3,2)]\\\hline
$\chi_1$ & 1 & 1 & 1 \\
$\chi_2$ & 1 & -1 & 1 \\
$\chi_3$ & 2 & 0 & 1 \\
\end{tabular}
\caption{The character table of C$_{3v}$.}
\end{table}

\begin{table}[h!]
\centering
\begin{tabular}{c|ccc}
         & [$e_A$=()] & [$a_A$=(1,2,3)]&[$a_A^{-1}$=(1,3,2)]\\\hline
$x_1$ & 1 & 1 & 1 \\
$x_2$ & 1 & $\omega$ & $\omega^2$ \\
$x_3$ & 1 & $\omega^2$ & $\omega$ \\
\end{tabular}
\caption{The character table of C$_{3}$, where $\omega=\exp(2\pi/3)$.}
\end{table}

\begin{table}[h!]
\centering
\begin{tabular}{c|cc}
         & [$e_G$=()] & [$g_G$=(1,2)]\\\hline
$y_1$ & 1 & 1 \\
$y_2$ & 1 & -1
\end{tabular}
\caption{The character table of I.}\
\end{table}

The automorphism on A by G is defined as:
\begin{eqnarray}
e_G(e_A)=e_A,\nonumber\\
e_G(a_A)=a_A,\nonumber\\
e_G(a_A^{-1})=a_A^{-1},\nonumber\\
g_G(e_A)=e_A,\nonumber\\
g_G(a_A)=a_A^{-1},\nonumber\\
g_G(a_A^{-1})=a_A.
\end{eqnarray}

As a trivial case, take $x_1$. Since $G_{x_1} = G$, $y_1$ can be chosen as $\chi_U$. Since $G$ is Abelian, i.e. $hgh^{-1}=g$, the formula is 
\begin{eqnarray}
\chi_V (a,g)&=&\frac{1}{|G_{x_1}|} \sum_{h\in G;hgh^{-1} \in G} x_1(h(a)) y_1(hgh^{-1})\\\nonumber     
&=&\frac{1}{2}    \sum_{h\in G;hgh^{-1} \in G_{y_1}} x_1(h(a))y_1(g)=\frac{1}{2}(1+1)=1,
\end{eqnarray}
for all elements in the semidirect product.

Again take $x_1$. Since $G_{x_1} = G$,$y_2$ can also be chosen as $\chi_U$. The formula is
\begin{eqnarray}
\chi_V (a,g)&=& \frac{1}{2}\sum_{h\in G;hgh^{(-1)} \in G_{x_1}} x_1(h(a)) y_2(hgh^{-1}).    
\end{eqnarray}
It gives
\begin{eqnarray}
\chi_V(e_A,e_G)&=&\frac{1}{2}(1+1)=1,\\\nonumber
\chi_V(a_A,e_G)&=&\chi_V(a_A^{-1},e_G)=\frac{1}{2}(1+1)=1,\\\nonumber
\chi_V(e_A,g_G)&=&\frac{1}{2}(-1-1)=-1,\\\nonumber
\chi_V(a_A,g_G)&=&\chi_V(a_A^{-1},g_G)=\frac{1}{2}(-1-1)=-1.
\end{eqnarray}

Take $x_2$. Since $G_{x_2} = \langle\rm{Id}\rangle$ (the trivial group of the identity), it must be $\chi_U=1$. The formula is 
\begin{eqnarray}
\chi_V (a,g)&=& \sum_{h\in G;hgh^{-1} \in G_{x_2}} x_2(h(a)) \chi_U (hgh^{-1})\\\nonumber    
&=&\sum_{h\in G;hgh^{-1} \in G_{x_2}} x_2(h(a))\cdot 1.
\end{eqnarray}
It gives
\begin{eqnarray}
\chi_V(e_A,e_G)&=&(1+1)=2,\\\nonumber
\chi_V(a_A,e_G)&=&\chi_V(a_A^{-1},e_G)=(\omega+\omega^2)=-1,\\\nonumber
\chi_V(e_A,g_G)&=&0,\\\nonumber
\chi_V(a_A,g_G)&=&\chi_V(a_A^{-1},g_G)=0.
\end{eqnarray}
The last two characters are zero, since for any $h$, $h g h^{-1}=g \notin \langle\rm{Id}\rangle$, no terms are summed up.

(N.B. When we take $x_1$, it seems plausible that we could also choose $G_{x_1}=\langle\rm{Id}\rangle$.Then the formula is 
\begin{eqnarray}
\chi_V (a,g)&=& \sum_{h\in G;hgh^{-1} \in G_{x_1}} x_1(h(a)) \chi_U (hgh^{-1})\\\nonumber    
&=&\sum_{h\in G;hgh^{-1} \in <{\rm Id}>} x_1(h(a))\cdot 1.
\end{eqnarray}
It gives
\begin{eqnarray}
\chi_V(e_A,e_G)&=&2,\\\nonumber
\chi_V(a_A,e_G)&=&\chi_V(a_A^{-1},e_G)=2,\\\nonumber
\chi_V(e_A,g_G)&=&0,\\\nonumber
\chi_V(a_A,g_G)&=&\chi_V(a_A^{-1},g_G)=0.
\end{eqnarray}
However, this result is not irreducible. It is essential to use the maximal set of $G_x$.)

There are other combinations of characters including $x_3$ of C$_{3v}$. However, they only produce the redundant results.   

The aforementioned computations are tedious and complicated; a naive implementation by means of GAP programming language could be given as in appendix \ref{PROGSEMID}.  
 
In GAP, the semidirect product can be constructed by a built-in function.
\begin{verbatim}
gap> A=Group((1,2,3));
Alt( [ 1 .. 3 ] )
gap> G=Group((1,2));
Sym( [ 1 .. 2 ] )
\end{verbatim}
The automorphism in A is evaluated as follows:
\begin{verbatim}
gap> au:=AutomorphismGroup(A);
<group of size 2 with 1 generators>
gap> auE:=Elements(au);
[ [ (1,2,3) ] -> [ (1,2,3) ], [ (1,2,3) ] -> [ (1,3,2) ] ]
gap> gens:=GeneratorsOfGroup(G);
[ (1,2) ]
\end{verbatim}

The automorphism induced by G is defined as follows: 
(The generator (1,2) in G causes the automorphism in A as (1,2,3)\verb!->!(1,3,2). Indeed 
(1,2,3)*(1,3,2)=1: this is the inversion.)

\begin{verbatim}
gap> hom:=GroupHomomorphismByImages(G,au,gens,[auE[2]]);
[ (1,2) ] -> [ [ (1,2,3) ] -> [ (1,3,2) ] ]
gap> p:=SemidirectProduct(G,hom,A);
Group([ (3,4,5), (1,2)(4,5) ])
gap> Irr(p);
[ Character( CharacterTable( Group([ (3,4,5), (1,2)(4,5) ]) ),
  [ 1, 1, 1 ] ), Character( CharacterTable( Group([ (3,4,5), (1,2)
  (4,5) ]) ), [ 1, 1, -1 ] ), Character( CharacterTable( Group([ (3,
   4,5), (1,2)(4,5) ]) ), [ 2, -1, 0 ] ) ]
\end{verbatim}
 

\subsection{The computation of the character table in the cubic diamond by GAP}
In this section, the character table in the cubic diamond structure is computed by GAP command for the semidirect product. The point group G of the basic diamond structure is extended by the Abelian group of fractional translations N. 

The point group G is given in the table in the appendix of this article, as the set of 48 three-dimensional matrices.(Although in the table accompanying fractional transformations are listed, the point group G is constructed solely from three-dimensional rotation matrices; the fractional translations are omitted in the point group construction; they are included in the Abelian group N.) The group N is the factor group F/T, where
 generators of F are
\begin{equation}
\tau_1:=(1/4,1/4,1/4),\tau_2:=(1/4,-1/4,-1/4),\tau_3:=(-1/4,-1/4,1/4)
\end{equation}
and those of T are
\begin{equation}
(1,0,0),(0,1,0),(0,0,1).
\end{equation}
Using these generators, the group A and T are given as the vector space (or module) over integers. 

For the convenience, the group F/T can be expressed as another factor group A/B, 
where
\begin{equation}
{\rm A:=Group((1,2,3,4),(5,6,7,8),(9,10,11,12));}
\end{equation}
and
\begin{eqnarray}
{\rm B} &:& = {\rm Group}((1,2,3,4)^\wedge 2*(5,6,7,8)^\wedge 2,(5,6,7,8)^\wedge 2*(9,10,11,12)^\wedge 2,\\\nonumber & &
(9,10,11,12)^\wedge 2*(1,2,3,4)^\wedge 2)\\\nonumber 
 ( &:& = {\rm Group}( (1,3)(2,4)(5,7)(6,8), (5,7)(6,8)(9,11)(10,12), (1,3)(2,4)(9,11)(10,12) );
\end{eqnarray}

This definition corresponds to the relations of the fractional translations in F/I;
\begin{equation}
4*\bar{\tau}_i:=(0,0,0)\:{\rm[Identity]}
\end{equation}
and
\begin{equation}
2*(\bar{\tau}_1+\bar{\tau}_2)=
2*(\bar{\tau}_2+\bar{\tau}_3)=
2*(\bar{\tau}_3+\bar{\tau}_1)
\equiv(0,0,0)\:{\rm[Identity]}
\end{equation}

The gap computation goes as follows:

\begin{verbatim}
gap> A:=Group((1,2,3,4),(5,6,7,8),(9,10,11,12));
Group([ (1,2,3,4), (5,6,7,8), (9,10,11,12) ])
gap> B:=Group((1,2,3,4)^2*(5,6,7,8)^2,(5,6,7,8)^2*(9,10,11,12)^2,
    (9,10,11,12)^2*(1,2,3,4)^2);
Group([ (1,3)(2,4)(5,7)(6,8), (5,7)(6,8)(9,11)(10,12), (1,3)(2,4)
(9,11)(10,12) ])
gap> N:=FactorGroup(A,B);
Group([ f1, f2, f3, f4 ])
gap> NaturalHomomorphismByNormalSubgroup(A,B);
[ (1,2,3,4), (5,6,7,8), (9,10,11,12) ] -> [ f3, f2, f1 ]
\end{verbatim}

The last command returns the homomorphism from A to another group whose kernel is B: in other words, gives the mapping of A onto A/B. The tree generators (1,2,3,4),(5,6,7,8),(9,10,11,12) in A are mapped to f3,f2,f1 in A/B. 

\begin{verbatim}
gap> N.1;
f1
gap> N.2;
f2
gap> N.3;
f3
gap> N.4;
f4
\end{verbatim}

There is a newcomer f4; we must identify it. For this purpose, it is enough to compute the multiplication table of for generators.

\begin{verbatim}
gap> NV:=[N.1,N.2,N.3,N.4];
[ f1, f2, f3, f4 ]
gap> TransposedMat([NV])*[NV];
[ [ f4, f1*f2, f1*f3, f1*f4 ], [ f1*f2, f4, f2*f3, f2*f4 ], 
  [ f1*f3, f2*f3, f4, f3*f4 ], 
  [ f1*f4, f2*f4, f3*f4, <identity> of ... ] ]
\end{verbatim}

We can see the relation f1*f1=f2*f2=f3*f3=f4, which means that the element f4 corresponds to elements \verb|(1,2,3,4)^2, (5,6,7,8)^2, (9,10,11,12)^2|, viz. the fractional transformation (1/2,1/2,1/2).

The size of the fractional group is given as
\begin{verbatim}
gap> Size(C);
16
\end{verbatim}
Concerning this, the volume of the parallelepiped spanned by $\tau_i$ is
\begin{verbatim}
gap> AbsoluteValue(Determinant(A));
1/16
\end{verbatim}
Thus, in the unit cubic cell, the translations by $\tau_i$ generates 16 discrete points, the number of which is equal to the order of F/I.    

It is sufficient to define the automorphisms in N induced by G only for the generators of G: the candidates of which are two matrices, those of No.36 and No.48 in the list of symmetry operations.
\begin{equation}
{\rm M36}:=\left(
\begin{array}{ccc}
 0 & 0  & 1  \\ 
 1     & 0 & 0 \\ 
 0 & -1 & 0
\end{array} 
\right),\: 
{\rm M48}:=\left(
\begin{array}{ccc}
 0 & 0  & -1  \\ 
 0 & -1 & 0 \\ 
 1 & 0 & 0
\end{array} 
\right)
\end{equation}

Let us check the operations of these two matrices on $\tau_i$.
\begin{verbatim}
gap> A:=[[1,1,1],[-1,-1,1],[-1,1,-1]]/4;
[ [ 1/4, 1/4, 1/4 ], [ -1/4, -1/4, 1/4 ], [ -1/4, 1/4, -1/4 ] ]
gap> A:=TransposedMat(A);
[ [ 1/4, -1/4, -1/4 ], [ 1/4, -1/4, 1/4 ], [ 1/4, 1/4, -1/4 ] ]
gap> M36:=[[0,0,1],[1,0,0],[0,-1,0]];
[ [ 0, 0, 1 ], [ 1, 0, 0 ], [ 0, -1, 0 ] ]
gap> M48:=[[0,0,-1],[0,-1,0],[1,0,0]];
[ [ 0, 0, -1 ], [ 0, -1, 0 ], [ 1, 0, 0 ] ]
\end{verbatim}

The generators are set to be the transposed forms of above matrices.
\begin{verbatim}
gap> TM36:=TransposedMat(M36);
[ [ 0, 1, 0 ], [ 0, 0, -1 ], [ 1, 0, 0 ] ]
gap> TM48:=TransposedMat(M48);
[ [ 0, 0, 1 ], [ 0, -1, 0 ], [ -1, 0, 0 ] ]
gap> (A*TM36*A^-1);
[ [ 0, -1, 0 ], [ 0, 0, -1 ], [ -1, 0, 0 ] ]
gap> (A*TM48*A^-1);
[ [ 0, 1, 0 ], [ 0, 0, 1 ], [ -1, -1, -1 ] ]
\end{verbatim}

The last two computations (such as $A\cdot M \cdot A^{-1}$) give the transformations on the row vectors ($\tau_i$) in $A$.

The operations by M36 and M48 are given as:
\begin{eqnarray}
M36\cdot\tau_1&=&-\tau_2\\\nonumber
M36\cdot\tau_2&=&-\tau_3\\\nonumber
M36\cdot\tau_3&=&-\tau_1\\\nonumber
M48\cdot\tau_1&=&\tau_2\\\nonumber
M48\cdot\tau_2&=&\tau_3\\\nonumber
M48\cdot\tau_3&=&-\tau_1-\tau_2-\tau_3
\end{eqnarray}

\begin{remark}
In the definition of GAP,
the semidirect product of a group N with a group G (acting on N via a homomorphism $\alpha$ from G into
the automorphism group of N) is the Cartesian product G $\times$ N with the multiplication $(g, n) \cdot (h, m) =
(gh, n^{h^\alpha} m)$, since in GAP all groups
act from the right. On the other hand, the familiar style for the physicists is $\{G|n\}\cdot\{H|m\}=\{G\cdot H|n+G*m\}$, as in the crystallography; the point group acts from the left. However, these two different definitions are equivalent. This is based on this: the crystallography, the symmetry operation $\{M|t\}$ is represented by 
an augmented matrix, which can take two forms. The one is the matrix of the form
\begin{verbatim}
                     [ M 0 ]
                     [ t 1 ]
\end{verbatim}
acting from the right on row vectors $(x,1)$. 
The another is 
\begin{verbatim}
                     [ M t ]
                     [ 0 1 ]
\end{verbatim}
acting from the left on column vectors $(x,1)$. 
These two definitions are convertible by transposition and one should interpret the GAP computations by the former definition. The generators used here, (M36, M48), should be replaced with their transposed forms.
\end{remark}

We can identify $\tau_i$ to three permutations:
\begin{eqnarray}
\tau_1&:=&(9,10,11,12) \equiv {\rm f1}\\\nonumber
\tau_2&:=&(5,6,7,8) \equiv {\rm f2}\\\nonumber
\tau_3&:=&(1,2,3,4) \equiv {\rm f3}
\end{eqnarray}

Other identifications will do. But we use the identification consistent with the result of the ``NaturalHomomorphismByNormalSubgroup'' command.

It suffices that the automorphism in N induced by G is defined at generators of groups. 
\begin{verbatim}
gap> gens:=GeneratorsOfGroup(N);;
gap> CC1:=[gens[2]^-1,gens[3]^-1,gens[1]^-1,gens[4]];;
gap> CC2:=[gens[2],gens[3],(gens[1]*gens[2]*gens[3])^-1,gens[4]];;
gap> hom1:=GroupHomomorphismByImages(N,N,gens,CC1);;
gap> hom2:=GroupHomomorphismByImages(N,N,gens,CC2);;
\end{verbatim}

The command ``GroupHomomorphismByImage(G, H, gens, imgs)'' defines a homomorphism from the group G to the group H through the mapping of the list ``gens'' of the generators of G to the list ``imgs'' of the images in H. If the definition of the mapping is irrelevant, the program returns a ``fail'' message.

\begin{verbatim}
gap> CC3:=[gens[3]^-1,gens[1]*gens[3]^-1,gens[2]*gens[3]^-1];;
gap> hom3:=GroupHomomorphismByImages(N,N,[gens[1],gens[2],gens[3]],CC3);
fail
\end{verbatim}

Let G be the point group of the minimal unit cell, generated from the matrices TM36 and TM48. By those preparations, we can make the semidirect product $G\ltimes N$.
\begin{verbatim}
gap> au:=AutomorphismGroup(N);;
gap> G:=Group(TM36,TM48);;
gap> GTON:=GroupHomomorphismByImages(G,au,GeneratorsOfGroup(G),[hom1,hom2]);;
gap> GPN:=SemidirectProduct(G,GTON,N);;
gap> irrep:=Irr(SemidirectProduct(G,GTON,N));;
\end{verbatim} 

Once the semidirect product ${\rm G}\ltimes {\rm N}$ is computed, the representation of the point group in the cubic diamond can be deduced from a suitable subgroup of this. For this purpose let us see the embedding of G and N into the semidirect product. The command ``Embedding'' creates the embedding from G and N into the semidirect product, and the command ``Projection'' creates the projection from the semidirect product to G. 
\begin{verbatim}
gap> embd1:=Embedding(GPN,1);
CompositionMapping( [ f1, f2, f3, f4, f5 ] -> [ f1, f2, f3, f4, f5 ],
 CompositionMapping( Pcgs([ (5,8)(6,7), (1,2)(3,5,7,4,6,8), (3,7,6)(4,8,5), 
  (1,6)(2,5)(3,7)(4,8), (1,3)(2,4)(5,8)(6,7) ]) -> [ f1, f2, f3, f4, f5 ],
 <action isomorphism> ) )
gap> embd2:=Embedding(GPN,2);
[ f1, f2, f3, f4 ] -> [ f6, f7, f8, f9 ]
gap> GeneratorsOfGroup(G);
[ [ [ 0, 1, 0 ], [ 0, 0, -1 ], [ 1, 0, 0 ] ], 
  [ [ 0, 0, 1 ], [ 0, -1, 0 ], [ -1, 0, 0 ] ] ]
gap> List(GeneratorsOfGroup(G),g->Image(embd1,g));
[ f2, f1*f3*f5 ]
gap> GeneratorsOfGroup(N);
[ f1, f2, f3, f4 ]
gap> List(GeneratorsOfGroup(N),g->Image(embd2,g));
[ f6, f7, f8, f9 ]
gap> Projection(GPN);
[ f1, f2, f3, f4, f5, f6, f7, f8, f9 ] -> 
[ [ [ 1, 0, 0 ], [ 0, 0, 1 ], [ 0, 1, 0 ] ], 
  [ [ 0, 1, 0 ], [ 0, 0, -1 ], [ 1, 0, 0 ] ], 
  [ [ 0, 0, -1 ], [ -1, 0, 0 ], [ 0, 1, 0 ] ], 
  [ [ -1, 0, 0 ], [ 0, 1, 0 ], [ 0, 0, -1 ] ], 
  [ [ 1, 0, 0 ], [ 0, -1, 0 ], [ 0, 0, -1 ] ], 
  [ [ 1, 0, 0 ], [ 0, 1, 0 ], [ 0, 0, 1 ] ], 
  [ [ 1, 0, 0 ], [ 0, 1, 0 ], [ 0, 0, 1 ] ], 
  [ [ 1, 0, 0 ], [ 0, 1, 0 ], [ 0, 0, 1 ] ], 
  [ [ 1, 0, 0 ], [ 0, 1, 0 ], [ 0, 0, 1 ] ] ]
\end{verbatim}

The correspondence of the generators is given in table\ref{tab:semidirectGn}.

\begin{table}[htbp]
\begin{tabular}{c|c}
\hline Elements in G and N& Elements in $G\ltimes N$\\
\hline TM36(in G) & f2 \\ 
\hline TM48(in G) & f1*f3*f5 \\ 
\hline f1(in N) (The image of (9,10,11,12))  & f6 \\ 
\hline f2(in N) (The image of (5,6,7,8))&  f7\\ 
\hline f3(in N) (The image of (1,2,3,4))&  f8 \\ 
\hline f4(in N) (The image of (1,3)(2,4) or (5,7)(6,8) or (1,3)(2,4))& f9 \\ 
\hline 
\end{tabular} 
\caption{The correspondence of the generators in groups G, N and $G\ltimes N$.} 
\label{tab:semidirectGn}
\end{table}

The generators of the point group of the cubic diamond structure are these elements:
\begin{eqnarray}
\{\rm m36|\tau_i\}\\\nonumber
\{\rm m48|0\}
\end{eqnarray}
($\tau_i$ can be any of three alternatives. The addition of pure fractional translations $\{E|\tau_i\}$ is superfluous in this case.) 

The corresponding subgroup can be constructed in this way:
\begin{verbatim}
Subgroup(GPN,[GPN.2*GPN.6,GPN.1*GPN.3*GPN.5])
\end{verbatim}
or
\begin{verbatim}
Subgroup(GPN,[GPN.2*GPN.7,GPN.1*GPN.3*GPN.5])
\end{verbatim}
or
\begin{verbatim}
Subgroup(GPN,[GPN.2*GPN.8,GPN.1*GPN.3*GPN.5])
\end{verbatim}
These three definitions generate the same group so that any choice can be used. 
(They are the subgroups of GPN, generated by the elements in the square brackets of the second argument.)
 
\subsection{Crystal lattice and Cohomology}
\subsubsection{A brief introduction to group cohomology}
In the modern style, the crystal group can be described by the theory of the cohomology, which is actually computable by the computer algebra. 

At first, we construct the one-cohomology. Let $K\in GL(n,\mathbb{Z})$ and define $V:=\mathbb{R}^n/\mathbb{Z}^n$.
The one-cocycle is defined as
\begin{equation}
Z^1(K,V):=\{\rho;k\rightarrow V|\rho(1)=0,\rho(gh)=\rho(g)+g\rho(h){\rm\,for\, all\,} g,h\in K\}.
\end{equation}
The one-coboundary is defined as
\begin{equation}
B^1(K,V):=\{\rho\in Z^1(K,V|\rho(g)=gv-v{\rm\,\, for\,\,some}\,\,v\in V\}.
\end{equation}
The one-cohomology is defined as their quotient:
\begin{equation}
H^1(K,V):=Z^1(K,V)/B^1(K,V)
\end{equation}
We can see the vector system in the space group, which satisfies the condition
\begin{equation}
\tau(gh)\equiv\tau(g)+g\cdot\tau(h) \,{\rm mod} \,\mathbb{Z}^n({\rm mod\, lattice \,translations})
\end{equation}
is a kind of cocycle.
Furthermore, the one-coboundary is interpreted as the shift of origin; we can observe this  from the conjugation operation:
\begin{equation}
\{1|s\}^{-1}\{g|t\}\{1|s\}=\{g|(gs-s)+t\}
\end{equation}
Thus, the vector system could be understood in the context of one-cohomology.

The above definition of the one-cohomology is derived from the general theory of group cohomology. Let $G$ be a finite group and $M$ be a module. Let $C^n(G,M)$ be the set functions from $G^n = (g_1,g_2,\dots,g_n)$ to $M$. The  differential ($d^n:C^n(G,M)\rightarrow C^{n+1}(G,M)$ is defined as
\begin{eqnarray}
d^n(f)(g_0,g_1,\dots,g_n)&=&g_0\cdot f(g_1,\dots,g_n)\\\nonumber
& &+\sum_{j}^{n}(-1)^j f(g_0,\dots,g_{j-2},g_{j-1}g_j,g_{j+1},\dots,g_n)\\\nonumber
& &+(-1)^{n+1}f(g_0,g_1,\dots,g_{n-1}).
\end{eqnarray}
It satisfies
\begin{equation}
d^n \circ d^{n-1}=0.
\end{equation}
The n-cocycle $Z^i(G,M)$ is defined as ${\rm Ker}(d^n)$, and the n-coboundary is as ${\rm Im}(d^{n-1})$. The n-th cohomology group is defined as $H^i(G,M):=Z^i(G,M)/B^i(G,M)$.

\subsubsection{The possible fractional translations with one point group}
We make use of following matrices (as explained elsewhere in this article) in order to define the crystallographic group in the diamond crystal lattice:
\begin{verbatim}
gap>M1:=[[0,0,1,1/4],[1,0,0,1/4],[0,-1,0,1/4],[0,0,0,1]];
gap>M2:=[[0,0,-1,0],[0,-1,0,0],[1,0,0,0],[0,0,0,1]];
gap>S:=AffineCrystGroupOnLeft([M1,M2]);
gap>P:=PointGroup(S);
\end{verbatim}

In this definition, the fractional translation is given a priori. With regard to this, there is a question: what is the possible fractional translation compatible to the point group of the diamond crystal? Is it uniquely determined? Let us inspect this problem. For example, take the face-centered lattice. For simplicity in the following computation, the symmetry operations are represented by the ``lattice axes coordinate'', by means of the vector system of the primitive lattice vectors. For this new coordinate, the new point group Q is made from the group P by the conjugation of the generators. (The symbol \verb!^! is used for the conjugation in GAP.)

\begin{verbatim}
gap> A:=[[0,1/2,1/2],[1/2,0,1/2],[1/2,1/2,0]];
gap> Q:=P^A;
gap> Q.1;Q.2;
[ [ -1, 0, 0 ], [ 0, 0, -1 ], [ 1, 1, 1 ] ]
[ [ 0, 1, 0 ], [ 0, 0, 1 ], [ -1, -1, -1 ] ]
\end{verbatim}

The generators of the space groups are defined now by means of unknown variables.(For the heuristic purpose, we venture to employ generators different to the first definition.)

\begin{verbatim}
a:=Indeterminate(Rationals,1);
b:=Indeterminate(Rationals,2);
c:=Indeterminate(Rationals,3);
N1:=[[-1,0,0,0],[0,0,-1,0],[1,1,1,0],[0,0,0,1]];
N2:=[[0,1,0,a],[0,0,1,b],[-1,-1,-1,c],[0,0,0,1]];
\end{verbatim}

Let us consider multiplications of two elements g$_i$ and g$_j$ in the point group, which yields the identity. If g$_i$ and g$_j$ are represented by the products of the generators, the multiplication gives a relation of the generators both in the point group and in the space group. For example, take (Q.2*Q.1)\verb!^!4=I. This relation is lifted up to the space group as (N2*N1)\verb!^!4$\equiv\{I|T\}$, where $T$ is a translation. The linear part of (N2*N1)\verb!^!4 is the identity matrix; and the translation part [2*a+2*b,2*a+2*b,-2*a-2*b] must be one of translation vectors. In general, the translation part in the relation of the generators in the space group gives three linear equations with respect to unknown variables ``a'',``b'', and ``c''. These equations should be equivalent to zero, modulo translation vectors; viz. they should be integer-valued. 

The relations in generators in Q shall be set up now. For this purpose, each elements in the group and its inverse are represented by generators respectively, so that the product of them gives the relation in generators which yields the unity. In the following, the symbols x1 and x2 indicate the two generators of the group Q. For simplicity, the trivial relation only representing the unity ( returned as \verb!<identity ...>! in the GAP result) is eliminated in the list.
\begin{verbatim}
gap> MQ:=Elements(Q);;
gap> rel:=List(MQ,i->Factorization(Q,i)*Factorization(Q,i^-1));;
gap> IDENTITY:=Factorization(Q,One(Q));;
gap> rel:=Filtered(rel,i->i<>IDENTITY);
[ (x2^-1*x1*x2*x1)^2, x1*(x2*x1^-1)^2*x1^-1, (x2*x1^2)^2,
 (x2^-1*x1)^2, x1*x2^2*x1*x2^-1*x1*x2*x1^-1, x1^6, 
  (x2*x1*x2^2)^2, (x2^2*x1*x2)^2, (x2*x1*x2^2*x1^2)^2, 
  (x2^-1*x1)^2, (x1*x2^2*x1^2)^2, x2^4, x2*x1^3*x2^-1*x1^3, 
  x1*x2^4*x1^-1, x1^-1*(x2^-1*x1)^2*x1, x2^-1*x1*x2^3*x1, 
  (x2^2*x1^3)^2, (x2*x1^-1)^2, x2^4, 
  x2^-1*x1^-1*(x2^-1*x1)^2*x1*x2, (x1*x2*x1)^2, (x2^-1*x1)^2, 
  x1^-1*x2^-1*x1*x2^3*x1^2, x2^-1*x1*x2^3*x1, 
  x1^-1*(x2^-1*x1)^2*x1, x2^2*x1*x2^-1*x1*x2, (x2*x1)^4, 
  (x1^2*x2)^2, (x2*(x2*x1)^2)^2, (x2*x1*x2^2*x1)^2, 
  (x1*x2^2*x1*x2)^2, x2^-1*x1^3*x2*x1^3, x2^2*x1*x2^-1*x1*x2, 
  (x2*x1^-1)^2, (x2^-1*x1^-1*x2^-1*x1^2)^2 ]
\end{verbatim}

The relations of the point group generators are lifted to the space group. The matrix relations are stored in the list ``relMat''; and the fractional translations are taken out in the list ``relv''. Now, if the fractional translation [a,b,c] is admissible, the list ``relv'' must be integer-valued by the substitutions to the indeterminates. (Apropos of this, we can use symbols of unknowns as ``a'',``b'',``c'' for the input, whereas the GAP results are represented by ``x\verb!_!1'',``x\verb!_!2'',``x\verb!_!3''. Be careful not to confuse indeterminates and generators of the group.) The GAP program cannot invert the symbolically represented matrices N1 and N2, so we must prepare the inverses for them (N1I and N2I) by the power products.

\begin{verbatim}
N1I:=N1^5;
N2I:=N2^3;
relMat:=[ (N2I^1*N1*N2*N1)^2, N1*(N2*N1I^1)^2*N1I^1, (N2*N1^2)^2, 
(N2I^1*N1)^2, N1*N2^2*N1*N2I^1*N1*N2*N1I^1, N1^6, 
  (N2*N1*N2^2)^2, (N2^2*N1*N2)^2, (N2*N1*N2^2*N1^2)^2, (N2I^1*N1)^2, 
  (N1*N2^2*N1^2)^2, N2^4, N2*N1^3*N2I^1*N1^3, 
  N1*N2^4*N1I^1, N1I^1*(N2I^1*N1)^2*N1, N2I^1*N1*N2^3*N1, 
  (N2^2*N1^3)^2, (N2*N1I^1)^2, N2^4, 
  N2I^1*N1I^1*(N2I^1*N1)^2*N1*N2, (N1*N2*N1)^2, (N2I^1*N1)^2,
  N1I^1*N2I^1*N1*N2^3*N1^2, N2I^1*N1*N2^3*N1, 
  N1I^1*(N2I^1*N1)^2*N1, N2^2*N1*N2I^1*N1*N2, (N2*N1)^4, 
  (N1^2*N2)^2, (N2*(N2*N1)^2)^2, (N2*N1*N2^2*N1)^2, 
  (N1*N2^2*N1*N2)^2, N2I^1*N1^3*N2*N1^3, N2^2*N1*N2I^1*N1*N2, 
  (N2*N1I^1)^2, (N2I^1*N1I^1*N2I^1*N1^2)^2 ];
relv:=List(relMat,i->[i[1][4],i[2][4],i[3][4]]);
[ [ 2*x_1+x_2+x_3, -2*x_1-x_2-x_3, -2*x_1-x_2-x_3 ], 
  [ -2*x_1-x_2-x_3, 0, 2*x_1+x_2+x_3 ], [ -x_2-x_3, 2*x_2, 2*x_3 ], 
  [ 2*x_1+x_2+x_3, -2*x_1-x_2-x_3, 0 ], [ 0, 2*x_1+x_2+x_3, 0 ], 
  [ 0, 0, 0 ], [ 0, 0, 2*x_1+x_2+x_3 ], 
  [ 0, 2*x_1+x_2+x_3, -2*x_1-x_2-x_3 ], [ 0, 2*x_1+4*x_2+2*x_3, 0 ],
  [ 2*x_1+x_2+x_3, -2*x_1-x_2-x_3, 0 ], 
  [ -2*x_1-2*x_2, 2*x_1+2*x_2, 2*x_2+2*x_3 ], [ 0, 0, 0 ], 
  [ 2*x_1, 2*x_2, 2*x_3 ], [ 0, 0, 0 ], 
  [ -2*x_1-x_2-x_3, 0, 2*x_1+x_2+x_3 ], 
  [ 2*x_1+x_2+x_3, -2*x_1-x_2-x_3, 0 ], 
  [ 2*x_1+2*x_2, 2*x_2+2*x_3, -2*x_1-2*x_2 ], 
  [ 2*x_1+x_2+x_3, 0, 0 ], [ 0, 0, 0 ], [ 0, -2*x_1-x_2-x_3, 0 ], 
  [ x_2+x_3, -2*x_3, x_2+x_3 ], [ 2*x_1+x_2+x_3, -2*x_1-x_2-x_3, 0 ]
    , [ -2*x_1-x_2-x_3, 0, 2*x_1+x_2+x_3 ], 
  [ 2*x_1+x_2+x_3, -2*x_1-x_2-x_3, 0 ], 
  [ -2*x_1-x_2-x_3, 0, 2*x_1+x_2+x_3 ], 
  [ 0, 2*x_1+x_2+x_3, -2*x_1-x_2-x_3 ], 
  [ 2*x_1+2*x_2, 2*x_1+2*x_2, -2*x_1-2*x_2 ], 
  [ -x_2-x_3, -x_2-x_3, 2*x_2 ], 
  [ 4*x_1+4*x_2+2*x_3, -x_2-x_3, -4*x_1-2*x_2 ], 
  [ 4*x_1+2*x_2, x_2+x_3, x_2+x_3 ], 
  [ -4*x_1-4*x_2-2*x_3, 4*x_1+2*x_2, x_2+x_3 ], 
  [ 2*x_1+2*x_2+2*x_3, -2*x_1, -2*x_2 ], 
  [ 0, 2*x_1+x_2+x_3, -2*x_1-x_2-x_3 ], [ 2*x_1+x_2+x_3, 0, 0 ], 
  [ 2*x_2+2*x_3, -2*x_1-2*x_3, 2*x_1+2*x_3 ] ]
\end{verbatim}

Since there are duplicates in the linear equations of the fractional translation, the unique ones should be extracted by 
\begin{verbatim}
gap> Unique(Flat(relv*a^0));
[ 2*x_1+x_2+x_3, -2*x_1-x_2-x_3, 0, -x_2-x_3, 2*x_2, 2*x_3, 
  2*x_1+4*x_2+2*x_3, -2*x_1-2*x_2, 2*x_1+2*x_2, 2*x_2+2*x_3, 2*x_1, 
  x_2+x_3, -2*x_3, 4*x_1+4*x_2+2*x_3, -4*x_1-2*x_2, 4*x_1+2*x_2, 
  -4*x_1-4*x_2-2*x_3, 2*x_1+2*x_2+2*x_3, -2*x_1, -2*x_2, 
  -2*x_1-2*x_3, 2*x_1+2*x_3 ]
\end{verbatim} 
It is necessary to use an input such as ``\verb!relv*a^0!'', so that all entries in the list should be polynomials, for the convenience of sorting, by dint of the fact \verb!a^0! is the unity as a polynomial. Now we have following four equations to be integer-valued: 
\begin{verbatim}
x_2+x_3
2*x_1
2*x_2
2*x_3
\end{verbatim}

In $\mathbb{Q}/\mathbb{Z}$, there are four representatives of the fractional translations, as are given below, with the primitive lattice vectors accompanied to them. (The translation vectors are computed by the substitution
in the list ``relv'', from which the basis vectors are extracted by the GAP built-in command ``BaseIntMat({\rm mat})''.)

\begin{table}[h!]
\begin{tabular}{c|c}
\hline
Fractional translation & Translation vectors \\\hline
[ 0, 1/2, 1/2 ]&[ [ 1, 0, 0 ], [ 0, 1, 0 ], [ 0, 0, 1 ] ]\\

[ 1/2, 0, 0 ] & [ [ 1, 0, 0 ], [ 0, 1, 0 ], [ 0, 0, 1 ] ]\\

[ 1/2, 1/2, 1/2 ] & [ [ 1, 1, 1 ], [ 0, 2, 0 ], [ 0, 0, 2 ] ]\\\hline
\end{tabular} 
\caption{The fractional translations and the translation vectors.}
\label{tab:dia_frac_trans}
\end{table}

To inspect the validity, by means of GAP built-in command, the possible space groups with a given point group can be computed. In order to check the computed fractional translations, it is enough to see the generators of the space group and the primitive translation vectors.
\begin{verbatim}
gap> SS:=SpaceGroupsByPointGroupOnLeft(Q);;
gap> Display(List(SS,GeneratorsOfGroup));
[ [ [ [ -1, 0, 0, 0 ], [ 0, 0, -1, 0 ], [ 1, 1, 1, 0 ], [ 0, 0, 0, 1 ] ], 
      [ [ 0, 1, 0, 0 ], [ 0, 0, 1, 0 ], [ -1, -1, -1, 0 ], [ 0, 0, 0, 1 ] ], 
      [ [ 1, 0, 0, 1 ], [ 0, 1, 0, 0 ], [ 0, 0, 1, 0 ], [ 0, 0, 0, 1 ] ], 
      [ [ 1, 0, 0, 0 ], [ 0, 1, 0, 1 ], [ 0, 0, 1, 0 ], [ 0, 0, 0, 1 ] ], 
      [ [ 1, 0, 0, 0 ], [ 0, 1, 0, 0 ], [ 0, 0, 1, 1 ], [ 0, 0, 0, 1 ] ] ], 
  [ [ [ -1, 0, 0, 0 ], [ 0, 0, -1, 0 ], [ 1, 1, 1, 0 ], [ 0, 0, 0, 1 ] ], 
      [ [ 0, 1, 0, 0 ], [ 0, 0, 1, 1/2 ], [ -1, -1, -1, 1/2 ], [ 0, 0, 0, 1 ] ], 
      [ [ 1, 0, 0, 1 ], [ 0, 1, 0, 0 ], [ 0, 0, 1, 0 ], [ 0, 0, 0, 1 ] ], 
      [ [ 1, 0, 0, 0 ], [ 0, 1, 0, 1 ], [ 0, 0, 1, 0 ], [ 0, 0, 0, 1 ] ], 
      [ [ 1, 0, 0, 0 ], [ 0, 1, 0, 0 ], [ 0, 0, 1, 1 ], [ 0, 0, 0, 1 ] ] ], 
  [ [ [ -1, 0, 0, 0 ], [ 0, 0, -1, 0 ], [ 1, 1, 1, 0 ], [ 0, 0, 0, 1 ] ], 
      [ [ 0, 1, 0, 1/2 ], [ 0, 0, 1, 0 ], [ -1, -1, -1, 0 ], [ 0, 0, 0, 1 ] ], 
      [ [ 1, 0, 0, 1 ], [ 0, 1, 0, 0 ], [ 0, 0, 1, 0 ], [ 0, 0, 0, 1 ] ], 
      [ [ 1, 0, 0, 0 ], [ 0, 1, 0, 1 ], [ 0, 0, 1, 0 ], [ 0, 0, 0, 1 ] ], 
      [ [ 1, 0, 0, 0 ], [ 0, 1, 0, 0 ], [ 0, 0, 1, 1 ], [ 0, 0, 0, 1 ] ] ], 
  [ [ [ -1, 0, 0, 0 ], [ 0, 0, -1, 0 ], [ 1, 1, 1, 0 ], [ 0, 0, 0, 1 ] ], 
      [ [ 0, 1, 0, 1/2 ], [ 0, 0, 1, 1/2 ], [ -1, -1, -1, 1/2 ], [ 0, 0, 0, 1 ] ], 
      [ [ 1, 0, 0, 1 ], [ 0, 1, 0, 0 ], [ 0, 0, 1, 0 ], [ 0, 0, 0, 1 ] ], 
      [ [ 1, 0, 0, 0 ], [ 0, 1, 0, 1 ], [ 0, 0, 1, 0 ], [ 0, 0, 0, 1 ] ], 
      [ [ 1, 0, 0, 0 ], [ 0, 1, 0, 0 ], [ 0, 0, 1, 1 ], [ 0, 0, 0, 1 ] ] ] ]
gap> Display(List(SS,TranslationBasis));
[ [ [ 1, 0, 0 ], [ 0, 1, 0 ], [ 0, 0, 1 ] ], 
  [ [ 1, 0, 0 ], [ 0, 1, 0 ], [ 0, 0, 1 ] ], 
  [ [ 1, 0, 0 ], [ 0, 1, 0 ], [ 0, 0, 1 ] ], 
  [ [ 1, 0, 0 ], [ 0, 1, 0 ], [ 0, 0, 1 ] ] ]
\end{verbatim}
There are four space groups with the point group Q. The last three of the GAP result is the same fractional translation in Table \ref{tab:dia_frac_trans}. However, their translational vectors are different: the latter has larger translational vectors. This is because the GAP computation implicitly includes the primitive translations of the unit cube in generators of the space groups.

Readers may question about the relationship between the fractional translation computed now and those of the definition of the space group given in the first part of this section. The generators defined earlier are rewritten in the lattice axes coordinate as:   
\begin{verbatim}
gap> MM1:=[[-1,0,0,1/4],[0,0,-1,1/4],[1,1,1,1/4],[0,0,0,1]];
gap> MM2:=[[0,1,0,0],[0,0,1,0],[-1,-1,-1,0],[0,0,0,1]];
gap> S2:=AffineCrystGroupOnLeft([MM1,MM2]);
\end{verbatim}
We can apply the shift of the origin to this space group by means of the conjugation ${\rm\{E|-s\}\{R|t\}\{E|s\}=\{R|t+(R-I)\cdot s\}}$. If the matrix $\rm R-I$ is normal, we can annihilate the fractional translation. Let us apply such a conjugation to the generators. (The command M\verb!^!T is the abbreviation of the conjugation ${\rm T^{-1}\cdot M\cdot T}$.)
\begin{verbatim}
gap> T:=[[1,0,0,1/8],[0,1,0,-3/8],[0,0,1,5/8],[0,0,0,1]];
gap> MM1^T;
[ [ -1, 0, 0, 0 ], [ 0, 0, -1, 0 ], [ 1, 1, 1, 0 ], [ 0, 0, 0, 1 ] ]
gap> MM2^T;
[ [ 0, 1, 0, -1/2 ], [ 0, 0, 1, 1 ], [ -1, -1, -1, -1 ], [ 0, 0, 0, 1 ] ]
\end{verbatim}
Thus we can see that the shift of the origin generates the space group, which is equivalent to that composed by matrices N1 and N2 with [a,b,c]=[ -1/2, 1, -1 ]$\cong$[1/2,0,0]. Incidentally, let us generate other representation of generators:
\begin{verbatim}
N1:=[[-1,0,0,a],[0,0,-1,b],[1,1,1,c],[0,0,0,1]];
N2:=[[0,1,0,0],[0,0,1,0],[-1,-1,-1,0],[0,0,0,1]];
\end{verbatim}
The linear equations for the fractional translation are given as
\begin{verbatim}
[ 0, 4*x_3, -2*x_2-2*x_3, -2*x_2+2*x_3, -x_2+x_3, x_2-x_3, 
2*x_2+2*x_3, -4*x_1-4*x_2-4*x_3, -2*x_1-2*x_3, -2*x_1+2*x_3, 
  -2*x_1-4*x_2-2*x_3, -2*x_1-2*x_2, -2*x_1-2*x_2-4*x_3,
   -2*x_1-x_2-x_3, -2*x_1+2*x_2, 2*x_1-2*x_3, 2*x_1+2*x_3, 
  2*x_1-2*x_2, 2*x_1+x_2+x_3, 2*x_1+2*x_2+4*x_3, 
  2*x_1+4*x_2+2*x_3, 4*x_1+4*x_2+4*x_3 ]
\end{verbatim}
From them, seven possible vectors of [a,b,c] (=\verb![x_1,x_2,x_3]!) are obtained as
\begin{verbatim}
[ 0, 1/2, 1/2 ], [ 1/4, 1/4, 1/4 ], [ 1/4, 3/4, 3/4 ], 
[ 1/2, 0, 0 ], [ 1/2, 1/2, 1/2 ], [ 3/4, 1/4, 1/4 ], 
[ 3/4, 3/4, 3/4 ] ]
\end{verbatim}
We can see one of the fractional translations is [1/4,1/4,1/4]. 

Whereas the GAP computation goes as
\begin{verbatim}
gap> Q2:=Group(Q.1,Q.2);
Group([ [ [ -1, 0, 0 ], [ 0, 0, -1 ], [ 1, 1, 1 ] ], 
  [ [ 0, 1, 0 ], [ 0, 0, 1 ], [ -1, -1, -1 ] ] ])
gap> Q2=Q;
true
gap> SS2:=SpaceGroupsByPointGroupOnLeft(Q2);
[ <matrix group with 5 generators>, <matrix group with 5 generators>
    , <matrix group with 5 generators>, 
  <matrix group with 5 generators> ]
gap> List(SS2,GeneratorsOfGroup);
[ [ [ [ 0, 1, 0, 0 ], [ 0, 0, 1, 0 ], [ -1, -1, -1, 0 ], [ 0, 0, 0, 1 ] ], 
      [ [ -1, 0, 0, 0 ], [ 0, 0, -1, 0 ], [ 1, 1, 1, 0 ], [ 0, 0, 0, 1 ] ], 
      [ [ 1, 0, 0, 1 ], [ 0, 1, 0, 0 ], [ 0, 0, 1, 0 ], [ 0, 0, 0, 1 ] ], 
      [ [ 1, 0, 0, 0 ], [ 0, 1, 0, 1 ], [ 0, 0, 1, 0 ], [ 0, 0, 0, 1 ] ], 
      [ [ 1, 0, 0, 0 ], [ 0, 1, 0, 0 ], [ 0, 0, 1, 1 ], [ 0, 0, 0, 1 ] ] ], 
  [ [ [ 0, 1, 0, 0 ], [ 0, 0, 1, 0 ], [ -1, -1, -1, 0 ], [ 0, 0, 0, 1 ] ], 
      [ [ -1, 0, 0, 0 ], [ 0, 0, -1, 1/2 ], [ 1, 1, 1, 1/2 ], [ 0, 0, 0, 1 ] ], 
      [ [ 1, 0, 0, 1 ], [ 0, 1, 0, 0 ], [ 0, 0, 1, 0 ], [ 0, 0, 0, 1 ] ], 
      [ [ 1, 0, 0, 0 ], [ 0, 1, 0, 1 ], [ 0, 0, 1, 0 ], [ 0, 0, 0, 1 ] ], 
      [ [ 1, 0, 0, 0 ], [ 0, 1, 0, 0 ], [ 0, 0, 1, 1 ], [ 0, 0, 0, 1 ] ] ], 
  [ [ [ 0, 1, 0, 0 ], [ 0, 0, 1, 0 ], [ -1, -1, -1, 0 ], [ 0, 0, 0, 1 ] ], 
      [ [ -1, 0, 0, 1/4 ], [ 0, 0, -1, 1/4 ], [ 1, 1, 1, 1/4 ], [ 0, 0, 0, 1 ] ], 
      [ [ 1, 0, 0, 1 ], [ 0, 1, 0, 0 ], [ 0, 0, 1, 0 ], [ 0, 0, 0, 1 ] ], 
      [ [ 1, 0, 0, 0 ], [ 0, 1, 0, 1 ], [ 0, 0, 1, 0 ], [ 0, 0, 0, 1 ] ], 
      [ [ 1, 0, 0, 0 ], [ 0, 1, 0, 0 ], [ 0, 0, 1, 1 ], [ 0, 0, 0, 1 ] ] ], 
  [ [ [ 0, 1, 0, 0 ], [ 0, 0, 1, 0 ], [ -1, -1, -1, 0 ], [ 0, 0, 0, 1 ] ], 
      [ [ -1, 0, 0, 1/4 ], [ 0, 0, -1, 3/4 ], [ 1, 1, 1, 3/4 ], [ 0, 0, 0, 1 ] ], 
      [ [ 1, 0, 0, 1 ], [ 0, 1, 0, 0 ], [ 0, 0, 1, 0 ], [ 0, 0, 0, 1 ] ], 
      [ [ 1, 0, 0, 0 ], [ 0, 1, 0, 1 ], [ 0, 0, 1, 0 ], [ 0, 0, 0, 1 ] ], 
      [ [ 1, 0, 0, 0 ], [ 0, 1, 0, 0 ], [ 0, 0, 1, 1 ], [ 0, 0, 0, 1 ] ] ] ]
\end{verbatim}
Four space groups are generated by GAP: non-zero fractional translations are three in number: 
\begin{verbatim}
[ 0, 1/2, 1/2 ], [ 1/4, 1/4, 1/4 ], [ 1/4, 3/4, 3/4 ]
\end{verbatim}
Apparently the GAP result is lacking in some of fractional translations computed just before by means of linear equations. The GAP result should rather be interpreted as the basis vectors for the possible fractional translations. The other possible fractional translations can be generated from this basis set.

It is well to remind readers of the point that the setting up of the admissible fractional translations (which is called ``vector system'' in the mathematical context) is nothing other than the computation of the one-cohomology. In the above example, the fractional translation in the generator N1 is chosen to be zero from the outset. It is possible to assume the non-zero translation in N1. However, a shift of the origin can cancel a nonzero fractional translation as we have seen. The shift of the origin could be caused by the conjugation, which corresponds to the choice of the representative of the elements in the map of the one-cohomology.

\subsubsection{Group extension}
The space group of the crystal is determined by the following short exact sequence (or the group extension):
\begin{equation}
1\rightarrow K \rightarrow E \xrightarrow{\pi} G \rightarrow 1
\end{equation}
with the condition that there exists a normal subgroup of $N$ in $E$ which is isomorphic to $K$ and that $E/K$ is isomorphic to $G$. The group $E$ is called an extension of $G$ by $K$. The equivalence classes of extension of $E$ are computed by a concept of the cohomological algebra, i.e. two-cohomology (denoted as $H^2(G, K)=Z^2(G, K)/B^2(G, K)$). In order to understand this concept, however, one should get acquainted with the language of commutative algebra, especially concerning homological algebra, which would necessitate long explanation. Thus if the reader would get interested in these topics, he should try advanced mathematical textbooks. However, the viewpoint of cohomological algebra might be useful in material designing, especially in studying the algebraic aspects of crystallography. For example, the Bravais lattice in two- or three-dimensional crystals can be classified by two-cohomology.

\begin{remark}
The construction of the two-cohomology is briefly stated here. For every $g \in G$ there are elements $e_g \in E$ with $\pi(e_g)=g$. However $e_g$ is not unique: another element $e'_g$ behaves similarly, if $e'_g(e_g)^{-1}\in\ker(\pi)=K$. Now we define a map $\varphi_g : k\rightarrow e_g k e_g^{-1}$, as an automorphism in $K$. The two-cocycle is defined by 
$$
f(g,h)=e_ge_he_{gh}^{-1}.
$$
For each g, let us give an element $k_g\in K$ (as a homomorphism $k(*)$ from $G$ to $K$). Then the two-coboundary is defined as 
$$
b(g,h)=k_g\varphi_g(k_h)k_{gh}^{-1}.
$$
(It will be represented as $k(g) + g\cdot k(h)- k(gh)$ when $K$ is a module.) 

Then we have
$$
f(g,h)f(gh,k)=\varphi_g(f(h,k))f(g,hk)
$$
and for another choice ($e'_g=k_ge_g$),
$$
f'(g,h)=e'_ge'_h(e'_{gh})^{-1}=b(h,g)f(g,h).
$$
In two-cohomology $f(*,*)$ and $f'(*,*)$ represent the same class.
The elements of $E$ is uniquely written in the form $ae_g$ for some $a \in K$ and $g \in G$. The product is given  as $(ae_g)(be_h)=a\varphi_g(b)f(g,h)e_{gh}$, which is the usual semidrect product when $f(g,h)=1$ for all $g,h$.
\end{remark}

For simplicity, consider the two-dimensional case. Let $P=\left\langle x,y | x^2=y^2=(xy)^2=1\right\rangle$, which acts on $T=\mathbb{Z}^2$ through 
$M_x=\left(\begin{array}{cc}
-1 & 0 \\ 0 & 1
\end{array} \right)$
and
$M_y=\left(\begin{array}{cc}
1 & 0 \\ 0 & -1
\end{array} \right)$.
The action on T is written as
$\{ ^xe_1=e_1^{-1},
\ ^xe_2=e_2,
\ ^ye_1=e_1,
\ ^ye_2=e_2^{-1} \}$.
($e_1=\left(\begin{array}{c}
1  \\ 0
\end{array} \right)$
and
$e_2=\left(\begin{array}{c}
 0 \\ 1
\end{array} \right)$.

Let us extend this group, and, if possible, let us construct space groups which have this group as their point groups. (This example is taken from the lecture notes of Webb\cite{COHOMGROUP}. For theoretical details, consult with that article and references therein.)

There is a general procedure (Zassenhauss algorithm) to compute two-cohomology, which is based on the following theorem.

\begin{theorem}
Let P be a finite group given by the representation

\begin{equation*}
P=\left\langle x_1,x_2,...,x_d | r_1,r_2,...,r_t\right\rangle.
\end{equation*}

Let n be the dimension of the space, and $T$ be a module. From the relators $r_i$, a matrix $\Lambda \in M_{nt,nd}(\mathbb{Z})$ is computed, and the elements $\Lambda_{ij}$ are defined as
\begin{equation*}
r_i-1=\sum_{j=1}^d\Lambda_{ij}(x_j-1) \ (i=1,...,t).
\end{equation*}

Then the two-cohomology is given by
\begin{equation}
H^2(P,T)\cong\{x \in (\mathbb{R}T)^d|\Lambda \cdot x\in T^t\}/({\rm Ker\ }\Lambda+T^d).
\end{equation}
This definition says that the representative elements in two-cohomology are given as $(v_1,\cdots,v_d)$ , the concatenation of the vectors in $\mathbb{R}T^n$, (from which the component in ${\rm Ker}\ \Lambda + T^d$ should be factored out),
so as to be
\begin{equation}
\Lambda \left(\begin{array}{c}
v_1 \\
\vdots \\ 
v_d  
\end{array} \right) \in T^t.
\end{equation}

The integer matrix $\Lambda$ can be converted into its Smith normal form
\begin{equation}
\left(\begin{array}{ccccccc}
b_1 &  &  &  &  &  &  \\ 
 & b_2 &  &  &  &  &  \\ 
 &  & \ddots &  &  &  &  \\ 
 &  &  & b_l &  &  &  \\ 
 &  &  &  & 0 &  &  \\ 
 &  &  &  &  & \ddots &  \\ 
 &  &  &  &  &  & 0
\end{array}  
\right)
\end{equation}
with $b_i\neq0 $, dividing $b_{j}$ for $i<j$. 
(A matrix $A \in M_{m,n} (\mathbb{Z})$ can be converted into the diagonal form as this, by means matrix operation $P\cdot A\cdot Q$ with $P \in GL(m,\mathbb{Z})$ and $Q \in GL(n,\mathbb{Z})$.) 
 
Then $H^2(P,T)\cong \mathbb{Z}/b_1\mathbb{Z}\oplus\cdots\oplus\mathbb{Z}/b_l\mathbb{Z}$. Now the relators in $P$ are lifted in the extension of $P$ by $T$ as
\begin{equation}
\left(\begin{array}{c}
r_1 \\
r_2 \\
\vdots \\ 
r_t  
\end{array} \right) = \Lambda \cdot v \ (\forall v \in H^2(P,T)).
\end{equation}
The vector systems are given as $\{x_i|v_i\}\ (i=1,...,d)$ in the crystallography.
\end{theorem}

\begin{remark}
This algorithm is a refined style of the computation of one-cohomology given in the previous section. The relation such as $xyxy=1 {\rm (\ in\ G)}$ is lifted to $e_{xyxy}=1_E$, as the relation in $E$. The latter is rewritten by two-cocycles
$$
1_E=e_{xyxy}=f(x,yxy)^{-1}\varphi_x(f(y,xy)^{-1})\varphi_x(\varphi_y(f(x,y)^{-1}))e_xe_ye_xe_y
$$
It is furthermore rewritten as
$$
e_xe_ye_xe_y=\varphi_x(\varphi_y(f(x,y)))\varphi_x(f(y,xy))f(x,yxy).
$$
The right side of this equation lies in $K$. The formula of $\hat{\Lambda}_{(xy)^2}$ in the algorithm represents the translational part in $e_xe_ye_xe_y$ by means of those of $e_x$ and $e_y$, and it must be in $K(=T)$. To find a $v$ such as $\Lambda\cdot v \in T$ is to fix a representative of cocycle $f(*,*)$ in two-cohomology.
\end{remark}

In the aforesaid question, the matrix $\Lambda$ is computed as
\begin{equation}
\Lambda=
\left( \begin{array}{ll}
\Lambda_{x^2,x} & \Lambda_{x^2,y} \\ 
\Lambda_{y^2,x} & \Lambda_{y^2,y} \\ 
\Lambda_{(xy)^2,x} & \Lambda_{(xy)^2,y}  
\end{array} \right)
=
\left( \begin{array}{cc}
x+1 & 0 \\ 
0 & y+1 \\ 
xy+1 & xyx+x
\end{array} \right)
=\left(
\begin{array}{cccc}
0 & 0 & 0 & 0 \\ 
0 & 2 & 0 & 0 \\ 
0 & 0 & 2 & 0 \\ 
0 & 0 & 0 & 0 \\ 
0 & 0 & 0 & 0 \\ 
0 & 0 & 0 & 0
\end{array} 
\right)
\end{equation}

From this, it can be concluded that $H^2(P,T)=\mathbb{Z}/2\mathbb{Z}\oplus\mathbb{Z}/2\mathbb{Z}$. The homomorphism representing the two-cohomology, which determines the extension of the group, is generated from these vectors: 

\begin{equation}
\left\{ \left(\begin{array}{c}
\frac{1}{2} e_2\\ 
0 
\end{array} \right),
\left(\begin{array}{c}
0 \\ 
\frac{1}{2}e_1 
\end{array} \right) \right\}
=
\left\{ \left(\begin{array}{c}
0 \\ 1/2 \\0  \\ 0 
\end{array} \right),
\left(\begin{array}{c}
0 \\ 0 \\ 1/2 \\0
\end{array} \right) \right\}
\end{equation}

The linear combination of them (with integer coefficients, of 0 or 1 ) admits four possible mappings. The relators in the extension E are obtained from the entries in $\Lambda (c_1\cdot e_1+c_2\cdot e_2)
=\left(\begin{array}{c}\hat{\Lambda}_{x^2}\\\hat{\Lambda}_{y^2}\\\hat{\Lambda}_{(xy)^2}\end{array}\right)$. They are listed as:
\begin{table}[h!]
\centering
\begin{tabular}{c|c|c|c|c}
\hline  & type 1 & type 2 & type 3 & type 4 \\ 
\hline 
$\hat{\Lambda}_{x^2}$    & $0\cdot e_1+ 0\cdot e_2$ & $0\cdot e_1+ 1\cdot e_2$ & $0\cdot e_1+ 0\cdot e_2$ & $0\cdot e_1+ 1\cdot e_2$ \\ 
$\hat{\Lambda}_{y^2}$    & $0\cdot e_1+ 0\cdot e_2$ & $0\cdot e_1+ 0\cdot e_2$ & $1\cdot e_1+ 0\cdot e_2$ & $1\cdot e_1+ 1\cdot e_2$ \\ 
$\hat{\Lambda}_{(xy)^2}$    & $0\cdot e_1+ 0\cdot e_2$ & $0\cdot e_1+ 0\cdot e_2$ & $0\cdot e_1+ 0\cdot e_2$ & $0\cdot e_1+ 0\cdot e_2$ \\ 
\hline 
\end{tabular} 
\caption{Four possible types of relators in the extensions.}
\end{table}
They generate four different extensions in table \ref{tab:fourextensionsin2D}. The extensions of type 2 and type 3 are isomorphic, since they are converted with each other by interchanging $x$ and $y$, and $e_1$ and $e_2$. 

\begin{table}[h!]
\centering
\begin{tabular}{c|c}
\hline
type 1 &
$\left\langle x,y,e_1,e_2|x^2=y^2=(xy)^2=[e_1,e_2]=1,
\ ^xe_1=e_1^{-1},
\ ^xe_2=e_2,
\ ^ye_1=e_1,
\ ^ye_2=e_2^{-1}\right\rangle$\\
type 2 &
$\left\langle x,y,e_1,e_2|x^2=e_2,y^2=(xy)^2=[e_1,e_2]=1,
\ ^xe_1=e_1^{-1},
\ ^xe_2=e_2,
\ ^ye_1=e_1,
\ ^ye_2=e_2^{-1}\right\rangle$
\\
type 3 &
$\left\langle x,y,e_1,e_2|y^2=e_1,x^2=(xy)^2=[e_1,e_2]=1,
\ ^xe_1=e_1^{-1},
\ ^xe_2=e_2,
\ ^ye_1=e_1,
\ ^ye_2=e_2^{-1}\right\rangle$
\\
type 4 &
$\left\langle x,y,e_1,e_2|x^2=e_2,y^2=e_1,(xy)^2=[e_1,e_2]=1,
\ ^xe_1=e_1^{-1},
\ ^xe_2=e_2,
\ ^ye_1=e_1,
\ ^ye_2=e_2^{-1}\right\rangle$
\\\hline
\end{tabular} 
\caption{Four types of the extension. $e_1$ and $e_2$ should be assumed to be translations $\{E|e_1\}$ and $\{E|e_2\}$, rather than vectors.}
\label{tab:fourextensionsin2D}
\end{table}

We can see the relators of the extensions provide the relations which determine the vector system in the non-symmorphic crystal, with the point group $P$.(To obtain the vector system, it is enough to re-express $x$,$y$ as the augmented matrix forms of the space groups, with indeterminate fractional translations. The $e_1$ and $e_2$ are also represented by matrices as translations. The equivalence relations in the entries of the matrix products in the group multiplications will give the vector systems, as was demonstrated elsewhere in this article. In that demonstration, the vector system was also derived from cohomological algebra, not from two-cohomology, but from one-cohomology.) From this simple example, we can observe some importance of homological algebra, which provides us the criterion of the existence of the non-symmorphic crystal, in accordance with the non-triviality of two-cohomology. On the other hand, if the two-cohomology is trivial, or of zero-mapping, the extension of the group simply becomes a semidirect product, which is always a symmorphic crystal. 
In this case of zero two-cohomology, the exact sequence splits, in the sense that there is a homomorphism $s: G\rightarrow H$ such as $\pi\circ s = {\rm id}_G$. It can be proved that an extension (or, the short exact sequence as above) is split if and only if the group H is the semidirect product of N and G. 
(We should remember this: elsewhere in this article, the point group of the non-symmorphic crystal is constructed by means of the semidirect product. In that computation, the point group of the non-symmorphic crystal is not directly obtained as a semidirect product but extracted as one of its subgroups. So there is no contradiction against the above statement.) 

Now the relationship is established among the symmorphic properties of crystals, the splitting extension of a group, and the two-cohomology, as is shown in the table below.

\begin{table}[h!]
\centering
\begin{tabular}{c|c}
\hline
type 1 & $\{M_x|0\},\{M_y|0\},\{E|e_1\},\{E|e_2\}$
\\
type 2 & $\{M_x|\frac{1}{2}e_2\},\{M_y|0\},\{E|e_1\},\{E|e_2\}$
\\
type 3 & $\{M_x|0\},\{M_y|\frac{1}{2}e_1\},\{E|e_1\},\{E|e_2\}$
\\
type 4 & $\{M_x|\frac{1}{2}e_2\},\{M_y|\frac{1}{2}e_1\},\{E|e_1\},\{E|e_2\}$
\\\hline
\end{tabular} 
\caption{The generators of the four types of the space groups classified by the two-cohomology.} 
\end{table}

In getting through this section, two necessary points for the computations in GAP are given. The first is how to derive the relators, or the finite representation: the second is the computation of the Smith Normal form. (The two-cohomology can be computed by GAP built-in command, for the case of a finite group $G$ and a module $M$, over \emph{a finite field}. Because of this assumption, the GAP built-in command is not applicable to the above example, in which the module $M$ is $\mathbb{Z}^n$.)

A definition is now needed: a group $G$ is polycyclic if there is a series of  subgroups
\[
 G=C_1>C_2>\dots>C_n>C_{n+1}=1,
\]
where $C_{i+1}$ is normal in $C_{i}$ and $C_{i}/C_{i+1}$ is a cyclic group. The generators of $G$ is given by a sequence $G:=(g_1,\dots,g_n)$ such that $C_{i}=\left\langle C_{i+1},g_i\right\rangle$. As a finite solvable group, the point group of the crystal is polycyclic, which is composed from the polycyclic generating system.(The point group of the crystal is always solvable, by means of a famous theorem of Burnside, which relates the order of the group to the solvability.) We have defined the point groups by means of matrix generators, or the multiplication table. These groups can be converted into isomorphic polycyclic groups, and then, polycyclic groups are converted into finitely presented groups with relators. The conversion goes as follows:

\begin{verbatim}
gap> A:=[ [ 0, 0, 1 ], [ 1, 0, 0 ], [ 0, -1, 0 ] ];
gap> B:=[ [ 0, 0, -1 ], [ 0, -1, 0 ], [ 1, 0, 0 ] ];
gap> G:=Group(A,B);
Group([ [ [ 0, 0, 1 ], [ 1, 0, 0 ], [ 0, -1, 0 ] ], 
  [ [ 0, 0, -1 ], [ 0, -1, 0 ], [ 1, 0, 0 ] ] ])
gap> p:=Pcgs(G);
Pcgs([ [ [ 0, 0, 1 ], [ 0, 1, 0 ], [ -1, 0, 0 ] ], 
  [ [ 1, 0, 0 ], [ 0, 1, 0 ], [ 0, 0, -1 ] ], 
  [ [ 0, 0, -1 ], [ -1, 0, 0 ], [ 0, 1, 0 ] ], 
  [ [ 1, 0, 0 ], [ 0, -1, 0 ], [ 0, 0, -1 ] ], 
  [ [ -1, 0, 0 ], [ 0, -1, 0 ], [ 0, 0, 1 ] ] ])
gap> iso:=IsomorphismFpGroupByPcgs(p,"g");
CompositionMapping( [ (1,4,6,8)(2,3,5,7), (1,4)(2,3)(5,7)(6,8), 
  (3,6,7)(4,5,8), (1,7)(2,8)(3,6)(4,5), (1,3)(2,4)(5,8)(6,7) ] -> 
[ g1, g2, g3, g4, g5 ], <action isomorphism> )
gap> fp:=Image(iso);
<fp group of size 48 on the generators [ g1, g2, g3, g4, g5 ]>
gap> RelatorsOfFpGroup(fp);
[ g1^2*g5^-1*g4^-1, g2^-1*g1^-1*g2*g1*g5^-1*g4^-1, 
  g3^-1*g1^-1*g3*g1*g5^-1*g4^-1*g3^-1, g4^-1*g1^-1*g4*g1*g5^-1*g4^-1, 
  g5^-1*g1^-1*g5*g1*g5^-1*g4^-1, g2^2, g3^-1*g2^-1*g3*g2*g4^-1, 
  g4^-1*g2^-1*g4*g2, g5^-1*g2^-1*g5*g2, g3^3, g4^-1*g3^-1*g4*g3*g5^-1*g4^-1, 
  g5^-1*g3^-1*g5*g3*g4^-1, g4^2, g5^-1*g4^-1*g5*g4, g5^2 ]
\end{verbatim}

The command ``Pcgs(G)'' returns the polycyclic generating system of $G$. (In this example, the generators are accessible through the list \verb!p[1],...p[5]!.) The command ``IsomorphismOfFpGroupByPcgs(p,"g")'' gives the map which converts the polycyclic generating system into a finitely represented group. As the image of this isomorphism, the polycyclic group is given. However, the polycyclic generating system is not minimal. To generate the minimal one, we should do as this:

\begin{verbatim}
gap> iso:=IsomorphismFpGroupByGenerators(G,GeneratorsOfGroup(G));
<composed isomorphism:[ [ [ 0, 0, 1 ], [ 1, 0, 0 ], [ 0, -1, 0 ] ], 
[ [ 0, 0, -1 ], [ 0, -1, 0 ], [ 1, 0, 0 ] ] ]->[ F1, F2 ]>
gap> fp:=Image(iso);
<fp group of size 48 on the generators [ F1, F2 ]>
gap> RelatorsOfFpGroup(fp);
[ F2^4, (F1*F2^-1)^2, F1^6, (F1*F2*F1)^2 ]
gap> [G.2^4,(G.1*G.2^-1)^2,G.1^6,(G.1*G.2*G.1)^2];
[ [ [ 1, 0, 0 ], [ 0, 1, 0 ], [ 0, 0, 1 ] ], 
  [ [ 1, 0, 0 ], [ 0, 1, 0 ], [ 0, 0, 1 ] ], 
  [ [ 1, 0, 0 ], [ 0, 1, 0 ], [ 0, 0, 1 ] ], 
  [ [ 1, 0, 0 ], [ 0, 1, 0 ], [ 0, 0, 1 ] ] ]
\end{verbatim}



 

The smith normal form is also computed by GAP:
\begin{verbatim}
gap> M:=[[0,0,0,0],[0,2,0,0],[0,0,0,0],[0,0,2,0],[0,0,0,0],[0,0,0,0]];
gap> SmithNormalFormIntegerMat(M);
[ [ 2, 0, 0, 0 ], [ 0, 2, 0, 0 ], [ 0, 0, 0, 0 ], [ 0, 0, 0, 0 ], 
  [ 0, 0, 0, 0 ], [ 0, 0, 0, 0 ] ]
\end{verbatim}

\subsection{Remarks to this section}
In this section, the technical details which would be important in the actual application of the computational group theory in the field of solid-state physics are explained. The chosen topics are as follows: the algorithm to determine the symmetry of the crystal; the computation of the compatibility relation, which is useful in the discussion of the splitting of the energy spectra with the reduction of the structural symmetry; the concept of the semidirect product, which is applicable to the preparation of the character table of the supercell structure;  the concept of the group extension, by means of cohomology,  which can be utilized to design the crystal lattice. The author hopes that the readers of this article would get familiar with these technical or advanced topics in order to deepen the knowledge of group theory, applicable to concrete problems in material science. The small programs presented in this article can be executed on a desktop PC. They are composed in accordance with the elementary formulas and definitions in the references and textbooks so that they should be as readable as possible, thus they are not are always effective in speed, and some part of them could be replaced by built-in functions in GAP. In the preceding sections of this article, the topics are taken from the solid-state physics. In the forthcoming sections, however, the application to the quantum chemistry (applicable to the isolated system, equipped with a certain symmetry, without periodicity) shall be discussed in a parallel way as the solid state physics, where the importance of the symmetry in the quantum dynamics of the molecule shall be illustrated in the different light.

﻿
\section{Symmetry in C$_{60}$}
The C$_{60}$ molecule is characterized by the peculiar structure, that of soccer-ball, composed of 60 atoms, 90 bonds, and 12 pentagons, and 60 hexagons. The group theoretical symmetry is of the icosahedral type, which contains 120 symmetric operations and 10 irreducible representations, according to which,  the eigenstates are classified and the origins of degeneracies in energy spectrum are explained. 
We begin the construction of the symmetric operations by means of permutations of atoms, generate the icosahedral group by GAP, compute the irreducible representations, and compose the projection operators. We execute the electronic structure calculations of the molecule and analyze the wave-functions from the viewpoint of their irreducible representation. 
Furthermore, some concepts of symmetries (to which physicists pays little notice) are presented also by the analysis of the computer algebra. Throughout the study, it is necessary that whole subgroups of icosahedral symmetry should be extracted and classified by equivalence, and we will see that the utilization of the computer algebra is effective for this purpose.  
 
\subsection{The initial set-up}

\begin{figure}
\centering
\includegraphics[width=0.5\linewidth]{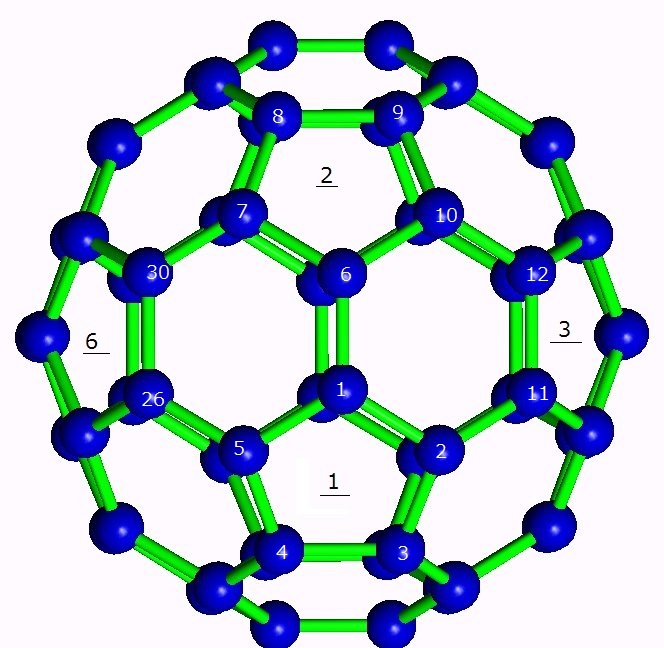}
\caption{A C$_{60}$ molecule. The numbering is given on some atoms and pentagons, in correspondence with the projected images of the molecule, in Fig. \ref{fig:C60VERTEX}.}
\label{fig:C60MOL}

\end{figure}

\begin{figure}
\centering
\includegraphics[width=0.65\linewidth]{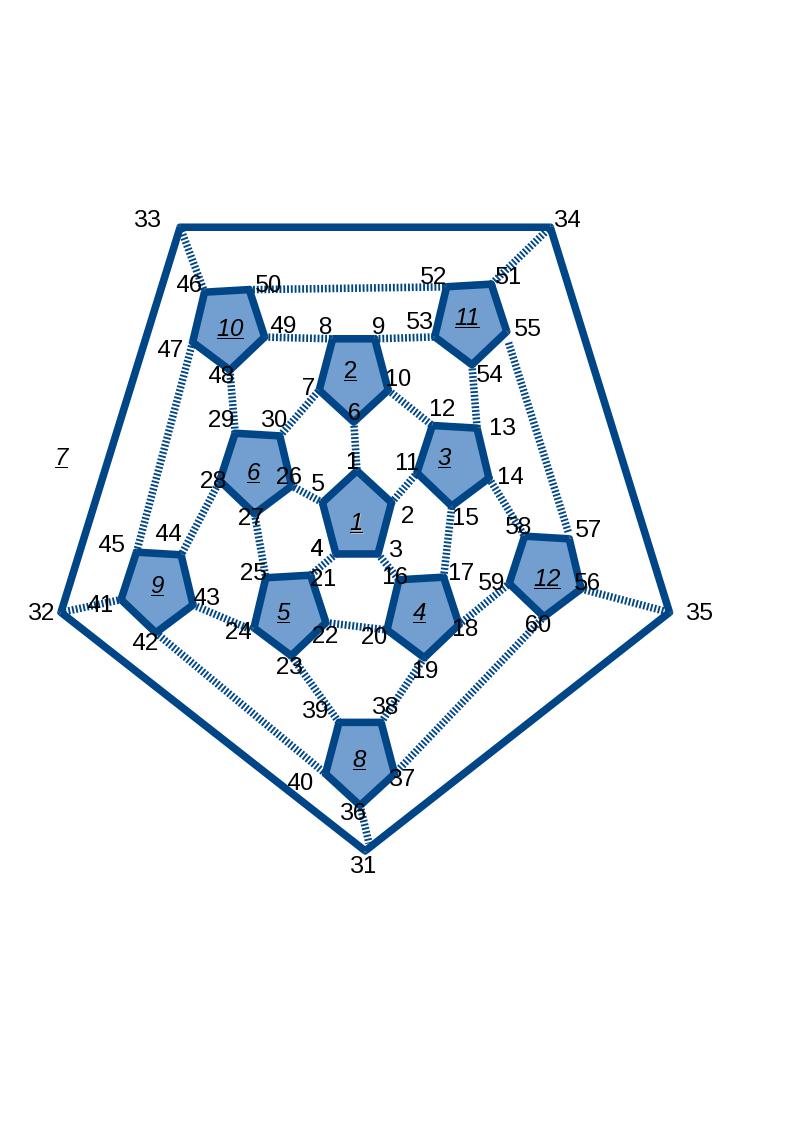}
\caption{The projected image of C$_{60}$ on a plane. The pentagon in the center of this figure (indexed as 1) is the one located in the lower half in Fig. \ref{fig:C60MOL}. (The pentagon at No.7, not shown in the projected image, is formed by five vertex at 31,32,33,34,35. )}
\label{fig:C60VERTEX}
\end{figure}

The C$_{60}$ molecule (in Fig. \ref{fig:C60MOL}) is projected on a plane as in the Fig. \ref{fig:C60VERTEX}. The vertexes are numbered from 1 to 60, where we follow the following rule: the two atoms numbered as I and I+30 (I=1,...,30) are located in the opposite poles in the molecular sphere. The pentagons are also numbered from 1 to 12 (with under-bars), and two pentagons numbered as J and J+6 (J=1,...,6) are opposed with each other.  There are two types of bonds, single and double, having slightly different lengths, distinguished by broad and broken segments in the figure. The icosahedral group are generated by three basic operations and represented by the permutations of atoms as follows. 
The group is the direct product of the fifth alternating group ``A5'' (the 5-th alternating group) and the group of the inversion ``I''. The generators are represented by permutations: 

\begin{verbatim}
gap> g1:=(1, 14, 20)(2, 15, 16)(3, 11, 17)(4, 12, 18)(5, 13, 19)
(6, 58, 22)(7, 57, 23)( 8, 56, 24)( 9, 60, 25)( 10, 59, 21)
( 31, 44, 50)( 32, 45, 46)( 33, 41, 47)( 34, 42, 48)( 35, 43, 49)
( 36, 28, 52)( 37, 27, 53)( 38, 26, 54)( 39, 30, 55)( 40, 29, 51);
gap> g2:=( 1, 5, 4, 3, 2)( 6,26,21,16,11)( 7,27,22,17,12)( 8,28,23,18,13)
( 9,29,24,19,14)(10,30,25,20,15)(31,35,34,33,32)(36,56,51,46,41)
(37,57,52,47,42)(38,58,53,48,43)(39,59,54,49,44)(40,60,55,50,45);
gap> g3:=( 1, 31)( 2, 32)( 3, 33)( 4, 34)( 5, 35)( 6, 36)( 7, 37)( 8, 38)
( 9, 39)( 10, 40)( 11, 41)( 12, 42)( 13, 43)( 14, 44)( 15, 45)( 16, 46)
( 17, 47)( 18, 48)( 19, 49)( 20, 50)( 21, 51)( 22, 52)( 23, 53)( 24, 54)
( 25, 55)( 26, 56)( 27, 57)( 28, 58)( 29, 59)( 30, 60);
\end{verbatim}

The elements g1 (the rotation by 120 degrees) and g2 (that by 72 degrees) generate ``A5`` and g3 generates ``I''. The permutation, denoted by (a, b, c, d,...,x), should be interpreted as the replacement of atoms in the following way: the atom located at the site ``a'' is replaced by that at the site ``b'', that at ``b'' by that at ``c'', then in turn and finally, the last entry ``x'' by the first ``a''. The usage of sixty symbols for the definition of the group is superfluous, and in fact, twelve symbols (representing the 12 vertexes of the icosahedron, or the 12 pentagons in the molecule) will do. If it is unnecessary to use the geometrical image, one can use 7 symbols, since, the icosahedral group is the direct product of the 5-th alternating group A5 and the inversion I, as the former requires five symbols ``1,2,3,4,5'' and the latter two symbols ``6,7''. The equivalence of these different definitions is assured by constructing isomorphisms among them, which is an easy task for the computer algebra.

\begin{remark}
This correspondence (between the group elements and the atoms) is consistent to the operation of the permutation in GAP. For example, the permutation (3,4,5) operates on the list [1,2,3,4,5] as follows:
\begin{verbatim}
 gap> ListPerm((3,4,5));
 [1,2,4,5,3];     
\end{verbatim}
Here the entries 3,4,5 are replaced by 4,5,3, as our interpretation assumes.
\end{remark}

The group is constructed as
\begin{verbatim}
gap> gp:=Group(g1,g2,g3);
\end{verbatim}

For the later utilization, it is more convenient to generate the group by means of matrices (which represents the permutations on sixty vertexes). We use the built-in function ``ListPerm()'' which computes the action on the permutation on the list [1,2,...,60].

\begin{verbatim}
gap> pl1:=ListPerm(g1);
[ 14, 15, 11, 12, 13, 58, 57, 56, 60, 59, 17, 18, 19, 20, 16, 2, 3, 4, 5, 1, 
  10, 6, 7, 8, 9, 54, 53, 52, 51, 55, 44, 45, 41, 42, 43, 28, 27, 26, 30, 29, 
  47, 48, 49, 50, 46, 32, 33, 34, 35, 31, 40, 36, 37, 38, 39, 24, 23, 22, 21, 
  25 ]
gap> pl2:=ListPerm(g2);;
gap> pl3:=ListPerm(g3);;
\end{verbatim}

These three lists are converted into three matrices by means of this user-defined function.
(The computed matrix element h[i][j] is nonzero when the symbol ``i'' goes to ``j'' by permutation given in the list ``l''.)

\begin{verbatim}
ListToMat:=function(l)
 local size,h,i,j;
 size:=Size(l);
 h:=List([1..size],i->List([1..size],j->0));
 for i in [1..size] do
  h[i][l[i]]:=1;
 od;
 return h;
end;

gap> m1:=ListToMat(pl1);;
gap> m2:=ListToMat(pl2);;
gap> m3:=ListToMat(pl3);;
\end{verbatim}

The irreducible representations are computed now. 

\begin{verbatim}
gap> g:=Group(m1,m2,m3);
gap> irr:=Irr(g);
gap> Size(irr);
10
\end{verbatim}

\begin{remark}
The computed irreducible representations by the group generated by the permutations and that by the matrices are listed by different orders (at least, in the result by GAP of the version 4.7.7.) This will cause some trouble in doing computations. In this article, to avoid confusion, the groups generated by different resources are given by different names even if they are isomorphic, and the generators are presented in each time of the definition.
\end{remark}

The characters for each element can be listed as numeral tables.
\begin{verbatim}
gap> elm:=Elements(g);
gap> A1:=List(elm,g->g^irr[1]);;
gap> A2:=List(elm,g->g^irr[2]);;
...............................
gap> A10:=List(elm,g->g^irr[10]);;
\end{verbatim}
The notation \verb!g^irr[X]! means the computation of the character value of the element ``g'' by the X-th representation in this context.

The characters of the irreducible representations are orthogonal and normalized.
\begin{verbatim}
gap> A1*A1;
120
gap> A2*A2;
120
gap> A1*A2;
0
\end{verbatim}
The virtual character is decomposed into the direct sum.
\begin{verbatim}
gap> (4*A1+2*A2)*TransposedMat([A1,A2,A3,A4,A5,A6,A7,A8,A9,A10])/Size(elm);
[4, 2, 0, 0, 0, 0, 0, 0, 0, 0]
\end{verbatim}
There are 10 conjugacy classes and 10 irreducible representations. They  are shown in the tables \ref{charactertbl} and \ref{representatives}. In table \ref{charactertbl}, the second row shows the number of elements in each conjugacy classes (CL1,...,CL10), and the ten rows bellow are characters for each of the ten irreducible representations (I.1,...,I.10). The last row (I.Tr) shows the trace of the group elements as matrices, which is a virtual character (not irreducible). The notation E(5) means a complex number, the fifth root of the unity $\exp(2\sqrt{-1}\pi/5)$, thus the characters are real-valued. Simply the computed result by GAP is tabulated here, as it is.
\begin{table}
\begin{tabular}{c|cccccccccc}\hline
&CL1&CL2&CL3&CL4&CL5&CL6&CL7&CL8&CL9&CL10\\\hline
&1&12&12&20&20&15&12&1&12&15\\\hline
I.1&1&1&1&1&1&1&1&1&1&1\\
I.2&1&-1&1&-1&1&-1&-1&-1&1&1\\
I.3&3&-E(5)-E(5)\verb!^!4&-E(5)\verb!^!2-E(5)\verb!^!3&0&0&-1&-E(5)\verb!^!2-E(5)\verb!^!3&3&-E(5)-E(5)\verb!^!4&-1\\
I.4&3&-E(5)\verb!^!2-E(5)\verb!^!3&-E(5)-E(5)\verb!^!4&0&0&-1&-E(5)-E(5)\verb!^!4&3&-E(5)\verb!^!2-E(5)\verb!^!3&-1\\
I.5&3&E(5)\verb!^!2+E(5)\verb!^!3&-E(5)-E(5)\verb!^!4&0&0&1&E(5)+E(5)\verb!^!4&-3&-E(5)\verb!^!2-E(5)\verb!^!3&-1\\
I.6&3&E(5)+E(5)\verb!^!4&-E(5)\verb!^!2-E(5)\verb!^!3&0&0&1&E(5)\verb!^!2+E(5)\verb!^!3&-3&-E(5)-E(5)\verb!^!4&-1\\
I.7&4&-1&-1&1&1&0&-1&4&-1&0\\
I.8&4&1&-1&-1&1&0&1&-4&-1&0\\
I.9&5&0&0&-1&-1&1&0&5&0&1\\
I.10&5&0&0&1&-1&-1&0&-5&0&1\\\hline
I.Tr&60&0&0&0&0&4&0&0&0&0\\\hline
\end{tabular}

\caption{The table of characters of the irreducible representation of the icosahedral group. Each row shows the characters of 10 conjugacy classes (CL1,...,CL10). }
\label{charactertbl}
\end{table}
\begin{table}
\centering
\begin{tabular}{c|c}\hline
      &  The representatives \\\hline
  CL1  & \verb!<identity ...>! \\
  CL2  & \verb!x1^-1*x2^-2*x3!\\
  CL3  & \verb!x2^-1*x1*x2*x1! \\
  CL4  & \verb!x3*x2*x1*x2^2!\\
  CL5  & \verb!x2^-2*x1*x2*x1 (also including x1)!\\
  CL6  & \verb!x3*x2^2*x1*x2^2!\\
  CL7  & \verb!x3*x2^2! \\
  CL8  & \verb!x3!\\
  CL9  & \verb!x1^-1*x2^-2 (also including x2)!\\
  CL10 & \verb!x1*x2*x1!\\\hline
\end{tabular}
\caption{The representatives in the conjugacy classes of the irreducible representations of the icosahedral group. The generators of the groups are represented by symbols x1...xN, as are numbered by GAP. In this case x1=m1,x2=m2,x3=m3.
}
\label{representatives}
\end{table}
The projector on the p-th irreducible representation is computed by the formula
$$
P^{(p)}=\frac{l(p)}{|G|}\sum_{T \in G}\chi^{(p)*}(T)\cdot O(T)
$$
by summing up the elements $T$ of the group $G$, where $l(p)$ is the dimension of the p-th irreducible representation, $g$ the order of the group,  $\chi$(p) the character, $O(T)$ the operation of the group (which would be represented as matrices and applied to the vector representation of the wave function. The irreducible representation of the wave function is determined numerically, by the condition that the application of the projector should keep the wave function as it is, or delete it.)
\subsection{The analysis of the eigenstates}
\label{TheAnalysisOfTheEigenstates}
Now that we have the irreducible representations and the symmetry operations, we can construct the projection operators.

At first, we assume a model of C$_{60}$, where, in each vertex, one s or $\pi$ orbital is located. The inter-atomic interactions are limited to those among three different neighboring sites, as are signified by single and double bonds. We permit two different strengths of interactions in the Fock matrix H, according to two types of the bonding. We neglect the overlap matrix elements between different sites, the overlapping matrix S being the unity. A computation is executed by the above assumption, with the values of H$_{\rm single}$=-2 (on the single bonds in the pentagons), H$_{\rm double}$=-1 (on the double bonds in the hexagons). 

Before the electronic structure computation, we can deduce a prediction from the group theory. The group elements given by the matrices which represent the replacement of the 60 vertexes provide us the trace representation I.Tr. The virtual representation is decomposed by the irreducible representations (as the sum of row vectors in the table): 
$$
\rm
I.Tr=1\times I.1+0\times I.2+1\times I.3+1\times I.4+2 \times I.5+2 \times I.6+2 \times I.7+2\times I.8+3 \times I.9+2\times I.10.
$$
This relation tells us that the 60-dimensional vector space (on which the matrix group acts) is divided into the ten irreducible components, given as the ten irreducible representations of the group.  The multiplicity of X-th component is equal to the coefficient to ``I.X''. One of these components,  belonging to ``I.2'', is a null space in this case. The wave functions, computed as the eigenvectors of the Fock matrix, are also subject to this division of the vector space. 
 
Let us verify this. The eigenstates and the corresponding irreducible representations are shown in the following list. The abbreviations as ``I.'',``D.'',``E'' denote the index of the irreducible representation, the degeneracy of the energy, and the energies.  The values of the energies are listed in the bracket. If the bracket is blank, the eigenstates in correspondence with the irreducible representations do not exist. The number of different eigenvalues in each irreducible representation is equal to the multiplicity in the above relation between I.Tr and I.1,..., I.10.

\begin{verbatim}
I.1  D.1 E.[ -5.0000 ]
I.2  D.1 E.[ None ]
I.3  D.3 E.[ 4.2361 ] 
I.4  D.3 E.[ -0.2361 ] 
I.5  D.3 E.[ -4.6542, -0.5819 ] 
I.6  D.3 E.[ -3.6765, 2.9126 ] 
I.7  D.4 E.[ -1.4495, 3.4495 ] 
I.8  D.4 E.[ -2.1623, 4.1623 ] 
I.9  D.5 E.[ -4.0642, -1.6946, 2.7588 ]
I.10 D.5 E.[ -1.0000, 3.0000 ]
\end{verbatim}    

For a realistic quantum chemistry computation of C$_{60}$, the similar analysis is possible. For simplicity, one can make use of the amplitudes of 1s orbitals, because those orbitals are transformed among themselves, without mixing with other types, 2s,2p,3s,3p. Except for one special case where the amplitudes of 1s orbital are null, the correspondence between the eigenstates and the irreducible representations can be well established. The result is listed in the following list. (The last entry of the list I.0 is for the case of the failures in the identification.) Even in the case of the failure of the analysis, the degeneracy is single, and one can conclude the corresponding irreducible representation is I.1 or I.2, that of 1-dimension. The numbers of the different eigenvalues in the each irreducible representations are predictable by means of the decomposition of the trace representation of the character of the transformation matrices of the 1s,2s,2px,2py,2pz,3s,3px,3py,3pz orbitals. (The p orbitals are transformed by the same matrices prepared for the vibrational mode of the molecule, as will be discussed later.)

\begin{verbatim}
I.1  D.1   E.[  7 different eigenvalues ]
I.2  D.1   E.[ ? ] 
I.3  D.3   E.[ 11 different eigenvalues ]
I.4  D.3   E.[ 11 different eigenvalues ]
I.5  D.3   E.[ 16 different eigenvalues ]
I.6  D.3   E.[ 16 different eigenvalues ]
I.7  D.4   E.[ 24 different eigenvalues ]
I.8  D.4   E.[ 24 different eigenvalues ]
I.9  D.5   E.[ 25 different eigenvalues ]
I.10 D.5   E.[ 20 different eigenvalues ]
I.0  D.1   E.[  2 different eigenvalues ]
\end{verbatim}

\subsection{The analysis again, against the failure in the identification}

As the degeneracy of the energy is one, one can see the levels denoted as I.0 in the previous section belong to the representation I.1 or I.2. The characters of the one-dimensional representation in table \ref{charactertbl} are shown here again (in table \ref{chatactertbl1D}). 


\begin{table}[h!] 
\centering
\begin{tabular}{c|cccccccccc}\hline
&CL1&CL2&CL3&CL4&CL5&CL6&CL7&CL8&CL9&CL10\\\hline
I.1&1&1&1&1&1&1&1&1&1&1\\
I.2&1&-1&1&-1&1&-1&-1&-1&1&1\\\hline
\end{tabular}
\caption{The one-dimensional irreducible representations.}
\label{chatactertbl1D}
\end{table}

To distinguish the representation I.1 and I.2 with ease, one may apply the element in the conjugacy class 8 (the inversion). By the inversion, one vertex moves to the opposite site with respect to the center of the geometry,  and the coefficients of p-orbitals change signs. The amplitudes of the wave-functions (in the two atoms located in the antipodes positions) in I.0 are typically obtained as in table \ref{TWOWFUNC}.

\begin{table}[h!] 
\centering
\begin{tabular}{c|c|c}
\hline
     CA& 1S       & 0\\
       & 2S       & 0\\
       & 2PX      & X \\
       & 2PY      & Y \\
       & 2PZ      & Z \\\hline
     CB& 1S       & 0\\
       & 2S       & 0\\
       & 2PX      & X\\
       & 2PY      & Y\\
       & 2PZ      & Z\\
\hline
\end{tabular}
\caption{The amplitudes of the wave-function (coefficients for each orbital) in two opposing atoms.}
\label{TWOWFUNC}
\end{table}
The inversion changes the sign of non-zero coefficients. We must recall that the wave-functions are basis functions of some irreducible representations, and the traces of their transformation matrices are the characters. Since the traces are -1 in the above cases, the wave-functions belong to the representation I.2. The reason of the failure in the analysis by the projection operators applied on s orbitals are fundamental one: the projection operator of I.2 over s orbitals is exactly the zero-matrix, which gives rise to exact zero amplitudes in s orbitals (i.e. the null space.) The reason of the existence of the orbitals of the I.2 is that the actual C$_{60}$ molecule has the symmetry of the inversion and the carbon atom has p-orbitals.

\subsection{The perturbation and the symmetry}

We often have to compute matrix elements in the perturbation theory such as
\begin{equation}
\rm
\left\langle\psi_i|V_p|\psi_j\right\rangle=[I_i|I_p|I_j]=[I_i|I_p*I_j].
\end{equation}
The product of the perturbation potential and the wave function  belongs to the direct product of the representation, which is decomposed into the direct sum
\begin{equation}
\rm
I_p*I_j=\sum_k A_k^{p,j}I_k.
\end{equation}
We can determine beforehand whether this matrix element would be zero or not, only from the group theoretical consideration: if the two irreducible representations in the inner product are the same, the matrix element will be non-zero. In the system with the high symmetry, the group theoretical constraint is so strong that it annihilates many matrix elements. However, the symmetry is reduced, there is some possibility where the matrix element will become non-zero. Indeed there is no symmetry, there is no group theoretical constraint which annihilates the matrix elements. And if one uses the group theory, one can see the least reduction of the symmetry where the matrix element will become non-zero. The irreducible representation of group g ramifies in the subgroups as follows.
\begin{equation}
\rm
I_p^g=C_{i,1}^{(g,s)}I_1^s+C_{i,2}^{(g,s)}I_2^s+\cdots+C_{i,M}^{(g,s)}I_M^s.
\end{equation}
One can find the subgroup where the non-zero matrix element will exist, by tracing the sequence of subgroups (in the subgroup lattice) and checking the possible combinations of the ramified irreducible representations. In this stage, the concrete style of the structural deformation is not discussed yet.


One of the important perturbation potentials is that of optical transition, such as $E\cdot r$. Now we determine the irreducible representation to which the perturbation $E\cdot r$ (or the spatial vector $r$) belongs. We now construct the icosahedral group from the rotation matrices in three-dimensional Euclidean space, acting on the twelve vertexes of one icosahedron. 


The twelve vertexes in one icosahedron are given by

\begin{verbatim}
1  [ 0, 0, 1 ] 
2  [ 2/5*E(5)-2/5*E(5)^2-2/5*E(5)^3+2/5*E(5)^4, 0, 
      1/5*E(5)-1/5*E(5)^2-1/5*E(5)^3+1/5*E(5)^4 ] 
3  [ -3/5*E(5)-2/5*E(5)^2-2/5*E(5)^3-3/5*E(5)^4, 
      -1/5*E(20)+1/5*E(20)^9+2/5*E(20)^13-2/5*E(20)^17, 
      1/5*E(5)-1/5*E(5)^2-1/5*E(5)^3+1/5*E(5)^4 ] 
4  [ 2/5*E(5)+3/5*E(5)^2+3/5*E(5)^3+2/5*E(5)^4, 
      -2/5*E(20)+2/5*E(20)^9-1/5*E(20)^13+1/5*E(20)^17, 
      1/5*E(5)-1/5*E(5)^2-1/5*E(5)^3+1/5*E(5)^4 ] 
5  [ 2/5*E(5)+3/5*E(5)^2+3/5*E(5)^3+2/5*E(5)^4, 
      2/5*E(20)-2/5*E(20)^9+1/5*E(20)^13-1/5*E(20)^17, 
      1/5*E(5)-1/5*E(5)^2-1/5*E(5)^3+1/5*E(5)^4 ], 
6  [ -3/5*E(5)-2/5*E(5)^2-2/5*E(5)^3-3/5*E(5)^4, 
      1/5*E(20)-1/5*E(20)^9-2/5*E(20)^13+2/5*E(20)^17, 
      1/5*E(5)-1/5*E(5)^2-1/5*E(5)^3+1/5*E(5)^4 ] 
7  [ 0, 0, -1 ] 
8  [ -2/5*E(5)+2/5*E(5)^2+2/5*E(5)^3-2/5*E(5)^4, 0, 
      -1/5*E(5)+1/5*E(5)^2+1/5*E(5)^3-1/5*E(5)^4 ] 
9  [ 3/5*E(5)+2/5*E(5)^2+2/5*E(5)^3+3/5*E(5)^4, 
      1/5*E(20)-1/5*E(20)^9-2/5*E(20)^13+2/5*E(20)^17, 
      -1/5*E(5)+1/5*E(5)^2+1/5*E(5)^3-1/5*E(5)^4 ] 
10  [ -2/5*E(5)-3/5*E(5)^2-3/5*E(5)^3-2/5*E(5)^4, 
      2/5*E(20)-2/5*E(20)^9+1/5*E(20)^13-1/5*E(20)^17, 
      -1/5*E(5)+1/5*E(5)^2+1/5*E(5)^3-1/5*E(5)^4 ] 
11  [ -2/5*E(5)-3/5*E(5)^2-3/5*E(5)^3-2/5*E(5)^4, 
      -2/5*E(20)+2/5*E(20)^9-1/5*E(20)^13+1/5*E(20)^17, 
      -1/5*E(5)+1/5*E(5)^2+1/5*E(5)^3-1/5*E(5)^4 ] 
12  [ 3/5*E(5)+2/5*E(5)^2+2/5*E(5)^3+3/5*E(5)^4, 
      -1/5*E(20)+1/5*E(20)^9+2/5*E(20)^13-2/5*E(20)^17, 
      -1/5*E(5)+1/5*E(5)^2+1/5*E(5)^3-1/5*E(5)^4 ] 
\end{verbatim}

The three generators of the group are as follows:

\begin{verbatim}
gap> f1:=(1,2,3)(4,6,11)(5,10,12)(7,8,9);; 
gap> f2:=(2,6,5,4,3)(8,12,11,10,9);;
gap> f3:=(1,7)(2,8)(3,9)(4,10)(5,11)(6,12);;
\end{verbatim}

These permutations are also represented by the rotation matrices as

\begin{verbatim}
gap> f1:=[ [ 3/10*E(5)+1/5*E(5)^2+1/5*E(5)^3+3/10*E(5)^4, 
           1/10*E(20)-1/10*E(20)^9-1/5*E(20)^13+1/5*E(20)^17,
           2/5*E(5)-2/5*E(5)^2-2/5*E(5)^3+2/5*E(5)^4 ], 
           [ -1/2*E(20)+1/2*E(20)^9, -1/2*E(5)-1/2*E(5)^4, 0 ], 
           [ -3/5*E(5)-2/5*E(5)^2-2/5*E(5)^3-3/5*E(5)^4, 
           -1/5*E(20)+1/5*E(20)^9+2/5*E(20)^13-2/5*E(20)^17, 
           1/5*E(5)-1/5*E(5)^2-1/5*E(5)^3+1/5*E(5)^4 ] ];
gap> f2:=[ [ 1/2*E(5)+1/2*E(5)^4, -1/2*E(20)+1/2*E(20)^9, 0 ], 
          [ 1/2*E(20)-1/2*E(20)^9, 1/2*E(5)+1/2*E(5)^4, 0 ], [ 0, 0, 1 ] ];;
gap> f3:=[ [ -1, 0, 0 ], [ 0, -1, 0 ], [ 0, 0, -1 ] ];;
\end{verbatim}


\begin{table}[h!] 
\begin{tabular}{lccll}
          &  Dimension &    Operator                            &     The characters of [f1,f2,f3]&\\
N.1       & 1    &     [ [ 0, 0, 0 ], [ 0, 0, 0 ], [ 0, 0, 0 ] ] &     [ 1, 1, 1 ]&\\
N.2       & 1    &     [ [ 0, 0, 0 ], [ 0, 0, 0 ], [ 0, 0, 0 ] ] &     [ 1, 1, -1 ]&\\
N.3       & 3    &     [ [ 0, 0, 0 ], [ 0, 0, 0 ], [ 0, 0, 0 ] ] &     [ 0,-E(5)\verb!^!2-E(5)\verb!^!3,  3 ]&       (I.4 in table \ref{charactertbl})\\
N.4       & 3    &     [ [ 0, 0, 0 ], [ 0, 0, 0 ], [ 0, 0, 0 ] ] &     [ 0,-E(5)-E(5)\verb!^!4,  3 ]  &       (I.3 in \ref{charactertbl})\\
N.5       & 3    &     [ [ 1, 0, 0 ], [ 0, 1, 0 ], [ 0, 0, 1 ] ] &     [ 0,-E(5)\verb!^!2-E(5)\verb!^!3, -3 ]&\\
N.6       & 3    &     [ [ 0, 0, 0 ], [ 0, 0, 0 ], [ 0, 0, 0 ] ] &     [ 0,-E(5)-E(5)\verb!^!4,  -3 ]&\\
N.7       & 4    &     [ [ 0, 0, 0 ], [ 0, 0, 0 ], [ 0, 0, 0 ] ] &     [ 1, -1, 4 ]&\\
N.8       & 4    &     [ [ 0, 0, 0 ], [ 0, 0, 0 ], [ 0, 0, 0 ] ] &     [ 1, -1, -4 ]&\\
N.9       & 5    &     [ [ 0, 0, 0 ], [ 0, 0, 0 ], [ 0, 0, 0 ] ] &      [ -1, 0, 5 ]&\\
N.10       & 5    &     [ [ 0, 0, 0 ], [ 0, 0, 0 ], [ 0, 0, 0 ] ] &      [ -1, 0, -5 ]&\\
\end{tabular}
\caption{The projection operators. Be careful that the computation by GAP returns the representations (N.1,...,N.10) in the slightly different order from those in table \ref{charactertbl}(I.1,...,I.10).}
\label{CHARTBLAGAIN}
\end{table}

The ten projection operators are computed as in table \ref{CHARTBLAGAIN}. The operators, except that of the fifth representation, are exactly zero. The symmetry operations on one vector, which is subject to the icosahedral symmetry, are only representable in the fifth irreducible representation, and the perturbation term $E\cdot r$ is represented by the fifth representation.  (The fifth representation can be distinguished from others, as such having characters \verb![0,-E(5)^2-E(5)^3,-3]! for three generators. Especially, the rotation of 72 degrees should have the character value \verb!-E(5)^2-E(5)^3!. The rotation by 72 degrees and its inverse (\verb!f2! and \verb!f2^-1!) have the same character value, because they are conjugate.)
\begin{remark}
In order to see whether the elements x and y are conjugate in group G, try the command. 
\begin{verbatim}
 gap> RepresentativeAction(G,x,y);
\end{verbatim}
\end{remark}

\subsection{The possible deformation in the reduced symmetry}
We can construct molecular geometry which is admissible in each subgroup. If the vertexes are dislocated as V(i) $\rightarrow$ V(i)+d(i), the set of d(i) should satisfy a certain geometrical constraint in the symmetry operation in the subgroup. We should take note of this : since the icosahedral group which includes 120 operations and acts on 60 vertexes in one C$_{60}$, there are two operations, both of which equally transfers one atomic site into same, but another site. If the dislocations of atoms are still subject to the icosahedral symmetry, there is a constraint. When the atom at the site V[i] moves to V[j] by the operation R, it must be 
$$
 \rm R\cdot V[i]= V[j],\hspace{0.5cm} and \hspace{0.5cm} \rm R \cdot (V[i]+d[i])= V[j]+d[j], 
$$ 
i.e.
$$
 \rm R \cdot d[i]= d[j].
$$
If there are two possible transformations R and Q, it must be 
$$
  \rm R \cdot d[i]=Q\cdot d[i].
$$
If we chose T=R$^{-1}$Q, then T$\cdot$d[i]=d[i], and the possible dislocation is the solution of the equation $\rm (T-E)\cdot d[i]=0$ (E: the unit matrix). In the case of the icosahedral symmetry in C$_{60}$, such a T is the reflection at the plain in which one double bond and its antipodes are located. 

In general, the dislocations d[i] on all of the vertex are generated as follows. For example, take V[1], and find the operation R, Q, T as above, and then find a possible solution d of $(T-E)\cdot d=0$ as the generator. The application of the 120 operations on V[1]+d shall generate 60 sites, which are represented by V[i]+d[i], and the set of d[i] is obtained. (This is equivalent to the projection on the trivial irreducible representation, in which the basis set is invariant under the group action.) 
In a subgroup, the symmetric operations do not act transitively, and the ``orbit'' of one vertex by the symmetry operations (which covers all of the vertexes in C$_{60}$ by the icosahedral symmetry) splits into several subsets, as we will see later. For each subset, we construct independent d, the generator of the dislocation in each subset. If the subgroups contain the reflection, the dislocations on the atoms have similar constraints to d (being fixed on one of the reflectional planes) as in the case of the icosahedral symmetry. On the contrary, when the reflection is missing in the subgroup, we can choose arbitrary generator of dislocations in each sub-orbits. One should note that the center of mass will not be kept since d[1]+d[2]+...+d[60] is not equal to 0 in general.  
   
Concerning this topic, we can classify the vibrational mode of C$_{60}$ according to the group theory. The procedure is to form the direct product of the virtual representation with respect to the replacement of atoms and the irreducible representation of the symmetry operations on the space vector, and then to decompose it into the direct sum. One will find that the vibrational mode in the icosahedral symmetry includes the duplicated trivial irreducible representation, to which the degrees of the freedom is two. The geometrical interpretation of this is that the possible distortion which keeps the icosahedral symmetry is confined in one plane as discussed above.

\subsection{The reduction of the symmetry, from the viewpoint of the orbit in the permutation}
The full symmetry of C$_{60}$ is the icosahedral group, generated by ``g1'', ``g2'', ``g3'' in the previous section, and it has 164 subgroups. All of the subgroups are computed as 

\begin{verbatim}
gap> gr:=Group(g1,g2,g3);
gap> AS:=AllSubgroups(gr);;
\end{verbatim}

The list of the subgroups is given in the descending order of the sizes of the subgroups, and the largest of them (No.164) is the icosahedral group. 
\begin{verbatim}
gap> irrA:=Irr(AS[164]);
\end{verbatim}

Take a subgroup No.162. This group is of the order 24, and the irreducible representation is also computed.
\begin{verbatim}
gap> irrB:=Irr(AS[162]);;
\end{verbatim}

The irreducible representations of these groups are given in tables \ref{CTBL164},\ref{CC164},\ref{CTBL162}, and \ref {CC162}.

\begin{table}[h!]
\begin{tabular}{l|llllllllll}\hline
   &CL1&CL2& CL3& CL4& CL5& CL6& CL7& CL8 &CL9 & CL10\\\hline
No.1&1&1&1&1&1&1&1&1&1&1\\
No.2&1&-1&1&1&1&-1&1&-1&-1&-1\\
No.3&3&-1&-E(5)-E(5)\verb!^!4&-E(5)\verb!^!2-E(5)\verb!^!3&-1&-E(5)\verb!^!2-E(5)\verb!^!3&0&0&3&-E(5)-E(5)\verb!^!4\\
No.4&3&-1&-E(5)\verb!^!2-E(5)\verb!^!3&-E(5)-E(5)\verb!^!4&-1&-E(5)-E(5)\verb!^!4&0&0&3&-E(5)\verb!^!2-E(5)\verb!^!3\\
No.5&3&1&-E(5)-E(5)\verb!^!4&-E(5)\verb!^!2-E(5)\verb!^!3&-1&E(5)\verb!^!2+E(5)\verb!^!3&0&0&-3&E(5)+E(5)\verb!^!4\\
No.6&3&1&-E(5)\verb!^!2-E(5)\verb!^!3&-E(5)-E(5)\verb!^!4&-1&E(5)+E(5)\verb!^!4&0&0&-3&E(5)\verb!^!2+E(5)\verb!^!3\\
No.7&4&0&-1&-1&0&-1&1&1&4&-1\\
No.8&4&0&-1&-1&0&1&1&-1&-4&1\\
No.9&5&1&0&0&1&0&-1&-1&5&0\\
No.10&5&-1&0&0&1&0&-1&1&-5&0\\\hline
\end{tabular}

\caption{The character table of the group AS[164].}
\label{CTBL164}
\end{table}
\begin{table}[h!]
\centering
\begin{tabular}{lll}\hline
      &  Representative &                               Size\\\hline
CL1& \verb!<identity ...>!&1\\
CL2& \verb!x1^-1*x2^-2*x3*x1*x2*x1!&15\\
CL3& \verb!x2^-1!&12\\
CL4& \verb!x2^-2!&12\\
CL5& \verb!x2^-2*x1*x2!&15\\ 
CL6& \verb!x2^2*x1*x2^2*x1^-1*x3!&12\\
CL7& \verb!x2^2*x1*x2!&20 (including \verb!g1!)\\
CL8& \verb!x2^-1*x1^-1*x2^-2*x1^-1*x3!&20\\
CL9& \verb!x3!&1\\
CL10& \verb!x2^-1*x3!&12\\\hline
\end{tabular}
\caption{The conjugacy classes of the group AS[164].}
\label{CC164}
\end{table}

\begin{table}[h!]
\centering
\begin{tabular}{l|llllllll}\hline
   &CL1&CL2& CL3& CL4& CL5& CL6& CL7& CL8 \\\hline
No.1& 1& 1& 1& 1& 1& 1& 1& 1 \\ 
No.2& 1& -1& -1& 1& 1& -1& 1& -1 \\  
No.3& 1& -1& -E(3)& E(3)& E(3)\verb!^!2& -E(3)\verb!^!2& 1& -1 \\ 
No.4& 1& -1& -E(3)\verb!^!2& E(3)\verb!^!2& E(3) & -E(3)& 1& -1 \\  
No.5& 1& 1& E(3)\verb!^!2& E(3)\verb!^!2& E(3)& E(3)& 1& 1 \\  
No.6& 1& 1& E(3)& E(3)& E(3)\verb!^!2& E(3)\verb!^!2& 1& 1 \\  
No.7& 3& 1& 0& 0& 0& 0& -1& -3 \\  
No.8& 3& -1& 0& 0& 0& 0& -1& 3 \\\hline 
\end{tabular}
\caption{The character table of the group AS[162].}
\label{CTBL162}
\end{table}

\begin{table}[h!]
\centering
\begin{tabular}{lll}\hline
      &  Representative &                   Size\\\hline
  CL1 & \verb!<identity ...>!              &    1\\        
  CL2 & \verb!x2^-1*x1^-1*x2^-2*x3*x1*x2*x1! &    3\\
  CL3 & \verb!x1^-1*x2^-2*x1^-1*x3!         &    4\\
  CL4 & \verb!x2^-1*x1*x2!                 &    4\\    
  CL5 & \verb!x2^-2*x1!                     &    4\\       
  CL6 & \verb!x1*x2^2*x1^-1*x2^-1*x3!         &    3\\
  CL7 & \verb!x2*x1^-1!                       &    3\\
  CL8 & \verb!x3!                          &    1\\\hline
\end{tabular}
\caption{The conjugacy classes of the group AS[162].}
\label{CC162}
\end{table}

The branching relation from the group of No.164 to that of No.162 is computed now. The characters of the elements of the group of No.162 are computed and listed in two ways: the one is by the irreducible representations of the group of No.164, and another by those of the group of No.162. The representations in the first list are decomposed as the summation of the entries in the second list, and the result is the branching relation. 

\begin{verbatim}
gap> LA:=List(irrA,a->List(Elements(AS[162]),g->g^a));
[ [ 1, 1, 1, 1, 1, 1, 1, 1, 1, 1, 1, 1, 1, 1, 1, 1, 1, 1, 1, 1, 1, 1, 1, 1 ], 
.............................................................................
.............................................................................
 [ 5, -1, 1, -1, -1, 1, -1, 1, 1, -1, 1, -1, -5, 1, -1, 1, 1, -1, 1, -1, -1, 
   1, -1, 1 ] ]
gap> LB:=List(irrB,a->List(Elements(AS[162]),g->g^a));
[ [ 1, 1, 1, 1, 1, 1, 1, 1, 1, 1, 1, 1, 1, 1, 1, 1, 1, 1, 1, 1, 1, 1, 1, 1 ], 
.............................................................................
.............................................................................
   0 ], [ 3, -1, 0, 0, 0, 0, -1, -1, 0, 0, 0, 0, 3, -1, 0, 0, 0, 0, -1, 
   -1, 0, 0, 0, 0 ] ]
gap> LA*TransposedMat(LB)/Size(AS[162]);
\end{verbatim}

The result returns as the matrix in the bellow: the irreducible representations are ramified as
\begin{verbatim}
                                 No.164 -> No.162
[ [ 1, 0, 0, 0, 0, 0, 0, 0 ],    1      -> 1
  [ 0, 1, 0, 0, 0, 0, 0, 0 ],    2      -> 2
  [ 0, 0, 0, 0, 0, 0, 0, 1 ],    3      -> 8
  [ 0, 0, 0, 0, 0, 0, 0, 1 ],    4      -> 8
  [ 0, 0, 0, 0, 0, 0, 1, 0 ],    5      -> 7 
  [ 0, 0, 0, 0, 0, 0, 1, 0 ],    6      -> 7 
  [ 1, 0, 0, 0, 0, 0, 0, 1 ],    7      -> 1 + 8
  [ 0, 1, 0, 0, 0, 0, 1, 0 ],    8      -> 2 + 7
  [ 0, 0, 0, 0, 1, 1, 0, 1 ],    9      -> 5 + 6 + 8
  [ 0, 0, 1, 1, 0, 0, 1, 0 ] ]  10      -> 3 + 4 + 7
\end{verbatim}

One can see the change in the geometry by checking the ``orbit'' of the vertex ``1'' and the bond ``[1,6]''. In the icosahedral symmetry, the vertex is transferred to other sixty vertexes transitively, and this double bond is transferred to all other thirty double bonds. 

\begin{verbatim}
gap> List(Elements(AS[164]),g->OnPoints(1,g));
[ 1, 1, 2, 2, 3, 3, 4, 4, 5, 5, 6, 6, 7, 7, 8, 8, 9, 9, 10, 10, 11, 11, 12, 
  12, 13, 13, 14, 14, 15, 15, 16, 16, 17, 17, 18, 18, 19, 19, 20, 20, 21, 21, 
  22, 22, 23, 23, 24, 24, 25, 25, 26, 26, 27, 27, 28, 28, 29, 29, 30, 30, 31, 
  31, 32, 32, 33, 33, 34, 34, 35, 35, 36, 36, 37, 37, 38, 38, 39, 39, 40, 40, 
  41, 41, 42, 42, 43, 43, 44, 44, 45, 45, 46, 46, 47, 47, 48, 48, 49, 49, 50, 
  50, 51, 51, 52, 52, 53, 53, 54, 54, 55, 55, 56, 56, 57, 57, 58, 58, 59, 59, 
  60, 60 ]
gap> BONDA:=Unique(List(Elements(AS[164]),g->OnSets([1,6],g)));
[ [ 1, 6 ], [ 2, 11 ], [ 3, 16 ], [ 4, 21 ], [ 5, 26 ], [ 7, 30 ], [ 8, 49 ], 
 [ 9, 53 ], [ 10, 12 ], [ 13, 54 ], [ 14, 58 ], [ 15, 17 ], [ 18, 59 ], 
 [ 19, 38 ], [ 20, 22 ], [ 23, 39 ], [ 24, 43 ], [ 25, 27 ], [ 28, 44 ], 
 [ 29, 48 ], [ 31, 36 ], [ 32, 41 ], [ 33, 46 ], [ 34, 51 ], [ 35, 56 ], 
 [ 37, 60 ], [ 40, 42 ], [ 45, 47 ], [ 50, 52 ], [ 55, 57 ] ]
\end{verbatim}

In the symmetry of the subgroup No.162, in contrast, the vertex ``1'' is transferred only to 24 other vertexes. The orbit of single vertexes (which covers the whole system in the case of the subgroup 164) splits into three subsets. Similarly the bond ``[1,6]'' is not transferred transitively, only to 24 of double bonds. The remaining six double bonds are transferred among themselves. This situation can be interpreted that one of the double bonds (one of these connecting one pentagon with five neighbors) slightly changes the length, while those of remaining four bonds are unchanged.
 
\begin{verbatim}
gap> List(Elements(AS[162]),g->OnPoints(1,g));
[ 1, 5, 9, 10, 11, 15, 18, 19, 21, 22, 27, 28, 31, 35, 39, 40, 41, 45, 48, 
  49, 51, 52, 57, 58 ]
gap> List(Elements(AS[162]),g->OnPoints(2,g));
[ 2, 4, 8, 6, 12, 14, 17, 20, 25, 23, 26, 29, 32, 34, 38, 36, 42, 44, 47, 50, 
  55, 53, 56, 59 ]
gap> List(Elements(AS[162]),g->OnPoints(3,g));
[ 3, 3, 7, 7, 13, 13, 16, 16, 24, 24, 30, 30, 33, 33, 37, 37, 43, 43, 46, 46, 
  54, 54, 60, 60 ]
gap> BONDB:=Unique(List(Elements(AS[162]),g->OnSets([1,6],g)));
gap> Sort(BONDB);
gap> BONDB;
[ [ 1, 6 ], [ 2, 11 ], [ 4, 21 ], [ 5, 26 ], [ 8, 49 ], [ 9, 53 ], 
 [ 10, 12 ], [ 14, 58 ], [ 15, 17 ], [ 18, 59 ], [ 19, 38 ], [ 20, 22 ], 
 [ 23, 39 ], [ 25, 27 ], [ 28, 44 ], [ 29, 48 ], [ 31, 36 ], [ 32, 41 ], 
 [ 34, 51 ], [ 35, 56 ], [ 40, 42 ], [ 45, 47 ], [ 50, 52 ], [ 55, 57 ] ]
\end{verbatim}

The splitting of the energy degeneracy with the symmetry reduction is verified numerically, for example, by a model computation, in which the matrix elements of these two split orbits of double bonds take slightly different values, in correspondence with the change in the symmetry. In this computation, the elements of the Fock matrix which belong to the one of the split orbit (including six bonds, the complementary part of BONDB in BONDA) are enhanced slightly, multiplied by 1.005. The energy spectra at HOMO and LUMO are computed as follows.(The positions of HOMO and LUMO are ambiguous in this case. We presently assume that the half of the 60 eigenstates are occupied by 60 electrons. The left and right columns show the spectra of the full and the reduced symmetries. The HOMO and LUMO belong to the irreducible representations of No.10 and No.5 of the character table of the group AS[164], and the splitting of the spectra is deduced from the above relation.) 

\begin{verbatim}
HOMO : Split from one fivefold degeneracy  
       to one threefold degeneracy and two single degeneracies.
26  -1.0000000  ->   -1.0000021     
27  -1.0000000  ->   -1.0000021     
28  -1.0000000  ->   -1.0000021     
29  -1.0000000  ->   -1.0000000     
30  -1.0000000  ->   -1.0000000     
\end{verbatim}

In this case, two single states still have the same energy levels. This is due to the fact that the two single states belong to two of the one-dimensional irreducible representation which are complex-conjugates with each other. As the eigenvalue problem is the real-valued one, these two eigenvalues should be equal owing to the invariance by the complex conjugation acting on it.  It could also be interpreted as of the time-reversal symmetry in the context of the quantum physics. 

\begin{verbatim}
LUMO: Without splitting.
31 -0.5818861  ->    -0.5810887     
32 -0.5818861  ->    -0.5810887     
33 -0.5818861  ->    -0.5810887     
\end{verbatim}

The symmetry reduction from the icosahedral group to one of the subgroups of order 24 can be realized clearly, as supposed in this computation. On the other hand, the difference between the icosahedral group and the maximal subgroup AS[163] (the 5-th alternating group: A5) is more subtle. The splitting of the ``orbits'' by the group action does not occur in the transposition of the bonds between nearest or second-nearest neighbors. The length of the single or double bonds is also unique in A5 symmetry, and the change in the symmetry could be seen in the ``orbit'' of the bonds among more distant neighbors. Take the bond of [1,11]. 

\begin{verbatim}
gap> lg:=List(Elements(AS[164]),i->OnSets([1,11],i));
[ [ 1, 11 ], [ 1, 26 ], [ 2, 6 ], [ 2, 16 ], [ 3, 11 ], [ 3, 21 ], [ 4, 16 ], 
 [ 4, 26 ], [ 5, 6 ], [ 5, 21 ], [ 6, 30 ], [ 6, 12 ], [ 1, 7 ], [ 7, 49 ], 
 [ 8, 30 ], [ 8, 53 ], [ 9, 49 ], [ 9, 12 ], [ 1, 10 ], [ 10, 53 ], 
 [ 10, 11 ], [ 11, 17 ], [ 2, 12 ], [ 12, 54 ], [ 10, 13 ], [ 13, 58 ], 
 [ 14, 54 ], [ 14, 17 ], [ 2, 15 ], [ 15, 58 ], [ 15, 16 ], [ 16, 22 ], 
 [ 3, 17 ], [ 17, 59 ], [ 15, 18 ], [ 18, 38 ], [ 19, 59 ], [ 19, 22 ], 
 [ 3, 20 ], [ 20, 38 ], [ 20, 21 ], [ 21, 27 ], [ 4, 22 ], [ 22, 39 ], 
 [ 20, 23 ], [ 23, 43 ], [ 24, 39 ], [ 24, 27 ], [ 4, 25 ], [ 25, 43 ], 
 [ 25, 26 ], [ 7, 26 ], [ 5, 27 ], [ 27, 44 ], [ 25, 28 ], [ 28, 48 ], 
 [ 29, 44 ], [ 7, 29 ], [ 5, 30 ], [ 30, 48 ], [ 31, 41 ], [ 31, 56 ], 
 [ 32, 36 ], [ 32, 46 ], [ 33, 41 ], [ 33, 51 ], [ 34, 46 ], [ 34, 56 ], 
 [ 35, 36 ], [ 35, 51 ], [ 36, 60 ], [ 36, 42 ], [ 31, 37 ], [ 19, 37 ], 
 [ 38, 60 ], [ 23, 38 ], [ 19, 39 ], [ 39, 42 ], [ 31, 40 ], [ 23, 40 ], 
 [ 40, 41 ], [ 41, 47 ], [ 32, 42 ], [ 24, 42 ], [ 40, 43 ], [ 28, 43 ], 
 [ 24, 44 ], [ 44, 47 ], [ 32, 45 ], [ 28, 45 ], [ 45, 46 ], [ 46, 52 ], 
 [ 33, 47 ], [ 29, 47 ], [ 45, 48 ], [ 8, 48 ], [ 29, 49 ], [ 49, 52 ], 
 [ 33, 50 ], [ 8, 50 ], [ 50, 51 ], [ 51, 57 ], [ 34, 52 ], [ 9, 52 ], 
 [ 50, 53 ], [ 13, 53 ], [ 9, 54 ], [ 54, 57 ], [ 34, 55 ], [ 13, 55 ], 
 [ 55, 56 ], [ 37, 56 ], [ 35, 57 ], [ 14, 57 ], [ 55, 58 ], [ 18, 58 ], 
 [ 14, 59 ], [ 37, 59 ], [ 35, 60 ], [ 18, 60 ] ]
\end{verbatim}

On the other hand, the ``orbit'' by A5 includes only 60 bonds.
\begin{verbatim}
gap> la5:=List(Elements(AS[163]),i->OnSets([1,11],i));
[ [ 1, 11 ], [ 2, 16 ], [ 3, 21 ], [ 4, 26 ], [ 5, 6 ], [ 6, 30 ], [ 7, 49 ], 
 [ 8, 53 ], [ 9, 12 ], [ 1, 10 ], [ 10, 11 ], [ 12, 54 ], [ 13, 58 ], 
 [ 14, 17 ], [ 2, 15 ], [ 15, 16 ], [ 17, 59 ], [ 18, 38 ], [ 19, 22 ], 
 [ 3, 20 ], [ 20, 21 ], [ 22, 39 ], [ 23, 43 ], [ 24, 27 ], [ 4, 25 ], 
 [ 25, 26 ], [ 27, 44 ], [ 28, 48 ], [ 7, 29 ], [ 5, 30 ], [ 31, 56 ], 
 [ 32, 36 ], [ 33, 41 ], [ 34, 46 ], [ 35, 51 ], [ 36, 42 ], [ 31, 37 ], 
 [ 38, 60 ], [ 19, 39 ], [ 23, 40 ], [ 41, 47 ], [ 32, 42 ], [ 40, 43 ], 
 [ 24, 44 ], [ 28, 45 ], [ 46, 52 ], [ 33, 47 ], [ 45, 48 ], [ 29, 49 ], 
 [ 8, 50 ], [ 51, 57 ], [ 34, 52 ], [ 50, 53 ], [ 9, 54 ], [ 13, 55 ], 
 [ 37, 56 ], [ 35, 57 ], [ 55, 58 ], [ 14, 59 ], [ 18, 60 ] ]
\end{verbatim}

The remaining 60 bonds are located in the separated orbit now.

\begin{verbatim}
gap> la5_2:=List(Elements(AS[163]),i->OnSets([2,6],i));
[ [ 2, 6 ], [ 3, 11 ], [ 4, 16 ], [ 5, 21 ], [ 1, 26 ], [ 1, 7 ], [ 8, 30 ], 
 [ 9, 49 ], [ 10, 53 ], [ 6, 12 ], [ 2, 12 ], [ 10, 13 ], [ 14, 54 ], 
 [ 15, 58 ], [ 11, 17 ], [ 3, 17 ], [ 15, 18 ], [ 19, 59 ], [ 20, 38 ], 
 [ 16, 22 ], [ 4, 22 ], [ 20, 23 ], [ 24, 39 ], [ 25, 43 ], [ 21, 27 ], 
 [ 5, 27 ], [ 25, 28 ], [ 29, 44 ], [ 30, 48 ], [ 7, 26 ], [ 35, 36 ], 
 [ 31, 41 ], [ 32, 46 ], [ 33, 51 ], [ 34, 56 ], [ 31, 40 ], [ 36, 60 ], 
 [ 19, 37 ], [ 23, 38 ], [ 39, 42 ], [ 32, 45 ], [ 40, 41 ], [ 24, 42 ], 
 [ 28, 43 ], [ 44, 47 ], [ 33, 50 ], [ 45, 46 ], [ 29, 47 ], [ 8, 48 ], 
 [ 49, 52 ], [ 34, 55 ], [ 50, 51 ], [ 9, 52 ], [ 13, 53 ], [ 54, 57 ], 
 [ 35, 60 ], [ 55, 56 ], [ 14, 57 ], [ 18, 58 ], [ 37, 59 ] ]
\end{verbatim}

The bonds, equivalent to ``[1,11]'', with the unique length in the icosahedral symmetry,  are permitted to have two different lengths in the A5 symmetry. The deformation from the icosahedral symmetry to that of A5 is realized not in the bond structures between the nearest or the second-nearest neighbors, but those between remoter atomic sites. By such deformation into A5 symmetry, the two pentagons, located in the antipodes positions, can rotate themselves in the opposite directions with each other.

\subsection{The symmetry in the eigenstates (hidden one)}

The irreducible representations of the same dimension show a certain similarity in the icosahedral group. For example, in table \ref{charactertbl}, the representations of I.3 and I.4 or I.5 and I.6 seem to be interchangeable, if we replace E(5), E(5)\verb!^!2, E(5)\verb!^!3, E(5)\verb!^!4 among themselves. This suggests a certain kind of affinity of the basis functions of the wave functions. This is the symmetry in the context of the Galois theory, especially of the cyclotomic field, and manifest itself in the secular equation of the energy spectrum. 

\begin{table}[h!]
\centering
\begin{tabular}{ccccccccccc}\hline
&CL1&CL2&CL3&CL4&CL5&CL6&CL7&CL8&CL9&CL10\\\hline
I.3&3&-E(5)-E(5)\verb!^!4&-E(5)\verb!^!2-E(5)\verb!^!3&0&0&-1&-E(5)\verb!^!2-E(5)\verb!^!3&3&-E(5)-E(5)\verb!^!4&-1\\
I.4&3&-E(5)\verb!^!2-E(5)\verb!^!3&-E(5)-E(5)\verb!^!4&0&0&-1&-E(5)-E(5)\verb!^!4&3&-E(5)\verb!^!2-E(5)\verb!^!3&-1\\
I.5&3&E(5)\verb!^!2+E(5)\verb!^!3&-E(5)-E(5)\verb!^!4&0&0&1&E(5)+E(5)\verb!^!4&-3&-E(5)\verb!^!2-E(5)\verb!^!3&-1\\
I.6&3&E(5)+E(5)\verb!^!4&-E(5)\verb!^!2-E(5)\verb!^!3&0&0&1&E(5)\verb!^!2+E(5)\verb!^!3&-3&-E(5)-E(5)\verb!^!4&-1\\\hline
\end{tabular}
\caption{The three dimensional irreducible representations in the icosahedral group.}
\end{table}

Let us start again at the model computation of the electronic structure of C$_{60}$. The eigenvalue problem is solved by one of the functions in the GAP package. By this, the secular equation is decomposed into polynomials with coefficients over the specified field, and the ``generalized'' eigenstates for these polynomials are obtained.(The eigenvalues are computed as the set of polynomials, each of which has a part of eigenvalues as the roots. The eigenstates of the matrix H are the solutions of the matrix equation p(H)$\cdot$x=0, where p(t) is one of these polynomials.)

Let ``hh'' be the Fock matrix, used in the model computation of the electronic structure of C$_{60}$, with the same parameter as in the previous section. The generators of the group g1,g2,g3 are the same as defined previously. The eigenvalue problem is solved now. The last part of the following computation is to extract the basis set of the eigenspaces. (The secular equation is representable as the product of polynomials equipped the parameter for the transfer integrals. See the appendix \ref{AnalyticSpectra}.)

\begin{verbatim}
Hgen:=function(GP,a)
 local i,h,tuples;
 h:=List([1..60],i->List([1..60],j->0));
 tuples:=Orbit(GP,a,OnTuples);
 for i in tuples
 do 
  h[i[1]][i[2]]:=1;
 od;
 return h;
end;
\end{verbatim}
This function sets non-zero matrix elements $h_{ij}$, when the symmetry of the molecule demand the bond between site ``i'' and ``j''. To do this, the bond defined in the list ``a'', in the argument of this function, is transferred to equivalent bonds in the symmetry of the group ``GP''.

\begin{verbatim}
gap> G:=Group(g1,g2,g3);
gap> h1:=Hgen(G,[1,2]);
gap> h2:=Hgen(G,[1,6]);
gap> hh:=-2*h1-h2;
gap> ge:=GeneralizedEigenvalues(Rationals,hh);
[ x_1-3,  x_1+1,  x_1+5,  x_1^2-4*x_1-1,  x_1^2-2*x_1-9, 
 x_1^2-2*x_1-5,  x_1^3+3*x_1^2-9*x_1-19,  
 x_1^4+6*x_1^3-4*x_1^2-54*x_1-29 ]
gap> gs:=GeneralizedEigenspaces(Rationals,hh);;
gap> gsbas:=List(gs,i->Basis(i));;
\end{verbatim}



The eight polynomials of the eigenvalues, decomposed up to the field of rational number are given as the table \ref{typeofsols}, and the basis vectors in the eigenspace are allotted into the irreducible representations. 

\begin{table}
\begin{tabular}{l l l l l}\hline
No.& Polynomial & Eigenspace  & Degeneracy   & Irreducible \\
   &      & dimension & of one eigenvalue & representation\\\hline
1&\verb!x_1-3!                 &     5 &  5 & I.10\\
2&\verb!x_1+1!                 &     5 &  5 & I.10\\
3&\verb!x_1+5!                 &     1 &  1 & I.1\\
4&\verb!x_1^2-4*x_1-1!         &     6 & 3 & I.3,I.4\\
5&\verb!x_1^2-2*x_1-9!         &     8 & 4 & I.8\\
6&\verb!x_1^2-2*x_1-5!         &     8 & 4 & I.7\\
7&\verb!x_1^3+3*x_1^2-9*x_1-19!&  15 & 3 & I.9\\
8&\verb!x_1^4+6*x_1^3-4*x_1^2-54*x_1-29!&  12 &  3 & I.5,I.6\\\hline
\end{tabular}
\caption{The classification of the solutions of the generalized eigenvalue problem. The irreducible representations are those presented in table \ref{charactertbl}.  }
\label{typeofsols}
\end{table}

The representations of I.3 and I.4 or I.5 and I.6 belong to the same eigenspace, specified by the same polynomial. The automorphism of the cyclotomic field (E(5)+E(5)\verb!^!4$\rightarrow$E(5)\verb!^!2 + E(5)\verb!^!3), which forms a group, interchanges these pairs of the irreducible representation, and consequently, also interchanges the vector solutions x of the equation p(H)$\cdot$x=0. From the viewpoint of the Galois theory, Let us review the situation. Take the last polynomial of the four degrees. The 8-th polynomial in the above table has C4 as the Galois group, and the four roots are transported with each other by the operation of this group( as $\rm 1 \rightarrow 2 \rightarrow 3 \rightarrow 4 \rightarrow 1$, by a certain sequential order of the roots.)
\begin{remark}
The Galois group of the polynomial is computed by GAP.
\begin{verbatim}
gap> ge[8];
x_1^4+6*x_1^3-4*x_1^2-54*x_1-29
gap> GaloisType(ge[8]);
1
gap> TransitiveGroup(4,1);
C(4) = 4
\end{verbatim}
The command Galoistype(f) computes the Galois Group of the polynomial f, and returns an integer i if the Galois group is isomorphic to TransitiveGroup(n, i), where n is the degree of f. 
\end{remark}

Let us represent this Galois group concretely as the operation of the eigenvectors. The polynomial splits in four polynomials of one degree in the 15-th cyclotomic field $\mathbb{Q}_{15}$.
The generalized eigenvalue problem is solved again in this field, for which CF(15) is the notation in GAP.

\begin{verbatim}
gap> ge2:=GeneralizedEigenvalues(CF(15),hh);
[ x_1-3, x_1+1, x_1+5, x_1+(E(5)+3*E(5)^2+3*E(5)^3+E(5)^4), 
  x_1+(3*E(5)+E(5)^2+E(5)^3+3*E(5)^4), 
  x_1+(E(15)+E(15)^2+E(15)^4+3*E(15)^7+3*E(15)^8+E(15)^11+E(15)^13+E(15)^14), 
  x_1+(E(15)+E(15)^2+3*E(15)^4+E(15)^7+E(15)^8+3*E(15)^11+E(15)^13+E(15)^14), 
  x_1+(E(15)+3*E(15)^2+E(15)^4+E(15)^7+E(15)^8+E(15)^11+3*E(15)^13+E(15)^14), 
  x_1+(3*E(15)+E(15)^2+E(15)^4+E(15)^7+E(15)^8+E(15)^11+E(15)^13+3*E(15)^14), 
  x_1^2-2*x_1-9, x_1^2-2*x_1-5, x_1^3+3*x_1^2-9*x_1-19 ]
gap> gs2:=GeneralizedEigenspaces(CF(15),hh);;
gap> gsbas2:=List(gs2,i->Basis(i));;
\end{verbatim}

\begin{table}
\begin{tabular}{l l l l }\hline
No.&Polynomial & Eigenspace   & Irreducible \\
    &      & dimension  & representation\\\hline
1 & \verb!x_1-3!                                     &     5 &   10\\
2 & \verb!x_1+1!                                      &    5 &   10\\
3 & \verb!x_1+5!                                      &    1 &     1\\
4 & \verb!x_1+(E(5)+3*E(5)^2+3*E(5)^3+E(5)^4)!        &    3 &     3\\
5 & \verb!x_1+(3*E(5)+E(5)^2+E(5)^3+3*E(5)^4)!        &    3 &     4\\
6 & \verb!x_1+(E(15)+E(15)^2+E(15)^4+3*E(15)^7!       &    3 &     6\\
  & \verb! +3*E(15)^8+E(15)^11+E(15)^13+E(15)^14)!          \\
7 & \verb!x_1+(E(15)+E(15)^2+3*E(15)^4+E(15)^7!  & 3  &    5\\
  & \verb! +E(15)^8+3*E(15)^11+E(15)^13+E(15)^14)!\\
8 & \verb!x_1+(E(15)+3*E(15)^2+E(15)^4+E(15)^7!  & 3 &     6\\
  & \verb! +E(15)^8+E(15)^11+3*E(15)^13+E(15)^14)!\\
9 & \verb!x_1+(3*E(15)+E(15)^2+E(15)^4+E(15)^7!  & 3 &     5\\ 
  & \verb! +E(15)^8+E(15)^11+E(15)^13+3*E(15)^14)!\\
10 & \verb!x_1^2-2*x_1-9!             & 8      &8\\
11 & \verb!x_1^2-2*x_1-5!             &8&      7\\
12 & \verb!x_1^3+3*x_1^2-9*x_1-19!     &      15&      9\\\hline
\end{tabular}
\caption{The classification of the solutions of the generalized eigenvalue problem in $\mathbb{Q}_{15}$.}
\label{typeofsols2}
\end{table}

\begin{remark}
In fact, the splitting field of the polynomial \verb!x_1^4+6*x_1^3-4*x_1^2-54*x_1-29! is not $\mathbb{Q}_{15}$ ( of the extension of eight degrees  to $\mathbb{Q}$), but $\mathbb{Q}$(a) (of the extension of four degrees  to $\mathbb{Q}$), the latter of which is obtained by adding one of the solution ``a'' of this polynomial to $\mathbb{Q}$. It is equal to $\mathbb{Q}$(\verb!E(15)+E(15)^14!) (denoted as NF(15,[1,14]) in GAP). To be exact, one should discuss in this field. 
\end{remark}


The Galois group of $\mathbb{Q}_{15}$ is computed as.
\begin{verbatim}
gap> GaloisGroup(CF(15));
<group with 2 generators>
gap> g:=GaloisGroup(CF(15));
<group with 2 generators>
gap> g.1;
ANFAutomorphism( CF(15), 11 )
gap> g.2;
ANFAutomorphism( CF(15), 7 )
\end{verbatim}

The generator g.1 is the map, replacing E(15)$\rightarrow$ E(15)\verb!^!11, of order 2, and g.2, E(15)$\rightarrow$ E(15)\verb!^!7, of order 4. The latter generator, in table \ref{typeofsols2}, transports the 6-th eigenspace to the 7-th, the 7-th to the 8-th, in turn, and the 10-th to the 6-th, generating the group C$_4$ ( the forth cyclic group.), realizing the Galois group of the polynomial
\verb!x_1^4+6*x_1^3-4*x_1^2-54*x_1-29! as the following computation verifies.
  
\begin{verbatim}
gap> ge2[6]*ge2[7]*ge2[8]*ge2[9]=ge[8];
true

gap> for j in [1..3] do;    
> Print(RootsOfPolynomial(ge2[6])[1]^g.2=RootsOfPolynomial(ge2[7])[1]);
> Print(RootsOfPolynomial(ge2[7])[1]^g.2=RootsOfPolynomial(ge2[8])[1]);
> Print(RootsOfPolynomial(ge2[8])[1]^g.2=RootsOfPolynomial(ge2[9])[1]);
> Print(RootsOfPolynomial(ge2[9])[1]^g.2=RootsOfPolynomial(ge2[6])[1]);
>od;
truetruetruetruetruetruetruetruetruetruetruetrue

gap> for j in [1..3] do;    
> Print(List(gsbas2[6][j],x->x^g.2)=gsbas2[7][j]);
> Print(List(gsbas2[7][j],x->x^g.2)=gsbas2[8][j]);
> Print(List(gsbas2[8][j],x->x^g.2)=gsbas2[9][j]);
> Print(List(gsbas2[9][j],x->x^g.2)=gsbas2[6][j]);
>od;
truetruetruetruetruetruetruetruetruetruetruetrue
\end{verbatim}

The first part of the above computation shows that the polynomial \verb!ge[8]! (the 8-th solution in the field of rational numbers) splits into \verb!ge2[6],...,ge2[9]!, (the four solutions, from the 6-th to 9-th, in $\mathbb{Q}_{15}$). The second part verifies the C$_4$ group property in the transportation of the eigenvalues (the root of \verb!ge2[6],...,ge2[9]!) , and the third part verifies it in the transportation of the eigenvectors, \verb!gsbas2[6],...,gsbas2[9]!, each of which includes three basis (\verb!gsbas2[X][1],gsbas2[X][2],gsbas2[X][3]!). The irreducible representations are transported as the following sequence by g.2, $\rm I.5 \rightarrow I.6 \rightarrow I.5 \rightarrow I.6$. In this sequence, the same types of the irreducible representations appear two times, but due to the duplicity of these representations in the whole eigenspace (as we have seen the decomposition of the trace representation I.Tr in the previous section) they are two distinct subspaces which belong to the same type of the irreducible representation. Therefore the Galois group has the order 4, running through the four distinct subspaces. The replacement of the eigenspaces by this group is given in table\ref{CF15REPR}, from which one could extract the Galois correspondence between sub-fields and subgroup. 

\begin{table}
\begin{tabular}{l | l l l l l l l l l l l l}\hline
\verb!<identity ...>!& 1& 2& 3& 4& 5& 6& 7& 8& 9& 10& 11& 12 \\
\verb!x2*x1!& 1& 2& 3& 5& 4& 9& 6& 7& 8& 10& 11& 12 \\
\verb!x2^2!& 1& 2& 3& 4& 5& 8& 9& 6& 7& 10& 11& 12 \\
\verb!x2!& 1& 2& 3& 5& 4& 7& 8& 9& 6& 10& 11& 12 \\
\verb!x2^-1*x1!& 1& 2& 3& 5& 4& 7& 8& 9& 6& 10& 11& 12 \\
\verb!x1!& 1& 2& 3& 4& 5& 8& 9& 6& 7& 10& 11& 12 \\
\verb!x2^-1!& 1& 2& 3& 5& 4& 9& 6& 7& 8& 10& 11& 12 \\
\verb!x2^2*x1!& 1& 2& 3& 4& 5& 6& 7& 8& 9& 10& 11& 12 \\\hline
\end{tabular}
\caption{The replacement of 12 eigenspaces by the automorphism in CF(15). The x1 and x2 are the generators of the Galois group, given as g.1 and g.2 in the article.}
\label{CF15REPR}
\end{table}

This example may be regarded as a hidden symmetry in the sense that the wave functions are transformed into each other by the operations of a group, which is different from that of the structural symmetry of the molecule.


We can approach the hidden symmetry in another direction: there is an automorphism on the icosahedral group which interchanges these two irreducible representations. For simplicity, we reduce the problem in A5. (The icosahedral group is constructed from this by the direct product with the inversion.) The generators of A5 are g1 and g2, (given as the permutations in the previous section.) 

\begin{verbatim}
gap> G:=Group(g1,g2);;
gap> Irr(G);;
\end{verbatim}


The character table is as follows (written briefly by a list in GAP).
\begin{verbatim}
[[ 1, 1, 1, 1, 1 ], 
 [ 3, -E(5)-E(5)^4, -E(5)^2-E(5)^3, -1, 0 ], 
 [ 3, -E(5)^2-E(5)^3, -E(5)-E(5)^4, -1, 0 ], 
 [ 4, -1, -1, 0, 1 ], 
 [ 5, 0, 0, 1, -1 ] ]
\end{verbatim}

The conjugacy classes are computed now. (As the list of the classes is given as the permutations, it is transformed to the representations by generators, to make the list as brief as possible. The symbols x1 and x2 represent g1 and g2 respectively.)

\begin{verbatim}
gap> A:=List(ConjugacyClasses(G),Elements);;
gap> B:=List(A,i->List(i,j->Factorization(G,j)));
[ [ <identity ...> ], 
  [ x2^-1, x2, x2*x1*x2, x1*x2, x2^2*x1, x2*x1, x2^-1*x1^-1, x2^-2*x1^-1, 
      x2^-1*x1^-1*x2^-1, x1^-1*x2^-1, x1*x2^2, x1^-1*x2^-2 ], 
  [ x2^-2, x2^2, x2^2*x1*x2^2*x1^-1, x1*x2^2*x1^-1, (x1*x2^2)^2, 
      (x2^-1*x1^-1*x2^-1)^2, (x2*x1*x2)^2, x2^-1*x1^-1*x2^-2*x1*x2, 
      (x2^2*x1)^2, x1^-1*x2^-2*x1, (x2*x1)^2, x2^-1*x1*x2*x1 ], 
  [ x2^-2*x1*x2, x2^-1*x1, x2*x1^-1, x2^2*x1^-1*x2^-1, x2^2*x1*x2^2, 
      x1^-1*x2^-2*x1*x2*x1, x1*x2*(x2*x1)^2, (x2*x1*x2)^2*x1, 
      (x2^2*x1)^2*x2*x1, x2^-1*x1^-1*x2^-2*x1*x2*x1, x2*x1*x2^2*x1^-1, 
      (x2*x1*x2)^2*x2, (x2^2*x1)^2*x2, x2^-1*x1^-1*x2^-2*x1, x1*x2*x1 ], 
  [ x2^2*x1*x2, x2^-1*x1*x2, x2^-2*x1, x1, x2^2*x1^-1, x1^-1, 
      x2^-2*x1^-1*x2^-1, x2*x1^-1*x2^-1, x2^-1*x1^-1*x2^-2, x2*x1*x2^2, 
      x2^-1*x1^-1*x2^-2*x1^-1, x1^-1*x2^-2*x1^-1, x1^-1*x2^-2*x1^-1*x2^-1, 
      x1*x2^2*x1^-1*x2^-1, x1*x2^2*x1*x2, x1^-1*x2^-2*x1*x2, x1*x2^2*x1, 
      x2*x1*x2^2*x1, x2*(x2*x1)^2, x2^-2*x1*x2*x1 ] ]
\end{verbatim}

The group of all automorphism is computed as
\begin{verbatim}
gap> AUALL:=AutomorphismGroup(G);
\end{verbatim}

Take first of the generator, AUALL.1.

\begin{verbatim}
gap> AUALL.1;
[ (1,14,20)(2,15,16)(3,11,17)(4,12,18)(5,13,19)(6,58,22)(7,57,23)
  (8,56,24)(9,60,25)(10,59,21)(26,54,38)(27,53,37)(28,52,36)(29,51,40)
  (30,55,39)(31,44,50)(32,45,46)(33,41,47)(34,42,48)(35,43,49), 
  (1,2,3,4,5)(6,11,16,21,26)(7,12,17,22,27)(8,13,18,23,28)(9,14,19,24,29)
  (10,15,20,25,30)(31,32,33,34,35)(36,41,46,51,56)(37,42,47,52,57)
  (38,43,48,53,58)(39,44,49,54,59)(40,45,50,55,60) ] -> 
[ (1,57,45)(2,56,44)(3,60,43)(4,59,42)(5,58,41)(6,55,47)(7,54,46)(8,53,50)
  (9,52,49)(10,51,48)(11,35,28)(12,34,29)(13,33,30)(14,32,26)
  (15,31,27)(16,37,24)(17,36,25)(18,40,21)(19,39,22)(20,38,23), 
  (1,59,9,16,55)(2,58,10,17,54)(3,57,6,18,53)(4,56,7,19,52)
  (5,60,8,20,51)(11,14,12,15,13)(21,35,30,38,50)(22,34,26,37,49)
  (23,33,27,36,48)(24,32,28,40,47)(25,31,29,39,46)(41,44,42,45,43) ]
\end{verbatim}

This automorphism maps g1 and g2 in the following way.

\begin{verbatim}
gap> Factorization(G,Image(AUALL.1,g1));
x2*(x2*x1)^2
gap> Factorization(G,Image(AUALL.1,g2));
x1^-1*x2^-2*x1
\end{verbatim}

The two conjugacy classes (of the second and of the third) are interchanged with each other, while the other classes are unchanged. This causes the exchange of the character values and that of the two irreducible representation (of No.2 and No.3). Let us verify this. The image of the automorphism on the elements of the conjugacy class 2 is computed and sorted as

\begin{verbatim}
gap> imgCl2:=List(A[2], a ->Image(AUALL.1,a));;
gap> SortedList(List(imgCl2,i->Factorization(G,i)));
[ x2^-2, x2^2, x1^-1*x2^-2*x1, x1*x2^2*x1^-1, x2^-1*x1*x2*x1, (x2*x1)^2, 
  (x1*x2^2)^2, (x2^-1*x1^-1*x2^-1)^2, x2^-1*x1^-1*x2^-2*x1*x2, (x2*x1*x2)^2, 
  x2^2*x1*x2^2*x1^-1, (x2^2*x1)^2 ]
\end{verbatim}

The sorted list of the conjugacy class 3 is: 
\begin{verbatim}
gap> SortedList(B[3]);
[ x2^-2, x2^2, x1^-1*x2^-2*x1, x1*x2^2*x1^-1, x2^-1*x1*x2*x1, (x2*x1)^2, 
  (x1*x2^2)^2, (x2^-1*x1^-1*x2^-1)^2, x2^-1*x1^-1*x2^-2*x1*x2, (x2*x1*x2)^2, 
  x2^2*x1*x2^2*x1^-1, (x2^2*x1)^2 ]
\end{verbatim}

These two lists are identical. As the automorphism AUALL.1 has the order 2, the conjugacy classes 2 and 3, or the irreducible representation (of No.2 and No.3), are interchanged with each other by this automorphism.

As the each element of the group A5 has a correspondence with one of the vertexes of the C$_{60}$ molecule, this automorphism can be interpreted as a geometrical operation, that of the replacement of vertex. But this replacement of atoms does not conserve the bonding structure between nearest neighbors. This means that this replacement of atoms is not the symmetry operation on the C$_{60}$.  

\subsection{The super-symmetries} 

In this section, we will see that we can construct a symmetry higher than the icosahedral one. The icosahedral group is the direct product of A5 and I. If one of the components, the group A5, is extended to a larger group, the direct product is a super-group of the icosahedral one. The bonding structure of the C$_{60}$ is representable by the pairs of the sequential numbers of the two vertexes, and it can be interpreted as the ``orbit'' of these pairs by the action of the icosahedral group. Then the orbit of the pairs of vertex, generated by the action of this new larger group, provides us with the bonding structure in the higher symmetry. The guiding principle is based the following observation: in the previous section, we have seen the existence of a certain group automorphism which interchanges two of the irreducible representation. If a group, including A5 and this group automorphism, could be constructed, the two three-dimensional irreducible representations will be merged into one. And if a molecule obeys this higher symmetry, the electronic structure will show different features, such as sixfold degeneracies. Now, concerning the extension of a group, the concept of the semi-direct product (with the automorphism group) is usable. For this purpose, we should construct a group composed from all of the automorphisms on the group A5. Taking this automorphism group or its subgroups, we can construct the semi-direct product with the group A5. The necessary computation is executed in the following way. As the full automorphism group of A5 is too large, we take a small subgroup of this (a subgroup generated by one of the generators of this group).

\begin{verbatim}
gap> G:=Group(g1,g2);
gap> AUALL:=AutomorphismGroup(G);
gap> p:=SemidirectProduct(Group(AUALL.1),G);
\end{verbatim}

The newly created group ``p'' is larger than A5 (the group ``G''), and contains A5 as a subgroup. The three generators are given as

\begin{verbatim}
gap> p.1;
(3,16,22)(4,17,18)(5,13,19)(6,14,20)(7,15,21)(8,60,24)(9,59,25)(10,58,26)
(11,62,27)(12,61,23)(28,56,40)(29,55,39)(30,54,38)(31,53,42)(32,57,41)
(33,46,52)(34,47,48)(35,43,49)(36,44,50)(37,45,51)
gap> p.2;
(3,7,6,5,4)(8,28,23,18,13)(9,29,24,19,14)(10,30,25,20,15)(11,31,26,21,16)(12,
32,27,22,17)(33,37,36,35,34)(38,58,53,48,43)(39,59,54,49,44)(40,60,55,50,
45)(41,61,56,51,46)(42,62,57,52,47)
gap> p.3;
(1,2)(4,61)(5,11)(6,18)(7,57)(8,21)(9,28)(10,42)(12,46)(13,54)(14,39)
(15,49)(16,59)(17,44)(19,51)(20,26)(22,47)(23,58)(25,34)(27,45)(29,36)
(31,55)(32,43)(33,50)(37,38)(40,48)(41,60)(52,53)
\end{verbatim}

The first and the second generators are inherited from A5, and the third is the new one. They are represented by 62 symbols. The symbols from 3 to 62 correspond to the 60 vertexes in one C$_{60}$. The extension of the group introduces extra symbols ``1'' and ``2''. We can give a geometrical interpretation to this group in the following way. From the 62 symbols, we can construct a bonding system which contains 120 atoms, located in the double C$_{60}$ molecules. The atoms will be indexed as [i,j] (i=3,..,62, and j=1,2) by the duplication of single C$_{60}$. This structure (C$_{120}$) is regulated by the symmetry of the new extended group. The operation p.1 and p.2 are the rotations by 120 and 72 degrees. The operation p.3 interchanges 60 vertexes between two C$_{60}$ shells. To see this, we have only to compute the orbit of the vertex at [I, J] by these operations. The permutation (1,2), applied to the latter element of the pairs, represents the exchange of vertexes between the two C$_{60}$.

By means of a function defined below, the bonding structure in this new symmetry can be set up in matrices (of 120 $\times$ 120 dimension). If one bond is bridged between the vertex [i[1],i[2]] and [i[3],i[4]], the corresponding element in the networking matrix is set to be the unity.

\begin{verbatim}
listforh:=function(A)
 local h,i1,i2,i3,i4,k,l,i;
 h:=List([1..120],i->List([1..120],j->0));
 for i in A do
  i1:=i[1];
  i2:=i[2];
  i3:=i[3];
  i4:=i[4];
  k:=(i1-2)+60*(i2-1);
  l:=(i3-2)+60*(i4-1);
  h[k][l]:=1;
  h[l][k]:=1;
 od;
 return h;
end; 
\end{verbatim}

 The orbits of the double bond [3,1,4,1] (read as from [3,1] to [4,1]), the single bond [3,1,8,1], and the bond between the first and the second C$_{60}$ ( such as [3,1,4,2], a possible example, taken par hazard) are computed into the lists ``A1'',``B1'',``C1'' as bellow. The bonding structures generated from them are represented by three matrices ``a1'',``b1'',``c1''. 
\begin{verbatim}
gap> A1:=Orbit(p,[3,1,4,1],OnTuples);
[ [ 3, 1, 4, 1 ], [ 16, 1, 17, 1 ], [ 4, 1, 5, 1 ], [ 3, 2, 61, 2 ], 
 [ 22, 1, 18, 1 ], [ 21, 1, 22, 1 ], [ 59, 2, 44, 2 ], [ 17, 1, 13, 1 ], 
 [ 5, 1, 6, 1 ], [ 61, 2, 11, 2 ], [ 16, 2, 23, 2 ], [ 4, 2, 41, 2 ],
...(omitted hereafter)...
gap> B1:=Orbit(p,[3,1,8,1],OnTuples);;
gap> C1:=Orbit(p,[3,1,4,2],OnTuples);;
gap> a1:=listforh(A1);;
gap> b1:=listforh(B1);;
gap> c1:=listforh(C1);;
\end{verbatim}

In the C$_{60}$, the symmetry operation of the icosahedral symmetry includes the inversion. The inversion itself is not included in the new group, but we can furnish this operation with first sixty vertexes (from the third to the sixty-second) in the same manner as the icosahedral case. And the inversion in the second 60 vertexes can be written in the following way. The second sixty vertex are transported from the first sixty, by the operation p.3. If the inversions on the first and second 60 vertexes are denoted as I1, I2, they are related by the conjugation $\rm I2:= (p.3)\cdot I1 \cdot (p.3)^{-1}$ (symbolically written here). The inversion on the total system is written by their union as
$$
\left(
\begin{tabular}{ll}
I1 & 0 \\
0  & I2 \\
\end{tabular}
\right)
$$
where the each entry is assumed to be 60$\times$60-dimensional matrices.

The destinations of the vertexes by the inversion ``g3'' in one C$_{60}$ shell are stored in the list bellow.
\begin{verbatim}
gap> ListInv:=List([1..60],i->OnPoints(i,g3)); 
\end{verbatim}
The destinations of the vertexes in the first C$_{60}$ shell to the second one, by the operation ``p.3'', is stored in the list bellow.
\begin{verbatim}
gap> mconjg:=List([3..62],i->OnPoints(i,p.3))-2;
\end{verbatim}
(In the last computation, we must extract 2 from the result, due to the numbering in the atoms, which are numbered from 3 to 62.)

The inversions in the first and the second C$_{60}$ shell are composed now (as matrices ``invrs'', ``invrs2'', and by uniting them, the inversion in the full system is computed, as in the matrix ``ivfull''.

\begin{verbatim}
gap> invrs:=List([1..60],i->List([1..60],j->0));
gap> mc:=List([1..60],i->List([1..60],j->0));
gap> for i in [1..60] do
    invrs[i][ListInv[i]]:=1;
    mc[i][mconjg[i]]:=1;
    od;
gap> invrs2:=mc*invrs*mc^-1;
gap> ivfull:=List([1..120],i->List([1..120],j->0));
gap> for i in [1..60] do
    for j in [1..60] do
    ivfull[i][j]:=invrs[i][j];
    ivfull[i+60][j+60]:=invrs2[i][j];
    od;
    od;    
\end{verbatim}

The generators of the group ``p'', as the symmetry operations of the 120 atoms, are also represented by matrices. At first, we prepare the lists to represent the permutation.

\begin{verbatim}
gap> l1:=List([1..2],j->List([3..62],i->(OnPoints(i,p.1)+60*(j-1))));
gap> l1:=Flat(l1);
gap> l2:=List([1..2],j->List([3..62],i->(OnPoints(i,p.2)+60*(j-1))));
gap> l2:=Flat(l2);
gap> l3:=List([1..2],j->List([3..62],i->(OnPoints(i,p.3)+60*(OnPoints(j,p.3)-1))));
gap> l3:=Flat(l3);
\end{verbatim}

The permutations are transformed to the matrix forms by this function.
\begin{verbatim}
ListToMat:=function(l)
 local size,h,i,j;
 size:=Size(l);
 h:=List([1..size],i->List([1..size],j->0));
 for i in [1..size] do
  h[i][l[i]]:=1;
 od;
 return h;
end;

gap> P1:=ListToMat(l1-2);
gap> P2:=ListToMat(l2-2);
gap> P3:=ListToMat(l3-2);
\end{verbatim}

We must examine the congruence of the computed data, by checking whether the matrices representing the bonding structure are subject to the symmetry of the extended group. For this purpose, we can compute the commutators between these matrices (``a1'',``b1'',``c1'') and the generators of this group (P1, P2, P3 and ivfull): the results should be zero. The matrices ``a1'' and ``b1'' pass the test, but ``c1'' not. As the matrix ``c1'' is not conserved by the matrix of the inversion ``ivfull'' (or not commutable), the symmetrization is necessary for the construction of the Fock matrix. One of the Fock matrices is given as, fixing the values of the transfer integrals,
\begin{verbatim}
gap> hh:=-2*a1-b1-1/2*(c1+ivfull*c1*ivfull);
\end{verbatim}

The generalized eigenvalue problems is solved up to in the field of rational numbers. 
\begin{verbatim}
gap> gev:=GeneralisedEigenvalues(Rationals,hh);
 [ x_1-3, x_1+1, x_1+3, x_1+7, x_1^2-4*x_1-13/4, x_1^2-4*x_1-2, 
 x_1^2-2*x_1-81/4, x_1^2-2*x_1-9/4, x_1^2-12, x_1^2-21/4, 
 x_1^3+5*x_1^2-4*x_1-24, x_1^4+6*x_1^3-6*x_1^2-60*x_1-36 ]
gap> ges:=GeneralisedEigenspaces(Rationals,hh);
 [ <vector space over Rationals, with 10 generators>, 
 <vector space over Rationals, with 15 generators>, 
 <vector space over Rationals, with 1 generators>, 
 <vector space over Rationals, with 1 generators>, 
 <vector space over Rationals, with 8 generators>, 
 <vector space over Rationals, with 12 generators>, 
 <vector space over Rationals, with 8 generators>, 
 <vector space over Rationals, with 8 generators>, 
 <vector space over Rationals, with 10 generators>, 
 <vector space over Rationals, with 8 generators>, 
 <vector space over Rationals, with 15 generators>, 
 <vector space over Rationals, with 24 generators> ]
\end{verbatim}

The group is composed and the irreducible representation is computed now.
\begin{verbatim}
gap> irrg:=Irr(Group(P1,P2,P3,ivfull));
 [ Character( CharacterTable( <matrix group of size 240 with 4 generators> ),
 [ 1, 1, 1, 1, 1, 1, 1, 1, 1, 1, 1, 1, 1, 1 ] ), 
 Character( CharacterTable( <matrix group of size 240 with 4 generators> ),
 [ 1, -1, -1, -1, -1, -1, 1, 1, 1, 1, 1, 1, 1, -1 ] ), 
 Character( CharacterTable( <matrix group of size 240 with 4 generators> ),
 [ 1, -1, -1, -1, 1, 1, 1, -1, 1, -1, 1, -1, -1, 1 ] ), 
 Character( CharacterTable( <matrix group of size 240 with 4 generators> ),
 [ 1, 1, 1, 1, -1, -1, 1, -1, 1, -1, 1, -1, -1, -1 ] ), 
 Character( CharacterTable( <matrix group of size 240 with 4 generators> ),
 [ 4, -2, 1, 0, 1, -2, 0, -1, -1, 1, 1, 0, 4, 0 ] ), 
 Character( CharacterTable( <matrix group of size 240 with 4 generators> ),
 [ 4, 2, -1, 0, -1, 2, 0, -1, -1, 1, 1, 0, 4, 0 ] ), 
 Character( CharacterTable( <matrix group of size 240 with 4 generators> ),
 [ 4, -2, 1, 0, -1, 2, 0, 1, -1, -1, 1, 0, -4, 0 ] ), 
 Character( CharacterTable( <matrix group of size 240 with 4 generators> ),
 [ 4, 2, -1, 0, 1, -2, 0, 1, -1, -1, 1, 0, -4, 0 ] ), 
 Character( CharacterTable( <matrix group of size 240 with 4 generators> ),
 [ 5, 1, 1, -1, 1, 1, 1, 0, 0, -1, -1, 1, 5, -1 ] ), 
 Character( CharacterTable( <matrix group of size 240 with 4 generators> ),
 [ 5, -1, -1, 1, -1, -1, 1, 0, 0, -1, -1, 1, 5, 1 ] ), 
 Character( CharacterTable( <matrix group of size 240 with 4 generators> ),
 [ 5, 1, 1, -1, -1, -1, 1, 0, 0, 1, -1, -1, -5, 1 ] ), 
 Character( CharacterTable( <matrix group of size 240 with 4 generators> ),
 [ 5, -1, -1, 1, 1, 1, 1, 0, 0, 1, -1, -1, -5, -1 ] ), 
 Character( CharacterTable( <matrix group of size 240 with 4 generators> ),
 [ 6, 0, 0, 0, 0, 0, -2, 1, 1, 0, 0, -2, 6, 0 ] ), 
 Character( CharacterTable( <matrix group of size 240 with 4 generators> ),
 [ 6, 0, 0, 0, 0, 0, -2, -1, 1, 0, 0, 2, -6, 0 ] ) ]
\end{verbatim}

\begin{table}[h!]
\begin{tabular}{l|l|l}\hline
Solution & Representation  & Dimension \\\hline
\verb!x_1-3! &   [ 11, 12 ]  & 5\\
\verb!x_1+1! &   [ 10, 11, 12 ] & 5\\
\verb!x_1+3! &   [ 2 ] & 1\\
\verb!x_1+7! &   [ 1 ] & 1\\
\verb!x_1^2-4*x_1-13/4! &   [ 6 ] & 4\\
\verb!x_1^2-4*x_1-2! &   [ 13 ] & 6\\
\verb!x_1^2-2*x_1-81/4! &   [ 8 ] & 4\\
\verb!x_1^2-2*x_1-9/4! &   [ 7 ] & 4 \\
\verb!x_1^2-12! &    [ 10 ] & 5 \\
\verb!x_1^2-21/4! &  [ 5 ]  & 4 \\
\verb!x_1^3+5*x_1^2-4*x_1-24! &   [ 9 ] & 5\\
\verb!x_1^4+6*x_1^3-6*x_1^2-60*x_1-36! &   [ 14 ] & 6\\\hline
\end{tabular}
\caption{The correspondence between the solutions of the generalized eigenvalue problem and the irreducible representations.
The left column shows the polynomial, the center the corresponding irreducible representation by sequential numbers,  and the right the dimension of the irreducible representation.  }
\label{SOLGENEIG}
\end{table}

The four eigenvalues, as the solutions of ``\verb!x_1^4+6*x_1^3-6*x_1^2-60*x_1-36!'', composed by the \verb!<vector space over Rationals, with 24 generators>!, are of the sixfold degeneracies, allotted to the 14th irreducible representation, which we have expected to see in this structure. Although this example is rather an imaginary one, it suggests us the importance of the concept of ``group extension'' in the artificial design of the electronic structure in molecular systems of high symmetries. Usually, we argue the electronic structure from the viewpoint of the splitting of the degeneracies in accordance with the reduction of the symmetry, but we will be able to proceed in the inverse direction by actually composing the ``super-group''.

\subsection{Remarks to this section}

In this section, with the viewpoint of the group theory, we analyze the electronic properties of C$_{60}$. The group theoretical properties of the icosahedral symmetry of C$_{60}$ are computed by a computer algebra program package GAP, and the obtained results are utilized to the analysis. We started in a traditional way in order to classify the eigenstates into irreducible representations. However, the possible applications of the computer algebra are not limited to this sort of the rather classical analysis, since we can construct the group and extract the group theoretical properties more freely and adroitly by the computer algebra than by the utilization of the group theoretical data given in some references. We can construct the all of the subgroups of the icosahedral symmetry. By means of the ``orbitals'' of the movement of the vertex by the symmetry operations, the correspondence between the subgroup and the possible deformation of the molecule can be established. The change in the energy spectra and the existence of the zero or nonzero transition moments could be discussed qualitatively, without massive computation. Furthermore, through the construction of the semi-direct group of the icosahedral group, the structure equipped with a symmetry higher than that of the icosahedral symmetry, can be generated. The interest with relation to the ``molecular design'' is as follows: the higher symmetry in the semidirect product merges some of the irreducible representations into one and gives rise to a higher degeneracy of the energy spectra. These examples suggest the importance of the group theoretical viewpoint in the material design, re-armored (or revived) by the modern computational tools, developed as ``computer algebra'' among mathematicians.

\section{Analysis of vibrational mode in C$_{60}$}
\label{vibrationalmodec60}
In this section, the computation of the vibrational modes of the C$_{60}$ by means of the computer algebra is demonstrated. The C$_{60}$ molecule has the unique symmetry, comparable to one soccer-ball (see Fig.~\ref{fig:C60MOL}). The symmetry of the molecule is described by the icosahedral group, and this group is represented by the permutation of 60 vertexes or the group of rotation matrices operating on them. Owing to this high symmetry, the molecule shows peculiar electronic and dynamic properties, both of scientific and industrial interest, and the vibrational mode might be one of the principal phenomena in governing the quantum dynamics in it. As a tool of the computation, we utilize the software GAP, developed in the field of the pure mathematics\cite{GAPSYSTEM}. This software can construct the symmetry group and compute the irreducible representations and other group-theoretical properties and enable us to put into practice the application in the material science, possibly in a modernized way. The computation presented hereafter will be useful to the working physicists and the students, who want to deepen the understanding of the group theory that is given in the textbooks\cite{DRESS, INUI, BURNSIDE, Ser, Fulton}.

\subsection{The vibrational modes of C$_{60}$}
The atoms and the pentagons in the molecule are provided with sequential numbers as in Fig.~\ref{fig:C60VERTEX} in the preceding section.
The symmetry operations on the spatial vectors on the vertexes of C60 are made by the direct products of the two types of matrices. The first type $R_i$ represents the replacement of 60 vertexes  and the second type  $G_i$ the rotation and the reflection of the spatial vectors. The symmetry operations on the spatial vectors placed into 60 vertexes are represented as this: 
\begin{equation}
O_i = R_i^f \otimes G_i, 
\end{equation}
where $R_i$ and $G_i$ should represent the equivalent operation, and the superscript $f$ is the group automorphism in the group $R$. The matrices generates the icosahedral group, and the trace of $O_i$ provides us with the trace representation $\chi^{tr}(O_i)$(not irreducible). The decomposition of the trace representation by means of the irreducible representations indicates the distinct modes of the vibration, and the projection operators composed by $O_i$ can generate the basis vectors for them. And to obtain the vibrational mode, the degrees of the freedoms of the translation and the rotation of the molecule should be removed:
\begin{equation}
\chi^{tr} - \chi^{Trans} - \chi^{Rot}.
\end{equation} 
The presence of the automorphism $f$ is inevitable. Consider the operation $r=(1,2,3,4,5)\cdots(\cdots)$. This operation, as the rotation $G$, acts on the dislocated vertexes ($ V([j]\rightarrow V[j]+ d[j]$):
\begin{eqnarray*}
   G(V[1]+d[1])&=&V[2]+G\cdot d[1]\\
   G(V[2]+d[2])&=&V[3]+G\cdot d[2]\\
   G(V[3]+d[3])&=&V[4]+G\cdot d[3]\\
   G(V[4]+d[4])&=&V[5]+G\cdot d[4]\\
   G(V[5]+d[5])&=&V[1]+G\cdot d[5]\\
   \cdots\cdots\cdots&\cdots&\cdots\cdots\cdots.
\end{eqnarray*}
The operation on $d[i]$ should be represented as   
\begin{eqnarray*}
\left(\begin{array}{cccccc} 0 & 0 & 0 & 0 & G & \cdots \\ G & 0 & 0 & 0 & 0 &\cdots \\ 0 & G & 0 & 0 & 0 & \cdots  
\\ 0 & 0 & G & 0 & 0 &\cdots \\ 0 & 0 & 0 & G & 0 & \cdots \\ 
\vdots & \vdots & \vdots & \vdots & \vdots \end{array}\right)
\left(\begin{array}{c} d[1] \\ d[2] \\ d[3] \\ d[4] \\ d[5] \\ \vdots \end{array}\right)
\end{eqnarray*}
 so that the rotation to $d[j]$ should be located at the correct position, while the permutation r is represented in the form which is the transposition of that operation:
\begin{eqnarray*}
\left(\begin{array}{cccccc} 0 & 1 & 0 & 0 & 0 & \cdots\\ 0 & 0 & 1 & 0 & 0 & \cdots \\ 0 & 0 & 0 & 1 & 0 & \cdots \\ 
0 & 0 & 0 & 0 & 1 &\cdots \\ 1 & 0 & 0 & 0 & 0 & \cdots \\ 
\vdots & \vdots & \vdots  & \vdots &\vdots & \end{array}\right)
\end{eqnarray*}
It happens that the permutation $r$ should act on $d[j]$ as its transposition, or its inverse. And the product of the operations is given by $(R_1,G_1)\circ(R_2,G_2)=(R_2\cdot R_1,G_1\cdot G_2)$, with inverse directions for $R_i$ and $G_i$. This observation shows the inadequacy of the simple definition of the direct product $R_i \otimes  G_i$ as the symmetry operation. However, the naive redefinition, such as $R_i^{-1}\otimes G_i$ for all $i$, does not work well. Now we can make use the fact the conjugation ($\rm A\rightarrow M \cdot A \cdot M^{-1}$) is an automorphism which preserves the sequential order of the group multiplication; we should find the elements which inverts the generators of the group by the conjugation in order to define the exact symmetry operations, the products of which go from right to left both for $R^f_i$ and $G_i$. Indeed this can be done in the final stage of the computation. So, before stepping into it, let us execute the naive computation (with the direct product $R_i\otimes G_j$). Concerning the dimension counting of the basis set, such a computation does not fail.  

The icosahedral group is defined in several ways. 
We prepare the twelve points, located at the centers of the pentagons in the molecule (or the vertexes in the icosahedron) as follows. E(5) is the fifth root of the unity, $\exp(2\pi i/5)$, and the symbolically presented data are all real. (The data and results are given by the expressions in GAP.)
\begin{verbatim}
1   [0,0,1],
2   [2/5*E(5)-2/5*E(5)^2-2/5*E(5)^3+2/5*E(5)^4,
     0,
     1/5*E(5)-1/5*E(5)^2-1/5*E(5)^3+1/5*E(5)^4],
3   [-3/5*E(5)-2/5*E(5)^2-2/5*E(5)^3-3/5*E(5)^4,
      1/5*E(20)-1/5*E(20)^9-2/5*E(20)^13+2/5*E(20)^17,
      1/5*E(5)-1/5*E(5)^2-1/5*E(5)^3+1/5*E(5)^4],
4   [2/5*E(5)+3/5*E(5)^2+3/5*E(5)^3+2/5*E(5)^4,
     2/5*E(20)-2/5*E(20)^9+1/5*E(20)^13-1/5*E(20)^17,
     1/5*E(5)-1/5*E(5)^2-1/5*E(5)^3+1/5*E(5)^4],
5   [2/5*E(5)+3/5*E(5)^2+3/5*E(5)^3+2/5*E(5)^4,
    -2/5*E(20)+2/5*E(20)^9-1/5*E(20)^13+1/5*E(20)^17,
     1/5*E(5)-1/5*E(5)^2-1/5*E(5)^3+1/5*E(5)^4],
6   [-3/5*E(5)-2/5*E(5)^2-2/5*E(5)^3-3/5*E(5)^4,
     -1/5*E(20)+1/5*E(20)^9+2/5*E(20)^13-2/5*E(20)^17,
      1/5*E(5)-1/5*E(5)^2-1/5*E(5)^3+1/5*E(5)^4],
7   [0,0,-1],
8   [-2/5*E(5)+2/5*E(5)^2+2/5*E(5)^3-2/5*E(5)^4,
      0,
     -1/5*E(5)+1/5*E(5)^2+1/5*E(5)^3-1/5*E(5)^4],
9   [3/5*E(5)+2/5*E(5)^2+2/5*E(5)^3+3/5*E(5)^4,
    -1/5*E(20)+1/5*E(20)^9+2/5*E(20)^13-2/5*E(20)^17,
    -1/5*E(5)+1/5*E(5)^2+1/5*E(5)^3-1/5*E(5)^4],
10 [-2/5*E(5)-3/5*E(5)^2-3/5*E(5)^3-2/5*E(5)^4,
    -2/5*E(20)+2/5*E(20)^9-1/5*E(20)^13+1/5*E(20)^17,
    -1/5*E(5)+1/5*E(5)^2+1/5*E(5)^3-1/5*E(5)^4],
11  [-2/5*E(5)-3/5*E(5)^2-3/5*E(5)^3-2/5*E(5)^4,
      2/5*E(20)-2/5*E(20)^9+1/5*E(20)^13-1/5*E(20)^17,
     -1/5*E(5)+1/5*E(5)^2+1/5*E(5)^3-1/5*E(5)^4],
12  [3/5*E(5)+2/5*E(5)^2+2/5*E(5)^3+3/5*E(5)^4,
     1/5*E(20)-1/5*E(20)^9-2/5*E(20)^13+2/5*E(20)^17,
    -1/5*E(5)+1/5*E(5)^2+1/5*E(5)^3-1/5*E(5)^4]
\end{verbatim}
These twelve points are transported among themselves by these three operations (g1,g2,g3). 
The three operations (g1,g2,g3) induces the three permutations (f1,f2,f3) on the 12 vertexes of the icosahedron, and(r1,r2,r3) on 60 vertexes in the C$_{60}$. 

\begin{verbatim}
gap> g1:=[ [ 3/10*E(5)+1/5*E(5)^2+1/5*E(5)^3+3/10*E(5)^4, 
 1/10*E(20)-1/10*E(20)^9-1/5*E(20)^13+1/5*E(20)^17,
 2/5*E(5)-2/5*E(5)^2-2/5*E(5)^3+2/5*E(5)^4 ], 
[ -1/2*E(20)+1/2*E(20)^9, -1/2*E(5)-1/2*E(5)^4, 0 ], 
[ -3/5*E(5)-2/5*E(5)^2-2/5*E(5)^3-3/5*E(5)^4, 
-1/5*E(20)+1/5*E(20)^9+2/5*E(20)^13-2/5*E(20)^17, 
1/5*E(5)-1/5*E(5)^2-1/5*E(5)^3+1/5*E(5)^4 ] ];

gap> g2:=[ [ 1/2*E(5)+1/2*E(5)^4, 1/2*E(20)-1/2*E(20)^9, 0 ], 
[ -1/2*E(20)+1/2*E(20)^9, 1/2*E(5)+1/2*E(5)^4, 0 ], [ 0, 0, 1 ] ];

gap> g3:=[ [ -1, 0, 0 ], [ 0, -1, 0 ], [ 0, 0, -1 ] ];

gap> G:=Group(g1,g2,g3);
\end{verbatim}

\begin{verbatim}
gap> f1:=(1,2,3)(4,6,11)(5,10,12)(7,8,9);
gap> f2:=(2,3,4,5,6)(8,9,10,11,12);
gap> f3:=(1,7)(2,8)(3,9)(4,10)(5,11)(6,12);
gap> F:=Group(f1,f2,f3);
\end{verbatim}

\begin{verbatim}
gap> r1:=(  1, 14, 20)(  2, 15, 16)(  3, 11, 17)(  4, 12, 18)(  5, 13, 19)
(  6, 58, 22)(  7, 57, 23)(  8, 56, 24)(  9, 60, 25)( 10, 59, 21)
( 31, 44, 50)( 32, 45, 46)( 33, 41, 47)( 34, 42, 48)( 35, 43, 49)
( 36, 28, 52)( 37, 27, 53)( 38, 26, 54)( 39, 30, 55)( 40, 29, 51);

gap> r2:=(  1,  2,  3,  4,  5)(  6, 11, 16, 21, 26)(  7, 12, 17, 22, 27)
(  8, 13, 18, 23, 28)(  9, 14, 19, 24, 29)( 10, 15, 20, 25, 30)( 31, 32, 33, 34, 35)
( 36, 41, 46, 51, 56)( 37, 42, 47, 52, 57)( 38, 43, 48, 53, 58)( 39, 44, 49, 54, 59)
( 40, 45, 50, 55, 60);

gap> r3:=(  1, 31)(  2, 32)(  3, 33)(  4, 34)(  5, 35)(  6, 36)(  7, 37)(  8, 38)
(  9, 39)( 10, 40)( 11, 41)( 12, 42)( 13, 43)( 14, 44)( 15, 45)( 16, 46)( 17, 47)
( 18, 48)( 19, 49)( 20, 50)( 21, 51)( 22, 52)( 23, 53)( 24, 54)( 25, 55)( 26, 56)
( 27, 57)( 28, 58)( 29, 59)( 30, 60);

gap> R:=Group(r1,r2,r3);
gap> Irr(R);
\end{verbatim}

The command ``Irr(R)'' returns the character table of R, a part of which is listed in table \ref{CHARACTERS}:

\begin{table}[H] 
\centering
\begin{tabular}{l|l|l|l}
Rep &  Dim   &     [f1,f2,f3] & \\\hline
N.1  &  1 &   \verb![ 1, 1, 1 ]!               &         A$_g$ \\
N.2  &  1 &   \verb![ 1, 1, -1 ]!              &         A$_u$  \\
N.3  &  3 &   \verb![ 0,-E(5)-E(5)^4,  3 ]!    &         F$_{2g}$  \\
N.4  &  3 &   \verb![ 0,-E(5)^2-E(5)^3,  3 ]!  &         F$_{1g}$ \\
N.5  &  3 &   \verb![ 0,-E(5)-E(5)^4,  -3 ]!   &         F$_{2u}$\\
N.6  &  3 &   \verb![ 0,-E(5)^2-E(5)^3, -3 ]! &          F$_{1u}$\\
N.7  &  4 &   \verb![ 1, -1, 4 ]!         &              H$_g$\\
N.8  &  4 &   \verb![ 1, -1, -4 ]!        &              H$_u$\\
N.9  &  5 &   \verb![ -1, 0, 5 ]!        &              G$_g$ \\
N.10 &  5 &   \verb![ -1, 0, -5 ]!       &               G$_u$\\
\end{tabular}
\caption{The characters of the generators in ten irreducible representations of the icosahedral group.}
\label{CHARACTERS}
\end{table}

We use the group R as the standard. The isomorphism from R to G is set up at the three generators (from [R.1,R.2,R.3] to [G.1,G.2,G.3].)
\begin{verbatim}
gap> hom:=GroupHomomorphismByImages(R,G,[R.1,R.2,R.3],[G.1,G.2,G.3]);
\end{verbatim}

It is necessary to judge the irreducible representations to which the translation belongs. The projectors (applicable to three-dimensional spatial vectors) are computed now for this purpose. They are defined as
$$
P^{(p)}=\frac{l(p)}{|G|}\sum_{T \in G}\chi^{(p)*}(T)\cdot O(T)
$$
by summing up the elements $T$ of the group $G$, where $l(p)$ is the dimension of the p-th irreducible representation, $|G|$ the order of the group,  $\chi$(p) the character, $O(T)$ the operation of the group. The gap script is given as:
\begin{verbatim}
Projopr:=function(irr,Em,hom)
#
# irr: Irreducible representation of the group.
# Em: Elements in the group.
# hom: Homomorphism of the permutation group to the matrix group
#
return List(irr,i->Sum(List(Em,j->j^i*Image(hom,j)))/Size(Em)*i[1]);
end;
\end{verbatim}
In this function, the elements ``j'' (in the list ``Em'') in the given group are mapped to another group by the homomorphism ``hom''. The characters are computed in the first group (with the irreducible representation ``irrep'') and multiplied to the images of ``j'' in the second group. The summation over them produces the projector. 
In this definition, the projectors of the matrix group G are composed through the isomorphism from the permutation group R to G, not directly from G. This is a precaution to avoid a possible confusion, since the character tables of R and G, computed by GAP, might be presented in the slightly different orders, even if these two groups are isomorphic.  
The result is given in table \ref{ProjVEC}. The non-zero one exists only in the representation N.6, so we conclude the spatial translation belongs to this representation.

\begin{table}[h] 
\begin{tabular}{lc}
          &      Operator                            \\
N.1       &      [ [ 0, 0, 0 ], [ 0, 0, 0 ], [ 0, 0, 0 ] ] \\
N.2       &      [ [ 0, 0, 0 ], [ 0, 0, 0 ], [ 0, 0, 0 ] ] \\
N.3       &      [ [ 0, 0, 0 ], [ 0, 0, 0 ], [ 0, 0, 0 ] ] \\
N.4       &      [ [ 0, 0, 0 ], [ 0, 0, 0 ], [ 0, 0, 0 ] ] \\
N.5       &      [ [ 0, 0, 0 ], [ 0, 0, 0 ], [ 0, 0, 0 ] ] \\
N.6       &      [ [ 1, 0, 0 ], [ 0, 1, 0 ], [ 0, 0, 1 ] ] \\
N.7       &      [ [ 0, 0, 0 ], [ 0, 0, 0 ], [ 0, 0, 0 ] ] \\
N.8       &      [ [ 0, 0, 0 ], [ 0, 0, 0 ], [ 0, 0, 0 ] ] \\
N.9       &      [ [ 0, 0, 0 ], [ 0, 0, 0 ], [ 0, 0, 0 ] ] \\
N.10       &      [ [ 0, 0, 0 ], [ 0, 0, 0 ], [ 0, 0, 0 ] ] \\
\end{tabular}
\caption{The projection operators for the spatial vectors.}
\label{ProjVEC}.
\end{table}

There is another necessity to judge the representation of the spatial rotations. The group operation T on the rotation matrix M is defined as the product from both sides, $\rm T \cdot M \cdot T^t$. This operation should be treated with care. This product is also a rotation matrix, but the summation of such products not. Thus the projection operator defined as above will not generate the proper rotation matrix if it is naively applied to the rotation matrices. Concerning this, the exponential map is useful, which is defined as
$$\rm
    M=\exp(A),
$$
$$\rm
    T \cdot M \cdot T^t  = \exp (T \cdot A \cdot T^t). 
$$
The pre-image A of the rotation M is the skew-symmetric matrix. The projection operator should be applied to the linear space of these pre-images. The gap computation for the determination of the irreducible representation is now presented. The projectors are computed by the formula
$$ 
P^{(p)}(A)=\frac{l(p)}{|G|}\sum_{T \in G} \chi^{(p)*}(T)  T\cdot A \cdot T^t
$$
by means of the function 
\begin{verbatim}
ProjoprR:=function(irr,Em,MM,hom)
 return List(irr,i->Sum(List(Em,j->j^i*Image(hom,j)
    *MM*TransposedMat(Image(hom,j))))/Size(Em)*i[1]);
end;
\end{verbatim}

The computation is done now.
\begin{verbatim}
gap> x:=Indeterminate(Rationals,"x");
gap> y:=Indeterminate(Rationals,"y");
gap> z:=Indeterminate(Rationals,"z");
gap> MM:=[[0*x,x,y],[-x,0*x,z],[-y,-z,0*x]]; 
gap> ProjoprR(Irr(R),Elements(R),MM,hom);
\end{verbatim}
\begin{remark}
The expressions such as 0 and 0*x signify different objects. The former is the zero as a number, and the latter the zero monomial. If one promiscuously uses 0 and 0*x, it might cause some errors.
\end{remark}
%

\begin{table}[h] 
\begin{tabular}{lc}
          &      Operator                            \\
N.1       &      [ [ 0, 0, 0 ], [ 0, 0, 0 ], [ 0, 0, 0 ] ] \\
N.2       &      [ [ 0, 0, 0 ], [ 0, 0, 0 ], [ 0, 0, 0 ] ] \\
N.3       &      [ [ 0, 0, 0 ], [ 0, 0, 0 ], [ 0, 0, 0 ] ] \\
N.4       &      [ [ 0, x, y ], [-x, 0, z ], [-y, -z, 0 ] ] \\
N.5       &      [ [ 0, 0, 0 ], [ 0, 0, 0 ], [ 0, 0, 0 ] ] \\
N.6       &      [ [ 0, 0, 0 ], [ 0, 0, 0 ], [ 0, 0, 0 ] ] \\
N.7       &      [ [ 0, 0, 0 ], [ 0, 0, 0 ], [ 0, 0, 0 ] ] \\
N.8       &      [ [ 0, 0, 0 ], [ 0, 0, 0 ], [ 0, 0, 0 ] ] \\
N.9       &      [ [ 0, 0, 0 ], [ 0, 0, 0 ], [ 0, 0, 0 ] ] \\
N.10       &      [ [ 0, 0, 0 ], [ 0, 0, 0 ], [ 0, 0, 0 ] ] \\
\end{tabular}
\caption{The projection operators for the rotations.}
\label{ProjROT}.
\end{table}

The non-zero projector is the fourth one. The spatial rotations belong to the fourth representation (N.4).

\begin{remark}
In the basis vectors of the vibrational mode computed hereafter, those which belong to the fourth irreducible representation, include the elements such as
\begin{eqnarray*}
\left(\begin{array}{cccc} I_i & 0 & 0 & \cdots\\ 0 & I_i & 0 & \cdots \\ 0 & 0 & I_i & \cdots \\ \vdots & \vdots & \vdots & \end{array}\right) \cdot v
\end{eqnarray*}
where $v$ are the basis vector in the trivial irreducible representation (invariant by the symmetry operation), and $I_i (i=1,2,3)$ are
\begin{eqnarray*}
I_1=\left(\begin{array}{ccc} 0 & 1 & 0  \\ -1 & 0 & 0   \\ 0 & 0 & 0  \end{array}\right),
I_2=\left(\begin{array}{ccc} 0 & 0 & 1  \\ 0 & 0 & 0   \\ -1 & 0 & 0  \end{array}\right),
I_3=\left(\begin{array}{ccc} 0 & 0 & 0  \\ 0 & 0 & 1   \\ 0 & -1 & 0  \end{array}\right). 
\end{eqnarray*}

These vectors represent the degrees of freedom of the rotation. They are set of 60 of three-dimensional vectors, defined on each vertex, and located in the tangent plane to the sphere where the 60 vertexes are placed.
\end{remark}

Now we construct the symmetry operation $(O_i = R_i \otimes G_i$) on the spatial vectors on the 60 vertexes. The following function converts a permutation $R_i$ to an integer matrix which represents the replacement of vertexes.
\begin{verbatim}
ListToMat:=function(l)
 local size,h,i,j;
 size:=Size(l);
 h:=List([1..size],i->List([1..size],j->0));
 for i in [1..size] do
  h[i][l[i]]:=1;
 od;
 return h;
end;
\end{verbatim}

The following function constructs the direct product of $R_i$ and $G_i$. 
The isomorphism ``iso'' from the group R to G convert an element ``rr'' (in R) to the equivalence in G exactly. In the generated matrix, the non-zero entries in $R_i$ are replaced by $G_i$ (the rotation), as this.
\begin{equation}
R_i = \left(\begin{array}{cccccc}
0 & 1 & 0 & \cdots &0\\\nonumber
1 & 0 & 0 & \cdots &0\\\nonumber
\vdots & & & & \\\nonumber
0 & 0 & 0 & \cdots &1
\end{array}\right)
\rightarrow
R_i \otimes G_i= \left(\begin{array}{cccccc}
0 & G_i & 0 & \cdots &0\\\nonumber
G_i & 0 & 0 & \cdots &0\\\nonumber
\vdots & & & & \vdots \\\nonumber
0 & 0 & 0 & \cdots & G_i
\end{array}\right)
\end{equation}

\begin{verbatim}
LargeMatrix:=function(rr,iso)
 local l,rmat,rrimg,i,i1,i2,n,m;
 l:=ListPerm(rr,60);
 rmat:=List([1..Size(l)*3],i->List([1..Size(l)*3],j->0));
 rrimg:=Image(iso,rr);
 for i in [1..Size(l)]
 do 
  i1:=i-1;i2:=l[i]-1;
  for n in [1..3]
  do
   for m in [1..3]
   do
    rmat[3*i1+n][3*i2+m]:=rrimg[n][m];
   od;
  od;
 od;
 return rmat;
end;
\end{verbatim}
This function at first prepares a $180\times180$ matrix, in order to represent the symmetry operation on the vectors at the vertex of C$_{60}$. The symmetry operation of the group element ``rr''  is transformed by  the isomorphism ``iso'' to $G_i$ and placed in the large matrix, by means of the list ``l'' which stores the data of the permutation $R_i$.

The projector function, composed from the direct product, is defined as:
\begin{verbatim}
ProjoprE:=function(irr,Em,hom)
#
# irr:Irreducible representation.
# Em :Elements of the group R.
# hom:Isomorphism from the group R to the group G.
#
return List(irr,i->Sum(List(Em,j->j^i*LargeMatrix(j,hom)))/Size(Em)*i[1]);
end;
\end{verbatim}

We can make now the basis set of the mode of deformation. The projectors are placed on the list of ten matrices (of dimension 180) (PRJE[1]...PRJE[10]), and each of them does the projection on each irreducible representations (N.1...N.10). The basis vectors for the each vibrational mode are generated from the column vectors of the corresponding projector. In the computation, the list ``BASSIZE'' gives us the dimensions of each mode.

\begin{verbatim}
gap> PRJE:=ProjoprE(Irr(R),Elements(R),hom);;
gap> BAS:=List(PRJE,i->BaseMat(TransposedMat(i)));;
gap> BASSIZE:=List(BAS,Size);
[ 2, 1, 12, 12, 15, 15, 24, 24, 40, 35 ]
gap> BAS;
[ [ [ 1, E(20)-E(20)^9+E(20)^13-E(20)^17, 0, 2*E(5)+E(5)^2+E(5)^3+2*E(5)^4, 
          -E(20)^13+E(20)^17, 0, -2*E(5)-2*E(5)^4, 0, 0, 
          2*E(5)+E(5)^2+E(5)^3+2*E(5)^4, E(20)^13-E(20)^17, 0, 1, 
          -E(20)+E(20)^9-E(20)^13+E(20)^17, 0, 
          -1/5*E(5)+1/5*E(5)^2+1/5*E(5)^3-1/5*E(5)^4, 
          1/5*E(20)-1/5*E(20)^9+3/5*E(20)^13-3/5*E(20)^17, 
          12/5*E(5)+8/5*E(5)^2+8/5*E(5)^3+12/5*E(5)^4, 
          -6/5*E(5)-4/5*E(5)^2-4/5*E(5)^3-6/5*E(5)^4, 
          ...........................................
          ...........................................
\end{verbatim}

\begin{remark}
One might demand numerical basis as well as symbolic one. The latter can be easily modified into quasi-Fortran form. For example, by replacements of strings, a formula such as
\begin{verbatim}
f:=-1/5*E(5)+1/5*E(5)^2+1/5*E(5)^3-1/5*E(5)^4; 
\end{verbatim}
is rewritten as
\begin{verbatim}
      FUNCTION E(I)
      INTEGER I
      COMPLEX*16 E
      REAL*8 TWOPI
      TWOPI=8.0D0*ATAN(1.0D0)
      E=DCMPLX(COS(TWOPI/I),SIN(TWOPI/I))
      RETURN
      END

      SUBROUTINE(f)
      COMPLEX*16 f,E
      f=-1./5*E(5)+1./5*E(5)**2+1./5*E(5)**3-1./5*E(5)**4
      RETURN
      END
\end{verbatim}

\end{remark}

At present projectors ``PRJE[i]'' are defined by the ``bare'' forms, using the direct product $R_i\otimes  G_i$. and they should be modified to the correct ones, such as $\rm M_f \cdot PRJE[i] \cdot M_f^{-1}$, composed by the direct product $R_i^f\otimes G_i$, where $f$ is a conjugation which inverts the generators of the permutation group R.  There are two candidates for such automorphisms.

\begin{verbatim}
gap>RPC:=Filtered(Elements(R),x->List([R.1,R.2,R.3],i->i^x)=[R.1^-1,R.2^-1,R.3^-1]);
[ (1,4)(2,3)(6,21)(7,25)(8,24)(9,23)(10,22)(11,16)(12,20)(13,19)(14,18)(15,
    17)(27,30)(28,29)(31,34)(32,33)(36,51)(37,55)(38,54)(39,53)(40,52)(41,
    46)(42,50)(43,49)(44,48)(45,47)(57,60)(58,59), 
  (1,34)(2,33)(3,32)(4,31)(5,35)(6,51)(7,55)(8,54)(9,53)(10,52)(11,46)(12,
    50)(13,49)(14,48)(15,47)(16,41)(17,45)(18,44)(19,43)(20,42)(21,36)(22,
    40)(23,39)(24,38)(25,37)(26,56)(27,60)(28,59)(29,58)(30,57) ]
gap>  RPC[1]*r3=RPC[2];
true
\end{verbatim}

The conjugation is computed as a matrix through this function.
\begin{verbatim}
ConjugMatrix:=function(rr)
 local l,rmat,vmat,i,i1,i2,n,m;
 l:=ListPerm(rr,60);
 vmat:=List([1..Size(l)*3],i->List([1..Size(l)*3],j->0));
 for i in [1..Size(l)]
 do 
  i1:=i-1;i2:=l[i]-1;
  for n in [1..3]
  do
   vmat[3*i1+n][3*i2+n]:=1;
  od;
 od;
 return vmat;
end;

gap> PCM:=ConjugMatrix(RPC[1]);
gap> PRJEC:=List(PRJE,i->i^PCM);;
gap> BASC:=List(PRJEC,i->BaseMat(TransposedMat(i)));;
gap> BASCSIZE:=List(BASC,Size);
[ 2, 1, 12, 12, 15, 15, 24, 24, 40, 35 ]
\end{verbatim}
In fact, in order to obtain the correct vibrational modes, it is enough to multiply the matrix ``PCM'' for the conjugation to the basis vectors ``BAS'' computed by the bare projectors. If we neglect such corrections, it  happens that the ``breathing mode'' would not belong to the trivial representation.

To check the validity, the characters of $\chi^{Tr}$ and those by the irreducible representations are computed in the two lists ``CTRC'' and ``CIRR''. The irreducible representation of $\chi^{Vec}$ is located at the sixth component of CIRR, and that of $\chi^{Rot}$ is at the fourth.
\begin{verbatim}
CTRC:=List(Elements(R),i->Trace(ListToMat(ListPerm(i,60))));
CIRR:=List(Irr(R),r->List(Elements(R),i->i^r));
\end{verbatim}

The $\chi^{Tr}\times\chi^{Vec}$ is computed and decomposed. 
\begin{verbatim}
gap> List([1..Size(R)],i->CTRC[i]*CIRR[6][i])*TransposedMat(CIRR)/Size(R);
[ 2, 1, 4, 4, 5, 5, 6, 6, 8, 7 ]
gap> List([1..10],i->BASSIZE[i]/Irr(R)[i][1]);
[ 2, 1, 4, 4, 5, 5, 6, 6, 8, 7 ]
\end{verbatim}
These results are the multiplicities to each irreducible representation (in theory and in computation) and shows the validity of the computation. The direct product of the characters (i.e. the vibrational mode plus extra terms) is  decomposed as $$\chi^{tr}\times\chi^{Vec} = 2\cdot{\rm N.1} + 1\cdot{\rm N.2} + 4\cdot{\rm N.3} + 4\cdot{\rm N.4} + 5\cdot{\rm N.5} + 5\cdot{\rm N.6} + 6\cdot{\rm N.7} + 6\cdot{\rm N.8} + 8\cdot{\rm N.9} + 7\cdot{\rm N.10}.$$

\subsection{One technical problem}
The modes of vibrations are decomposed into irreducible representations, some of which appear with multiplicities. Take one of the basis vectors, computed in the previous section, and apply the symmetry operations on it: for example, one of those which belong to the tenth irreducible representation. In the following computation the two lists ``EMR'' and ``AR'' contains the group elements in R and the whole of the symmetry operation, ``vs'' is the set of the generated vectors, ``vsb'', the basis set to it, and the command RankMat(...) computes the dimension of the vector space.
\begin{verbatim}
gap> EMR:=Elements(R) ;;
gap> AR:=List(Elements(R),j->LargeMatrix(j,hom));;
gap> vs:=List(AR,x->x*BAS[10][1]);
gab> vsb:=BaseMat(vs);
gap> RankMat(vs);
25
\end{verbatim}

The chosen vector generates the vector space of dimension 25 by the symmetry operation.(The basis vectors generated in this way are ``vertex-wise'' orthogonal to those in the trivial representation: the basis vectors are the set of 60 small three-dimensional vectors, and the small vectors of the former and the latter are orthogonal at each vertex. The vectors, ``vertex-wise'' orthogonal to those in the trivial representation, make the subspace that includes several copies of the proper irreducible space.) But the tenth irreducible representation is of dimension 5. The generated vectors span the copies of this irreducible space, not confined to one of them. For the practical purpose, we should remove the degrees of freedoms which correspond to those of the translation and the rotation. And it is enough to the reset the dislocated origin or the disorientated axes at the proper positions when we make the structures altered by vibrations. Thus the technical problem of this kind might not be of importance. However, we can divide the multiplied irreducible spaces, so that the each component is proper minimal one, separated with each other, kept invariant under the symmetry operation. In this section, two kinds of such constructions are demonstrated.

The first is to use the projectors composed from matrix representations $D^{(p)}_{kl}(O_T)$, not from characters, as was explained in the previous section. The formula is given again:
\begin{equation*}
P^{(p)}_{kl}=\frac{l_p}{|G|}\sum_{T \in G} D^{(p)*}_{kl}(O_T) \cdot O_T.
\end{equation*}
The basis vectors are transformed by the matrix representation, forming the subspace which has the proper dimension of the irreducible representation. We can make the plural irreducible spaces from the column vectors of the projector one of the $P^{(p)}_{kk}$, the diagonal part of the projectors. Unfortunately, even in this construction, the generated irreducible spaces overlap with each other, not being separated. Thus we should make the separation, projecting out overlapping subspaces. At first the matrix representation is computed in the following way. (The GAP, in each run, returns the different results concerning the matrix representation, since we can determine them up to the isomorphism, but not furthermore; there are plural possibilities.)

\begin{verbatim}
gap> irrR:=Irr(R);;
gap> irrRM:=IrreducibleRepresentations(R);;
gap> EMR:=Elements(R);;
gap> AM:=List(EMR,j->LargeMatrix(j,hom));;
\end{verbatim}

The projector $P^{(p)}_{ij}$ is defined as a function
\begin{verbatim}
PRJR:=function(i,j,p)
 local C,REP,dim;
 REP:=irrRM[p];
 dim:=Trace(()^REP);
 C:=List([1..Size(AM)],k->AsList(EMR[k]^REP)[i][j]*AM[k]);
 return ComplexConjugate(Sum(C))/Size(AM)*dim;
end;
\end{verbatim}

Take the $P^{(10)}_{1,1}$ component of the projector. The basis vectors are generated from column vectors of this. As the GAP software defines the vector set by row vectors, The basis sets are computed from the transposed matrix. The basis set of the vector space is listed in ``vtb''. The projectors $P^{(10)}_{1,1},...,P^{(10)}_{5,1}$ are stored now for the afterward usage.(As for the matrix representations, see Appendix \ref{FOURDIM}.)

\begin{verbatim}
gap> vt:=TransposedMat(PRJR(1,1,10));;
gap> vtb:=BaseMat(vt);;
gap> PRJRS:=List([1..5],i->PRJR(i,1,10));;
\end{verbatim}

The validity of the transformation of the partner functions is checked now.
\begin{verbatim}
gap> PRJR(2,2,10)*PRJR(2,1,10)*vtb[1]=PRJR(2,1,10)*vtb[1];
true
\end{verbatim}
This function at first puts p-th irreducible matrix representation in ``REP''. (The irreducible matrix representations are already stored in ``irrRM''.) For a group element ``x'', the matrix representation $D^{(p)}({\rm x})$ is computed by ``x$^\wedge$REP''. The dimension of the irreducible representation ``dim'' is computed from the trace of the matrix representation of the unit of the group ( represented by ``(~)''). Each group element ``x'' (stored in the list ``EMR'') is transformed to $D^{(p)}_{ij}({\rm x})$, and multiplied to the equivalent matrix operation on the spatial vectors on the vertex of C$_{60}$  (stored in the list ``AM'').  The summation over them, through the complex conjugation, makes the projector.

Each of the basis vector $x$ generates the five-dimensional, irreducible vector spaces under the symmetry operation, and the vector space is composed of $P^{(10)}_{1,1} \cdot x,...,P^{(10)}_{5,1}\cdot x$. In the following computations, the irreducible vector space for each of basis vector is stored in the list ``zz'' (in the seven component zz[1],...,zz[7]). However, these irreducible spaces are overlapped with each other. The computations of the inner products show this.

\begin{verbatim}
gap> zz:=List(vtb,y->List(List([1..5],i->PRJRS[i]*y)));;
gap> List(zz,a->List(zz,b->RankMat(a*TransposedMat(b))));
[ [ 5, 0, 5, 0, 0, 0, 0 ], [ 0, 5, 0, 0, 0, 0, 0 ], [ 5, 0, 5, 5, 0, 0, 0 ], 
  [ 0, 0, 5, 5, 0, 5, 0 ], [ 0, 0, 0, 0, 5, 0, 0 ], [ 0, 0, 0, 5, 0, 5, 5 ], 
  [ 0, 0, 0, 0, 0, 5, 5 ] ]
\end{verbatim}

We now project out the overlapping between irreducible subspaces. We start from the one of the irreducible space, say zz[1], and construct the orthogonal space to it. One of the basis vectors in the latter space, if the projector $P^{(10)}_{1,1}$ is applied to it, belongs to the irreducible space invariant by $P^{(10)}_{1,1}$. This vector is chosen as the generator of another irreducible space. Similarly, we construct the orthogonal space to the set of the irreducible spaces obtained until now, from this, we extract the generator of another irreducible space and proceed so on.

\begin{verbatim}
gap> B:=ShallowCopy(zz[1]);;
gap> D:=[ShallowCopy(zz[1])];;
gap> for n in [2..Size(zz)]
 do
 OV:=BaseOrthogonalSpaceMat(B);;
 xx:=BaseMat(List(OV,x->PRJRS[1]*x))[1];
 vv:=List(List([1..5],i->PRJRS[i]*xx));
 Append(B,vv);
 Append(D,[vv]);
 od;
\end{verbatim}

We generated seven separated irreducible spaces; let us check this (by computing the inner products of vectors):
\begin{verbatim}
gap> List(D,a->List(D,b->RankMat(a*TransposedMat(b))));;
[ [ 5, 0, 0, 0, 0, 0, 0 ], [ 0, 5, 0, 0, 0, 0, 0 ], [ 0, 0, 5, 0, 0, 0, 0 ], 
  [ 0, 0, 0, 5, 0, 0, 0 ], [ 0, 0, 0, 0, 5, 0, 0 ], [ 0, 0, 0, 0, 0, 5, 0 ], 
  [ 0, 0, 0, 0, 0, 0, 5 ] ]
\end{verbatim}

The vector space generated by the symmetry operations from one of the separated irreducible spaces should be orthogonal to others; let us check this (using the list ``AM'' of symmetry operations prepared in the preceding computation):
\begin{verbatim}
gap> OO:=List(AM,o->o*D[1][1]);;
gap> List(D,d->RankMat(OO*TransposedMat(d)));
[ 5, 0, 0, 0, 0, 0, 0 ]
gap> OO:=List(AM,o->o*D[2][1]);;
gap> List(D,d->RankMat(OO*TransposedMat(d)));
[ 0, 5, 0, 0, 0, 0, 0 ]
. . . etc . . .
\end{verbatim}

We can see that all is done well.

The second method is to divide the multiplied irreducible spaces W (as computed in the previous section, by the projectors defined by the characters) into proper minimal components, invariant under the symmetry operation. This method utilizes the coset decomposition of the group G by the subgroup H, G/H. In general, one vector v$_1$ in W, under the operation of the elements g$_i$ in G, generates the set of vectors, the total number of which is equal to the order of G. Now let us take another vector w$_1$, also in W, generated by the operation of the elements h$_i$ in H, such as $\rm w_1=\sum_i h_i \cdot v_1$. Under the symmetry operation of G, it generates the set of vectors, the total number of which is equal to the order of G/H. The number of the generated vectors decreased in the latter case. We might expect the dimension of the basis vectors of the generated vectors would also be diminished in the latter case. And if w$_1$ would generate the vector space of the proper dimension of the irreducible representation, we adopt it, as the generator of the separated, proper, irreducible space. (If not, with another chose of H, this step is repeated.) Then we project out the vector space generated by w$_1$ from W. We take another vector v$_2$ from W, make an H-invariant vector w$_2$ with the suitable H (if necessary, H should be chosen again) to generate the another proper irreducible space. We proceed so on until W is exhausted. This method will be applicable to the present problem in order to separate each irreducible component, even if it involves the trial-and-error, in the choice of the suitable subgroup H. At the start, the vector v$_1$ is taken from the multiplied irreducible spaces. The candidate vector w$_i$ is generated by the projection of v$_i$ into the trivial irreducible representation of the subgroup H. (Indeed, it could also be prepared by the projection of v into other irreducible representations of H, for the purpose of diminishing the number of the generated vectors by the symmetry operation.)

In order to execute this computation, the function ``Cutout'' (given in appendix \ref{Cutout}) is prepared. The function cuts out the subspace (as minimal as possible) from the given vector space, making use of the coset decomposition of R by the subgroups H. In projecting out the proper irreducible component from the multiplied copies, we employ slightly different step from the above description. Let W be the initial starting place, that is, the multiplied components of the irreducible representations, from which the separated proper irreducible components $\rm W_1,..., W_i$ are computed. The next candidate of the generator v$_{i+1}$ is not taken directly from the complementary space ($W\setminus (W_1 \cup W_2 \cup \cdots \cup W_i)$); instead, in the whole vector space S (of 180 dimension in this case), the orthogonal space to $W_1 \cup W_2 \cup \cdots \cup W_i$, biz, $S\setminus(W_1 \cup W_2 \cup \cdots \cup W_i)$ is constructed. The basis set in the latter space is projected into the irreducible representation under the consideration, and the candidate v$_{i+1}$ is taken from this projected space (the renewed W). This treatment is to quicken the symbolic computation. 


In our problem, the latter method works well, too. We can separate out seven distinct irreducible spaces.
In the following computation the function ``Cutout'' tries to divide out the vector space ``BAS[10]'' with respect to the tenth irreducible representation. In this vector space seven copies of irreducible spaces exist. The divided subspaces are given in list ``H1''. Each subspace is of dimension 5 and orthogonal to others. The symmetry operation (given in list ``AM'', in the preceding computation) to each vector in each subspace generates the 5 dimensional vector space,  and the generated vector space exits exactly in the same subspace of that vector.
(As for the other irreducible representations, it is successful, with the suitable choice of the coset decomposition. For example, the subgroups of order 5 operate well at the first, third, fourth, fifth, sixth, ninth and tenth irreducible representations; that of order 8 does at the seventh; that of order 4 does at the eighth. )
\begin{verbatim}
gap> H1:=Cutout(BAS[10],PRJE[10],R,hom,5,5);;
gap> List(H1,a->List(H1,b->RankMat(a*TransposedMat(b))));
[ [ 5, 0, 0, 0, 0, 0, 0 ], [ 0, 5, 0, 0, 0, 0, 0 ], [ 0, 0, 5, 0, 0, 0, 0 ], 
  [ 0, 0, 0, 5, 0, 0, 0 ], [ 0, 0, 0, 0, 5, 0, 0 ], [ 0, 0, 0, 0, 0, 5, 0 ], 
  [ 0, 0, 0, 0, 0, 0, 5 ] ]
gap> OO:=List(AM,o->o*H1[1][1]);;
gap> List(H1,d->RankMat(OO*TransposedMat(d)));
[ 5, 0, 0, 0, 0, 0, 0 ]
gap> OO:=List(AM,o->o*H1[2][1]);;
gap> List(H1,d->RankMat(OO*TransposedMat(d)));
[ 0, 5, 0, 0, 0, 0, 0 ]
. . . etc . . .
\end{verbatim}

A warning is given: in the above discussion, the symmetry operations are defined in the bare forms, using the direct product $R_i\otimes  G_i$. To be precise one should use $R_i^f \otimes G_i$, even if the argument on the dimension counting on the vector space is not affected. To obtain the proper vector system one should dutifully operate $\rm M_f$ ( the matrix prepared in the previous section for the conjugation) to any vector $\rm v$, such as $\rm M_f \cdot v$.

\subsection{Remark to this section}
In this section, the computation of the vibrational mode of C$_{60}$ by means of the computer algebra system GAP is demonstrated. The computation can be executed by a standard desktop personal computer. Though the topic in the present article is limited to the C$_{60}$ molecule, the way of the computation shall be applicable to other systems, such as crystal. The small programs in the article are composed as brief as possible and as comprehensible as possible but without considerations of the optimization, such as memory usage, or the avoidance of the re-computation of same objects. So there is the possibility to improve the efficiency, although they are computable even by the standard desktop personal computer. The computed basis of the vibrational modes are obtained as the symbolic formulas, and they are convertible to the numerical program functions simply by the rewriting of strings. Some of the formulas might be so lengthy that the generated program functions might be beyond the capacity of the compiler. However, the computations presented in this article can be done by the floating-point computation, started from the numerical representation of the rotation matrices, with a certain toleration to the numerical error under the threshold. 

I must add this: there are various studies on the electronic structure and the deformation of C$_{60}$ in the quantitative electronic structure computations (not of the group theoretical view) such as in the references\cite{HADDON, GREEN, COULON}, or in the empirical model (of the group theoretical view)\cite{MOOIJ}. The computations in this article, of this and the previous section, only classifies geometrically possible deformations with respect to irreducible representations and generates the basis set of deformations in the symbolic formulas. The obtained results, however, could be applied to more quantitative studies as those.


\section{Final Remarks}
The theme of this article is how to simulate the property of the material by means of electronic structure calculations with the aid of computer algebra. The collaboration of the electronic structure calculation and the computer algebra is not novel one; rather being regarded as the modernization of the union of the quantum dynamics and the group theory, the success of which was brilliantly displayed in the former half of last century. In this article, the treated topics in the group theory are rather classical ones concerning the character theory. However, there can be various applications of the myriad topics of mathematics. As for the combinatorics, the Polya counting theory is a typical example; it is applicable to determine the number of configurations, equivalent by some symmetry operations, for example, to determine the total number of possible molecular formations indicated by a single chemical formula\cite{POLYA}. This type of computations in combinatorics can also be executed by the computer algebra. As for the application of combinatorics in quantum physics, a sort of extension of Polya counting theorem can be useful in computing the spin weight in molecules\cite{MOLSPEC, JONAS, LANDAU}. In a recent study, it is shown that the framework of the electronic structure could be reconstructed by means of a concept in the algebraic geometry, so-called Groebner basis theory, through the utilization of symbolic computation\cite{AKIHITO}. 

In my article, I have omitted several important application of the group theory and the algebra. One of this is the treatment of magnetic or color groups. Also, I omitted the treatment of Lie algebra, important in the spin system or the study of angular momentum. As for the former, a package of color groups is implemented in GAP\cite{GAPSYSTEM}. As for the latter, the Lie algebra is computable by GAP, although it is not of the digested style, suitable for the practical usage of physicists. (Some of the implemented GAP packages are intended to study the model of spin systems, proposed in the statistical- or mathematical physics.) I must admit that I have not yet developed my skills to compute them through GAP. But I believe that the readers if they are acquainted with GAP through my article,  can construct their own application in these topics.

In the present circumstances of the material physics, indeed, there is an awkward inclination in which one solely relies on computer powers; only by trial and errors, or, only by varying the input data, one tries to execute ``material design'', without any effective guiding principle. We must admit that the computational packages have developed so complex and so vast that a common researcher, if not being an expert, would be obliged to utilize them as black boxes, without knowing the detail of the programs; fortunately, those packages are so powerful and effective that one can obtain reliable results with ease. The blind reliance on these packages, however, should be avoided; the mere enumeration of results of computations, with the slight modification of input data, will be the waste of computational resources. Certainly, by gathering and comparing computed results we could surmise vague tendencies, but may not go anymore, if we do not know the physical or mathematical origin. Instead, it is necessary for us to have another utensil, which will compensate for such a shortcoming; the group theory, or its modernization, the computer algebra, will be of great use for this purpose. With the aid of computer algebra, advanced concepts in group theory are easily accessible and applicable, so that we can execute systematic research of material design from the mathematical first principles, as the group theoretical view illuminates for us the mathematical principles lurking behind observable phenomena. I hope that the computer algebra will bring forth the blessed reunion of physics and mathematics in the research of materials; the available tool for us is not limited to the classics such as Courant-Hilbert\cite{COUR-HILB}, although they had been powerful in the early stages of the development of quantum dynamics. The mathematics and physics have progressed greatly since then; the forefront of these developments seems to be so esoteric that non-experts feel dismayed when urged to put it into practice. Nevertheless, one should not be timid; for the sake of ardent researchers, the computer algebra will be a most suitable guide for their efforts, and there they will find a most speedily passable bridge over the ever-deepening chasm between physics and mathematics.        
﻿\appendix
\section{Symmetry operations in the diamond structure (real space)}
 The symmetry operations in the diamond structure (in Cartesian coordinates) are given in the table. The operation on the atomic coordinate is 
given by $x\rightarrow R\cdot x + T1\cdot a_1 + T2\cdot a_2 +T3\cdot a_3$, where $(a_1,a_2,a_3)$ are primitive translations.There are two types of fractional translations, according to the location of the origin of the unit cell.
The first, denoted as (T1,T2,T3) corresponds to the case of atomic positions of $\pm(1/8\cdot a_1 +1/8\cdot a_3 +1/8\cdot a_3)$:the second,
(T1(*),T2(*),T3(*)) corresponds to positions of $(0\cdot a_1 +0\cdot a_3 +0\cdot a_3)$ and  $(1/4\cdot a_1 +1/4\cdot a_3 +1/4\cdot a_3)$. Although the latter case is more conventional, the former has a merit such that the coefficients of the plane-wave-expression of the wavefunctions are real-valued.
\begin{longtable}{r|rrrrrrrrr|rrr|rrr}
    \hline
          & {R11} & {R12} & {R13} & {R21} & {R22} & {R23} & {R31} & {R32} & {R33} & {T1} & {T2} & {T3} & {T1(*)} & {T2(*)} & {T3(*)} \\
          \hline
    1     & 1     & 0     & 0     & 0     & 1     & 0     & 0     & 0     & 1     & 0     & 0     & 0     & 0     & 0     & 0 \\
    2     & 1     & 0     & 0     & 0     & -1    & 0     & 0     & 0     & -1    & 0.5   & 0     & 0     & 0     & 0     & 0 \\
    3     & -1    & 0     & 0     & 0     & 1     & 0     & 0     & 0     & -1    & 0     & 0.5   & 0     & 0     & 0     & 0 \\
    4     & -1    & 0     & 0     & 0     & -1    & 0     & 0     & 0     & 1     & 0     & 0     & 0.5   & 0     & 0     & 0 \\
    5     & 0     & 1     & 0     & 0     & 0     & 1     & 1     & 0     & 0     & 0     & 0     & 0     & 0     & 0     & 0 \\
    6     & 0     & 1     & 0     & 0     & 0     & -1    & -1    & 0     & 0     & 0.5   & 0     & 0     & 0     & 0     & 0 \\
    7     & 0     & -1    & 0     & 0     & 0     & 1     & -1    & 0     & 0     & 0     & 0.5   & 0     & 0     & 0     & 0 \\
    8     & 0     & -1    & 0     & 0     & 0     & -1    & 1     & 0     & 0     & 0     & 0     & 0.5   & 0     & 0     & 0 \\
    9     & 0     & 0     & 1     & 1     & 0     & 0     & 0     & 1     & 0     & 0     & 0     & 0     & 0     & 0     & 0 \\
    10    & 0     & 0     & 1     & -1    & 0     & 0     & 0     & -1    & 0     & 0.5   & 0     & 0     & 0     & 0     & 0 \\
    11    & 0     & 0     & -1    & 1     & 0     & 0     & 0     & -1    & 0     & 0     & 0.5   & 0     & 0     & 0     & 0 \\
    12    & 0     & 0     & -1    & -1    & 0     & 0     & 0     & 1     & 0     & 0     & 0     & 0.5   & 0     & 0     & 0 \\
    13    & 0     & -1    & 0     & -1    & 0     & 0     & 0     & 0     & -1    & 0     & 0     & 0     & 0.25  & 0.25  & 0.25 \\
    14    & 0     & -1    & 0     & 1     & 0     & 0     & 0     & 0     & 1     & -0.5  & 0     & 0     & 0.25  & 0.25  & 0.25 \\
    15    & 0     & 1     & 0     & -1    & 0     & 0     & 0     & 0     & 1     & 0     & -0.5  & 0     & 0.25  & 0.25  & 0.25 \\
    16    & 0     & 1     & 0     & 1     & 0     & 0     & 0     & 0     & -1    & 0     & 0     & -0.5  & 0.25  & 0.25  & 0.25 \\
    17    & -1    & 0     & 0     & 0     & 0     & -1    & 0     & -1    & 0     & 0     & 0     & 0     & 0.25  & 0.25  & 0.25 \\
    18    & -1    & 0     & 0     & 0     & 0     & 1     & 0     & 1     & 0     & -0.5  & 0     & 0     & 0.25  & 0.25  & 0.25 \\
    19    & 1     & 0     & 0     & 0     & 0     & -1    & 0     & 1     & 0     & 0     & -0.5  & 0     & 0.25  & 0.25  & 0.25 \\
    20    & 1     & 0     & 0     & 0     & 0     & 1     & 0     & -1    & 0     & 0     & 0     & -0.5  & 0.25  & 0.25  & 0.25 \\
    21    & 0     & 0     & -1    & 0     & -1    & 0     & -1    & 0     & 0     & 0     & 0     & 0     & 0.25  & 0.25  & 0.25 \\
    22    & 0     & 0     & -1    & 0     & 1     & 0     & 1     & 0     & 0     & -0.5  & 0     & 0     & 0.25  & 0.25  & 0.25 \\
    23    & 0     & 0     & 1     & 0     & -1    & 0     & 1     & 0     & 0     & 0     & -0.5  & 0     & 0.25  & 0.25  & 0.25 \\
    24    & 0     & 0     & 1     & 0     & 1     & 0     & -1    & 0     & 0     & 0     & 0     & -0.5  & 0.25  & 0.25  & 0.25 \\
    25    & -1    & 0     & 0     & 0     & -1    & 0     & 0     & 0     & -1    & 0     & 0     & 0     & 0.25  & 0.25  & 0.25 \\
    26    & -1    & 0     & 0     & 0     & 1     & 0     & 0     & 0     & 1     & -0.5  & 0     & 0     & 0.25  & 0.25  & 0.25 \\
    27    & 1     & 0     & 0     & 0     & -1    & 0     & 0     & 0     & 1     & 0     & -0.5  & 0     & 0.25  & 0.25  & 0.25 \\
    28    & 1     & 0     & 0     & 0     & 1     & 0     & 0     & 0     & -1    & 0     & 0     & -0.5  & 0.25  & 0.25  & 0.25 \\
    29    & 0     & -1    & 0     & 0     & 0     & -1    & -1    & 0     & 0     & 0     & 0     & 0     & 0.25  & 0.25  & 0.25 \\
    30    & 0     & -1    & 0     & 0     & 0     & 1     & 1     & 0     & 0     & -0.5  & 0     & 0     & 0.25  & 0.25  & 0.25 \\
    31    & 0     & 1     & 0     & 0     & 0     & -1    & 1     & 0     & 0     & 0     & -0.5  & 0     & 0.25  & 0.25  & 0.25 \\
    32    & 0     & 1     & 0     & 0     & 0     & 1     & -1    & 0     & 0     & 0     & 0     & -0.5  & 0.25  & 0.25  & 0.25 \\
    33    & 0     & 0     & -1    & -1    & 0     & 0     & 0     & -1    & 0     & 0     & 0     & 0     & 0.25  & 0.25  & 0.25 \\
    34    & 0     & 0     & -1    & 1     & 0     & 0     & 0     & 1     & 0     & -0.5  & 0     & 0     & 0.25  & 0.25  & 0.25 \\
    35    & 0     & 0     & 1     & -1    & 0     & 0     & 0     & 1     & 0     & 0     & -0.5  & 0     & 0.25  & 0.25  & 0.25 \\
    36    & 0     & 0     & 1     & 1     & 0     & 0     & 0     & -1    & 0     & 0     & 0     & -0.5  & 0.25  & 0.25  & 0.25 \\
    37    & 0     & 1     & 0     & 1     & 0     & 0     & 0     & 0     & 1     & 0     & 0     & 0     & 0     & 0     & 0 \\
    38    & 0     & 1     & 0     & -1    & 0     & 0     & 0     & 0     & -1    & 0.5   & 0     & 0     & 0     & 0     & 0 \\
    39    & 0     & -1    & 0     & 1     & 0     & 0     & 0     & 0     & -1    & 0     & 0.5   & 0     & 0     & 0     & 0 \\
    40    & 0     & -1    & 0     & -1    & 0     & 0     & 0     & 0     & 1     & 0     & 0     & 0.5   & 0     & 0     & 0 \\
    41    & 1     & 0     & 0     & 0     & 0     & 1     & 0     & 1     & 0     & 0     & 0     & 0     & 0     & 0     & 0 \\
    42    & 1     & 0     & 0     & 0     & 0     & -1    & 0     & -1    & 0     & 0.5   & 0     & 0     & 0     & 0     & 0 \\
    43    & -1    & 0     & 0     & 0     & 0     & 1     & 0     & -1    & 0     & 0     & 0.5   & 0     & 0     & 0     & 0 \\
    44    & -1    & 0     & 0     & 0     & 0     & -1    & 0     & 1     & 0     & 0     & 0     & 0.5   & 0     & 0     & 0 \\
    45    & 0     & 0     & 1     & 0     & 1     & 0     & 1     & 0     & 0     & 0     & 0     & 0     & 0     & 0     & 0 \\
    46    & 0     & 0     & 1     & 0     & -1    & 0     & -1    & 0     & 0     & 0.5   & 0     & 0     & 0     & 0     & 0 \\
    47    & 0     & 0     & -1    & 0     & 1     & 0     & -1    & 0     & 0     & 0     & 0.5   & 0     & 0     & 0     & 0 \\
    48    & 0     & 0     & -1    & 0     & -1    & 0     & 1     & 0     & 0     & 0     & 0     & 0.5   & 0     & 0     & 0 \\
\hline
  \caption{Symmetry operations in the point group in the diamond unit cell. If fractional translations are set to be zero, these operations are applicable to the cubic lattice.}
  \label{tab:symopr_fcc_dia}%
\end{longtable}%

\section{Symmetry operations in the hexagonal lattice (real space)}
 The symmetry operations in graphite (in Cartesian coordinates)are given in the table:

\begin{longtable}{r|rrrrrrrrr|rrr}
\hline
 & R11 & R12 & R13 & R21 & R22 & R23 & R31 & R32 & R33 & T1 & T2 & T3\\\hline
    1     & 1     & 0     & 0     & 0     & 1     & 0     & 0     & 0     & 1     & 0     & 0     & 0 \\
    2     & 0.5   & -$\sqrt{3}/2$ & 0     & $\sqrt{3}/2$ & 0.5   & 0     & 0     & 0     & 1     & 0     & 0     & -0.5 \\
    3     & -0.5  & -$\sqrt{3}/2$ & 0     & $\sqrt{3}/2$ & -0.5  & 0     & 0     & 0     & 1     & 0     & 0     & 0 \\
    4     &  -1    & 0     & 0     & 0     & -1    & 0     & 0     & 0     & 1     & 0     & 0     & -0.5 \\
    5     & -0.5  & $\sqrt{3}/2$ & 0     & -$\sqrt{3}/2$ & -0.5  & 0     & 0     & 0     & 1     & 0     & 0     & 0 \\
    6     & 0.5   & $\sqrt{3}/2$ & 0     & -$\sqrt{3}/2$ & 0.5   & 0     & 0     & 0     & 1     & 0     & 0     & -0.5 \\
    7     & -0.5  & -$\sqrt{3}/2$ & 0     & -$\sqrt{3}/2$ & 0.5   & 0     & 0     & 0     & -1    & 0     & 0     & -0.5 \\
    8     & 0.5   & -$\sqrt{3}/2$ & 0     & -$\sqrt{3}/2$ & -0.5  & 0     & 0     & 0     & -1    & 0     & 0     & 0 \\
    9     & 1     & 0     & 0     & 0     & -1    & 0     & 0     & 0     & -1    & 0     & 0     & -0.5 \\
    10    & 0.5   & $\sqrt{3}/2$ & 0     & $\sqrt{3}/2$ & -0.5  & 0     & 0     & 0     & -1    & 0     & 0     & 0 \\
    11    & -0.5  & $\sqrt{3}/2$ & 0     & $\sqrt{3}/2$ & 0.5   & 0     & 0     & 0     & -1    & 0     & 0     & -0.5 \\
    12    & -1    & 0     & 0     & 0     & 1     & 0     & 0     & 0     & -1    & 0     & 0     & 0 \\
    13    & -1    & 0     & 0     & 0     & -1    & 0     & 0     & 0     & -1    & 0     & 0     & -0.5 \\
    14    & -0.5  & $\sqrt{3}/2$ & 0     & -$\sqrt{3}/2$ & -0.5  & 0     & 0     & 0     & -1    & 0     & 0     & 0 \\
    15    & 0.5   & $\sqrt{3}/2$ & 0     & -$\sqrt{3}/2$ & 0.5   & 0     & 0     & 0     & -1    & 0     & 0     & -0.5 \\
    16    & 1     & 0     & 0     & 0     & 1     & 0     & 0     & 0     & -1    & 0     & 0     & 0 \\
    17    & 0.5   & -$\sqrt{3}/2$ & 0     & $\sqrt{3}/2$ & 0.5   & 0     & 0     & 0     & -1    & 0     & 0     & -0.5 \\
    18    & -0.5  & -$\sqrt{3}/2$ & 0     & $\sqrt{3}/2$ & -0.5  & 0     & 0     & 0     & -1    & 0     & 0     & 0 \\
    19    & 0.5   & $\sqrt{3}/2$ & 0     & $\sqrt{3}/2$ & -0.5  & 0     & 0     & 0     & 1     & 0     & 0     & 0 \\
    20    & -0.5  & $\sqrt{3}/2$ & 0     & $\sqrt{3}/2$ & 0.5   & 0     & 0     & 0     & 1     & 0     & 0     & -0.5 \\
    21    & -1    & 0     & 0     & 0     & 1     & 0     & 0     & 0     & 1     & 0     & 0     & 0 \\
    22    & -0.5  & -$\sqrt{3}/2$ & 0     & -$\sqrt{3}/2$ & 0.5   & 0     & 0     & 0     & 1     & 0     & 0     & -0.5 \\
    23    & 0.5   & -$\sqrt{3}/2$ & 0     & -$\sqrt{3}/2$ & -0.5  & 0     & 0     & 0     & 1     & 0     & 0     & 0 \\
    24    & 1     & 0     & 0     & 0     & -1    & 0     & 0     & 0     & 1     & 0     & 0     & -0.5 \\
 \hline
  \caption{The symmetry operations in graphite in real space.}
  \label{tab:symopr_graphite}%
\end{longtable}%

\section{Symmetry operations in the diamond structure (wave-number space)} 
 The symmetry operations in the diamond structure, as the rotation matrices in the reciprocal spaces are given in the table:
\begin{longtable}{r|rrrrrrrrr|rrr}
\hline
          & R'11   & R'12   & R'13   & R'21   & R'22   & R'23   & R'31   & R'32   & R'33   & T1    & T2    & T3 \\
\hline
    1     & 1     & 0     & 0     & 0     & 1     & 0     & 0     & 0     & 1     & 0     & 0     & 0 \\
    2     & -1    & 0     & 0     & -1    & 0     & 1     & -1    & 1     & 0     & 0.5   & 0     & 0 \\
    3     & 0     & -1    & 1     & 0     & -1    & 0     & 1     & -1    & 0     & 0     & 0.5   & 0 \\
    4     & 0     & 1     & -1    & 1     & 0     & -1    & 0     & 0     & -1    & 0     & 0     & 0.5 \\
    5     & 0     & 1     & 0     & 0     & 0     & 1     & 1     & 0     & 0     & 0     & 0     & 0 \\
    6     & 0     & -1    & 0     & 1     & -1    & 0     & 0     & -1    & 1     & 0.5   & 0     & 0 \\
    7     & 1     & 0     & -1    & 0     & 0     & -1    & 0     & 1     & -1    & 0     & 0.5   & 0 \\
    8     & -1    & 0     & 1     & -1    & 1     & 0     & -1    & 0     & 0     & 0     & 0     & 0.5 \\
    9     & 0     & 0     & 1     & 1     & 0     & 0     & 0     & 1     & 0     & 0     & 0     & 0 \\
    10    & 0     & 0     & -1    & 0     & 1     & -1    & 1     & 0     & -1    & 0.5   & 0     & 0 \\
    11    & -1    & 1     & 0     & -1    & 0     & 0     & -1    & 0     & 1     & 0     & 0.5   & 0 \\
    12    & 1     & -1    & 0     & 0     & -1    & 1     & 0     & -1    & 0     & 0     & 0     & 0.5 \\
    13    & 0     & -1    & 0     & -1    & 0     & 0     & 0     & 0     & -1    & 0     & 0     & 0 \\
    14    & 0     & 1     & 0     & 0     & 1     & -1    & -1    & 1     & 0     & -0.5  & 0     & 0 \\
    15    & 1     & 0     & -1    & 1     & 0     & 0     & 1     & -1    & 0     & 0     & -0.5  & 0 \\
    16    & -1    & 0     & 1     & 0     & -1    & 1     & 0     & 0     & 1     & 0     & 0     & -0.5 \\
    17    & -1    & 0     & 0     & 0     & 0     & -1    & 0     & -1    & 0     & 0     & 0     & 0 \\
    18    & 1     & 0     & 0     & 1     & -1    & 0     & 1     & 0     & -1    & -0.5  & 0     & 0 \\
    19    & 0     & -1    & 1     & 0     & 0     & 1     & -1    & 0     & 1     & 0     & -0.5  & 0 \\
    20    & 0     & 1     & -1    & -1    & 1     & 0     & 0     & 1     & 0     & 0     & 0     & -0.5 \\
    21    & 0     & 0     & -1    & 0     & -1    & 0     & -1    & 0     & 0     & 0     & 0     & 0 \\
    22    & 0     & 0     & 1     & -1    & 0     & 1     & 0     & -1    & 1     & -0.5  & 0     & 0 \\
    23    & -1    & 1     & 0     & 0     & 1     & 0     & 0     & 1     & -1    & 0     & -0.5  & 0 \\
    24    & 1     & -1    & 0     & 1     & 0     & -1    & 1     & 0     & 0     & 0     & 0     & -0.5 \\
    25    & -1    & 0     & 0     & 0     & -1    & 0     & 0     & 0     & -1    & 0     & 0     & 0 \\
    26    & 1     & 0     & 0     & 1     & 0     & -1    & 1     & -1    & 0     & -0.5  & 0     & 0 \\
    27    & 0     & 1     & -1    & 0     & 1     & 0     & -1    & 1     & 0     & 0     & -0.5  & 0 \\
    28    & 0     & -1    & 1     & -1    & 0     & 1     & 0     & 0     & 1     & 0     & 0     & -0.5 \\
    29    & 0     & -1    & 0     & 0     & 0     & -1    & -1    & 0     & 0     & 0     & 0     & 0 \\
    30    & 0     & 1     & 0     & -1    & 1     & 0     & 0     & 1     & -1    & -0.5  & 0     & 0 \\
    31    & -1    & 0     & 1     & 0     & 0     & 1     & 0     & -1    & 1     & 0     & -0.5  & 0 \\
    32    & 1     & 0     & -1    & 1     & -1    & 0     & 1     & 0     & 0     & 0     & 0     & -0.5 \\
    33    & 0     & 0     & -1    & -1    & 0     & 0     & 0     & -1    & 0     & 0     & 0     & 0 \\
    34    & 0     & 0     & 1     & 0     & -1    & 1     & -1    & 0     & 1     & -0.5  & 0     & 0 \\
    35    & 1     & -1    & 0     & 1     & 0     & 0     & 1     & 0     & -1    & 0     & -0.5  & 0 \\
    36    & -1    & 1     & 0     & 0     & 1     & -1    & 0     & 1     & 0     & 0     & 0     & -0.5 \\
    37    & 0     & 1     & 0     & 1     & 0     & 0     & 0     & 0     & 1     & 0     & 0     & 0 \\
    38    & 0     & -1    & 0     & 0     & -1    & 1     & 1     & -1    & 0     & 0.5   & 0     & 0 \\
    39    & -1    & 0     & 1     & -1    & 0     & 0     & -1    & 1     & 0     & 0     & 0.5   & 0 \\
    40    & 1     & 0     & -1    & 0     & 1     & -1    & 0     & 0     & -1    & 0     & 0     & 0.5 \\
    41    & 1     & 0     & 0     & 0     & 0     & 1     & 0     & 1     & 0     & 0     & 0     & 0 \\
    42    & -1    & 0     & 0     & -1    & 1     & 0     & -1    & 0     & 1     & 0.5   & 0     & 0 \\
    43    & 0     & 1     & -1    & 0     & 0     & -1    & 1     & 0     & -1    & 0     & 0.5   & 0 \\
    44    & 0     & -1    & 1     & 1     & -1    & 0     & 0     & -1    & 0     & 0     & 0     & 0.5 \\
    45    & 0     & 0     & 1     & 0     & 1     & 0     & 1     & 0     & 0     & 0     & 0     & 0 \\
    46    & 0     & 0     & -1    & 1     & 0     & -1    & 0     & 1     & -1    & 0.5   & 0     & 0 \\
    47    & 1     & -1    & 0     & 0     & -1    & 0     & 0     & -1    & 1     & 0     & 0.5   & 0 \\
    48    & -1    & 1     & 0     & -1    & 0     & 1     & -1    & 0     & 0     & 0     & 0     & 0.5 \\
\hline
      \caption{Symmetry operations in the diamond crystal in the reciprocal space.}
    \label{tab:symopr_fcc_dia_reciprocal}
\end{longtable}%

\section{Symmetry operations in the hexagonal lattice (wave-number space)}
The crystal axes are chosen to be $a_1=(1,0,0),a_2=(-1/2,\sqrt(3)/2,0),a_3=(0,0,c)$. Then
The symmetry operations as the operations in the wave-number space are given as:
    \begin{longtable}{r|rrrrrrrrr|rrr}
\hline
     & R'11 & R'12 & R'13 & R'21 & R'22 & R'23 & R'31 & R'32 & R'33 & T1 & T2 & T3\\\hline
    1     & 1     & 0     & 0     & 0     & 1     & 0     & 0     & 0     & 1     & 0     & 0     & 0 \\
    2     & 0     & -1    & 0     & 1     & 1     & 0     & 0     & 0     & 1     & 0     & 0     & -0.5 \\
    3     & -1    & -1    & 0     & 1     & 0     & 0     & 0     & 0     & 1     & 0     & 0     & 0 \\
    4     & -1    & 0     & 0     & 0     & -1    & 0     & 0     & 0     & 1     & 0     & 0     & -0.5 \\
    5     & 0     & 1     & 0     & -1    & -1    & 0     & 0     & 0     & 1     & 0     & 0     & 0 \\
    6     & 1     & 1     & 0     & -1    & 0     & 0     & 0     & 0     & 1     & 0     & 0     & -0.5 \\
    7     & -1    & -1    & 0     & 0     & 1     & 0     & 0     & 0     & -1    & 0     & 0     & -0.5 \\
    8     & 0     & -1    & 0     & -1    & 0     & 0     & 0     & 0     & -1    & 0     & 0     & 0 \\
    9     & 1     & 0     & 0     & -1    & -1    & 0     & 0     & 0     & -1    & 0     & 0     & -0.5 \\
    10    & 1     & 1     & 0     & 0     & -1    & 0     & 0     & 0     & -1    & 0     & 0     & 0 \\
    11    & 0     & 1     & 0     & 1     & 0     & 0     & 0     & 0     & -1    & 0     & 0     & -0.5 \\
    12    & -1    & 0     & 0     & 1     & 1     & 0     & 0     & 0     & -1    & 0     & 0     & 0 \\
    13    & -1    & 0     & 0     & 0     & -1    & 0     & 0     & 0     & -1    & 0     & 0     & -0.5 \\
    14    & 0     & 1     & 0     & -1    & -1    & 0     & 0     & 0     & -1    & 0     & 0     & 0 \\
    15    & 1     & 1     & 0     & -1    & 0     & 0     & 0     & 0     & -1    & 0     & 0     & -0.5 \\
    16    & 1     & 0     & 0     & 0     & 1     & 0     & 0     & 0     & -1    & 0     & 0     & 0 \\
    17    & 0     & -1    & 0     & 1     & 1     & 0     & 0     & 0     & -1    & 0     & 0     & -0.5 \\
    18    & -1    & -1    & 0     & 1     & 0     & 0     & 0     & 0     & -1    & 0     & 0     & 0 \\
    19    & 1     & 1     & 0     & 0     & -1    & 0     & 0     & 0     & 1     & 0     & 0     & 0 \\
    20    & 0     & 1     & 0     & 1     & 0     & 0     & 0     & 0     & 1     & 0     & 0     & -0.5 \\
    21    & -1    & 0     & 0     & 1     & 1     & 0     & 0     & 0     & 1     & 0     & 0     & 0 \\
    22    & -1    & -1    & 0     & 0     & 1     & 0     & 0     & 0     & 1     & 0     & 0     & -0.5 \\
    23    & 0     & -1    & 0     & -1    & 0     & 0     & 0     & 0     & 1     & 0     & 0     & 0 \\
    24    & 1     & 0     & 0     & -1    & -1    & 0     & 0     & 0     & 1     & 0     & 0     & -0.5 \\
 \hline
    \caption{The symmetry operations in graphite in reciprocal space.}
  \label{tab:symopr_graphite_reciprocal}%
\end{longtable}%

\section{The computation of Wyckoff positions by GAP}
Computations for space groups can be executed by means of ``Cryst'' package in GAP. Though the manual of this package is provided, the description is too curt and the computed result is too unfriendly for beginners to utilize. So in this section, the usage of this package is illustrated.

The crystallographic group is defined at first (in this example, with minimal generators):

\begin{verbatim}
gap> M1:=[[0,0,1,0],[1,0,0,0],[0,-1,0,0],[1/4,1/4,1/4,1]];
gap> M2:=[[0,0,-1,0],[0,-1,0,0],[1,0,0,0],[0,0,0,1]];
gap> S:=AffineCrystGroup([M1,M2]);
\end{verbatim}

The point group is obtained by:

\begin{verbatim}
gap> P:=PointGroup(S);
\end{verbatim}

The Wyckoff positions are computed now: 

\begin{verbatim}
gap> W:=WyckoffPositions(S);
[ < Wyckoff position, point group 4, translation := [ 1/8, 3/8, 7/8 ], 
    basis := [  ] >
    , < Wyckoff position, point group 4, translation := [ 1/8, 3/8, 3/8 ], 
    basis := [  ] >
    , < Wyckoff position, point group 5, translation := [ 1/4, 1/4, 3/4 ], 
    basis := [  ] >
    , < Wyckoff position, point group 5, translation := [ 1/4, 1/4, 1/4 ], 
    basis := [  ] >
    , < Wyckoff position, point group 3, translation := [ 0, 0, 0 ], 
    basis := [ [ 1, -1, -1 ] ] >
    , < Wyckoff position, point group 6, translation := [ 1/4, 1/4, 1/2 ], 
    basis := [ [ 0, 0, 1 ] ] >
    , < Wyckoff position, point group 7, translation := [ 0, 1/4, 7/8 ], 
    basis := [ [ 1/2, 1/2, 0 ] ] >
    , < Wyckoff position, point group 2, translation := [ 0, 0, 0 ], 
    basis := [ [ 1/2, -1/2, 0 ], [ 0, 0, 1 ] ] >
    , < Wyckoff position, point group 1, translation := [ 0, 0, 0 ], 
    basis := [ [ 1/2, 0, 1/2 ], [ 0, 1/2, 1/2 ], [ 0, 0, 1 ] ] >
     ]
\end{verbatim}

The returned result is the list of the generating set of the Wyckoff positions. The expression such as

\begin{verbatim}
    , < Wyckoff position, point group 3, translation := [ 0, 0, 0 ], 
    basis := [ [ 1, -1, -1 ] ] >
\end{verbatim}
    
indicates the information of a generator for the Wyckoff positions, which is located at the segment [0,0,0]+t[1,-1,-1], where t is arbitrary real number inasmuch that the segment is confined in the unit cell. Now we get nine generators. The computed result, therefore, should be read as, 

\begin{verbatim}
[ 1/8, 3/8, 7/8 ]
[ 1/8, 3/8, 3/8 ]
[ 1/4, 1/4, 3/4 ]
[ 1/4, 1/4, 1/4 ]
[ 0, 0, 0 ]+t[ 1, -1, -1 ]
[ 1/4, 1/4, 1/2 ]+t[ 0, 0, 1 ]
[ 0, 1/4, 7/8 ]+t[ 1/2, 1/2, 0 ] 
[ 0, 0, 0 ]+t1[ 1/2, -1/2, 0 ]+t2[ 0, 0, 1 ]
[ 0, 0, 0 ]+t1[ 1/2, 0, 1/2 ]+t2[ 0, 1/2, 1/2 ]+t3[ 0, 0, 1 ] 
\end{verbatim}

The first four are single points, but the remaining five are one-dimensional segments, two-dimensional planes, or, a three-dimensional shape. (The last of them is the arbitrary position in the unit cell.) Let us pay attention to the fourth of them. It is the position of one carbon atom in the minimal diamond unit cell; which is moved to [0,0,0] (the position of another Carbon atom) by means of symmetry operations. In order to obtain whole of the Wyckoff positions, the equivalent coordinate points (orbits by the symmetry operations) can be evaluated in the following way. 

The total numbers of the equivalent coordinate points are given by
\begin{verbatim}
gap> List(WyckoffPositions(S),x->Size(WyckoffOrbit(x)));
[ 4, 4, 2, 2, 8, 12, 24, 24, 48 ]
\end{verbatim}

For each generating points,  the equivalents coordinates can be computed separately:
\begin{verbatim}
gap> WyckoffOrbit(WyckoffPositions(S)[1]);
[ < Wyckoff position, point group 4, translation := [ 1/8, 3/8, 7/8 ], 
    basis := [  ] >
    , < Wyckoff position, point group 4, translation := [ 3/8, 1/8, 7/8 ], 
    basis := [  ] >
    , < Wyckoff position, point group 4, translation := [ 3/8, 3/8, 5/8 ], 
    basis := [  ] >
    , < Wyckoff position, point group 4, translation := [ 1/8, 1/8, 5/8 ], 
    basis := [  ] >
     ]
gap> WyckoffOrbit(WyckoffPositions(S)[2]);
[ < Wyckoff position, point group 4, translation := [ 1/8, 3/8, 3/8 ], 
    basis := [  ] >
    , < Wyckoff position, point group 4, translation := [ 3/8, 1/8, 3/8 ], 
    basis := [  ] >
    , < Wyckoff position, point group 4, translation := [ 3/8, 3/8, 1/8 ], 
    basis := [  ] >
    , < Wyckoff position, point group 4, translation := [ 1/8, 1/8, 1/8 ], 
    basis := [  ] >
     ]
gap> ...................................................................
gap> WyckoffOrbit(WyckoffPositions(S)[8]);
[ < Wyckoff position, point group 2, translation := [ 0, 0, 0 ], 
    basis := [ [ 1/2, -1/2, 0 ], [ 0, 0, 1 ] ] >
    , < Wyckoff position, point group 2, translation := [ 1/4, 1/4, 1/4 ], 
    basis := [ [ -1/2, 0, 1/2 ], [ 0, -1, 0 ] ] >
    , < Wyckoff position, point group 2, translation := [ 0, 0, 0 ], 
    ..................................................................
    ..................................................................
     ]
\end{verbatim}

\section{Subgroup lattice}
In this article, the sequence of crystalline symmetry reduction is displayed by means of the mathematical concept of the lattice of subgroups, in which the inclusion relations among them are illustrated as a graph. The GAP computation can be done as follows:
\begin{verbatim}
gap> M1:=[[0,1,0],[0,0,1],[1,0,0]];
[ [ 0, 1, 0 ], [ 0, 0, 1 ], [ 1, 0, 0 ] ]
gap> M2:=[[0,1,0],[-1,0,0],[0,0,1]];
[ [ 0, 1, 0 ], [ -1, 0, 0 ], [ 0, 0, 1 ] ]
gap> G:=Group(M1,M2);
Group([ [ [ 0, 1, 0 ], [ 0, 0, 1 ], [ 1, 0, 0 ] ], 
  [ [ 0, 1, 0 ], [ -1, 0, 0 ], [ 0, 0, 1 ] ] ])
gap> Size(G);
24
gap> l:=LatticeSubgroups(G);
<subgroup lattice of Group(
[ [ [ 0, 1, 0 ], [ 0, 0, 1 ], [ 1, 0, 0 ] ], 
  [ [ 0, 1, 0 ], [ -1, 0, 0 ], [ 0, 0, 1 ] ] ]), 11 classes, 
30 subgroups>
gap> ConjugacyClassesSubgroups(l);
[ Group([],[ [ 1, 0, 0 ], [ 0, 1, 0 ], [ 0, 0, 1 ] ])^G, 
  Group([ [ [ 1, 0, 0 ], [ 0, -1, 0 ], [ 0, 0, -1 ] ] ])^G, 
  Group([ [ [ -1, 0, 0 ], [ 0, 0, -1 ], [ 0, -1, 0 ] ] ])^G, 
  Group([ [ [ 0, 1, 0 ], [ 0, 0, 1 ], [ 1, 0, 0 ] ] ])^G, 
  Group([ [ [ -1, 0, 0 ], [ 0, 1, 0 ], [ 0, 0, -1 ] ], 
      [ [ 1, 0, 0 ], [ 0, -1, 0 ], [ 0, 0, -1 ] ] ])^G, 
  Group([ [ [ -1, 0, 0 ], [ 0, 0, -1 ], [ 0, -1, 0 ] ], 
      [ [ 1, 0, 0 ], [ 0, -1, 0 ], [ 0, 0, -1 ] ] ])^G, 
  Group([ [ [ 1, 0, 0 ], [ 0, 0, 1 ], [ 0, -1, 0 ] ], 
      [ [ 1, 0, 0 ], [ 0, -1, 0 ], [ 0, 0, -1 ] ] ])^G, 
  Group([ [ [ -1, 0, 0 ], [ 0, 0, -1 ], [ 0, -1, 0 ] ], 
      [ [ 0, 1, 0 ], [ 0, 0, 1 ], [ 1, 0, 0 ] ] ])^G, 
  Group([ [ [ -1, 0, 0 ], [ 0, 1, 0 ], [ 0, 0, -1 ] ], 
      [ [ 1, 0, 0 ], [ 0, -1, 0 ], [ 0, 0, -1 ] ], 
      [ [ -1, 0, 0 ], [ 0, 0, -1 ], [ 0, -1, 0 ] ] ])^G, 
  Group([ [ [ -1, 0, 0 ], [ 0, 1, 0 ], [ 0, 0, -1 ] ], 
      [ [ 1, 0, 0 ], [ 0, -1, 0 ], [ 0, 0, -1 ] ], 
      [ [ 0, 1, 0 ], [ 0, 0, 1 ], [ 1, 0, 0 ] ] ])^G, 
  Group([ [ [ -1, 0, 0 ], [ 0, 1, 0 ], [ 0, 0, -1 ] ], 
      [ [ 1, 0, 0 ], [ 0, -1, 0 ], [ 0, 0, -1 ] ], 
      [ [ 0, 1, 0 ], [ 0, 0, 1 ], [ 1, 0, 0 ] ], 
      [ [ -1, 0, 0 ], [ 0, 0, -1 ], [ 0, -1, 0 ] ] ])^G ]
gap> DotFileLatticeSubgroups(l,"paradigm.dot");
\end{verbatim}

The last command ``DotfileLatticeSubgroups'' is used to generate an output file for the visualization.

\section{A short program to compute the semidirect product of groups}
\label{PROGSEMID}
In the main article, the characters of $C_{3v}={\rm Group}((1,2,3),(1,2))\equiv S_3$ are computed by the semidirect product between $C_3={\rm Group}((1,2,3))$ and $I={\rm Group}((1,2))$. The computations are tedious and complicated; a naive implementation by means of GAP programming language could be given as follows:  

\begin{verbatim}
# The definition of the groups G and A is done.
# The irreducible representations and the lists  
# of group elements are prepared.

G:=Group((1,2));
A:=Group((1,2,3));
repG:=Irr(G);
repA:=Irr(A);
Gmem:=Elements(G);
Amem:=Elements(A);

#
# The definition of the automorphism on A induced by G
# is given in list "au".
# The elements of G and A are indexed
#  G[1],G[2],...
#    and
#  A[1],A[2],....
# The elements in A are mapped as
#
#  A[i]^G[j]=A[au[i,j]];
#
au:=[[1,1],[2,3],[3,2]];
#  This list now means the following automorphism:
#  x(in A)^E(in G)  : E -> E, a -> a,    a^-1 -> a^-1 
#  x(in A)^g(in_G)  : E -> E, a -> a^-1, a^-1 -> a    
#  Here a=(1,2,3), 
#  and the elements in A are listed [e, a, a^-1].

#  The following is the list of elements 
#  in the semidirect product [A,G],
#  given by the Cartesian products of sequential numbers.
#
Elist:=[[1,1],[1,2],[2,2],[3,2],[2,1],[3,1]];

GAction:=function(a,g)
#
# A function to compute the automorphism  (a^g)
# on an element "a" in A
# induced by an element "g" in G.
# This function returns the group element,
# according to the list "au".
#
 local apos,gpos,ha;
 apos:=Position(Amem,a);
 gpos:=Position(Gmem,g);
 ha:=au[apos][gpos];
 return Amem[ha];
end;

InvarianceCHK:=function(iA,g,au)
#
# This function checks whether the character x in A 
# (that by the iA-th representation) is
# fixed by the automorphism induced by the element g in G.
#
# iA: index to the irreducible representation of A
# au: the automorphism on A by G, defined above.
#
 local ol,tl;
#
# The list of characters, before the automorphism.
 ol:=List(Amem,x->x^repA[iA]);
# The list of characters, after the automorphism.
 tl:=List(Amem,x->Amem[au[Position(Amem,x)][Position(Gmem,g)]]);
 tl:=List(tl,x->x^repA[iA]);
 return ol=tl;
end;

GetGx:=function(iA,au)
#
# Gx is the set of elements in G, 
# such that the character x in A 
# (that by the iA-th representation)
# is fixed by the automorphism induced 
# by the elements in Gx.
#
# iA: the index to the irreducible representation of A
# au: the automorphism on A by G, defined above.
# 
#
 local Gx;
 Gx:=Filtered(G,g->InvarianceCHK(iA,g,au));
 return Group(Gx);
end;

ProSumma:=function(a,g,Gx,iA,iG)
# This is a subordinate function to compute the character 
# of the elements (a,g),
# when the representations in group A and G, 
# and Gx (defined above) are specified. 
#
# iA: index to the irreducible representations of A
# iG: index to the irreducible representations of G
#
 local k,sum,l1,l2,l3,l4,repGx;
 l1:=List(Gmem,x->x*g*x^-1);
 l2:=List(Gmem,x->GAction(a,x));
 l3:=List(l2,x->x^repA[iA]);
 repGx:=Irr(Gx);
 sum:=0;
 for k in [1..Size(l1)]
 do 
  if l1[k] in Gx then   
    l4:=l1[k]^repGx[iG];
    sum:=sum+l3[k]*l4;
  fi;
 od;
 sum:=sum/Size(Gx);
 return sum;
end;

ProChr:=function(au,iA)
#
# au  : the automorphism, defined in the list as above.
# iA  : the index to the irreducible representations in A
#
# This function computes the characters in the semidirect product
# (for each element in Elist, defined above).
#
 local Gx,repGx,index,retTBL;
 Gx:=GetGx(iA,au);
 repGx:=Irr(Gx);
 retTBL:=[];
 for index in [1..Size(repGx)] 
 do;
   Append(retTBL,
   [List(Elist,x->ProSumma(Amem[x[1]],Gmem[x[2]],Gx,iA,index))]
   ) ;
 od;
 return retTBL;
end;

gap> for iA in [1..Size(repA)]
do;
  Display(ProChr(au,iA));
od;
[ [   1,   1,   1,   1,   1,   1 ],
  [   1,  -1,  -1,  -1,   1,   1 ] ]
[ [   2,   0,   0,   0,  -1,  -1 ] ]
[ [   2,   0,   0,   0,  -1,  -1 ] ]
\end{verbatim}

\section{The analytical representation of the energy spectra of C$_{60}$.}
\label{AnalyticSpectra}

In the model of C$_{60}$ in section \ref{TheAnalysisOfTheEigenstates}, if we chose the transfer integral as H$_{\rm single}$=-1 (on the single bonds in the pentagons) and H$_{\rm double}$=-y (on the double bonds in the hexagons), the secular equation is decomposed into the polynomial with some multiplicity in each representation, as in  table \ref{POLYNOMIALSC60}. The irreducible representations I.1,...,I.10 are the same as given in the main article, at table \ref{charactertbl}. The energy spectra of in the main article can be constructed from this table, setting $\rm E=2x$ and y=-1/2. 

\begin{table}[h!]
\begin{tabular}{l l l }\hline
  & Polynomial  & multiplicity \\\hline
I.1&\verb![ x+y+2 ]!& 1\\
I.2&\verb![  ]!&   \\
I.3&\verb![ x-y+(E(5)^2+E(5)^3) ]!& 3\\
I.4&\verb![ x-y+(E(5)+E(5)^4) ]!& 3\\
I.5&\verb![ x^2-y^2+(-E(5)-2*E(5)^2-2*E(5)^3-E(5)^4)*x !& 3\\
   &\verb!   +(-E(5)-E(5)^4)*y+(2*E(5)+2*E(5)^4) ]!& \\
I.6&\verb![ x^2-y^2+(-2*E(5)-E(5)^2-E(5)^3-2*E(5)^4)*x !& 3\\
   &\verb!   +(-E(5)^2-E(5)^3)*y+(2*E(5)^2+2*E(5)^3) ]!& \\
I.7&\verb![ x^2-y^2-x-1 ]!& 4\\
I.8&\verb![ x^2-y^2-x-2*y-1 ]!& 4\\
I.9&\verb![ x^3+x^2*y-x*y^2-y^3+x^2+2*x*y+y^2-3*x-y-2 ]!& 5\\
I.10&\verb![ x^2-y^2-x+y-1 ]!& 5\\\hline
\end{tabular}
\caption{The decomposition of the secular equation.}
\label{POLYNOMIALSC60}
\end{table}

\section{Comment on the irreducible representations in the symmetry group of C$_{60}$ }
\label{FOURDIM}
In section \ref{vibrationalmodec60} we have computed the irreducible matrix representations of the symmetry group of C$_{60}$. 
Since the tenth irreducible matrix representation is a real representation, the generated bases are real and we can work in the real number. Meanwhile, the four-dimensional irreducible matrix representations (computed by GAP) are complex and the generated basis are complex. However, the complex irreducible representation, in this case, can be remade into real one. The complex-valued matrix representation of a generator ($M_i$) can be replaced by a real matrix:
$$
\left(\begin{array}{cc} Re(M_i) & -Im(M_i) \\ Im(M_i) & Re(M_i) \end{array}\right).
$$
This representation is reducible, including two copies of the irreducible representation, but as the matrix group, it is isomorphic to the original group. We extract the basis set from the columns of the projector of this matrix group and divide it into two independent irreducible vector spaces by means of the procedure explained thereafter in the article. Each of the separated vector space provides us the real-valued basis set and the real-valued matrix representation. One set of such real matrix representations (for the generators r1 and r2) is given as follows:
\begin{verbatim}
r1:=[[1,
      3/2*E(5)+1/2*E(5)^2+1/2*E(5)^3+3/2*E(5)^4,
     -3*E(5)-7*E(5)^2-7*E(5)^3-3*E(5)^4,
     -1081/2131*E(5)-1797/2131*E(5)^2-1797/2131*E(5)^3-1081/2131*E(5)^4],
     [0,
     -1/2*E(5)-E(5)^2-E(5)^3-1/2*E(5)^4,
     -5*E(5)-14*E(5)^2-14*E(5)^3-5*E(5)^4,
     -1681/2131*E(5)-3510/2131*E(5)^2-3510/2131*E(5)^3-1681/2131*E(5)^4],
     [0,
     -2*E(5)-3/2*E(5)^2-3/2*E(5)^3-2*E(5)^4,
     -2*E(5)-7*E(5)^2-7*E(5)^3-2*E(5)^4,
    -600/2131*E(5)-1713/2131*E(5)^2-1713/2131*E(5)^3-600/2131*E(5)^4],
     [0,
      113/44*E(5)+92/11*E(5)^2+92/11*E(5)^3+113/44*E(5)^4,
      623/22*E(5)+807/11*E(5)^2+807/11*E(5)^3+623/22*E(5)^4,
      5/2*E(5)+8*E(5)^2+8*E(5)^3+5/2*E(5)^4]];

r2:=[[-4/11*E(5)-3/11*E(5)^2-3/11*E(5)^3-4/11*E(5)^4,
      7/22*E(5)+19/22*E(5)^2+19/22*E(5)^3+7/22*E(5)^4,
      36/11*E(5)+104/11*E(5)^2+104/11*E(5)^3+36/11*E(5)^4,
      383/2131*E(5)+2127/2131*E(5)^2+2127/2131*E(5)^3+383/2131*E(5)^4],
     [12/11*E(5)-2/11*E(5)^2-2/11*E(5)^3+12/11*E(5)^4,
     -16/11*E(5)-1/11*E(5)^2-1/11*E(5)^3-16/11*E(5)^4,
      57/11*E(5)+128/11*E(5)^2+128/11*E(5)^3+57/11*E(5)^4,
     -1516/2131*E(5)+1596/2131*E(5)^2+1596/2131*E(5)^3-1516/2131*E(5)^4],
      [32/11*E(5)+13/11*E(5)^2+13/11*E(5)^3+32/11*E(5)^4,
      -39/11*E(5)-31/22*E(5)^2-31/22*E(5)^3-39/11*E(5)^4,
       53/11*E(5)+92/11*E(5)^2+92/11*E(5)^3+53/11*E(5)^4,
      -1784/2131*E(5)+703/2131*E(5)^2+703/2131*E(5)^3-1784/2131*E(5)^4],
      [0,0, 
      -623/22*E(5)-807/11*E(5)^2-807/11*E(5)^3-623/22*E(5)^4,
      -2*E(5)-7*E(5)^2-7*E(5)^3-2*E(5)^4]];
\end{verbatim}
We can use them in common in the two four-dimensional representations, with 
$\rm r3 := \pm I$ (the four dimensional unit matrix).
\section{The GAP script in the computation of deformation in C$_{60}$}
\label{Cutout}

As is pointed in the main article, some of the modes of deformation in C$_{60}$ computed by the projectors includes several copies of proper irreducible representations. To decompose them into separated irreducible spaces, the following function is prepared. The function tries to find the irreducible subspace of dimension ``REPSIZ'' in the operation of the group ``R''. The function cuts out the subspace (as minimal as possible) from the given vector space, making use of the coset decomposition of R by the subgroups H of order ``NMULTI''. In projecting out the proper irreducible component from the multiplied copies, we employ these steps. Let W be the initial starting place, that is, the multiplied components of the irreducible representations, from which the separated proper irreducible components $\rm W_1,...,W_i$ are computed. The next candidate of the generator v$_{i+1}$ is not taken directly from the complementary space ($W\setminus (W_1 \cup W_2 \cup \cdots \cup W_i)$); instead, in the whole vector space S (of 180 dimension in this case), the orthogonal space to $W_1 \cup W_2 \cup \cdots \cup W_i$, biz, $S\setminus(W_1 \cup W_2 \cup \cdots \cup W_i)$ is constructed. The basis set in the latter space is projected into the irreducible representation under the consideration, and the candidate v$_{i+1}$ is taken from this projected space (the renewed W). This treatment is to quicken the symbolic computation. 
\begin{verbatim} 
gopr:=function(O,v,i,hom)
#
# When i=1, This function applies the symmetry operations of Group O
# to a vector v. And the generated vectors are summed up.
# (In fact, this function is the operation of the projector to the vector v.)
#
 local irr;
 irr:=Irr(O);
 return Sum(List(Elements(O),j->j^irr[i]*LargeMatrix(j,hom)*v));
end;

Cutout:=function(BAS,PROJ,R,hom,NMULTI,REPSIZ)
#
# BAS    : THE VECTOR SPACE INCLUDING THE COPIES OF IRREDUCIBLE SPACES
# PROJ   : THE PROJECTORS
# R      : THE GROUP IN THE PROBLEM.
# hom    : THE HOMOMORPHISM FROM <R> TO THE MATRIX GROUP. 
# NMULTI : THE SIZE OF THE SUBGROUP TO FORM THE COSET.
# REPSIZ : THE SIZE OF THE IRREDUCIBLE REPRESENTATION.
#
 local C,D,BASN,EA,O,l,P,p,B,OV,SI,AS,OREP,ww,OT,ith,ifind;
 C:=[];
 D:=[];
 BASN:=BAS;
 EA:=List(Elements(R),j->LargeMatrix(j,hom));;
#
# The subgroups with the order of "NMULTI" are extracted.
#
 AS:=AllSubgroups(R);
 while (BASN<>[]) 
 do
   Print("\n",Size(BASN),"\n");
   O:=Filtered(AS,g->Order(g)=NMULTI);;
   Print(O);
#
# The subgroups which generate the vector space of 
# the correct dimension of the irreducible representation 
# are chosen.
#
  if (O=[]) then
   O:=[Group(One(R))];
  fi;
  l:=List([1..Size(O)],i->0);
  ifind:=0;
  ith:=1;
  while (ifind=0 and ith<=Size(O))
  do
   ww:=gopr(O[ith],BASN[1],1,hom);
   l[ith]:=RankMat(List(EA,g->g*ww));
   Print(l);  
   if (l[ith]=REPSIZ) then
    ifind:=1;
    OT:=O[ith];
   fi;
   ith:=ith+1;
  od;
  if (ifind=0) then
   Print("CHANGE THE SUPPOSITION!\n");
   return 0;
  fi;
#
#  By means of the chosen subgroup, the irreducible 
#  vector space "B" is generated. 
#  It is stored in the vector space "D".
#
  B:=BaseMat(List(EA,g->g*gopr(OT,BASN[1],1,hom)));;
  Append(C,[B]);
  Append(D,B);
#
# The orthogonal space "OV" to the vector space "D" is prepared.
#
  OV:=BaseOrthogonalSpaceMat(D);
#
# The vector space "BASN" to be processed is renewed now, 
# by the projection "PROJ" on the vectors in "OV".
#
  BASN:=BaseMat(List(OV,o->PROJ*o));
#
# Instead, it is possible that "BASN" is renewed 
# as the intersection between "BASN" and "OV". 
#
#  SI:=SumIntersectionMat(BASN,OV);
#  BASN:=SI[2];
#
 od;
#
# Return C, the separated subspaces.
#
 return C;
end;
\end{verbatim}

\section*{Supplement: matrix generators of the point group}
In this supplement I give the matrix generators, in order to define the point group. They are written in the GAP language. 
\begin{verbatim}
#
# THE MATRIX IN THE POINT GROUP OF CUBIC LATTICE.
#
# THE OPERATIONS IN THE REAL SPACE.
#
CMR:=[[[1,0,0],[0,1,0],[0,0,1]],[[1,0,0],[0,-1,0],[0,0,-1]],
[[-1,0,0],[0,1,0],[0,0,-1]],[[-1,0,0],[0,-1,0],[0,0,1]],
[[0,1,0],[0,0,1],[1,0,0]],[[0,1,0],[0,0,-1],[-1,0,0]],
[[0,-1,0],[0,0,1],[-1,0,0]],[[0,-1,0],[0,0,-1],[1,0,0]],
[[0,0,1],[1,0,0],[0,1,0]],[[0,0,1],[-1,0,0],[0,-1,0]],
[[0,0,-1],[1,0,0],[0,-1,0]],[[0,0,-1],[-1,0,0],[0,1,0]],
[[0,-1,0],[-1,0,0],[0,0,-1]],[[0,-1,0],[1,0,0],[0,0,1]],
[[0,1,0],[-1,0,0],[0,0,1]],[[0,1,0],[1,0,0],[0,0,-1]],
[[-1,0,0],[0,0,-1],[0,-1,0]],[[-1,0,0],[0,0,1],[0,1,0]],
[[1,0,0],[0,0,-1],[0,1,0]],[[1,0,0],[0,0,1],[0,-1,0]],
[[0,0,-1],[0,-1,0],[-1,0,0]],[[0,0,-1],[0,1,0],[1,0,0]],
[[0,0,1],[0,-1,0],[1,0,0]],[[0,0,1],[0,1,0],[-1,0,0]],
[[-1,0,0],[0,-1,0],[0,0,-1]],[[-1,0,0],[0,1,0],[0,0,1]],
[[1,0,0],[0,-1,0],[0,0,1]],[[1,0,0],[0,1,0],[0,0,-1]],
[[0,-1,0],[0,0,-1],[-1,0,0]],[[0,-1,0],[0,0,1],[1,0,0]],
[[0,1,0],[0,0,-1],[1,0,0]],[[0,1,0],[0,0,1],[-1,0,0]],
[[0,0,-1],[-1,0,0],[0,-1,0]],[[0,0,-1],[1,0,0],[0,1,0]],
[[0,0,1],[-1,0,0],[0,1,0]],[[0,0,1],[1,0,0],[0,-1,0]],
[[0,1,0],[1,0,0],[0,0,1]],[[0,1,0],[-1,0,0],[0,0,-1]],
[[0,-1,0],[1,0,0],[0,0,-1]],[[0,-1,0],[-1,0,0],[0,0,1]],
[[1,0,0],[0,0,1],[0,1,0]],[[1,0,0],[0,0,-1],[0,-1,0]],
[[-1,0,0],[0,0,1],[0,-1,0]],[[-1,0,0],[0,0,-1],[0,1,0]],
[[0,0,1],[0,1,0],[1,0,0]],[[0,0,1],[0,-1,0],[-1,0,0]],
[[0,0,-1],[0,1,0],[-1,0,0]],[[0,0,-1],[0,-1,0],[1,0,0]]];;
#
# THE OPERATIONS IN THE RECIPROCAL SPACE.
# 
CMT:=[[[1,0,0],[0,1,0],[0,0,1]],[[-1,0,0],[-1,0,1],[-1,1,0]],
[[0,-1,1],[0,-1,0],[1,-1,0]],[[0,1,-1],[1,0,-1],[0,0,-1]],
[[0,1,0],[0,0,1],[1,0,0]],[[0,-1,0],[1,-1,0],[0,-1,1]],
[[1,0,-1],[0,0,-1],[0,1,-1]],[[-1,0,1],[-1,1,0],[-1,0,0]],
[[0,0,1],[1,0,0],[0,1,0]],[[0,0,-1],[0,1,-1],[1,0,-1]],
[[-1,1,0],[-1,0,0],[-1,0,1]],[[1,-1,0],[0,-1,1],[0,-1,0]],
[[0,-1,0],[-1,0,0],[0,0,-1]],[[0,1,0],[0,1,-1],[-1,1,0]],
[[1,0,-1],[1,0,0],[1,-1,0]],[[-1,0,1],[0,-1,1],[0,0,1]],
[[-1,0,0],[0,0,-1],[0,-1,0]],[[1,0,0],[1,-1,0],[1,0,-1]],
[[0,-1,1],[0,0,1],[-1,0,1]],[[0,1,-1],[-1,1,0],[0,1,0]],
[[0,0,-1],[0,-1,0],[-1,0,0]],[[0,0,1],[-1,0,1],[0,-1,1]],
[[-1,1,0],[0,1,0],[0,1,-1]],[[1,-1,0],[1,0,-1],[1,0,0]],
[[-1,0,0],[0,-1,0],[0,0,-1]],[[1,0,0],[1,0,-1],[1,-1,0]],
[[0,1,-1],[0,1,0],[-1,1,0]],[[0,-1,1],[-1,0,1],[0,0,1]],
[[0,-1,0],[0,0,-1],[-1,0,0]],[[0,1,0],[-1,1,0],[0,1,-1]],
[[-1,0,1],[0,0,1],[0,-1,1]],[[1,0,-1],[1,-1,0],[1,0,0]],
[[0,0,-1],[-1,0,0],[0,-1,0]],[[0,0,1],[0,-1,1],[-1,0,1]],
[[1,-1,0],[1,0,0],[1,0,-1]],[[-1,1,0],[0,1,-1],[0,1,0]],
[[0,1,0],[1,0,0],[0,0,1]],[[0,-1,0],[0,-1,1],[1,-1,0]],
[[-1,0,1],[-1,0,0],[-1,1,0]],[[1,0,-1],[0,1,-1],[0,0,-1]],
[[1,0,0],[0,0,1],[0,1,0]],[[-1,0,0],[-1,1,0],[-1,0,1]],
[[0,1,-1],[0,0,-1],[1,0,-1]],[[0,-1,1],[1,-1,0],[0,-1,0]],
[[0,0,1],[0,1,0],[1,0,0]],[[0,0,-1],[1,0,-1],[0,1,-1]],
[[1,-1,0],[0,-1,0],[0,-1,1]],[[-1,1,0],[-1,0,1],[-1,0,0]]];;
#
# THE PRIMITIVE TRANSLATIONS.
#
A:=[[0,1,1],[1,0,1],[1,1,0]]/2;;
#
# TRANSFORMATION FROM MR TO MT;
#
CMT=List(CMR,m->TransposedMat(A^-1*TransposedMat(m)*A));
#
# THE MULTIPLICATION TABLE.
# 
TBL:=List(CMR,i->List(CMR,j->Position(CMR,i*j)));
#
#
# THE MATRIX IN THE POINT GROUP OF HEXAGONAL LATTICE.
#
# THE PRIMITIVE TRANSLATIONS HA:=[A1,A2,A3]; 
#    [ Sqrt(3)/2=-1/2*E(12)^7+1/2*E(12)^11 ] 
#
HA:=[[1,-1/2,0],[0,-1/2*E(12)^7+1/2*E(12)^11,0],[0,0,1]];;
#
# THE OPERATIONS IN THE REAL SPACE.
#
HMR:=[[[1,0,0],[0,1,0],[0,0,1]],
[[1/2,1/2*E(12)^7-1/2*E(12)^11,0],[-1/2*E(12)^7+1/2*E(12)^11,1/2,0],[0,0,1]],
[[-1/2,1/2*E(12)^7-1/2*E(12)^11,0],[-1/2*E(12)^7+1/2*E(12)^11,-1/2,0],[0,0,1]],
[[-1,0,0],[0,-1,0],[0,0,1]],
[[-1/2,-1/2*E(12)^7+1/2*E(12)^11,0],[1/2*E(12)^7-1/2*E(12)^11,-1/2,0],[0,0,1]],
[[1/2,-1/2*E(12)^7+1/2*E(12)^11,0],[1/2*E(12)^7-1/2*E(12)^11,1/2,0],[0,0,1]],
[[-1/2,1/2*E(12)^7-1/2*E(12)^11,0],[1/2*E(12)^7-1/2*E(12)^11,1/2,0],[0,0,-1]],
[[1/2,1/2*E(12)^7-1/2*E(12)^11,0],[1/2*E(12)^7-1/2*E(12)^11,-1/2,0],[0,0,-1]],
[[1,0,0],[0,-1,0],[0,0,-1]],
[[1/2,-1/2*E(12)^7+1/2*E(12)^11,0],[-1/2*E(12)^7+1/2*E(12)^11,-1/2,0],[0,0,-1]],
[[-1/2,-1/2*E(12)^7+1/2*E(12)^11,0],[-1/2*E(12)^7+1/2*E(12)^11,1/2,0],[0,0,-1]],
[[-1,0,0],[0,1,0],[0,0,-1]],[[-1,0,0],[0,-1,0],[0,0,-1]],
[[-1/2,-1/2*E(12)^7+1/2*E(12)^11,0],[1/2*E(12)^7-1/2*E(12)^11,-1/2,0],[0,0,-1]],
[[1/2,-1/2*E(12)^7+1/2*E(12)^11,0],[1/2*E(12)^7-1/2*E(12)^11,1/2,0],[0,0,-1]],
[[1,0,0],[0,1,0],[0,0,-1]],
[[1/2,1/2*E(12)^7-1/2*E(12)^11,0],[-1/2*E(12)^7+1/2*E(12)^11,1/2,0],[0,0,-1]],
[[-1/2,1/2*E(12)^7-1/2*E(12)^11,0],[-1/2*E(12)^7+1/2*E(12)^11,-1/2,0],[0,0,-1]],
[[1/2,-1/2*E(12)^7+1/2*E(12)^11,0],[-1/2*E(12)^7+1/2*E(12)^11,-1/2,0],[0,0,1]],
[[-1/2,-1/2*E(12)^7+1/2*E(12)^11,0],[-1/2*E(12)^7+1/2*E(12)^11,1/2,0],[0,0,1]],
[[-1,0,0],[0,1,0],[0,0,1]],
[[-1/2,1/2*E(12)^7-1/2*E(12)^11,0],[1/2*E(12)^7-1/2*E(12)^11,1/2,0],[0,0,1]],
[[1/2,1/2*E(12)^7-1/2*E(12)^11,0],[1/2*E(12)^7-1/2*E(12)^11,-1/2,0],[0,0,1]],
[[1,0,0],[0,-1,0],[0,0,1]]];;
#
# 
# THE OPERATIONS IN THE RECIPROCAL SPACE.
#
HMT:=[[[1,0,0],[0,1,0],[0,0,1]],[[0,-1,0],[1,1,0],[0,0,1]],
[[-1,-1,0],[1,0,0],[0,0,1]],[[-1,0,0],[0,-1,0],[0,0,1]],
[[0,1,0],[-1,-1,0],[0,0,1]],[[1,1,0],[-1,0,0],[0,0,1]],
[[-1,-1,0],[0,1,0],[0,0,-1]],[[0,-1,0],[-1,0,0],[0,0,-1]],
[[1,0,0],[-1,-1,0],[0,0,-1]],[[1,1,0],[0,-1,0],[0,0,-1]],
[[0,1,0],[1,0,0],[0,0,-1]],[[-1,0,0],[1,1,0],[0,0,-1]],
[[-1,0,0],[0,-1,0],[0,0,-1]],[[0,1,0],[-1,-1,0],[0,0,-1]],
[[1,1,0],[-1,0,0],[0,0,-1]],[[1,0,0],[0,1,0],[0,0,-1]],
[[0,-1,0],[1,1,0],[0,0,-1]],[[-1,-1,0],[1,0,0],[0,0,-1]],
[[1,1,0],[0,-1,0],[0,0,1]],[[0,1,0],[1,0,0],[0,0,1]],
[[-1,0,0],[1,1,0],[0,0,1]],[[-1,-1,0],[0,1,0],[0,0,1]],
[[0,-1,0],[-1,0,0],[0,0,1]],[[1,0,0],[-1,-1,0],[0,0,1]]];;
#
#
# THE OPERATIONS (ROTATIONS AND TRANSLATIONS), 
# AS ARE GIVEN IN THE TABLES IN THE ARTICLE.
#
# CUBIC LATTICE.(DIAMOND)
#
OPRCUB:=[[1,1,0,0,0,1,0,0,0,1,0,0,0,0,0,0],
[2,1,0,0,0,-1,0,0,0,-1,1/2,0,0,0,0,0],
[3,-1,0,0,0,1,0,0,0,-1,0,1/2,0,0,0,0],
[4,-1,0,0,0,-1,0,0,0,1,0,0,1/2,0,0,0],
[5,0,1,0,0,0,1,1,0,0,0,0,0,0,0,0],
[6,0,1,0,0,0,-1,-1,0,0,1/2,0,0,0,0,0],
[7,0,-1,0,0,0,1,-1,0,0,0,1/2,0,0,0,0],
[8,0,-1,0,0,0,-1,1,0,0,0,0,1/2,0,0,0],
[9,0,0,1,1,0,0,0,1,0,0,0,0,0,0,0],
[10,0,0,1,-1,0,0,0,-1,0,1/2,0,0,0,0,0],
[11,0,0,-1,1,0,0,0,-1,0,0,1/2,0,0,0,0],
[12,0,0,-1,-1,0,0,0,1,0,0,0,1/2,0,0,0],
[13,0,-1,0,-1,0,0,0,0,-1,0,0,0,1/4,1/4,1/4],
[14,0,-1,0,1,0,0,0,0,1,-1/2,0,0,1/4,1/4,1/4],
[15,0,1,0,-1,0,0,0,0,1,0,-1/2,0,1/4,1/4,1/4],
[16,0,1,0,1,0,0,0,0,-1,0,0,-1/2,1/4,1/4,1/4],
[17,-1,0,0,0,0,-1,0,-1,0,0,0,0,1/4,1/4,1/4],
[18,-1,0,0,0,0,1,0,1,0,-1/2,0,0,1/4,1/4,1/4],
[19,1,0,0,0,0,-1,0,1,0,0,-1/2,0,1/4,1/4,1/4],
[20,1,0,0,0,0,1,0,-1,0,0,0,-1/2,1/4,1/4,1/4],
[21,0,0,-1,0,-1,0,-1,0,0,0,0,0,1/4,1/4,1/4],
[22,0,0,-1,0,1,0,1,0,0,-1/2,0,0,1/4,1/4,1/4],
[23,0,0,1,0,-1,0,1,0,0,0,-1/2,0,1/4,1/4,1/4],
[24,0,0,1,0,1,0,-1,0,0,0,0,-1/2,1/4,1/4,1/4],
[25,-1,0,0,0,-1,0,0,0,-1,0,0,0,1/4,1/4,1/4],
[26,-1,0,0,0,1,0,0,0,1,-1/2,0,0,1/4,1/4,1/4],
[27,1,0,0,0,-1,0,0,0,1,0,-1/2,0,1/4,1/4,1/4],
[28,1,0,0,0,1,0,0,0,-1,0,0,-1/2,1/4,1/4,1/4],
[29,0,-1,0,0,0,-1,-1,0,0,0,0,0,1/4,1/4,1/4],
[30,0,-1,0,0,0,1,1,0,0,-1/2,0,0,1/4,1/4,1/4],
[31,0,1,0,0,0,-1,1,0,0,0,-1/2,0,1/4,1/4,1/4],
[32,0,1,0,0,0,1,-1,0,0,0,0,-1/2,1/4,1/4,1/4],
[33,0,0,-1,-1,0,0,0,-1,0,0,0,0,1/4,1/4,1/4],
[34,0,0,-1,1,0,0,0,1,0,-1/2,0,0,1/4,1/4,1/4],
[35,0,0,1,-1,0,0,0,1,0,0,-1/2,0,1/4,1/4,1/4],
[36,0,0,1,1,0,0,0,-1,0,0,0,-1/2,1/4,1/4,1/4],
[37,0,1,0,1,0,0,0,0,1,0,0,0,0,0,0],
[38,0,1,0,-1,0,0,0,0,-1,1/2,0,0,0,0,0],
[39,0,-1,0,1,0,0,0,0,-1,0,1/2,0,0,0,0],
[40,0,-1,0,-1,0,0,0,0,1,0,0,1/2,0,0,0],
[41,1,0,0,0,0,1,0,1,0,0,0,0,0,0,0],
[42,1,0,0,0,0,-1,0,-1,0,1/2,0,0,0,0,0],
[43,-1,0,0,0,0,1,0,-1,0,0,1/2,0,0,0,0],
[44,-1,0,0,0,0,-1,0,1,0,0,0,1/2,0,0,0],
[45,0,0,1,0,1,0,1,0,0,0,0,0,0,0,0],
[46,0,0,1,0,-1,0,-1,0,0,1/2,0,0,0,0,0],
[47,0,0,-1,0,1,0,-1,0,0,0,1/2,0,0,0,0],
[48,0,0,-1,0,-1,0,1,0,0,0,0,1/2,0,0,0]];;
#
#
# HEXAGONAL LATTICE.
#
OPRHEXA:=[[1,1,0,0,0,1,0,0,0,1,0,0,0],
[2,1/2,-Sqrt(3)/2,0,Sqrt(3)/2,1/2,0,0,0,1,0,0,-1/2],
[3,-1/2,-Sqrt(3)/2,0,Sqrt(3)/2,-1/2,0,0,0,1,0,0,0],
[4,-1,0,0,0,-1,0,0,0,1,0,0,-1/2],
[5,-1/2,Sqrt(3)/2,0,-Sqrt(3)/2,-1/2,0,0,0,1,0,0,0],
[6,1/2,Sqrt(3)/2,0,-Sqrt(3)/2,1/2,0,0,0,1,0,0,-1/2],
[7,-1/2,-Sqrt(3)/2,0,-Sqrt(3)/2,1/2,0,0,0,-1,0,0,-1/2],
[8,1/2,-Sqrt(3)/2,0,-Sqrt(3)/2,-1/2,0,0,0,-1,0,0,0],
[9,1,0,0,0,-1,0,0,0,-1,0,0,-1/2],
[10,1/2,Sqrt(3)/2,0,Sqrt(3)/2,-1/2,0,0,0,-1,0,0,0],
[11,-1/2,Sqrt(3)/2,0,Sqrt(3)/2,1/2,0,0,0,-1,0,0,-1/2],
[12,-1,0,0,0,1,0,0,0,-1,0,0,0],
[13,-1,0,0,0,-1,0,0,0,-1,0,0,-1/2],
[14,-1/2,Sqrt(3)/2,0,-Sqrt(3)/2,-1/2,0,0,0,-1,0,0,0],
[15,1/2,Sqrt(3)/2,0,-Sqrt(3)/2,1/2,0,0,0,-1,0,0,-1/2],
[16,1,0,0,0,1,0,0,0,-1,0,0,0],
[17,1/2,-Sqrt(3)/2,0,Sqrt(3)/2,1/2,0,0,0,-1,0,0,-1/2],
[18,-1/2,-Sqrt(3)/2,0,Sqrt(3)/2,-1/2,0,0,0,-1,0,0,0],
[19,1/2,Sqrt(3)/2,0,Sqrt(3)/2,-1/2,0,0,0,1,0,0,0],
[20,-1/2,Sqrt(3)/2,0,Sqrt(3)/2,1/2,0,0,0,1,0,0,-1/2],
[21,-1,0,0,0,1,0,0,0,1,0,0,0],
[22,-1/2,-Sqrt(3)/2,0,-Sqrt(3)/2,1/2,0,0,0,1,0,0,-1/2],
[23,1/2,-Sqrt(3)/2,0,-Sqrt(3)/2,-1/2,0,0,0,1,0,0,0],
[24,1,0,0,0,-1,0,0,0,1,0,0,-1/2]];;
\end{verbatim}

\section*{Supplement:Some functions used in the computations at "Symmetry in C$_{60}$"}
\begin{verbatim}

#
# This file contains some functions used in the computations
# in VII. "SYMMETRY IN C60".
#

listtoh:=function(A)
 local h,i1,i2,i3,i4,k,l,i;
 h:=List([1..120],i->List([1..120],j->0));
 for i in A do
 i1:=i[1];
 i2:=i[2];
 i3:=i[3];
 i4:=i[4];
 k:=(i1-2)+60*(i2-1);
 l:=(i3-2)+60*(i4-1);
 h[k][l]:=1;
 h[l][k]:=1;
 od;
 return h;
end;

#
# We construct an extended group by the direct product 
# between A5 and one of the automorphism group.
#
# The new group contains 120 elements, represented by 62 symbols.
# The symbols from the 3rd to the 62nd are descendants from A5.
# The symbols of the first and the second are the newcomers
# in the extension.
#
# From the 62 symbols, we construct a bonding system 
# which contains 120 atoms, regulated by the symmetry of the
# new group. (C120)  
# The atoms will be indexed as (i,j) (i=3,..,62, and j=1,2)
# by the duplication of single C60.
#

#
# The generators of the group. (A5 and I)
#
g1:=(  1, 14, 20)(  2, 15, 16)(  3, 11, 17)(  4, 12, 18)(  5, 13, 19)
(  6, 58, 22)(  7, 57, 23)(  8, 56, 24)(  9, 60, 25)( 10, 59, 21)
( 31, 44, 50)( 32, 45, 46)( 33, 41, 47)( 34, 42, 48)( 35, 43, 49)
( 36, 28, 52)( 37, 27, 53)( 38, 26, 54)( 39, 30, 55)( 40, 29, 51);

g2:=(  1,  2,  3,  4,  5)(  6, 11, 16, 21, 26)(  7, 12, 17, 22, 27)
(  8, 13, 18, 23, 28)(  9, 14, 19, 24, 29)( 10, 15, 20, 25, 30)
( 31, 32, 33, 34, 35)( 36, 41, 46, 51, 56)( 37, 42, 47, 52, 57)
( 38, 43, 48, 53, 58)( 39, 44, 49, 54, 59)( 40, 45, 50, 55, 60);

g3:=(  1, 31)(  2, 32)(  3, 33)(  4, 34)(  5, 35)(  6, 36)(  7, 37)
(  8, 38)(  9, 39)( 10, 40)( 11, 41)( 12, 42)( 13, 43)( 14, 44)
( 15, 45)( 16, 46)( 17, 47)( 18, 48)( 19, 49)( 20, 50)( 21, 51)
( 22, 52)( 23, 53)( 24, 54)( 25, 55)( 26, 56)( 27, 57)( 28, 58)
( 29, 59)( 30, 60);


# I define g2 in this way in the article.
g2:=g2^-1;
#

G:=Group(g1,g2);
AUALL:=AutomorphismGroup(G);
p:=SemidirectProduct(Group(AUALL.1),G);
N:=Image(Embedding(p,2));

#
# The orbits of the double bond [(3,1),(4,1)]
# and the single bonds [3,1,8,1]
# and the bond between the first and the second C60
# are computed as A1,B1,C1
#
# The bonding structures from them are represented by
# three matrices a1,b1,c1. 
#
A1:=Orbit(p,[3,1,4,1],OnTuples);
B1:=Orbit(p,[3,1,8,1],OnTuples);
C1:=Orbit(p,[3,1,4,2],OnTuples);
a1:=listtoh(A1);
b1:=listtoh(B1);
c1:=listtoh(C1);

#
# The symmetry operations of the 120 atoms are written as:  
#
l1:=List([1..2],j->List([3..62],i->(OnPoints(i,p.1)+60*(j-1))));
l1:=Flat(l1);
l2:=List([1..2],j->List([3..62],i->(OnPoints(i,p.2)+60*(j-1))));
l2:=Flat(l2);
l3:=List([1..2],j->List([3..62],i->(OnPoints(i,p.3)+60*(OnPoints(j,p.3)-1))));
l3:=Flat(l3);

#
#
inv1:=List([1..60],i->OnPoints(i,g3)+2);

inv2:=inv1+60;
inv:=Flat([inv1,inv2]);

ListToMat:=function(l)
 local size,h,i,j;
 size:=Size(l);
 h:=List([1..size],i->List([1..size],j->0));
 for i in [1..size] do
  h[i][l[i]]:=1;
 od;
 return h;
end;

#
# Now generators (p.1,p.2,p.3) are represented by matrix P1,P2,P3.
#
P1:=ListToMat(l1-2);
P2:=ListToMat(l2-2);
P3:=ListToMat(l3-2);

#
# In C60, of the icosahedral symmetry, the symmetry operation also includes
# the inversion. The inversion itself is not includes in the new group,
# but we can write the operation of this operation on first 60 vertexes
# (from the 3rd to the 62nd)
# The inversion in the second 60 vertexes can be written in the following way.
# The second 60 vertexes are transmuted from the first 60, by the operation p.3.
#
# If the inversions on the first and second 60 vertexes are denoted as I1,I2,
# they are related by the conjugation I2:= (p.3)*I1*(p.3)^(-1). 
# The inversion on the total system is written as
# [[ I1,  0]
#  [  0, I2]].
#
#

# ListInv: the operation of "g3" in the first shell.
ListInv:=List([1..60],i->OnPoints(i,g3));
# mconjg : the operation of "p.3" from the first to the second shell.
mconjg:=List([3..62],i->OnPoints(i,p.3))-2;

#
# These operations are put into matrix, "invrs", "mc".
#
invrs:=List([1..60],i->List([1..60],j->0));
mc:=List([1..60],i->List([1..60],j->0));
for i in [1..60] do
 invrs[i][ListInv[i]]:=1;
 mc[i][mconjg[i]]:=1;
od;
invrs2:=mc*invrs*mc;

# "invrs": the inversion in the first shell.
# "invrs2": the inversion in the second shell.
#
# Set the inversions in two shells in a matrix "ivfull".
#
ivfull:=List([1..120],i->List([1..120],j->0));
for i in [1..60] do
 for j in [1..60] do
  ivfull[i][j]:=invrs[i][j];
  ivfull[i+60][j+60]:=invrs2[i][j];
 od;
od;	

#
# As the matrix c1 are not conserved by the inversion "ivfull",
# (as ivfull*c1*ivfull^-1 is not equal to c1)
# it is modified to have inversion symmetry.
#

hh:=-2*a1-b1-1/2*(c1+ivfull*c1*ivfull);

#
# The generalized eigenvalue problems are solved in 
# the field of rational numbers. 
#
gev:=GeneralisedEigenvalues(Rationals,hh);
ges:=GeneralisedEigenspaces(Rationals,hh);

#
# The irreducible representation is computed.
#
# "irrg" is the group of the symmetry in the composed structure.
# "irrg2" is the subgroup.
# 
irrg:=Irr(Group(P1,P2,P3,ivfull));
irrg2:=Irr(Group(P1,P2,ivfull));
Em:=Elements(Group(P1,P2,P3,ivfull));;
Em2:=Elements(Group(P1,P2,ivfull));;

Projopr:=function(irr,Em)
#
# The projector, by means of characters.
#
return List(irr,i->Sum(List(Em,j->j^i*j))/Size(Em)*i[1]);
end;

prjs:=Projopr(irrg,Em);;
prjs2:=Projopr(irrg2,Em2);;

#
# The basis vectors are computed and allotted to the irreducible representation.
#
bases:=List(ges,i->Basis(i));

for b in bases do;Display(List(b,j->List( prjs,p->(p*j)^2 ))); od;

for b in bases do;Display(List(b,j->List(prjs2,p->(p*j)^2 )));od;
\end{verbatim}

\section*{Supplement:some functions used in the computations at "Analisys of vibrational mode in C$_{60}$"}

\begin{verbatim}
#
# THIS FILE CONTAINS SMALL FUNCTIONS FOR THE COMPUTATION CONCERNING C60.
# (VIII."ANALYSIS OF VIBRATIONAL MODE IN C60")
#
cc:=(E(5)+E(5)^4)/2;
ss:=(E(5)-E(5)^4)/2/E(4);
cc10:=(E(10)+E(10)^4)/2;
ss10:=(E(10)-E(10)^4)/2/E(4);

x:=Indeterminate(Rationals,"x");
z:=Indeterminate(Rationals,"z");
y:=Indeterminate(Rationals,"y");


#
# MATRIX OF ROTATION OF 72 DEGREES (A GENERATOR OF GROUP A5).
#
M72D:=[[cc,-ss,0],[ss,cc,0],[0,0,1]];
#

#
# HEAFTER, VERY TRIVIAL COMPUTATIONS ARE DONE
# IN ORDER TO GIVE 12 VERTEX ON ICOSAHEDRON.
#
v:=[[0,0,0],[0,0,0],[0,0,0]];
v[1]:=[0*x,0*x,x^0];
v[2]:=[x,0*x,z];
v[3]:=M72D^-1*v[2];
v21:=v[2]-v[1];
v32:=v[3]-v[2];
L21:=v21*v21;
L32:=v32*v32;
B:=[L21-L32,v[2]*v[2]-1];

GB:=GroebnerBasis(B,MonomialLexOrdering());
RTPS:=RootsOfPolynomial(GB[3]);
zval:=RTPS[2];
xval:=Sqrt(1-zval*zval);
vval:=List(v,i->List(i,j->Value(j,[x,z],[xval,zval])));

#
# NOW 12 VERTEX ARE PREPARED.
#
sommets:=List([1..12],i->0);
sommets[1]:=vval[1];
sommets[2]:=        vval[2];
sommets[3]:=M72D^-1*vval[2];
sommets[4]:=M72D^-2*vval[2];
sommets[5]:=M72D^-3*vval[2];
sommets[6]:=M72D^-4*vval[2];
sommets[7]:=-vval[1];
sommets[8]:=-        vval[2];
sommets[9]:=-M72D^-1*vval[2];
sommets[10]:=-M72D^-2*vval[2];
sommets[11]:=-M72D^-3*vval[2];
sommets[12]:=-M72D^-4*vval[2];

#
Tv:=TransposedMat(vval);
Tv2:=TransposedMat([vval[2],vval[3],vval[1]]);
Inv:=[[-1,0,0],[0,-1,0],[0,0,-1]];

#  
# THE THREE VECTORS, STORED IN THE COLUMNS in "Tv" AND "Tv2"
# ARE RELATED BY THE ROTATION, WHICH IS OTHER GENERATOR OF A5.
#
# THE GROUP (A5 x I) , DEFINED IN 3D-EUCLID SPACE.(THE GROUP G)
#
# THE GENERATORS OF THIS GROUP (G.1,G.2,G.3) ARE (g1,g2,g3) IN THE ARTICLE.
#
G:=Group(Tv2*Tv^-1,M72D^-1,Inv);
irrg:=Irr(G);
elm:=Elements(G);

#
# THE PROJECTORS, EXPLANED IN THE ARTICLE.
#
Projopr:=function(irr,Em,hom)
#
return List(irr,i->Sum(List(Em,j->j^i*Image(hom,j)))/Size(Em)*i[1]);
end;


ProjoprR:=function(irr,Em,MM,hom)
#
return 
List(irr,i->Sum(List(Em,j->j^i*Image(hom,j)*MM*TransposedMat(Image(hom,j))))
/Size(Em)*i[1]);
end;


#
# THE DEFINITION OF GROUP (A5 x I) BY THE PERMUTATION OF 12 VERTEX IN ICOSAHEDRON
# (GROUP F)
#
f1:=(1,2,3)(4,6,11)(5,10,12)(7,8,9);
f2:=(2,3,4,5,6)(8,9,10,11,12);
f3:=(1,7)(2,8)(3,9)(4,10)(5,11)(6,12);
F:=Group(f1,f2,f3);

#
# THE DEFINITION OF GROUP (A5 x I) BY THE PERMUTATION OF 60 VERTEX IN C60
# (GROUP R)
#
r1:=(  1, 14, 20)(  2, 15, 16)(  3, 11, 17)(  4, 12, 18)(  5, 13, 19)
(  6, 58, 22)(  7, 57, 23)(  8, 56, 24)(  9, 60, 25)( 10, 59, 21)
( 31, 44, 50)( 32, 45, 46)( 33, 41, 47)( 34, 42, 48)( 35, 43, 49)
( 36, 28, 52)( 37, 27, 53)( 38, 26, 54)( 39, 30, 55)( 40, 29, 51);
r2:=(  1,  2,  3,  4,  5)(  6, 11, 16, 21, 26)(  7, 12, 17, 22, 27)
(  8, 13, 18, 23, 28)(  9, 14, 19, 24, 29)( 10, 15, 20, 25, 30)
( 31, 32, 33, 34, 35)( 36, 41, 46, 51, 56)( 37, 42, 47, 52, 57)
( 38, 43, 48, 53, 58)( 39, 44, 49, 54, 59)( 40, 45, 50, 55, 60);
r3:=(  1, 31)(  2, 32)(  3, 33)(  4, 34)(  5, 35)(  6, 36)(  7, 37)
(  8, 38)(  9, 39)( 10, 40)( 11, 41)( 12, 42)( 13, 43)( 14, 44)
( 15, 45)( 16, 46)( 17, 47)( 18, 48)( 19, 49)( 20, 50)( 21, 51)
( 22, 52)( 23, 53)( 24, 54)( 25, 55)( 26, 56)( 27, 57)( 28, 58)
( 29, 59)( 30, 60);
R:=Group(r1,r2,r3);

#
# THE ISOMORPHISM BETWEEN GROUPS
#
iso:=IsomorphismGroups(F,R);
iso2:=IsomorphismGroups(F,G);
hom:=GroupHomomorphismByImages(F,G,[F.1,F.2,F.3],[G.1,G.2,G.3]);
hom2:=GroupHomomorphismByImages(R,G,[R.1,R.2,R.3],[G.1,G.2,G.3]);

Display(List(Irr(R),r->List([R.1,R.2,R.3],i->i^r)));
MM:=[[0*x,x,y],[-x,0*x,z],[-y,-z,0*x]];
PRJ:=Projopr(Irr(R),Elements(R),hom2);
PRJR:=ProjoprR(Irr(R),Elements(R),MM,hom2);


#
# THE FUNCTIONS DEFINED AND EXPLAINED IN THE ARTICLE.
#
#
ListToMat:=function(l)
 local size,h,i,j;
 size:=Size(l);
 h:=List([1..size],i->List([1..size],j->0));
 for i in [1..size] do
  h[i][l[i]]:=1;
 od;
 return h;
end;

LargeMatrix:=function(rr,iso)
 local l,rmat,rrimg,i,i1,i2,n,m;
 l:=ListPerm(rr,60);
 rmat:=List([1..Size(l)*3],i->List([1..Size(l)*3],j->0));
 rrimg:=Image(iso,rr);
 for i in [1..Size(l)]
 do 
  i1:=i-1;i2:=l[i]-1;
  for n in [1..3]
  do
   for m in [1..3]
   do
    rmat[3*i1+n][3*i2+m]:=rrimg[n][m];
   od;
  od;
 od;
 return rmat;
end;


#
# THE DEFINITION OF PROJECTOR, DEFINED IN THE ARTICLE
# IN ORDER TO COMPUTE THE MODE OF DEFORMATION IN C60.
#
ProjoprE:=function(irr,Em,hom2)
#
local A,B;
#A:=List(Em,j->LargeMatrix(j,hom2));
#B:=List(irr,i->List(Em,j->j^i));
return List(irr,i->Sum(List(Em,j->j^i*LargeMatrix(j,hom2)))/Size(Em)*i[1]);
end;

PRJE:=ProjoprE(Irr(R),Elements(R),hom2);
BAS:=List(PRJE,i->BaseMat(TransposedMat(i)));
BASSIZE:=List(BAS,Size);


CROT:=List(Elements(R),i->Trace(ListToMat(ListPerm(i,60))));
CIRR:=List(Irr(R),r->List(Elements(R),i->i^r));

# In the trace representation the basis set composed from column vectors 
# of the projector may include the multiplicated spaces of 
# the representation. The basis vectors might be transmuted 
# with each other transitively, not confined in the one of 
# the component of the multiplicated irreducible spaces.


# The basis set of the decomposion into the irreduciple 
# representations and the conjugacy classes are prepared. 

cng:=ConjugacyClasses(R);;
rcng:=List(cng,Representative);;
rcng2:=List(rcng,j->LargeMatrix(j,hom2));;

charact:=function(g,B)
return Trace(B*g*TransposedMat(B)*(B*TransposedMat(B))^-1);
end;


gopr:=function(O,v,i,hom)
local irr;
irr:=Irr(O);
return Sum(List(Elements(O),j->j^irr[i]*LargeMatrix(j,hom)*v));
end;

Cutout:=function(BAS,PROJ,R,hom,NMULTI,REPSIZ)
#
# BAS    : THE VECTOR SPACE INCLUDING THE COPIES OF IRREDUCIBLE SPACES
# PROJ   : THE PROJECTORS
# R      : THE GROUP IN THE PROBLEM.
# hom    : THE HOMOMORPHISM FROM <R> TO THE MATRIX GROUP. 
# NMULTI : THE SIZE OF THE SUBGROUP TO FORM THE COSET.
# REPSIZ : THE SIZE OF THE IRREDUCIBLE REPRESENTATION.
#
 local C,D,BASN,EA,O,l,P,p,B,OV,SI,AS,OREP,ww,OT,ith,ifind;
 C:=[];
 D:=[];
 BASN:=BAS;
 EA:=List(Elements(R),j->LargeMatrix(j,hom));;
#
# The subgroups with the order of "NMULTI" are extracted.
#
 AS:=AllSubgroups(R);
 while (BASN<>[]) 
 do
   Print("\n",Size(BASN),"\n");
   O:=Filtered(AS,g->Order(g)=NMULTI);;
   Print(O);
#
# The subgroups which generate the vector space of 
# the correct dimension of the irreducible representation 
# are chosen.
#
  if (O=[]) then
   O:=[Group(One(R))];
  fi;
  l:=List([1..Size(O)],i->0);
  ifind:=0;
  ith:=1;
  while (ifind=0 and ith<=Size(O))
  do
   ww:=gopr(O[ith],BASN[1],1,hom);
   l[ith]:=RankMat(List(EA,g->g*ww));
   Print(l);  
   if (l[ith]=REPSIZ) then
    ifind:=1;
    OT:=O[ith];
   fi;
   ith:=ith+1;
  od;
  if (ifind=0) then
   Print("CHANGE THE SUPPOSITION!\n");
   return 0;
  fi;
#
#  By means of the chosen subgroup, the irreducible 
#  vector space "B" is generated. 
#  It is stored in the vector space "D".
#
  B:=BaseMat(List(EA,g->g*gopr(OT,BASN[1],1,hom)));;
  Append(C,[B]);
  Append(D,B);
#
# The orthogonal space "OV" to the vector space "D" is prepared.
#
  OV:=BaseOrthogonalSpaceMat(D);
#
# The vector space "BASN" to be processed is renewed now, 
# by the projection "PROJ" on the vectors in "OV".
#
  BASN:=BaseMat(List(OV,o->PROJ*o));
#
# Instead, it is possible that "BASN" is renewed 
# as the intersection between "BASN" and "OV". 
#
#  SI:=SumIntersectionMat(BASN,OV);
#  BASN:=SI[2];
#
 od;
#
# Return C, the separated subspaces.
#
 return C;
end;

H1:=Cutout(BAS[9],PRJE[9],R,hom2,5,5);


#
# TO MAKE THE MATRIX NEEDED IN THE COMPUTATION OF MATRIX CONJUGATION
# EXPLAINED IN THE ARTICLE.
#
ConjugMatrix:=function(rr)
 local l,rmat,vmat,i,i1,i2,n,m;
 l:=ListPerm(rr,60);
 vmat:=List([1..Size(l)*3],i->List([1..Size(l)*3],j->0));
 for i in [1..Size(l)]
 do 
  i1:=i-1;i2:=l[i]-1;
  for n in [1..3]
  do
   vmat[3*i1+n][3*i2+n]:=1;
  od;
 od;
 return vmat;
end;

\end{verbatim}

\end{document}